\documentclass[a4paper,fontsize=12pt,twoside=true,numbers=noenddot]{kaobook}

\usepackage[T1]{fontenc}

\usepackage[english]{babel} 
\usepackage[english=british]{csquotes}
\usepackage{blindtext}
\usepackage{kaobiblio}
\usepackage[framed=true]{kaotheorems}
\usepackage{physics}

\newcommand{\Sec}{Section~}

\newcommand{\eq}{Eq.~}
\newcommand{\eqs}{Eqs.~}
\newcommand{\fig}{Fig.~}

\newcommand{\cf} {cf.~}
\newcommand{\ug} {\!=\!}

\newcommand{\piu} {\!+\!}
\newcommand{\meno} {\!-\!}
\renewcommand{\ie} {i.e.~}
\renewcommand{\eg} {e.g.~}

\newcommand{\rref} {Ref.~}
\newcommand{\rrefs} {Refs.~}

\newcommand{\se}{Schr\"{o}dinger equation}

\renewcommand{\U}{\hat{\mathcal{U}}}
\newcommand{\V}{\hat{V}}

\newcommand{\dt}{\Delta t}

\newcommand{\VNS}[1]{{\rm Tr}\{#1\}}
\newcommand{\de}{\Delta}
\renewcommand{\etal}{\textit{et al.}}
\renewcommand{\rm}{\textrm}

\makeatletter
\let\@fnsymbol\@arabic
\newcommand*\smashcaption{
	\def\FR@makecaption##1##2{%
		\vbox to\z@{%
			\vss
			\captionfont
			{\captionlabelfont##1}\caption@lsep##2%
			\par
			\vss
		}%
	}%
	\caption
}
\makeatother

\usepackage{geometry}
\renewcommand{\marginlayout}{%
	\newgeometry{
		top=27.4mm,             
		bottom=27.4mm,          
		inner=24.8mm,           
		textwidth=127mm,        
		marginparsep=8.2mm,     
		marginparwidth=39.4mm,  
	}%
}

\usepackage{xcolor}
\usepackage{graphicx}
\usepackage{lipsum}
\definecolor{titlepagecolor}{cmyk}{1,.60,0,.40}
\definecolor{namecolor}{cmyk}{1,.50,0,.10} 

\usepackage{cleveref}
\usepackage{floatrow}

\begin{document}
\begin{titlepage}
	\newgeometry{left=5cm}
	\pagecolor{titlepagecolor}
	\noindent
	\color{white}
	\center
	{\fontsize{32}{32}\selectfont \textsf{Quantum Collision Models}}
	\par
	\par
	\noindent
	{\fontsize{16}{32}\selectfont \textsf{Open system dynamics from repeated interactions}}
	\vspace{2.5cm}
	

	\includegraphics[width=0.8\linewidth]{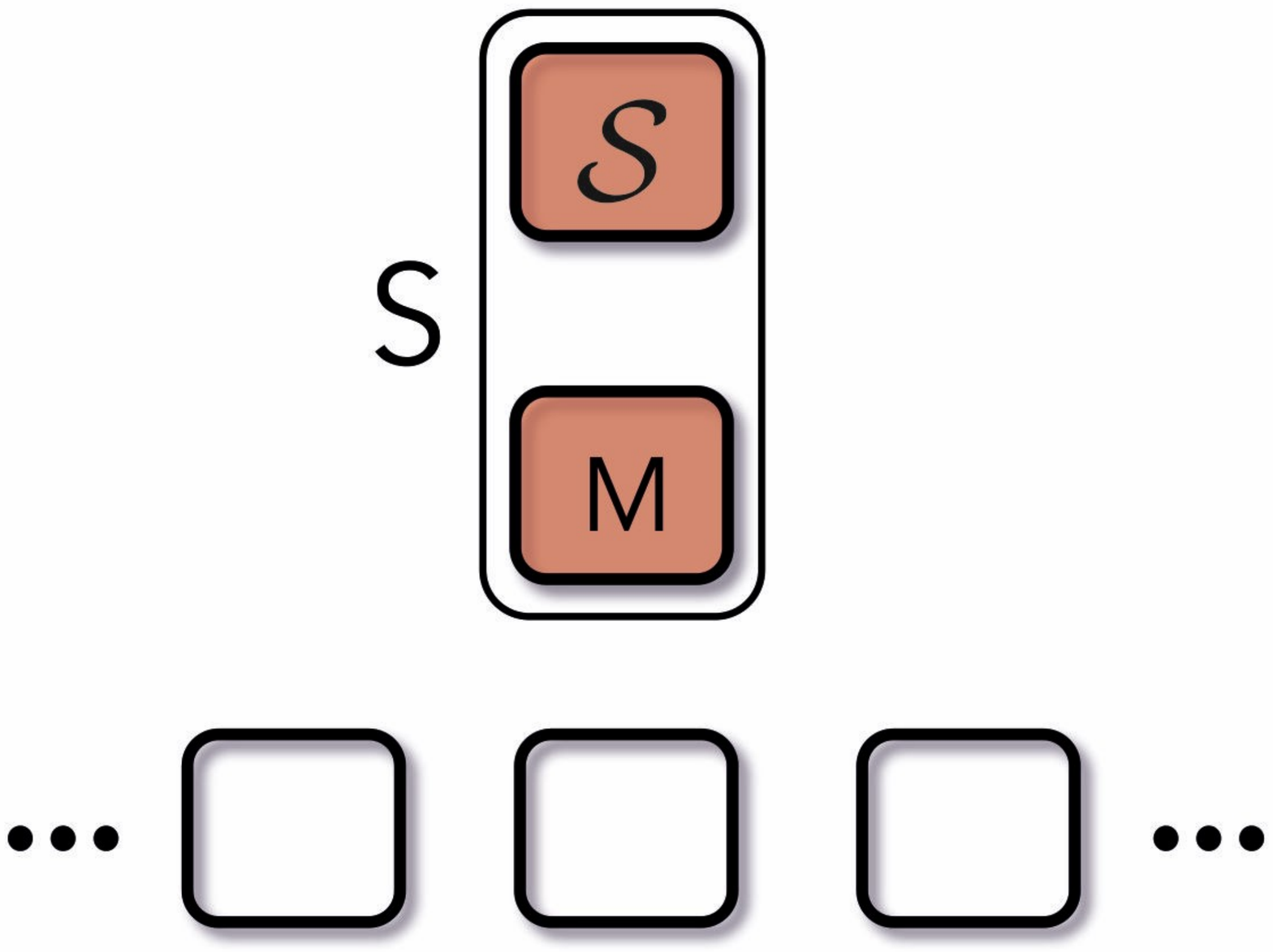}
	
	\vspace{2.0cm}
	\noindent
	\center
	{\fontsize{18}{32}\selectfont Francesco Ciccarello}\\
	\vspace{0.3cm}
	{\fontsize{18}{32}\selectfont Salvatore Lorenzo}\\
	\vspace{0.3cm}
	{\fontsize{18}{32}\selectfont Vittorio Giovannetti}\\
	\vspace{0.3cm}
	{\fontsize{18}{32}\selectfont G. Massimo Palma}
	\vskip\baselineskip
\end{titlepage}
\restoregeometry
\pagecolor{white}


\subject{}
\title[Quantum Collision Models]{Quantum Collision Models}
\subtitle{Open system dynamics from repeated interactions}

\author{Francesco Ciccarello
	\footnote{Università~degli~Studi~di~Palermo~Dipartimento~di~Fisica~e~Chimica-Emilio~Segrè,\protect\\~via~ Archirafi~36,~I-90123~Palermo,~Italy}\;\,\footnote{NEST,~Istituto~Nanoscienze-CNR, \protect\\~Piazza~S.~Silvestro~12,~56127~Pisa,~Italy}\\
	Salvatore Lorenzo\footnotemark[1]\\
	Vittorio Giovannetti
	\footnote{NEST,~Scuola~Normale~Superiore~and~Istituto~Nanoscienze,~Consiglio~Nazionale~delle~Ricerche,\protect\\~Piazza~dei~Cavalieri~7,~IT-56126~Pisa,~Italy}\\
	Massimo Palma\footnotemark[1]\;\,\footnotemark[2]}
\date{\today}%


\frontmatter 

\makeatletter
\uppertitleback{\@titlehead} 

\lowertitleback{
	\textbf{No copyright}\\
	\cczero\ This book is released into the public domain using the CC0 code. To the extent possible under law, I waive all copyright and related or neighbouring rights to this work.
	
	To view a copy of the CC0 code, visit: \\\url{http://creativecommons.org/publicdomain/zero/1.0/}
	
	\medskip
	
	This document was typeset with the help of \href{https://sourceforge.net/projects/koma-script/}{\KOMAScript} and \href{https://www.latex-project.org/}{\LaTeX} using the \href{https://github.com/fmarotta/kaobook/}{kaobook} class.
	
	The source code of this book is available at:\\\url{https://github.com/fmarotta/kaobook}
	
	\medskip
}
\makeatother

\maketitle
\chapter*{Preface}
We present an extensive introduction to quantum collision models (CMs), also known as repeated interactions schemes: a class of microscopic system-bath models for investigating open quantum systems dynamics whose use is currently spreading in a number of research areas. Through dedicated sections and a pedagogical approach, we discuss the CMs definition and general properties, their use for the derivation of master equations, their connection with quantum trajectories, their application in non-equilibrium quantum thermodynamics, their non-Markovian generalizations, their emergence from conventional system-bath microscopic models and link to the input-output formalism. The state of the art of each involved research area is reviewed through dedicated sections. The article is supported by several complementary appendices, which review standard concepts/tools of open quantum systems used in the main text with the goal of making the material accessible even to readers possessing only a basic background in quantum mechanics.

The paper could also be seen itself as a friendly, physically intuitive, introduction to fundamentals of open quantum systems theory since most main concepts of this are treated such as quantum maps, Lindblad master equation, steady states, POVMs, quantum trajectories and stochastic Schr\"odinger equation.

\index{preface}

\setlength{\textheight}{230\hscale}
\tableofcontents
\listoffigures
	
\mainmatter
\setchapterstyle{kao}

\setchapterpreamble[u]{\margintoc}
\chapter{Introduction and historical notes}

The last two decades or so have seen the compelling emergence and subsequent
consolidation of a set of research areas that today usually go under the joint name
of \textit{quantum technologies}~\cite{dowling2003quantum}. This is the idea of
taking advantage of some distinctive features of quantum mechanics -- such as the
superposition principle and entanglement -- for devising a plethora of novel,
potentially groundbreaking, applications. These include tasks such as quantum
computing, quantum cryptography, quantum sensing, quantum metrology, quantum
simulation, quantum imaging. As a paradigmatic instance (also in light of our goals
here), harnessing ``quantumness" in order to challenge longstanding thermodynamics
bounds such as the Carnot efficiency so as to engineer more efficient thermal
machines is a possibility that is being more and more investigated these days in the
lively field of \textit{quantum
thermodynamics}~\cite{e15062100,goold_role_2016,vinjanampathy_quantum_2016,binder2018thermodynamics,deffner2019quantum}. 

The above scenario in particular gave momentum to the study of an old, but always
topical, quantum mechanics problem: the dynamics of a system in contact with an
external environment, namely a so called \textit{open quantum
system} \cite{breuer2007,gardiner2004quantum,rivas_open_2012}. In some respects, this
problem arises from the hope to find an irreversible-dynamics analogue of the
Schr\"odinger equation that governs quantum systems coupled to a large bath
(\textit{master equation}). No truly general master equation is known to date except
for a restricted, although conceptually prominent, class of dynamics known as
\textit{Markovian dynamics}; and a very few others. It is likely that this formidable
problem may even be unsolvable as in general the system's degrees of freedom can get
entangled with the bath in such a way that one cannot give up keeping track of the
environment dynamics, or at least a portion of it. In various contexts such as
quantum thermodynamics, this may even be desirable \eg in order to study energy or
entropy exchange between system and bath, which requires describing the latter as
well. In practice, especially when running experiments, ``looking" at some
environment is inevitable. A measurement on the system of interest, for instance,
requires to make it interact with an external probe which is then
analyzed~\cite{wiseman2009quantum,jacobs2014quantum}. 

On a methodological ground, tackling system--bath dynamics at a microscopic level is
in general a very hard task, which necessarily demands for appropriate
\textit{models}. Traditionally, the standard scheme is to decompose the bath
$B$ into a \textit{continuum} of normal modes (defined by its free
Hamiltonian) and let them interact with the system $S$ according to some
physically-motivated coupling model~\cite{breuer2007,gardiner2004quantum}. 


\begin{figure}[!h] 
	\raggedright
	\begin{floatrow}[1]
		\ffigbox[\FBwidth]{\caption[Collision model versus conventional system--bath model]{\textit{Collision model versus conventional system--bath model}. In a collision model (a) the bath is made out of a large, discrete, collection of smaller units (ancillas) with which the open system $S$ interacts (collides) one at a time. In a conventional system--bath model (b), instead, the bath typically comprises a continuum of normal modes and $S$ interacts with (generally) all of them at any time.}\label{fig-intro}}%
		{\includegraphics[width=\textwidth]{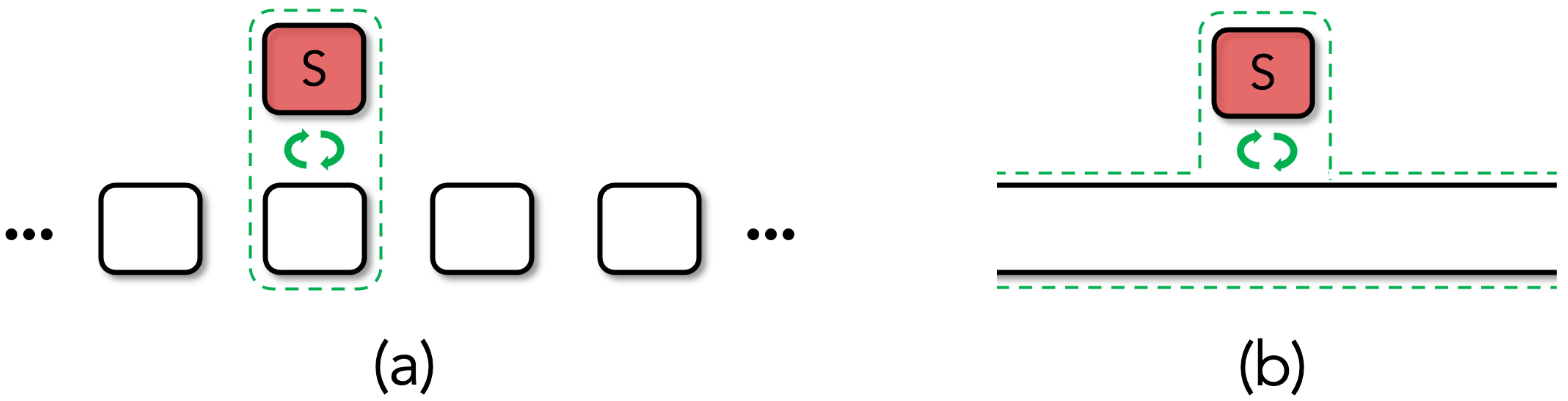}}
	\end{floatrow}
\end{figure}

The last few years have yet seen a growing use of a less conventional class of
system--bath models known as quantum \textit{collision models} (CMs) or
\textit{repeated interaction schemes}.
In its most basic formulation [see \cref{fig-intro}(a)], a CM model imagines the bath $B$ as a large collection
of smaller subunits (\textit{ancillas}) with which the open system $S$
interacts -- one at a time -- through a sequence of pairwise, short unitary
interactions (\textit{collisions}). 
Arguably inspired by the famous Boltzmann's
\textit{Stosszahlansatz}~\cite{brown2008boltzmann} and first adopted in the study of
optical masers and weak continuous measurements, quantum CMs are currently spreading
across research fields such as quantum non-Markovian dynamics, quantum optics and
quantum thermodynamics (where they have become now a standard approach).\marginnote{Many authors use the name ``collision\textit{al} models". Occasionally, it was used ``refreshing models"~\cite{modi_positivity_2012}.} 

Compared to the conventional system--bath modeling mentioned before, CMs differ in many respects. Two hallmarks in particular stand out. First, they are intrinsically \textit{discrete}: continuous time is effectively replaced by a step number (although the continuous-time limit is often taken in the end) and the bath is thought as a discrete collection of elementary subsystems instead of a continuum as usual. Second, as schematically pictured in  \fig. \ref{fig-intro}, in contrast to standard models where $S$ at each time interacts with (generally) all the normal modes, in CMs (at least memoryless ones) $S$ crosstalks with a single little portion of bath at a time. This in a way decomposes the extremely complex system--bath dynamics into simple elementary contributions, a traditionally effective strategy in Physics.

To our knowledge, the first appearance of a quantum CM in the literature dates back
to the 60s through a paper by J. Rau~\cite{rau_relaxation_1963}. Later on in the 80s,
CMs appeared in seminal works on \textit{weak measurements} by C. M. Caves and G. J.
Milburn~\cite{caves_quantum_1986,caves_quantum_1987,caves_quantum-mechanical_1987}.
CMs are indeed a natural microscopic framework for introducing this important class
of weak quantum measurements~\cite{wiseman2009quantum,jacobs2014quantum} because,
taking a metrological viewpoint, ancillas can be seen as a large collection of
``meters" each of which being measured after the collision. More or less in the same
years, Javanainen and
Meystre~\cite{filipowicz_theory_1986,filipowicz1986quantum,filipowicz1986microscopic}
developed the theory of \textit{micromaser} whose basic setup features flying atoms
that one at a time interact with a lossy cavity mode. This can be seen as a
physically intuitive implementation of a CM with atoms embodying ancillas which
undergo collisions with $S$ (the cavity mode).

A hallmark of the CM approach is viewing the system--bath dynamics as a sequence of
\textit{two-body} unitary collisions. This is very similar in spirit to a cornerstone
of quantum information processing (and generally quantum
technologies)~\cite{nielsen2002quantum}, namely that two-qubit gates (assisted by
one-qubit gates) are sufficient to carry out universal quantum computation, and was 
probably the reason why CMs gained renewed attention in the early 2000s. V.~Scarani
\etal in 2002 approached the thermalization of a qubit (two-level system) due to
collisions with a bath of qubits as a quantum task whose goal is taking $S$
to a Gibbs state no matter what state it started from (``quantum thermalizing
machine")~\cite{ziman_diluting_2002,scarani_thermalizing_2002}. At about the same
time, A.~Brun~\cite{brun_simple_2002} used a CM made out of qubits and the language
of quantum information to study basic concepts of quantum trajectories, including the
stochastic \se, connecting as well to the aforementioned weak measurements.

Around the beginning of 2010s, a strong (still ongoing) interest arose in attacking
quantum non-Markovian dynamics and defining on a firm basis the meaning of
(non-)Markovian evolution in quantum
mechanics~\cite{breuer_foundations_2012,breuer_colloquium:_2016,de_vega_dynamics_2017,li2018concepts}.
CMs are an ideal playground in this respect as was shown by Rybar, Filippov, Ziman
and Buzek~\cite{rybar_simulation_2012}, who demonstrated that CMs can simulate any
indivisible dynamics of a qubit, and by Ciccarello, Palma and
Giovannetti~\cite{ciccarello_collision-model-based_2013} who added ancilla--ancilla
collisions to a basic CM to derive a completely-positive non-Markovian master
equation. 

In the same years, the field of quantum thermodynamics was emerging, prompted by a
number of questions calling for manageable microscopic models. Due to their
simplicity and the possibility to describe system--bath coupling non-perturbatively,
CMs are a quite natural tool in this framework so that it is hard establishing when
they were used for the first time. A comprehensive quantum thermodynamics theory of
CMs (repeated interaction schemes) was presented in 2017 by Strasberg, Schaller,
Brandes and Esposito~\cite{Esp17}. In this context, CMs are actually seen mostly as a
resource to harness in order to design machines with enhanced thermodynamic
performances, possibly powered by genuinely quantum features. A paradigm in this
respect came from an influential 2003 paper by Scully, Zubairy, Agarwal and
Walther~\cite{scully2003Science}, which considered a single-bath thermal machine made
out of a stream of three-level atoms flying through a cavity.

Having in mind a readership of physicists, even those armed with only a basic background in quantum mechanics, here we present a self-contained introduction to quantum CMs theory, including overviews of the state of the art and recent developments.

While to our knowledge this is the first, fully dedicated, comprehensive review on
CMs, we note that there are some papers and PhD dissertations which introduce to
certain aspects of
CMs~\cite{brun_simple_2002,rodriguez_theory_2008,ziman_open_2010,layden2016indirect,altamirano_unitarity_2017,ciccarello_collision_2017,gross_qubit_2018,grimmer2020interpolated}.
Dedicated sections on CMs can be found in the review on non-Markovian dynamics in
\rref~\cite{de_vega_dynamics_2017} and the review on irreversible entropy production
in \rref~\cite{landi2020irreversible}.
We also quote a recent perspective on the topic~\cite{campbell2021collision}.

Finally a disclaimer. The present review does not cover mathematical aspects, for
which we point the interested reader to \rref~\cite{bruneau2014repeated} and
references therein.

\chapter{Outline and structure of the paper}

\begin{figure}[!h] 
	\raggedright
	\begin{floatrow}[1]
		\ffigbox[\FBwidth]{\caption[Structure of the paper]{\textit{Structure of the paper}. The body of the paper comprises six big sections, numbered from 4 to 9. \cref{section-1,section-eqs} are the central ones, to which all the others are directly linked to}.\label{fig-struc}}%
		{\includegraphics[width=0.7\linewidth]{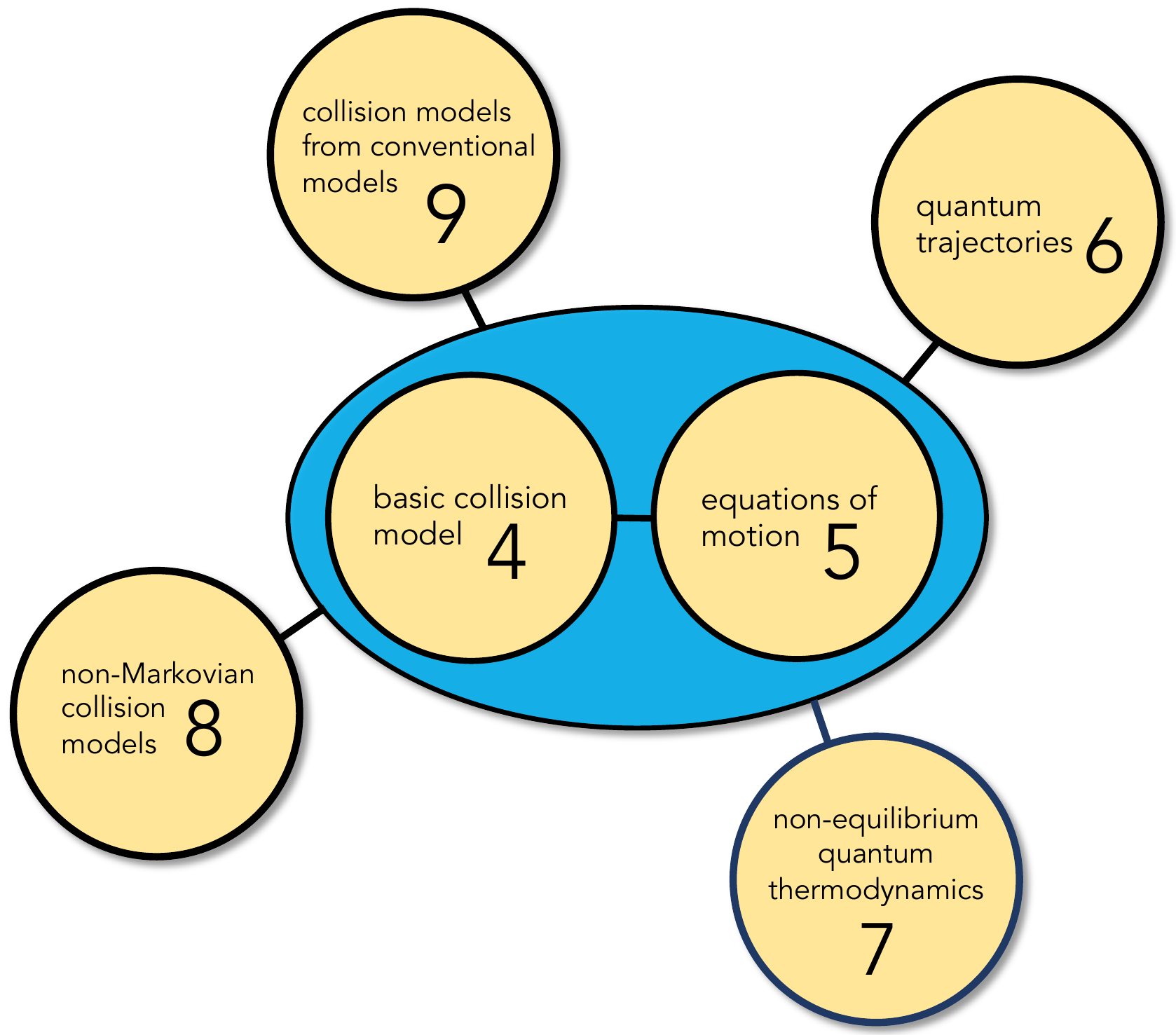}}
	\end{floatrow}
\end{figure}

The body of the paper is organized into six big sections (each in turn structured in a number of subsections): \textit{Basic collision model} (\cref{section-1}), \textit{Equations of motion} (\cref{section-eqs}), \textit{Quantum trajectories} (\cref{section-qtraj}), \textit{Non-equilibrium quantum thermodynamics} (\cref{section-thermo}), \textit{Non-Markovian collision models} (\cref{section-NM}) and, finally, \textit{Collision models from conventional models} (\cref{section-CMQO}). As sketched in  \cref{fig-struc}, the paper's central Sections are  \cref{section-1,section-eqs} with which each of the other sections is directly connected. 

Each of these six big sections is written with a quite pedagogical attitude. In particular,  we note that -- similarly to a textbook -- there intentionally appear very \textit{few references} in order not to distract the reader from the main line of discussion.\marginnote{A general criterion is that a reference is given if a certain property is used in the main text but not proven (nor in the appendices).} In the same spirit, in order not to interfere with the main discussion, references to previous equations or sections often appear between brackets like ``(see Section xxx)" or ``[see \eq(xxx)]". Also, a large use of \textit{footnotes} is made, which supply extra details, explanations, comments and disclaimers. Each big section, from 4 to 9, ends with a dedicated \textit{State of the art} subsection, reviewing relevant literature related to the topic of the corresponding section.

We begin with a preliminary technical section (\cref{section-not}), which is intended to provide a sort of reading guide. The main conventions underpinning the notation we use are explained along with the (relatively few) acronyms appearing throughout the paper.

\cref{section-1} defines the most basic CM \cref{section-def} focusing first on the open dynamics of $S$ \cref{section-open} and then also that of ancillas in \ref{AD-section}. Next, in  \ref{markovianity-section}, we discuss Markovianity, a property of utmost importance for CMs and open quantum systems theory in general. Thereafter (\cref{section-inhomCP}) after introducing inhomogeneous CMs, we discuss a generalized notion of Markovian behavior called CP divisibility (where CP stands for ``completely positive"). Some paradigmatic CMs are presented in  \cref{section-all-qubit} (all-qubit CM) and  \cref{section-cascaded} (cascaded CMs). A major issue when dealing with open dynamics, \ie the convergence to a steady state (if any), is discussed in  \cref{section-homo}. We close with  \cref{section-MPS} which studies the tensor-network structure of the joint system--bath state at each step.

\cref{section-eqs} deals with the derivation of equations of motion for both states and expectation values of observables. The basis is the second-order expansion of the collision unitary operator with respect to the collision time (\cref{section-EMs}), resulting in finite-difference equations of motion having the structure of discrete Lindblad master equations and ensuing dynamical equations for expectation values (see  \cref{section-EVs}). The Lindblad structure can be proven based on the spectral decomposition of the ancilla's initial state (see  \cref{section-lind}) or solely in terms of bath moments (see  \cref{section-moments}), the analogous of the latter being next worked out for expectation values as well in  (\cref{section-ev-moments}). We then show in  (\cref{sec-coarse2}) how the intrinsically discrete dynamics can be turned into a continuous-time one through coarse graining. The prominent example of micromaser is then discussed in the extensive (\cref{section-maser}). The possibility to define a strict continuous-time limit $\de t\,{\rightarrow}\, 0$ by introducing a diverging coupling strength is studied in (\cref{sec-div}). We close with a section devoted to multiple baths (\cref{section-multiB}) and one deriving the master equation of cascaded CMs (\cref{sec-cascME}).

\cref{section-qtraj} discusses quantum trajectories and weak measurements, these being important general topics that are naturally introduced in the CM language. The starting point (\cref{section-unrav}) is to view ancillas as probes and study how measurements of these condition the dynamics of $S$. This framework is used in the following  \cref{section-povm} to introduce the important general concept of POVM (Positive Operator-Valued Measure). We then focus on a specific dynamics in the all-qubit CMs, which is used to introduce the concept of quantum jumps (\cref{section-qt-qubits}), the stochastic Schr\"odinger equation~(\cref{section-stoca}) and, finally, how averaging over all trajectories returns the Lindblad master equation~(\cref{section-uncond}). We conclude in (\cref{section-SE-general}) by deriving the stochastic Schr\"odinger equation for a general interaction Hamiltonian, at the same time highlighting the role of the bath's first and second moments.

\cref{section-thermo} is dedicated to non-equilibrium quantum thermodynamics, beginning with the definition of quantum thermalization \cref{section-QT1} and discussing next the important instance of a system thermalizing with a bath of quantum harmonic oscillators (\cref{section-harm-th}) and the connection between thermalization and energy conservation (\cref{section-balance}). There follow instances of non-equilibrium steady states with baths at different temperatures (\cref{section-QT1T2}). The intrinsic time dependence of the system--bath Hamiltonian, a major distinctive feature of CMs, is analyzed in  \cref{section-HSBt}. Following this, we present one by one the computation of relevant thermodynamic quantities like: the change of system free energy \cref{section-free}, that of ancillas or heat \cref{section-heat} and work \cref{section-work}. We then derive the non-equilibrium version of the 1st and 2nd law of thermodynamics (\cref{section-1st,section-2nd}, respectively) and discuss the Landauer's principle in  \cref{section-Land}. The energy balance of some of the previously introduced instances is studied in  \ref{qubit-sect}.

\cref{section-NM} deals with non-Markovian CMs. Three basic classes are introduced, where each arises from the introduction of a memory mechanism into the basic memoryless CM of  \cref{section-1}: ancilla--ancilla collisions (\cref{section-aacu}), initially-correlated ancillas \cref{section-ica}, multiple system--ancilla collisions \cref{section-multiple}.  \cref{section-NMME} shows the derivation of a fully CPT non-Markovian master equation based on the class in  \cref{section-aacu}. A further class, the so called composite CMs, is presented in  \cref{section-comp} and illustrated in a paradigmatic instance. We close with the demonstration that, so long as the open dynamics is concerned, ancilla--ancilla collisions can be mapped into a composite CM (\cref{section-mapCCM}).

The last  \cref{section-CMQO} deals with the relationship between CMs and conventional system--bath models (see  \ref{fig-intro}). The two descriptions are shown to be equivalent pictures in the case (recurrent in quantum optics) that $S$ is weakly coupled to a continuum of bosonic modes (field). All the steps of the mapping are illustrated in detail in  Sections \ref{wn}, \ref{section-tm}, \ref{section-int-pic}, \ref{section-tdcg} and \ref{section-emer}. Occurrence of Markovian behavior depends on the field's initial state (see  \ref{in-field}). This is then specifically illustrated for the vacuum state leading to spontaneous emission (see  \cref{section-vacuum}), thermal states yielding Einstein coefficients (\cref{section-thermal}), coherent states and optical Bloch equations~\cref{section-coh}. These are all special cases of a general master equation, fully defined by the field's first and second moments, which holds for Gaussian white-noise field states \cref{section-gaussian}. Non-Markovian dynamics can occur for non-Gaussian initial states such as single-photon wavepackets \cref{section-singleph}. Finally, we explain the link to the input--output formalism that is broadly used in quantum optics (\ref{link-io}).
We conclude in  \cref{section-concl} with a discussion of future prospects and open questions about quantum CMs.

In order to help the reader, in addition to the aforementioned reading guide in  \cref{section-not}, there are a number of appendices. Those from A to F have a tutorial scope and recall basic notions: density matrices (\ref{app-rho}), various entropic quantities (\ref{app-ent}), quantum maps (\ref{app-qmaps}), the dynamical map (\ref{app-DM}), the Stinespring dilation theorem (\ref{app-stine}) and the Lindblad master equation (\ref{app-lind}). There follow technical sections featuring mostly proofs of properties used in the main text (\cref{appendix-sse,lin-sys}).

\setchapterpreamble[u]{\margintoc}
\chapter{Reading guide}
\label{section-not}

Hats are used throughout to identify all the operators, except density operators and the identity operator $\mathbb{I}$. When appearing, the identity operator is frequently used without subscripts, the (sub)system it refers to being often clear from the context. If an operator acts trivially on a subsystem, \eg $\hat O_A\otimes \mathbb{I}_B$, then the identity operator is generally omitted.

Usually, \textit{joint} states (generally represented by density operators) of the system plus bath are denoted with letter $\sigma$, while $\rho$ and $\eta$ are used for the \textit{reduced} state of the open system $S$ and a single bath ancilla, respectively. 

Letter $\sigma$ but with a hat is also used for spin operators such as $\hat\sigma_{\pm}$ and $\hat \sigma_z$. 

The eigenstates of $\hat \sigma_z$ are denoted with $\ket{0}$ and $\ket{1}$, having respectively eigenvalues $-1$ and $+1$. We point out that this choice does \textit{not} follow the usual convention in the quantum information literature, where $\ket{0}$ ($\ket{1}$) corresponds to eigenvalue $+1$ ($-1$). The rationale of this choice is that, in many cases, we deal with a ground and an excited state so that $\ket{0}$ and $\ket{1}$ are understood as the state with zero and one excitations, respectively.

Superoperators, including quantum maps, are denoted with capital (usually calligraphic) letters and the argument is shown between square brackets, \eg $\mathcal M[\varrho]$. The symbol of composition of quantum maps is most of the times omitted, thus 
\[
\mathcal M \mathcal M'[\varrho]=(\mathcal M\circ \mathcal M')[\varrho]=\mathcal M [\mathcal{M}'[\varrho]].
\]

Arguments of partial traces are shown between curly brackets.

Anti-commutators are denoted as $[\hat A,{\hat B}]_+=\hat A{\hat B}+{\hat B}\hat A\,$.

We use units such that $\hbar=1$ throughout.

In some contexts such as  \Sec\ref{section-CMQO}, in order to avoid making the notation cumbersome, dependencies on a continuous variable are shown through a subscript (as is usually done with discrete time variables), thus \eg $f_t=f(t)$.

The tensor product symbol $\otimes$ is often omitted.

The ancilla index usually appears as a subscript, \eg $\eta_n$ stands for a state of ancilla $n$.

Time dependencies, where time is often embodied by the (discrete) number of steps, can appear as subscripts or superscripts, which generally depends on the quantity or subsystem they refer to or on the considered context.

We generally do not use the same symbol for different purposes depending on the context/section. Some exceptions are yet unavoidable (given the considerable size of the paper), \eg ``$M$" stands for the number of baths in  \Sec\ref{section-multiB} while in  \Sec\ref{section-comp} it denotes the memory coupled to $S$.

\section{Acronyms and some terminology}
\label{section-acro}

\begin{itemize}
	\item [] CM $=$ ``Collision model"
	\item [] ME $=$ ``Master equation"
	\item [] CPT $=$ ``Completely positive and trace-preserving"
	\item [ ]NM $=$ ``Non-Markovian"
	\item [] ``Qubit" $=$ Two-level system (quantum bit), formally equivalent to a spin-1/2 particle.
\end{itemize}
\setchapterpreamble[u]{\margintoc}
\chapter{Basic collision model}
\label{section-1}

\section{Definition}
\label{section-def}

\begin{figure}[!h] 
	\raggedright
	\begin{floatrow}[1]
		\ffigbox[\FBwidth]{\caption[Basic collision model]{\textit{Basic collision model}. (a): First collision: the pairwise unitary $\hat U_1$ is applied on $S$ and ancilla 1 (initially in state $\rho_0$ and $\eta$, respectively). At the end of the collision, $S$ is in state $\rho_1$. (b): Second collision: unitary $\hat U_2$ is applied on $S$ and ancilla 2  (initially in state $\rho_1$ and $\eta$, respectively). (c): Quantum circuit representation of the first two CM steps. Each wire represents a subsystem ($S$ or an ancilla), while each rectangular box is a two-body quantum gate (collision unitary).  
		(d): Correlations: $S$ and all of the ancillas it already collided with are jointly correlated, while each ancilla which still has to collide with $S$ is yet in the initial state $\eta$ (hence uncorrelated with $S$ and all the remaining ancillas)}.\label{fig-CMdyn}}%
		{\includegraphics[width=\textwidth]{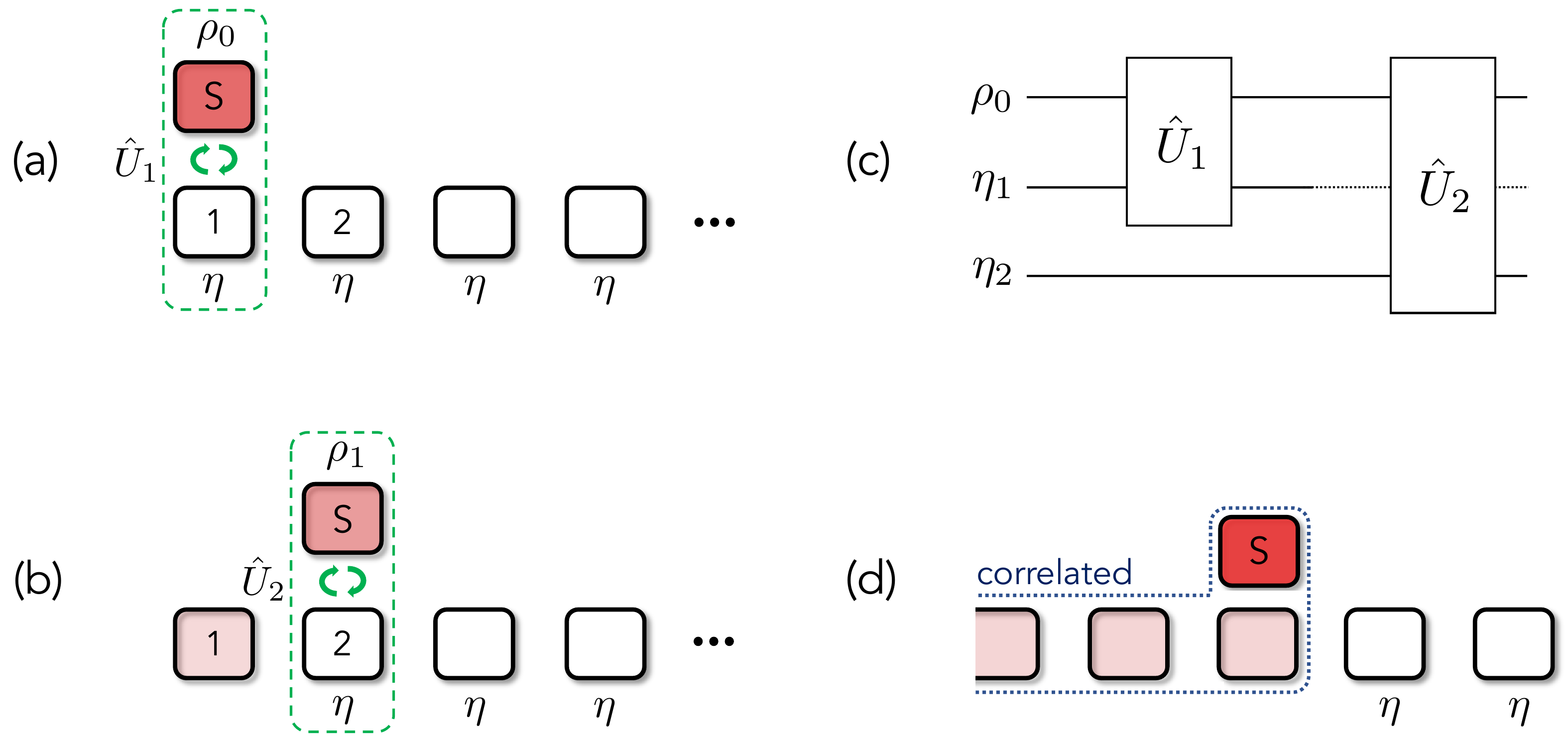}}
	\end{floatrow}
\end{figure}

Consider a quantum system with unspecified Hilbert-space dimension called $S$ (\textit{open syste}m) coupled to a quantum \textit{bath} $B$. 
By hypothesis, the bath is made out of a large collection of smaller identical subunits (\textit{ancillas}) labeled by an integer number $n$.   It is assumed that $S$ and $B$ start in the joint state
\begin{equation}
\sigma_{0}=\rho_{0}\otimes\,\eta\otimes\eta\otimes\cdots\,\label{sigma0}   
\end{equation}
with $\rho_0$ the initial state of $S$ and $\eta$ the initial state common to all ancillas [see  \fig \ref{fig-CMdyn}(a)]. Here, $\sigma_{0}$, $\rho_{0}$ and $\eta$ are (generally mixed) density matrices (see \ref{app-rho}).
Note that $\sigma_{0}$ is a \textit{product} state: neither correlations between $S$ and $B$ nor between ancillas are present. 

By hypothesis, as sketched in  \fig\ref{fig-CMdyn}(a) and (b), the entire dynamics takes place through successive {collisions}, namely pairwise short interactions each involving $S$ and one ancilla of $B$. Collision $S$-$1$ (i.e.,~between $S$ and ancilla 1) occurs at step $n=1$, then $S$-$2$ at step $n=2$, then $S$-$3$ and so on. Importantly, each ancilla $n$ collides with $S$ only \textit{once} (at the corresponding step $n$). 

The dynamics of an elementary collision is described by the time evolution unitary operator
\begin{equation}
\hat {U}_{n}=e^{-i\left(\hat H_{S}+\hat H_n+\hat{V}_{n}\right) \Delta t }\,,\label{USn1}
\end{equation} 
with $\Delta t$ the collision duration, $\hat H_{S}$ ($\hat H_{n}$) the free Hamiltonian of $S$ ($n$th ancilla) and $\hat V_n$ the $S$-$n$ interaction Hamiltonian. Note that, although only the ancilla subscript appears, $\hat V_n$ acts on \textit{both} $S$ and $n$, and so does $\hat U_n$.

\subsection{Conditions for Markovian behavior}
\label{conds}

Among the series of assumptions introduced so far that define the CM, three in particular stand out:
\begin{itemize}
\item[(1)] Ancillas do not interact with each other\,;
\item[(2)] Ancillas are initially uncorrelated\,;
\item[(3)] Each ancilla collides with $S$ only once\,.
\end{itemize}
Hypotheses (1)--(3) crucially underpin many major properties of CMs, in particular those related to Markovian behavior. Is worth pointing out that the essential meaning of (3) is to rule out sequences of collisions such as $S$-$1$, $S$-$2$, $S$-$3$, $S$-$1$, $S$-$2$, \ldots \,, while dynamics like $S$-$1$, $S$-$1$, $S$-$2$, $S$-$2$, \ldots \, can be seen as not violating (3) provided that one simply redefines the collision as a double collision with the same ancilla.

\section{Open dynamics and collision map}
\label{section-open}

After $n$ collisions, i.e.~at step $n$, the joint system--bath state reads
\begin{equation}
\sigma_n=\hat U_{n} \cdots\hat{ U}_{1}\,\sigma_{0}\,\hat U_{1}^\dag\cdots\hat{ U}_{n}^\dag\,\,.\label{sigman}
\end{equation}
In passing, we note that this dynamics can be represented [see \
\fig \ref{fig-CMdyn}(c)] as a quantum circuit~\cite{nielsen2002quantum} where each wire
stands for a subsystem ($S$ or an ancilla) while each rectangular box is a
two-body quantum gate (collision unitary $\hat U_n$).

By replacing $\sigma_{0}$ with the uncorrelated state \ref{sigma0}, \eq\ref{sigman} can be expressed as\footnote{Note that subscript $n$ is used here with different although related meanings. For $S-B$ and $S$ states, such as $\sigma_n$ and $\rho_n$, it refers to the time step. For single-ancilla states, such as $\eta_n$, it indicates which ancilla the state refers to.}
\begin{align}
\sigma_n&=\hat U_{n} \cdots\hat{ U}_{1}\,\rho_{0}\,\eta_1\cdots \eta_n\,\hat U_{1}^\dag\cdots\hat{ U}_{n}^\dag\,\eta_{n{+}1}\eta_{n+2}\cdots\nonumber\\
&=\left(\hat U_n\cdots\,\left(\hat{U}_2\left(\hat{U}_1\rho_{0}\,\eta_1\hat U_1^\dagger\right)\eta_2\hat U_2^\dag\right)\, \ldots\,\eta_n\hat{U}_n^\dagger\right)\eta_{n{+}1}\cdots.
\end{align}
In the last identity we used that $\hat U_n$ acts on $S$ and $n$, hence it commutes with any $\eta_m$ with $m\neq n$. We see that, up to step $n$, ancillas $\eta_m$ with $m>n$ play no role and we will thus ignore them in the following.

Tracing off the bath, the corresponding state of the open system $S$ is
\begin{align}
\rho_{n}=&{\rm Tr}_{B} \{\sigma_n\}={\rm Tr}_{n}\cdots{\rm Tr}_1 \{\sigma_n\}\label{eps-0}\nonumber\\
=&{\rm Tr}_n\left\{\hat U_n\cdots\,{\rm Tr}_2\left\{\hat{U}_2\,{\rm Tr}_1\left\{\hat{U}_1\rho_{0}\,\eta_1\hat U_1^\dagger\right\}\eta_2\hat U_2^\dag\right\}\, \ldots\eta_n\hat{U}_n^\dagger\right\}\,,
\end{align} 
with ${\rm Tr}_m$ the partial trace [see \eq\ref{trace}] over the $m$th ancilla and where we used that ${\rm Tr}_n$ and $\hat U_n$ do not involve ancillas different from $n$. We next express \cref{eps-0} in the compact form
\begin{equation}
\rho_{n}=\mathcal E \left[\cdots\left[\mathcal E\left[\mathcal E\left[\rho_0\right]\right]\right]\right]={\mathcal{E}^n}[\rho_{0}]\,,\label{conc}  
\end{equation}
where we defined the quantum map (see \ref{app-qmaps}) on $S$
\begin{equation}
\rho'={\mathcal{E}}[\rho]={\rm Tr}_{n}\left\{\hat {U}_ {n} \left(\rho\otimes\eta_n\right)\,\hat {U}_{ {n} }^\dag\right\}\,\,\,\,\,\,\,(\rm{collision map})\,. \label{coll-map}
\end{equation}
We will henceforth refer to \ref{coll-map} as the \textit{collision map}\marginnote{Map ${\mathcal{E}}$ does not depend on $n$ since we are assuming a fully homogeneous model (same initial state for all ancillas and same collision unitary $\hat U_n$). This assumption will be relaxed in  \Sec \ref{section-inhomCP}.} : the knowledge of map $\mathcal E$ allows to determine the state of $S$ at the end of a collision, $\rho'$, for any state $\rho$ prior to the collision. Note that map $\mathcal E$ depends on the unitary \ref{USn1} describing each collision (in turn depending on $H_S$, $H_n$ and $\hat V_n$) as well as on the  ancilla's initial state $\eta$. As a property of utmost importance in CMs theory, the form of \ref{coll-map} ensures that $\mathcal E$ is a \textit{completely positive and trace-preserving} (CPT) map (see \ref{app-qmaps}). The essential reason behind this property is that, \textit{before} the $S$-$n$ collision starts, $S$ is still {uncorrelated} with ancilla $n$ [see  \fig\ref{fig-CMdyn}(a), (b) and (d)] this being a consequence of assumptions (1)--(3) in  \ref{conds}. The breaking of even only one of these generally brings about that the initial and final states of $S$ in a collision are no longer connected to one another by a CPT map as we will discuss extensively in  \Sec \ref{section-NM}.

Thus \eq\ref{conc} states that, when looking only at the open system $S$, $n$ collisions correspond to $n$ applications of collision map ${\mathcal{E}}$ on $\rho_{0}$ (initial state of $S$).
We immediately see that \eq\ref{conc} entails\marginnote{The converse holds as well, i.e.,~\eq\ref{eps} implies $\rho_n=\mathcal E^n [\rho_{0}]$. \eqs\cref{conc,eps} are thus equivalent.}
\begin{equation}
\rho_n=\mathcal E [\rho_{n-1}]\,.\label{eps}
\end{equation}
\eq\ref{eps} in fact governs the entire open dynamics and, as will become clearer in  \Sec \ref{section-eqs}, it can be regarded as the discrete analogue of a time-local master equation (see \ref{app-lind}). In particular, it shows that the state of $S$ at the current step depends only on that at the previous step. This means that the dynamics keeps \textit{no memory} of past history: if we are given state $\rho_{n-1}$ but we do not know what happened to $S$ up to step $n{-}1$, the entire future evolution at any step $m\ge n$ can be predicted from \ref{eps}. 
This property no more holds for non-Markovian CMs to be discussed in  \Sec \ref{section-NM}. Yet, the Markovian nature of a basic CM has a tremendous conceptual relevance for all CMs, including non-Markovian ones, as will become clear in the development of this paper.

\section{Ancilla dynamics}
\label{AD-section}

While, as shown above, the open dynamics of $S$ is relatively easy to work out, deriving the full bath dynamics is generally far more challenging (see however  \Sec \ref{section-MPS}). Although not directly coupled, indeed, ancillas that \textit{already} collided with $S$ get correlated with each other [see  \fig \ref{fig-CMdyn}(d)].
However, if one is concerned only with the \textit{single} ancilla dynamics (as is often the case) this is simply obtained from \eq\ref{sigman} by tracing off $S$ and all the remaining ancillas. The result is similar to the collision map \ref{coll-map} except that the partial trace is now over $S$ (instead of $n$)
\begin{equation}
\eta'_n={\rm Tr}_{S}\{\hat U_n\rho_{n-1}\eta_n\hat U_n^\dag\}\,,\label{etapn}
\end{equation}
Thereby, the collision with $S$ transforms the ancilla state as
\begin{equation}
\eta'_n={\mathcal{A}}_{\rho_{n-1}}[\eta_n]\label{ancilla-map}
\end{equation}
with
\begin{equation}
\eta'={\mathcal{A}}_{\rho}[\eta]={\rm Tr}_{S}\left\{\hat {U}_ {n} \left(\rho\otimes \eta\right)\,\hat {U}_{ {n} }^\dag\right\}\,. \label{ancilla-map-def}
\end{equation}
\eq\ref{ancilla-map-def} defines a CPT map on the ancilla, showing that this evolves at each collision somewhat similarly to $S$. Yet, at variance with \ref{eps} which can be determined once for all given $\hat U_{n}$ and $\eta$, map \ref{ancilla-map-def} does depend parametrically on the current state of $S$. Thereby, to work out $\eta'_n$ we need to keep track of the open dynamics of $S$ (\ie $\rho_{n}$).\marginnote{The ancilla's reduced dynamics~\ref{ancilla-map}  can be equivalently described as a CPT map connecting \textit{different} Hilbert spaces in that its input is a state of $S$ while its output is the final state of the $n$th ancilla after colliding with $S$. This shows even more directly that recording the whole dynamics of $S$ is required in order to determine the ancilla evolution.}
Note that after colliding with $S$ the ancilla's state no longer changes [due to conditions (1)--(3) in  \ref{conds}], hence \ref{ancilla-map} fully specifies the single-ancilla dynamics. At the next steps, however, the correlations between the ancilla and the rest of the system (both $S$ and other ancillas) generally change [see  \fig \ref{fig-CMdyn}(d)].

\section{Markovianity}
\label{markovianity-section}

It is convenient to introduce the \textit{dynamical map} (see \ref{app-DM}))
\begin{equation}
\rho_n=\Lambda_n [\rho_0]\,,\label{DM}
\end{equation}
which, given the initial state $\rho_{0}$, returns the state of $S$ at any step $n$. The dynamical map (in fact the propagator of the open dynamics) depends on the collision unitary \ref{USn1} and the initial state of ancillas. It is ensured to be CPT since $S$ shares no initial correlations with the bath. 

In terms of the dynamical map $\Lambda_{n}$, \eq\ref{conc} is simply expressed as
\begin{equation}
\Lambda_n={\mathcal{E}}^n\,, \label{DMCM}
\end{equation}
showing the exponential dependence of $\Lambda_n$ on the collision map $\mathcal E$.
It immediately follows that, for any integer $m$ between 0 and $n$,
\begin{equation}
\Lambda_{n}=\Lambda_{n-m} \,\Lambda_m\,.\label{SG}
\end{equation}
This is the discrete-dynamics version of the so called \textit{semigroup property} (see \ref{app-lind}). It states that, like for any group (in the mathematical sense), the composition of dynamical maps produces another legitimate dynamical map. Here, the prefix ``semi" comes from the fact that dynamical maps are generally non-unitary and thus cannot be inverted [see \ref{app-qmaps}] (physically this means that they describe irreversible dynamics).

The semi-group property is another equivalent way to express the memoryless nature of the dynamics [already stressed below \eq\ref{eps}]: if we know the state at an intermediate step $m$, $\rho_{m}$, no matter what happened at previous steps, we can determine the following evolution up to a final time $n$.

\section{Inhomogeneous collision model and CP divisibility}
\label{section-inhomCP}

So far we assumed that the entire model is homogeneous: the ancilla's initial state $\eta$ was assumed to be the same for all ancillas [\cf\cref{sigma0}] and so was the collision unitary \ref{USn1}. Accordingly, the open dynamics resulted from repeated applications of the same map \ref{coll-map} [recall \cref{conc,DMCM}]. This homogeneity assumption, made mostly for the sake of argument to better highlight general properties, can be relaxed straightforwardly. By still maintaining the assumption of initially uncorrelated ancillas (a constraint which we will relax in  \Sec \cref{section-ica,section-singleph}), the initial state \ref{sigma0} can be generalized as
\begin{equation}
\sigma_{0}=\rho_{0}\bigotimes_n \eta_n\,\label{sigma0-2}
\end{equation}
with $\eta_n$ not necessarily the same state for all ancillas, while in the collision unitary \ref{USn1} $\hat H_S$, $\hat H_n$, $\hat V_{n}$ can be different for different values of $n$. Accordingly, the system's collision map \ref{coll-map} is generally step-dependent and we thus rename it $\mathcal E^{(n)}$.\marginnote{This kind of integer subscript between brackets will usually denote the step number (discrete time).} Correspondingly, the dynamical map \ref{DMCM} is generalized as
\begin{equation}
\Lambda_n={\mathcal{E}}^{(n)}{\mathcal{E}}^{(n-1)}\cdots \,{\mathcal{E}}^{(0)}\,.\label{DMCM-2}
\end{equation}
The semigroup property \ref{SG} is replaced by the more general
\begin{equation}
\Lambda_{n}=\Phi_{n,m} \,\Phi_{m,0}\,,\label{SG-2}
\end{equation}
holding for any integer $0\le m\le n$. Here, we defined map
\begin{equation}
\Phi_{n_2,n_1}={\mathcal{E}}^{(n_2)}{\mathcal{E}}^{(n_2-1)}\cdots \,{\mathcal{E}}^{(n_1)}\,\,\,\,\,\,\,\,\,\,\,(n_2\ge n_1)\,\,,
\end{equation}
which, being a composition of CPT maps, is also CPT (this can be easily shown).

\eq\ref{SG-2} is the discrete version of a property called ``CP
divisibility"~\cite{rivas_open_2012}. This is in fact the statement that the open
dynamics can be divided into a sequence of elementary CPT maps which generally need
not be the same. In the special case that they are the same, we recover the semigroup
property \ref{SG}. Fulfillment of {CP} divisibility has been proposed as a
quantitative definition of Markovian behavior that is more general than the
traditional Markovianity associated with the semigroup
property~\cite{breuer_colloquium:_2016,RHP}. In this sense, the inhomogeneous CM
defined above can still be considered to be Markovian, an assessment in agreement
with the fact that conditions (1)--(3) of  \ref{conds} are still matched. 

\section{All-qubit collision model}
\label{section-all-qubit}

To illustrate more concretely some of the concepts introduced so far, we next present one the simplest instances of CM which we will refer to repeatedly in this paper as the ``all-qubit CM". 
The open system $S$ is a \textit{``qubit"} (two-level system), whose Hilbert space is spanned by the orthonormal basis $\left\{\ket{0}, \ket{1}\right\}$ with $\hat\sigma_z\ket{0}{=}{-}\ket{0}$ and $\hat\sigma_z\ket{1}{=}\ket{1}$, where $\hat\sigma_\alpha$ ($\alpha=x,y,z$) are the usual Pauli spin operators.
Ancillas are also qubits, each with orthonormal basis $\left\{\ket{0}_n, \ket{1}_n\right\}$ (eigenstates of $\hat\sigma_{nz}$, \ie the $z$-component of the $n$th ancilla spin operator). We assume for simplicity no free Hamiltonian for both $S$ and ancillas, \ie $\hat H_S=\hat H_n=0$, and consider the (homogeneous) system--ancilla coupling Hamiltonian
\begin{equation}
\hat V_n= g \left(\hat\sigma_{+}\otimes\hat\sigma_{-}+\hat\sigma_{-}\otimes\hat\sigma_{+}\right)+g_z \,\hat\sigma_{z}\otimes\hat\sigma_{z} \,,\label{Vn-qubits}
\end{equation}
where in each term the first (second) operator acts on $S$ ($n$th ancilla) and with
\begin{equation}
\hat\sigma_-=\hat\sigma_+^\dag=\tfrac{1}{\sqrt{2}}\left(\hat\sigma_x+i \hat\sigma_y\right)=\vert 0\rangle \langle 1\vert  \,,\label{ladder}
\end{equation}
the usual spin ladder operators. The eigenstates of $\hat V_n$ are $\ket{00}$, $\ket{11}$ and
\begin{equation}
\vert \Psi^{\pm}\rangle =\tfrac{1}{\sqrt{2}}\left(\ket{10}\pm \ket{01}\right)\label{Psipm}
\end{equation}
with eigenvalues $g_z$, $g_z$ and $\pm g-g_z$, respectively (we use the short notation $\vert ab\rangle =\vert a\rangle _S\vert b\rangle _n$). 

Hence, the collision unitary \ref{USn1} for this class of CMs explicitly reads\marginnote{Using eigenstates and eigenvalues of $\hat V_n$, $\hat U_n$ is spectrally decomposed as $\hat U_{n}=e^{-i g_z \Delta t}\left(\ket{00}\bra{00}+\ket{11}\bra{11}\right)+e^{-i (g-g_z) \Delta t} \ket{\Psi^+}\bra{\Psi^+}+e^{i(g+g_z)\Delta t}\ket{\Psi^-}\bra{\Psi^-}$.  We next expand $\ket{\Psi^\pm}$ and then use that $\hat\sigma_{+}\otimes\hat\sigma_{-}=\ket{10}\bra{01}$}
\begin{align}
\hat U_{n}&=e^{-i 2g_z \Delta t}\left(\ket{00}\bra{00}+\ket{11}\bra{11}\right)\label{Un-qubits}\nonumber\\
&+\cos (g\Delta t) \left(\ket{10}\bra{10}+\ket{01}\bra{01}\right)-i\sin (g\Delta t) \left(\hat\sigma_{+}\hat\sigma_{-}+\hat\sigma_{-}\hat\sigma_{+}\right)\,
\end{align}
where we omitted an irrelevant phase factor $e^{i g_z \Delta t}$ and all tensor product symbols.

Although the all-qubit CM at first may appear somewhat artificial, there are realistic physical scenarios (see  \Sec \ref{section-CMQO}) where it provides an accurate description of the dynamics (especially in the case $g_z=0$).

\subsection{Partial swap unitary collision}
\label{section-swap}

An important special case is when $\hat U_{n}$ reduces to a \textit{partial swap}, this being a recurrent collision unitary in the CM literature. 

Let us first define the SWAP unitary operator $\hat S_{n}$ as the operator such that
\begin{equation}
\label{part-s}
\hat S_{n}\ket{\psi}_S\ket{\chi}_n=\ket{\chi}_S\ket{\psi}_n
\end{equation}
for any pair of states $\ket{\psi}$ and $\ket{\chi}$. In line with the notation used for $\hat V_n$ and $\hat U_n$, only the ancilla index appears in the subscript of $\hat S_{n}$ (yet recall that it acts on both $S$ and ancilla). Note that
\begin{equation}
\hat S_{n} \,\hat S_{n}=\mathbb{ I}\,.\label{S2}
\end{equation}
Thus operator $\hat S_{n}$ is both Hermitian and unitary.

Operator $\hat S_{n}$ thus swaps the states of $S$ and the ancilla. Note that definition \ref{part-s} applies even if $S$ and ancilla $n$ are not qubits, the essential requirement being that they have the same Hilbert space dimension.

A partial swap is a generalization of the SWAP defined as\marginnote{The conversion from the exponential to the trigonometric form straightforwardly follows from $\hat S_n^2=\mathbb{I}$ (the analogous property holds for any spin-1/2 operator).}
\begin{equation}
\label{p-swap}
\hat U_n=e^{-i \theta \hat S_{n} }=\cos\theta\,\mathbb{I}-i \sin\theta \hat S_{n}\,,   
\end{equation}
where angle $\theta$ (such that $0\le \theta\le \pi/2$) measures the amount of swapping. For $\theta=0$, $\hat U_n$ reduces to the identity corresponding to a fully ineffective collision. For $\theta=\pi/2$, instead, the collision has the maximum effect, swapping the states of the interacting systems. 

\subsection{Partial swap in the all-qubit CM}

In the case of the all-qubit CM ($S$ and ancilla $n$ are both qubits), $\hat S_{n}$ leaves $\ket{00}$ and $\ket{11}$ unchanged while $\ket{01}$ and $\ket{10}$ are turned into one another. It is then easily shown that the SWAP operator can be expressed in terms of the identity and Pauli operators as
\begin{equation}
\label{Snq}
\hat S_{n}=\tfrac{1}{2}\left({\mathbb{ I}+\hat{{\boldsymbol \sigma}}\cdot\hat{{\boldsymbol \sigma}}_n}\right) \,.
\end{equation}
The partial swap unitary \ref{p-swap} occurs in the all-qubit model for $g_z=g/2$ [\cf\eq\ref{Vn-qubits}], corresponding to the \textit{Heisenberg exchange} interaction Hamiltonian
\begin{equation}
\label{heis}
\hat V_n= \frac{g}{2} \,\hat{{\boldsymbol \sigma}}\cdot\hat{{\boldsymbol \sigma}}_n\,\,.
\end{equation}
Indeed, up to an irrelevant phase factor, the corresponding unitary \ref{Un-qubits} has the form \ref{p-swap} with $\hat S_{n}$ given by \ref{Snq} and $\theta=g\de t$.\marginnote{Replacing $\hat S_{n}{=}1/2\,({\mathbb{ I}}{+}\hat{{\boldsymbol \sigma}}{\cdot}\hat{{\boldsymbol \sigma}}_n)$, we get $\hat U_n{=}e^{-i g/2\hat{{\boldsymbol \sigma}}\cdot\hat{{\boldsymbol \sigma}}_n \de t}{=}e^{i \frac{g}{2} \Delta t}e^{-i g \hat S_{n} \Delta t}$. Thus, $\theta=g\de t$ up to phase factor $e^{i \frac{g}{2} \Delta t}$.}

\subsection{Reduced dynamics of $S$ and ancilla}

Take all ancillas initially in the same state $\eta_n=\ket{0}_n\bra{0}$ [\cf\eq\ref{sigma0}]. Using basis $\left\{\ket{0}_n, \ket{1}_n\right\}$ to carry out the partial trace over each ancilla, the collision map \ref{coll-map} is worked out from \ref{Un-qubits}  as
\begin{equation}
\rho'=\mathcal{E}[\rho]=\hat K_0 \rho\hat K_0^\dagger+\hat K_1 \rho\hat K_1^\dagger\label{eps-qubits}\,,
\end{equation}
where the Kraus operators $\hat K_k={\mbox{}_n}\bra{k}\hat U_n\vert 0\rangle _n$ (see \ref{app-qmaps}) explicitly
read\marginnote{For $g_z=0$, the collision map reduces to a so called amplitude
damping channel~\cite{nielsen2002quantum}.}
\begin{align}
&\hat K_0=e^{-i g_z\Delta t}\ket{0}_S\bra{0}+e^{i g_z \Delta t}\cos(g\Delta t)\ket{1}_S\bra{1}\label{kraus-qubits}\\
&\hat K_1=-i\,e^{i g_z \Delta t}\sin(g\Delta t)\,\hat \sigma_{-}\,. \nonumber 
\end{align}
Any qubit state has the general form
\begin{equation}
\label{qubit-state}
\rho=
\begin{pmatrix}
\bra{1}\rho\ket{1}&\bra{0}\rho\ket{1}\\
\bra{1}\rho\ket{0}&\bra{0}\rho\ket{0}
\end{pmatrix}
=
\begin{pmatrix}
p& c\\ c^{*} &1{-}p
\end{pmatrix}
\end{equation}
with $0\le p\le 1$ and $(1-2p)^2+4\vert c\vert ^2\le 1$ (to ensure that eigenvalues of $\rho$ are positive). Entries $c$ and $p$ are routinely called ``coherences" and ``populations", respectively.

Plugging \ref{qubit-state} into \ref{eps-qubits} yields
\begin{equation}
\label{map-all}
\rho'=\mathcal{E}[\rho]=
\begin{pmatrix}
\cos ^2(g\Delta t)\,p  & e^{2i g_z\Delta t}\cos (g\Delta t)\,c \\
e^{-2i g_z\Delta t}\cos (g\Delta t)\,c^{*} &(1{-}p)+ \sin^2(g\Delta t)\,p  
\end{pmatrix}
\,,
\end{equation}
which is an alternative way to represent the collision map \ref{eps-qubits}. The effect of the map, as can be seen, is to multiply the coherences $c$ by a factor $e^{2i g_z\Delta t}\cos(g\Delta t)$ and populations $p$ by $\cos^2(g\Delta t)$. In light of \eq\ref{conc}, the state of $S$ at step $n$ is thus given by
\begin{equation}
\label{rhon-qubits}
\rho_n=\mathcal{E}^n[\rho]=
\begin{pmatrix}
p_n  & c_n \\
c_n^{*} &1{-}p_n  
\end{pmatrix}
\,
\end{equation}
with
\begin{equation}
p_n=\cos ^{2n}(g\Delta t)\,p\,,\,\,\,c_n=e^{2i g_z n \Delta t} \cos^n (g\Delta t)\,c\,\,.\label{pncn}
\end{equation}
This entails the following: provided that $\vert \cos(g\Delta t)\vert <1$, no matter what state $S$ started from (\ie regardless of $c$ and $p$), the coherences and populations vanish for $n\rightarrow\infty$ meaning that $S$ asymptotically ends up in state $\ket{0}_S$. This is a rather extreme example of non-unitary, \ie irreversible, open dynamics, which can be pictured as a transformation mapping the entire Hilbert space of $S$ into a single point representing the asymptotic states $\ket{0}_S$.

By replacing \cref{Un-qubits,rhon-qubits} into \ref{ancilla-map} for $\eta_n=\ket{0}_n\bra{0}$, we get that the state of ancilla $n$ after colliding with $S$ is given by
\begin{align}
\label{ancilla-maps-qubits}
\eta'_n=
\begin{pmatrix}
\pi_n & d_n \\ d_n^{*}& 1{-}\pi_n\,
\end{pmatrix}
\,.
\end{align}
with
\begin{align}
&\pi_n=\sin ^2(g \Delta t) \cos ^{2(n-1)}(g \Delta t)\,p\\
&d_n= -i \sin (g\Delta t)\cos ^{n-1}(g \Delta t) e^{2i g_z n\Delta t}\,c\,\,.\nonumber\label{pncn-eta}
\end{align}
Note that, as $n$ grows up and for $\vert \cos (g\Delta t)\vert <1$, $\eta'_n\rightarrow \ket{0}_n\bra{0}=\eta_n$, Namely, after a sufficient number of steps, ancillas basically no longer change their state after colliding with $S$. This is consistent with the convergence of $S$ to $\ket{0}_S$ since the collision leaves state $\ket{0}_S\ket{0}_n$ unaffected, \ie $\hat U_n \ket{00}_{Sn}\bra{00} \hat U_n^\dag=\ket{00}_{Sn}\bra{00}$.

\section{Steady states}
\label{section-homo}

As discussed, {\eq\ref{rhon-qubits} shows that for, $\vert \cos (g\Delta t)\vert <1$, $S$ eventually ends up in state $\ket{0}_S$,\marginnote{The rigorous mathematical statement is that, for any $\varepsilon>0$, there exists $n_\varepsilon$ such that $\Vert \rho_n-\ket{0}_S\bra{0}\Vert <\varepsilon$ for any $n>n_\varepsilon$, where $\Vert \ldots \Vert $ is some distance measure between quantum states (\eg trace distance).} \ie $\rho_n\rightarrow \rho^{*}=\ket{0}_S\bra{0}$.
Once $S$ reaches this state, this will be not be affected by collisions with ancillas. In these cases, we say that $\rho^{*}$ is a \textit{steady} or \textit{stationary} state for $S$.

In the language of quantum maps (see \ref{app-qmaps}), a steady state $\rho^{*}$ is a \textit{fixed point} of the collision map, \ie
\begin{equation}
\mathcal{E}[\rho^{*}]=\rho^{*}\,\,\,\,\,\,({\rm steady}\,\,\mathrm{state}).\label{fixedE}
\end{equation}
This expresses the fact that $\rho^{*}$ is unchanged by the collisions, no matter how many (since we also have $\mathcal{E}^n[\rho^{*}]=\rho^{*}$ for any $n$). 
Note that, in general, map $\mathcal E$ could admit more than one fixed point, \ie many steady states can exist. When only one steady state is possible (as in the previous instance), \ie 
there is a \textit{unique} fixed point, we say that the collision map is  \textit{ergodic}.\marginnote{More formally, since a fixed point is an eigenstate of ${\mathcal{E}}$ with eigenvalue 1 [\cf\eq\ref{fixedE}], the map is ergodic when 1 is a non-degenerate eigenvalue of ${\mathcal{E}}$.}

Actually, the instance in the previous subsection fulfills a stronger property in that $\rho_n\rightarrow \rho^{*}$ for \textit{any} initial state $\rho_{0}$.
In the language of quantum maps, in such cases map ${\mathcal{E}}$ is said to be \textit{mixing}. A paradigm of mixing processes is thermalization (which will be discussed in  \Sec \ref{section-QT1}), enforcing $S$ to end up in the Gibbs state at the reservoir temperature no matter what state it started from.  Importantly, note that a necessary -- but not sufficient -- condition for a map to be mixing with respect to a steady state $\rho^{*}$ is that this be a fixed point \ie fulfill \ref{fixedE}]. A more stringent necessary condition, although still insufficient, is that $\rho^{*}$ be the \textit{only} fixed point, namely (see above) the collision map must be ergodic (if there were two or more fixed points, mixingness clearly could not occur).

A simple paradigmatic instance where ergodicity, hence mixingness, does not take place is the all-qubit CM for $g\Delta t=\pi$ and $g_z=0$. The corresponding collision map \ref{map-all} then is
\begin{equation}
\label{map-all2}
\rho'=\mathcal{E}[\rho]=
\begin{pmatrix}
	p  & -c \\
	-c^{*} & 1{-}p  
\end{pmatrix}
\,.
\end{equation}
This leaves populations unaffected, while coherences change sign. Clearly, any mixture of $\ket{0}_S\bra{0}$ and $\ket{1}_S\bra{1}$ (zero coherences) is a fixed point of map ${\mathcal{E}}$ [\cf\eq\ref{fixedE}]. Notably, there exist initial states giving rise to a dynamics where $S$ never reaches a steady state. For instance, observing that ${\mathcal{E}}\left[\ket{\pm}\bra{\pm}\right]=\ket{\mp}\bra{\mp}$ with $\ket{\pm}=\tfrac{1}{\sqrt{2}}(\ket{0}\pm \ket{1})$ (eigenstates of $\hat \sigma_x$), we see that if $\rho_0=\ket{+}\bra{+}$ then $S$ will indefinitely oscillate between states $\ket{+}$ ($n$ even) and $\ket{-}$ ($n$ odd).

We finally mention a special type of mixing dynamics called \textit{quantum homogenization}, which occurs when ${\mathcal{E}}$ is mixing with steady state $\rho^{*}$ such that $\rho^{*}=\eta$ for any initial state $\eta$ of ancillas. Note how this definition poses the constraint that $S$ and each ancilla have the same Hilbert space dimension (and moreover that all ancillas start in the same state). Physically, the intuitive idea behind quantum homogenization is that, since the bath is made out of a huge number of identical subsystems, if $S$ ``talks" long enough with them then its state will more and more look like that of ancillas until becoming homogeneous with these. It can be shown~\cite{ziman_diluting_2002} that in the
all-qubit CM quantum homogenization occurs when the collision unitary $\hat U_n$ is
a partial swap [\cf\eq\ref{p-swap}] corresponding to the Heisenberg exchange
interaction \ref{heis}.

\section{Cascaded collision model}
\label{section-cascaded}

The basic CM of  \fig \ref{fig-CMdyn} comes with an intrinsic \textit{unidirectionality}: $S$ explores the bath along a specific direction (say from left to right as in  \fig \ref{fig-CMdyn}). Remarkably, if we let $S$ be \textit{multipartite} in such a way that each ancilla collides with one subsystem of $S$ at a time, then the above unidirectionality yields an interesting effect.

To see this, let $S$ comprise a pair of subsystems, $S_1$ and $S_2$. 
By hypothesis, the collision with each ancilla consists of two \textit{cascaded} sub-collisions (see  \fig \ref{fig-chiral}): $n$ collides \textit{first} with $S_1$ and only \textit{afterward} with $S_2$. Accordingly, the collision unitary reads
\begin{equation}
\hat U_n=\hat U_{2,n}\hat U_{1,n}\,.\label{Un-casc}
\end{equation}
The remaining hypotheses of the basic CM in  \Sec \ref{section-def} are unchanged. Already at this stage, it is clear that there exists an asymmetry between $S_1$ and $S_2$ since $\hat U_n$ does change if 1 and 2 are swapped (as $\hat U_{1,n}$ and $\hat U_{2,n}$ generally do not commute). The open dynamics of $S_1$ is indeed quite different from that of $S_2$, as we show next.

We first note that, just like in the basic collision model, the \textit{joint} system $S$ undergoes a fully Markovian dynamics according to [\cf\eq\ref{coll-map}]
\begin{equation}
\rho_n={\mathcal{E}}[\rho_{n-1}]={\rm Tr}_n \left\{\hat U_{2,n}\hat U_{1,n}\,\rho_{n-1}\eta_n\, \hat U_{1,n}^\dag\hat U_{2,n}^\dag\right\}\,.\label{map-chir}
\end{equation}

\begin{figure}[!h] 
	\raggedright
	\begin{floatrow}[1]
		\ffigbox[\FBwidth]{\caption[Cascaded collision model]{\textit{Cascaded collision model}. The open system $S$ is made out of two subsystems, $S_1$ and $S_2$. At each collision, the ancilla interacts \textit{first} with $S_1$ (unitary $\hat U_{1,n}$) and only \textit{afterward} with $S_2$ (unitary $\hat U_{2,n}$). Thus the collision consists of two time-ordered sub-collisions according to the collision unitary $\hat U_n=\hat U_{2,n}\hat U_{1,n}$. Subsystem $S_1$ always collides with ``fresh" ancillas (still in the initial state $\eta$), while $S_2$ collides with ``recycled" ancillas that already interacted (and got correlated) with $S_1$}.\label{fig-chiral}}%
		{\includegraphics[width=\textwidth]{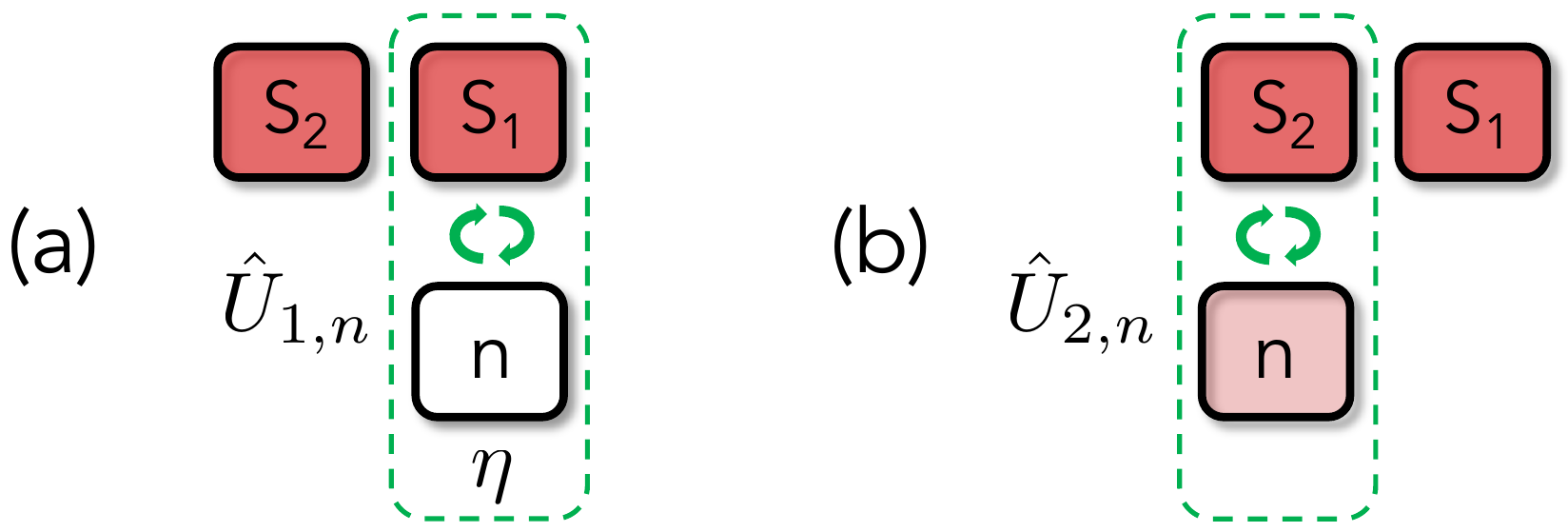}}
	\end{floatrow}
\end{figure}

We next ask whether or not the same statement holds for the reduced dynamics of $S_1$ and $S_2$ (whose reduced states will be respectively denoted as $\rho_{1,n}$ and $\rho_{2,n}$). Let us start with $S_1$: tracing off $S_2$ in \ref{map-chir} yields\footnote{We use that ${\rm Tr}_2{\rm Tr}_n \left\{\hat U_{2,n}\sigma\hat U_{2,n}^\dag\right\}\equiv{\rm Tr}_2{\rm Tr}_n \left\{\sigma\right\}$ since if $\{\vert k_2, k_n\rangle \}$ is an orthonormal basis of system $2$-$n$ used to compute ${\rm Tr}_2{\rm Tr}_n\{\ldots \}$ another legitimate basis to perform the trace is $\{\hat U_{2,n}^\dag\vert k_2, k_n\rangle \}$ (recall that the trace can be carried out in any basis).}
\begin{align}
\rho_{1,n}=&{\rm Tr}_2\{\rho_n\}=
{\rm Tr}_2{\rm Tr}_n \left\{\hat U_{2,n}\hat U_{1,n}\,\rho_{n-1}\eta_n\, \hat U_{1,n}^\dag\hat U_{2,n}^\dag\right\}\nonumber\\
=&{\rm Tr}_2{\rm Tr}_n \left\{\hat U_{1,n}\,\rho_{n-1}\eta_n\, \hat U_{1,n}^\dag\right\}\,.
\end{align}
Since ${\rm Tr}_2\{\ldots \}$ does not act on either $S_1$ or ancilla $n$, it can be moved to the right of $\hat U_{1,n}$
\[
\\
\rho_{1,n}={\rm Tr}_n \left\{\hat U_{1,n}\,{\rm Tr}_2\{\rho_{n-1}\}\eta_n\, \hat U_{1,n}^\dag\right\}={\rm Tr}_n \left\{\hat U_{1,n}\,\rho_{1,n-1}\eta_{n}\hat U_{1,n}^\dag\right\}\equiv {\mathcal{E}}[\rho_{1,n-1}]\,,\label{map-chir4}\]
where we introduced the usual collision map \ref{coll-map}.
Thus $S_1$ evolves exactly as if $S_2$ were absent, entailing in particular that its dynamics is Markovian. This occurs because $S_1$ always collides with ``fresh" ancillas that are still in the initial state $\eta$ [see  \fig \ref{fig-chiral}(a)]. Once the ancilla has collided with $S_1$, the following collision with $S_2$ cannot affect the reduced state of $S_1$. 

In contrast, since it collides with ``recycled" ancillas that already collided with $S_1$, the dynamics of $S_2$ does depend on that of $S_1$. Indeed, if we now trace off subsystem $S_1$ from \eq\ref{map-chir} we get
\begin{align}
\rho_{2,n}&={\rm Tr}_1{\rm Tr}_n \left\{\hat U_{2,n}\hat U_{1,n}\,\rho_{n-1}\eta_n\, \hat U_{1,n}^\dag\hat U_{2,n}^\dag\right\}\nonumber\\
&={\rm Tr}_n \left\{\hat U_{2,n}\,{\rm Tr}_1\left\{\hat U_{1,n}\,\rho_{n-1}\eta_n\, \hat U_{1,n}^\dag\right\}\hat U_{2,n}^\dag\right\}\,.
\end{align}
At least two features stand out. First, the state of $S_2$ is affected by the previous subcollision (the one involving $S_1$). Second, upon comparison with $\rho_{1,n}$, we see that $\rho_{2,n}$ does \textit{not} evolve according to a CPT map. This is because, after the first sub-collision but before the second one starts, $S_2$ is in general already correlated with ancilla $n$. Indeed, even if $S_1$ and $S_2$ start in a product state, very soon they will get correlated during the collisional dynamics due to their interaction with the common bath of ancillas. Thus, as soon as ancilla $n$ has finished colliding with $S_1$, it establishes correlations with both $S_1$ and $S_2$.

To summarize, in a cascaded CM, the two subsystems jointly undergo a Markovian evolution. The reduced dynamics of $S_1$ is Markovian as well and completely insensitive to the presence of $S_2$. Instead, the reduced dynamics of $S_2$ depends on that of $S_1$ and is generally \textit{non}-Markovian since it cannot be divided into a sequence of CPT maps. This \textit{asymmetry} in the mutual dependence of the two reduced dynamics reflects the intrinsic CM unidirectionality (causal order) that we discussed above.

The next subsection (connecting CMs with matrix product states theory) is not indispensable to access the remainder of the paper. As such, it could be skipped by the uninterested reader.

\section{Collision models and Matrix Product States}
\label{section-MPS}

We have previously focused on the reduced dynamics of either $S$ or an ancilla. Here, we will consider the joint dynamics of $S$ and all ancillas showing that it enjoys interesting properties. 

Starting from state \ref{sigma0}, as the collisional dynamics proceeds, multipartite correlations are established so that the joint system evolves at step $n$ into a generally \textit{entangled} state\marginnote{An entangled state is a state which is not separable, \ie such that the corresponding density matrix cannot be expressed as a mixture of product states. For bipartite systems, a separable state reads $\sigma_{12}=\sum_j p_j \rho_{1}^{(j)} \otimes \rho_2^{(j)}$ with $\sum_j p_j=1$ (this naturally generalizes to $N$ systems).} having the generic form
\begin{equation}
\ket{\Psi_n}=\sum_{\alpha,k_1, \ldots , k_n}c_{\alpha,k_1,k_2, \ldots ,k_n}\ket{\alpha,k_1,k_2, \ldots ,k_n}\,\,\label{exp-1}
\end{equation}
with $\{\ket{\alpha}\}$ denoting a basis of $S$ and $\{\ket{k_n}\}$ a basis of the $n$th ancilla.\marginnote{Here, $\ket{\alpha}$ and $\ket{k_n}$ respectively stand for $\ket{\alpha}_S$ and $\ket{k_n}_n$, a light notation that will be used again later on in the paper.}

We will show next that state \ref{exp-1} can be expressed in a computationally advantageous form. The basic idea is to view the expansion coefficients $c_{\alpha,k_1,k_2, \ldots ,k_n}$ (each labeled by $n+1$ indexes) as a rank-$(n+1)$ tensor and decompose it into $n$ tensors each with the smallest possible rank.

For the sake of argument, we will refer to the basic CM of  \Sec \ref{section-def} and assume that $\alpha=1,2, \ldots ,d_S$ with $d_S$ the Hilbert space dimension of $S$, while $k_n$ takes integer values from 1 to $d_A$ with $d_A$ the dimension of each ancilla. We consider an initial pure state $\sigma_0=\ket{\Psi_0}\bra{\Psi_0}$, which without loss of generality can be written as
\begin{equation}
\ket{\Psi_0}=\ket{1,1_1,1_2, \ldots }.   
\end{equation}
At the end of the first collision, the joint state reads
\begin{align}
\ket{\Psi_1}=\hat U_1\ket{\Psi_0}&=\sum_{\alpha,k_1}\ket{\alpha,k_1}\bra{\alpha,k_1}\hat U_1\ket{1,1_1}\otimes\ket{1_2,1_3, \ldots }\\
&=\sum_{\alpha,k_1}{\rm U}_{1 \alpha}^{(k_1)}\,\ket{\alpha,k_1,1_2,1_3, \ldots }\,\nonumber,
\end{align}
where we plugged in the identity operator $\mathbb{I}_S\otimes \mathbb{I}_1$ expressed in terms of basis $\ket{\alpha,k_1}$ and defined
\begin{equation}
{\rm U}_{\alpha\alpha'}^{(k_1)}=\bra{\alpha',k_1}\hat U_1\ket{\alpha,1_1}\,.\label{mat-U}
\end{equation}
This is a rank-3 tensor of dimension $d_S\times d_S\times d_A$ due to dependence on the three indexes $\alpha$, $\alpha' $ and $k$.

\begin{figure}[!h] 
	\raggedright
	\begin{floatrow}[1]
		\ffigbox[\FBwidth]{\caption[Tensor-network representation of the joint CM dynamics]{\textit{Tensor-network representation of the joint CM dynamics}. The joint state at step $n$ is generally defined by the rank-$(n+1)$ tensor $c_{\alpha,k_1, \ldots ,k_n}$ [\cf\eq\ref{exp-1}]. This can be decomposed into one rank-2 tensor of dimension $d_S\times d_A$ (leftmost square) and $n-1$ rank-3 tensors each of dimension $d_S\times d_S\times d_A$ (squares with three legs) with $d_S$ ($d_A$) the Hilbert space dimension of $S$ (ancilla). A joined leg (each link between nearest-neighbor squares)  represents an index contraction. The steps $n=1$ and $n=2$ are also shown for comparison.}\label{fig-MPS}}%
		{\includegraphics[width=\textwidth]{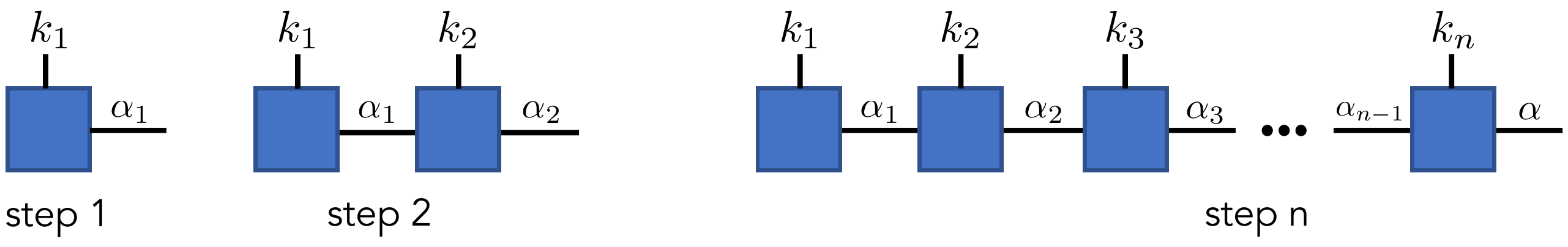}}
	\end{floatrow}
\end{figure}

Using this, the joint state at the end of the second collision, $\ket{\Psi_2}=\hat{ U}_2\ket{\Psi_1}$, can be worked out as
\begin{align}
\ket{\Psi_2}&=\sum_{\alpha, k_1}{\rm U}_{1 \alpha}^{(k_1)}\,\hat U_2\ket{\alpha, k_1,1_2, \ldots }\nonumber\\
&=\sum_{\alpha,k_1}{\rm U}_{1 \alpha}^{(k_1)}\sum_{\alpha',k_2}\ket{\alpha',k_2}\bra{\alpha',k_2}\hat U_2\ket{\alpha,1_2}\otimes\ket{k_1}\otimes\ket{1_3, \ldots }\nonumber\\
&=\sum_{\alpha, k_1}\sum_{\alpha',k_2}{\rm U}_{1\alpha}^{(k_1)}\,{\rm U}_{\alpha \alpha'}^{(k_2)}\ket{\alpha',k_1,k_2,1_3, \ldots }\nonumber\\
&=\sum_{\alpha',k_1,k_2}\left(\sum_{\alpha}{\rm U}_{1 \alpha}^{(k_1)}\,{\rm U}_{\alpha \alpha'}^{(k_2)}\right)\ket{\alpha',k_1,k_2,1_3, \ldots }\,\,.
\end{align}
For $n=3$, analogous steps lead to
\begin{equation}
\ket{\Psi_3}{=}\sum_{\alpha'',k_1,k_2,k_3}\left(\sum_{\alpha \alpha'}{\rm U}_{1,\alpha}^{(k_1)}\,{\rm U}_{\alpha\alpha'}^{(k_2)}{\rm U}_{\alpha',\alpha''}^{(k_3)}\right)\ket{\alpha'',k_1,k_2,k_3,1_4, \ldots }\,.   
\end{equation}
Finally, at the $n$th step, we get \ref{exp-1} with each coefficient given by
\begin{equation}
\label{c-dec}
c_{\alpha,k_1, \ldots ,k_n}=\sum_{\alpha_1, \ldots , \alpha_{n-1}}{\rm U}_{1 \alpha_1}^{(k_1)}\,{\rm U}_{\alpha_1\alpha_2}^{(k_2)}\dots{\rm U}_{\alpha_{n-1} \alpha}^{(k_n)}\,\,.
\end{equation}
Thus, as schematically sketched in  \fig \ref{fig-MPS}, we get that the rank-$(n+1)$ tensor $c_{\alpha,k_1, \ldots ,k_n}$ can be decomposed in terms of $n-1$ rank-3 tensors of dimension $d_S\times d_S\times d_A$ and one rank-2 tensor of dimensions $d_S\times d_A$.\marginnote{The rank-2 tensor is ${\rm U}_{1 \alpha_1}^{(k_1)}$, which derives from the rank-3 tensor \ref{mat-U} by fixing one $\alpha$ index.}
Interestingly, each of these low-rank tensors [\cf\eq\ref{mat-U}] corresponds to a single collision: \eq\ref{c-dec} thus reflects the decomposition of the overall complex system--bath dynamics in terms of elementary two-body unitaries. 
This way of expressing the multipartite $S$-bath state is very close to
the so called \textit{matrix product states}
decomposition~\cite{orus2014practical,gawatz2017matrix,biamonte2017tensor}. The idea
is that reducing to low-rank tensors with \textit{small} enough dimension (if
possible) allows to limit the computational complexity of the problem (with clear
advantages for numerical simulations of the dynamics). A collisional dynamics
typically has such features in that, as shown, the dimension of each rank-three
tensor is bounded in terms of the Hilbert space dimensions of the open system
$S$ and a single ancilla these being often small.

\section{Basic collision model: state of the art}
\label{soa-def}

Throughout we considered each collision to be described by a well-defined unitary.
One can yet consider \textit{random unitary collisions}. These were investigated in
\rref~\cite{gennaro_relaxation_2009}, where it was shown that $S$ reaches
the same asymptotic state which would be attained for repeated random collisions with
a single effective ancilla of suitable dimension.

A more general and formal treatment than  \ref{AD-section} of the  ancilla
dynamics was carried out in the context of so called \textit{non-anticipatory quantum
channels with memory}~\cite{CarusoRMP}. Similarly to cascaded CMs (see
\Sec \ref{section-cascaded}), this dynamics features an explicit causal ordering of the
ancillas, which reflects
the different times at which these interact with $S$.

The Markovianity notion based on divisibility discussed in  \Sec \ref{section-inhomCP}
is featured in the review paper \rref~\cite{rivas_quantum_2014}, where CMs are used
as an effective way to visualize the memoryless properties characteristic of quantum
Markovian processes.

A thorough treatment of mixing channels and fixed points mentioned in
\Sec \ref{section-homo} can be found in \rref~\cite{burgarth_ergodic_2013}. Note that the
properties of mixing CPT maps which we referred to are directly connected with the
concept of \textit{forgetful channels} introduced in 2005 by Kretschmann and
Werner~\cite{KW2005} within the general framework of memory quantum communication
lines (reviewed in \rref~\cite{CarusoRMP}). These models describe the evolution of an
ordered collection of quantum information carriers, which sequentially interact with
a common reservoir. In this context, if the reservoir asymptotically loses track of
its initial state for a growing number of carriers then the resulting transformation
is said to be ``forgetful". Accordingly, from the point of view of the bath ancillas,
any CM featuring a collision map ${\mathcal E}$ that is mixing can be seen as a special
instance of forgetful channel.

\textit{Quantum homogenization} (see  \Sec \ref{section-homo}) was first considered by
Ziman, Stelmachovic, Buzek, Scarani and Gisin in \rref~\cite{ziman_diluting_2002},
where they introduced a so called ``universal quantum homogenizer" this being in fact
an all-qubit CM such that $\rho_n\rightarrow \eta$ for any $\rho_0$ and $\eta$. A
related paper~\cite{ziman_saturation_2003} carried out a detailed analysis of the
nature of correlations (in the form of entanglement) between $S$ and the
bath of ancillas that are established during the collisional dynamics [\cf\
\fig \ref{fig-CMdyn}(d)]. 
We also note that quantum homogenization was studied also in the more general case
that $S$ is a composite system (spin chain) colliding locally with a bath of
ancillas~\cite{burgarth_mediated_2007}.

\textit{Cascaded CMs} (\Sec \ref{section-cascaded}) were first introduced in 2012 by
Giovannetti and Palma~\cite{giovannetti_master_2012,giovannetti_master_2012-1} mostly
with the goal of defining a simple microscopic framework underpinning cascaded master
equations (which will be discussed in detail in  \Sec \ref{sec-cascME}).

Connections between CMs and \textit{matrix product states} (for a friendly
introduction see \eg \rrefs~\cite{gawatz2017matrix,biamonte2017tensor}) can be found
in papers dealing with the more general framework of non-Markovian dynamics, see \eg
\rref~\cite{pichler_photonic_2016,regidor2020modelling} (which we will discuss in
\ref{soa-nm}) and \rref~\cite{filippovPRL}.

\setchapterpreamble[u]{\margintoc}
\chapter{Equations of motion}
\label{section-eqs}

A hallmark of CMs is their \textit{discrete} nature, which is indeed a major reason why these models are useful. Yet, most dynamics in Physics are intrinsically continuous or, better to say, conveniently approached through a continuous-time description, allowing to write down an associated differential equation of motion. 

When it comes to open quantum systems, a relevant equation of motion is the so called \textit{master equation} (ME) governing the time evolution of the open system state $\rho$ (much like the Schr\"odinger equation does for closed systems). In some applications, such as quantum thermodynamics (see  \cref{section-thermo}, it is yet often convenient working with a specific dynamical equation for the expectation value of an observable of concern (\eg energy). Accordingly, in this section we will introduce both kinds of equations of motion (although they are tightly connected to one another of course).

In the last part of the present section, we will in particular revisit the instances of CMs introduced in the previous section with the aim of providing the corresponding ME for each.

\section{Equations of motion for small collision time: states}
\label{section-EMs}

In light of an eventual conversion of the discrete collisional dynamics into a continuous-time one, such that $t_n=n \Delta t$ is turned into the continuous time variable $t$, the collision duration $\Delta t$ must approach zero.

With this in mind, we are interested in the approximated expression of the
collision unitary in the regime of \textit{small collision time}. We thus consider
the basic CM in  \cref{section-def} and replace $\hat U_n$ with the
small-$\Delta t$ approximation
\begin{equation}
	\hat {U}_{n}\,\simeq\, \hat{\mathbb I}-i (\hat{H}_{0}+{\hat V_{n} })\Delta t -\tfrac{1}{2}\,{\hat V}_{n}^2\Delta t ^2\,,\label{USn2}
\end{equation} 
where $\hat H_0$ is the total system--ancilla free Hamiltonian\footnote{Note that this lowest-order expansion
	relies on treating the $S$-ancilla Hamiltonian as time-independent. If not, an additional 2nd-order term would appear in the expansion as is the case of
	\cref{Un-app1} in  \cref{section-CMQO} (see also
	\rref~\cite{altamirano_unitarity_2017}).}
\begin{equation}
\hat H_0=\hat H_S+\hat H_n\label{H0-coll}\,.
\end{equation}
Note that \ref{USn2} is of the 2nd-order in $\hat V_{n}$ but of the 1st order in $\hat{H}_{S}$ and $\hat{H}_{n}$. This is in fact due to a hypothesis of the CM that we are making: \textit{second-order terms in $\Delta t$ that are not quadratic in $\hat V_n$ are negligible}. The rationale of this assumption, requiring in fact that $\hat H_S$ and $\hat H_n$ be much weaker than $\hat V_n$, will become clear later on. \marginnote{This hypothesis will be partially relaxed in  \cref{section-work}.}

Accordingly, at each collision, the joint $S$-bath state $\sigma_n$ evolves according to
\begin{equation}
	{\Delta \sigma_n}=-i \,[\hat H_0 +\hat V_n,\sigma_{n}]\,\Delta t+\left({\hat V}_{n}\,\sigma_n \,{\hat V}_{n}-\tfrac{1}{2}[{\hat V}_{n}^2,\sigma_n]_+\right)\Delta t^2\label{central}
\end{equation}
with $\Delta \sigma_n=\sigma_{n}-\sigma_{n-1}$ and $[\,\,\,,\,\,\,]_+$ the anti-commutator. We dropped third-order terms in $\Delta t$ and, in line with the aforementioned hypothesis, all second-order terms but those having a quadratic dependence on $\hat{V}_{n}$. \cref{central} has a central role in CM theory.\marginnote{Note that \cref{central} is not restricted to the memoryless CMs specified by assumptions (1)--(3) in  \ref{conds} (\ie the CM which we refer to in the present section). In particular, it remains valid for initially correlated ancillas (see  \cref{section-ica}).}

\cref{central} can be equivalently arranged solely in terms of commutators as
\begin{equation}
	{\Delta \sigma_n}=-i \,[\hat H_0+\hat V_n,\sigma_{n}]\,\Delta t-\tfrac{1}{2}\,[\hat V_n,[\hat V_n,\sigma_n]]\Delta t^2\label{central2}\,,
\end{equation}
an alternative expression which is useful in some contexts.

We next focus on $S$ and the $n$th ancilla. Before colliding, they are in the \textit{product} state $\rho_{n-1}\otimes \eta_n$ (see  \cref{section-def}). The collision changes their joint state according to
\begin{equation}
	\frac{\Delta \varrho_{Sn}}{\Delta t}=-i \,[\hat H_0 +\hat V_n,\rho_{n-1} \,\eta_n]+\Delta t\,({\hat V}_{n}\,\rho_{n-1} \,\eta_n \,{\hat V}_{n}-\tfrac{1}{2}[{\hat V}_{n}^2,\rho_{n-1} \,\eta_n]_+)\,,\label{central-M}
\end{equation}
with $\Delta \varrho_{Sn}=\varrho_{Sn}-\rho_{n-1} \,\eta_n$.
This equation, which simply follows from \ref{central} by tracing off all ancillas not involved in the collision and  dividing either side by $\Delta t$, underpins memoryless CMs.

Note that \cref{central,central-M} also hold for the general inhomogeneous CM in  \cref{section-inhomCP}, in which case $\hat H_S$, $\hat H_n$, $\hat V_n$ and $\eta_n$ are understood as generally dependent on step $n$.

To get a closed equation for the reduced dynamics of $S$ we trace off the $n$th ancilla in \ref{central-M}, obtaining
\begin{equation}
	\frac{\Delta \rho_n}{\Delta t}=-i \,[\hat H_S+{\rm Tr}_{n}\{\hat{V}_{n}\,\eta_n\},\rho_{n-1}]+\Delta t\, {\rm Tr}_{n}\{{\hat V}_{n}\,\rho_{n-1}\eta_n \,{\hat V}_{n}-\tfrac{1}{2}[{\hat V}_{n}^2,\rho_{n-1}\eta_n]_+\}\,\label{drho}
\end{equation}
with $\Delta \rho_n=\rho_n-\rho_{n-1}={\rm Tr}_n\{{\Delta \varrho_{Sn}}\}$. Note that, since $\Delta \rho_n=\left(\mathcal{E}-\mathcal{I}\right)[\rho_{n-1}]$ with $\mathcal I$ the identity map on $S$, \cref{drho} in fact represents the short-time expression of the collision map $\mathcal{E}$ [\cf\cref{coll-map}].

By tracing off $S$ (instead of ancilla $n$) in \cref{central-M}, a similar discrete-time equation of motion can be obtained for the change of ancilla's state $\Delta \eta_n=\eta_n'-\eta_n$ due to the collision with $S$ (see  \ref{AD-section}). This reads
\begin{equation}
	\frac{\Delta \eta_n}{\Delta t}=-i \,[\hat H_n+{\rm Tr}_{S}\{\hat{V}_{n}\,\rho_{n-1}\},\eta_{n}]+\Delta t \,{\rm Tr}_{S}\{{\hat V}_{n}\rho_{n-1}\eta_n \,{\hat V}_{n}-\tfrac{1}{2}[{\hat V}_{n}^2,\rho_{n-1}\eta]_+\}\,.\label{deta}
\end{equation}
\eq \ref{drho} (discrete-time \textit{master equation}) and \cref{deta} are finite-difference equations that govern the reduced dynamics of $S$ and ancilla $n$, respectively, at the discrete times $t_n=n \Delta t$.\marginnote{This is sometimes called ``stroboscopic evolution" in that we are not interested in the dynamics at any possible instant but only at regular intervals of duration $\Delta t$.} We will show shortly (\cref{section-lind}) that these equations are in the so called Lindblad form (see \ref{app-lind}).

\section{Equations of motion for small collision time: expectation values}
\label{section-EVs}

While all the above equations of motion describe the evolution of states, one may be interested in the evolution of the expectation value of a given observable $\hat O$, denoted as $\langle \hat O\rangle_{n}={\rm Tr}_{SB}\{\hat O\sigma_{n}\}$ (at this stage we allow $\hat O$ to generally act on the joined system, \ie $S$ plus all the ancillas).
The general change of the expectation value $\Delta \langle \hat O\rangle_{n}{=}\langle \hat O\rangle_{n}- \langle \hat O\rangle_{n-1}$ at each time step reads
\begin{equation}
	\Delta \langle \hat O\rangle_{n}={\rm Tr}_{SB}\{ \Delta \hat O_n\,\sigma_{n-1}\}+
	{\rm Tr}_{SB}\{  \hat O_n\,\Delta\sigma_n\}\,\label{dOn1}
\end{equation}
with $\Delta \hat O_n=\hat O_{n}-\hat O_{n-1}$.
The former and latter terms respectively describe the contribution due to an intrinsic time dependence of operator $\hat O$ (if any) and that due to the evolution of state $\sigma_n$. For a time-independent observable, only the second term can contribute.

Plugging \cref{central} in \ref{dOn1} and exploiting the cyclic property of trace, we find that the rate of change of $\langle \hat O\rangle_{n}$ is given by
\begin{equation}
	\frac{\Delta \langle \hat O\rangle_n}{\Delta t}=\Bigl\langle \frac{\de \hat O_n}{\Delta t}\Bigr\rangle+i\,\langle[\hat H_0 +\hat V_n,\hat O]\rangle+\Delta t\,\langle{\hat V}_{n}\,\hat O \,{\hat V}_{n}-\tfrac{1}{2}[{\hat V}_{n}^2,\hat O]_+\rangle\label{centralO}\,,   
\end{equation}
where on the right hand side $\langle\ldots \rangle={\rm Tr}_{SB}\{\ldots \,\sigma_{n-1}\}$.

Alternatively, expressing $\de \sigma_n$ in the form \ref{central2}, we get
\begin{equation}
	\frac{\Delta \langle \hat O\rangle_n}{\Delta t}=\Bigl\langle \frac{\de \hat O_n}{\Delta t}\Bigr\rangle+i \,\langle [\hat H_0+\hat V_n,\hat O]\rangle-\tfrac{1}{2}\de t\,\langle[\hat V_n,[\hat V_n,\hat O]]\rangle\label{centralx}\,.
\end{equation}

\section{Lindblad form}
\label{section-lind}

The initial ancilla's density operator state $\eta_n$ can be spectrally decomposed (see \ref{app-rho}) as
\begin{equation}
	\eta_n=\sum_k\,p_k\,\vert k\rangle _n\langle k\rvert  \label{SD-mt}\,.
\end{equation}
with $\sum_k p_k=1$. Replacing this in \ref{drho} yields
\begin{equation}
	\frac{\Delta \rho_n}{\Delta t}=-i \,[\hat H_S+{\rm Tr}_{n}\{\hat{V}_{n}\,\eta_n\},\rho_{n-1}]+\sum_{kk'} \left(\hat{L}_{kk'}\rho_{n-1}\hat{L}_{kk'}^\dag-\tfrac{1}{2}[\hat L_{kk'}^\dag\hat L_{kk'},\rho_{n-1}]_+\right)\,\label{drho-lind}
\end{equation}
with jump operators $\hat{L}_{kk'}$ given by
\begin{equation}
	\label{Lkkp}
	\hat{L}_{kk'}=\sqrt{p_k}\,\,_n\langle k'\rvert  \hat{V}_{n}\vert k\rangle _n\,\sqrt{\Delta t}\,\,.   
\end{equation}
Here, $\vert k\rangle  $ and $\vert k'\rangle$ are eigenstates of $\eta$ [\cf\cref{SD}]. Note that operator ${\rm Tr}_{n}\{\hat{V}_{n}\,\eta_n\}$ appearing in the commutator is Hermitian. 

\cref{drho-lind} has the form of a discrete Lindblad master equation (see \ref{app-lind}). An analogous reasoning, this time based on the spectral decomposition of $\rho_{n-1}$, shows that \cref{deta} is also in Lindblad form. 

The Lindblad form essentially arises because both $S$ and the ancilla evolve at each step according to a CPT map that can be expanded in Kraus operators [see \cref{coll-map,ancilla-map}]. We stress that this crucially relies on the fact that $S$ and each ancilla are uncorrelated before colliding (their initial state $\rho_{n-1} \,\eta_n$ is factorized), which is guaranteed by assumptions (1)--(3) in  \ref{conds}.

\section{Reduced equations of motion in terms of moments}
\label{section-moments}

\cref{drho-lind} relies on the spectral decomposition \ref{SD} of the ancilla's state, whose calculation could be impractical in some cases. 
We derive next an alternative form of \cref{drho,deta} in terms of first and second moments of the bath/system operator entering the coupling Hamiltonian $\hat V_n$, which is both technically advantageous and conceptually important in that it pinpoints the essential quantities controlling the reduced dynamics of either subsystem.

The system--ancilla coupling Hamiltonian can be always decomposed as
\begin{equation}
	\hat V_{n}= \sum_\nu g_\nu \hat A_\nu  {\hat B}_{\nu} \,,\label{AB}
\end{equation}
with $g_\nu$ generally complex coefficients and $\hat A_\nu$(${\hat B}_{\nu}$) a set of (generally non-Hermitian) operators on $S$ (ancilla) subject to the constraint $\hat V_{n}=\hat V_{n}^\dag$ (index $n$ is omitted in ancilla operators).\marginnote{It can be shown that there always exists a decomposition such that $\hat A_\nu=\hat A_\nu^\dag$, $\hat B_\nu=\hat B_\nu^\dag$ and $g_\nu=g_\nu^{*}$. Yet, we prefer allowing for generally non-Hermitian operators since this is the natural form of  many usual interactions (\eg atom-field interactions, in which case $\hat A_\nu$ and $\hat B_\nu$ are ladder operators).}

Let us define first and second moments of $S$ and ancilla as
\begin{align}
	&\langle{\hat A_\nu}\rangle={\rm Tr}_n\{\hat A_\nu\rho_{n-1}\},\,\,&\langle\hat A_\nu\hat A_\mu\rangle={\rm Tr}_n\{\hat A_\nu\hat A_\mu\rho_{n-1}\},\nonumber\\
	&\langle{{\hat B}_\nu}\rangle={\rm Tr}_n\{{\hat B}_\nu\eta_n\},\,\,&\langle{\hat B}_\nu{\hat B}_\mu\rangle={\rm Tr}_n\{{\hat B}_\nu{\hat B}_\mu\eta_n\}\label{moments}\,.
\end{align}
Note that the moments of $\hat A_\nu$ are calculated on the current state of $S$, $\rho_{n-1}$, to be updated after each collision. Regardless, both moments of $S$ and ancilla have an intrinsic dependence on step $n$ when the CM is inhomogeneous (see  \cref{section-inhomCP}; \eg when ancillas are prepared in different states).

In terms of the moments just defined, the contributions to the discrete ME \ref{drho} can be decomposed as
\begin{align}\label{avs}
	&{\rm Tr}_{n}\{\hat{V}_{n}\,\eta_n\}=\sum_\nu  g_\nu \langle{{\hat B}_\nu}\rangle\hat A_\nu\,,\\
	&{\rm Tr}_{n}\{{\hat V}_{n}\rho_n\eta_n {\hat V}_{n}\}=\sum_{\nu\mu}g_{\nu} g_{\mu}\langle {\hat B}_\mu {\hat B}_\nu\rangle\,\hat A_\nu\rho_n\hat A_\mu\,,\nonumber \\
	&{\rm Tr}_{n}\{[\hat V_{n}^2,\rho_{n-1}\,\eta_n]_+ \}=\sum_{\nu\mu}g_{\nu} g_{\mu}\langle {\hat B}_\nu{\hat B}_\mu\rangle\,[\hat A_\nu A_\mu\,\rho_{n-1}]_+\,.\nonumber
\end{align}
Analogous expressions are worked out for \cref{deta} in terms of moments of ${\hat A}_\nu$'s calculated on state $\rho_{n-1}$.

To summarize, \eq \ref{drho} can be written as
\begin{equation}
	\frac{\Delta \rho_n}{\Delta t}=-i \,[\hat H_S+\hat H'_S,\rho_{n-1}]+ \,{\mathcal{D}}_S[\rho_{n-1}]\label{drho2}
\end{equation}
with
\begin{align}
	&\hat H'_S={\rm Tr}_{n}\{\hat{V}_{n}\eta_n\}= \sum_\nu\,g_\nu\, \langle{{\hat B}_\nu}\rangle  \hat A_\nu\,,\label{HS-diss}\\
	&{\mathcal{D}}_S[\rho_{n-1}]=\sum_{\nu\mu}\,\gamma_{\nu\mu}\langle {\hat B}_\mu {\hat B}_\nu\rangle(\hat A_\nu\rho_{n-1}\hat A_\mu-\tfrac{1}{2}[\hat A_\mu\hat A_\nu,\rho_{n-1}]_+)\,\nonumber,
\end{align}
while \cref{deta} as
\begin{equation}
	\frac{\Delta \eta_n}{\Delta t}=-i \,[\hat H_n+\hat H'_n,\eta_{n}]+ \,{\mathcal{D}}_n[\eta_n]\label{deta2}
\end{equation}
with
\begin{align}
	&\hat H'_n={\rm Tr}_{S}\{\hat{V}_{n}\rho_{n-1}\}= \sum_\nu\,g_\nu\, \langle{\hat A_\nu}\rangle  {\hat B}_\nu\,,\\
	&{\mathcal{D}}_n[\eta_n]=\sum_{\nu\mu}\,\gamma_{\nu\mu}\langle \hat A_\mu \hat A_\nu\rangle({\hat B}_\nu\rho_n{\hat B}_\mu-\tfrac{1}{2}[{\hat B}_\mu{\hat B}_\nu,\rho_n]_+)\,,\label{Hn-diss}
\end{align}
and where the rates appearing in the dissipators ${\mathcal{D}}_S$ and ${\mathcal{D}}_n$ are given by
\begin{equation}
	\gamma_{\nu\mu}=g_{\nu} g_{\mu} \,\Delta t\,\,.\label{rates}
\end{equation}
We see that the $S$-bath interaction brings about two main effects on the reduced dynamics. One is the appearance of an extra Hamiltonian term ($\hat H'_S$ and $\hat H'_n$) that adds to the free Hamiltonian ($\hat H_S$ and $\hat H_n$, respectively). Hamiltonian $\hat H'_S$, taken alone, would change the reduced dynamics of $S$ without yet affecting its \textit{unitary} nature, despite the $S$-bath coupling (and similarly $\hat H'_n$ with respect to ancilla $n$). The other effect, embodied by dissipator ${\mathcal{D}}_S$ (${\mathcal{D}}_n$), instead causes \textit{non-unitary} dynamics. 

Finally, note the explicit appearance of a $\Delta t$ factor in the rates \ref{rates}, which will be shown later to have consequences on the passage to the continuous-time limit.

\section{Equations of motion for expectation values in terms of moments}
\label{section-ev-moments}

In the (frequent) case of observables acting only on $S$ or ancilla, also the equations of motion in  \cref{section-EVs} can be simply decomposed in terms of simple moments.

For an operator on $S$, in  \cref{dOn1} $\sigma$ can be replaced with $\rho$ so that, in light of \cref{drho2,HS-diss}, we get
\begin{align} 
	\frac{\Delta \langle \hat O_S\rangle}{\Delta t}&=\Bigl\langle \frac{\de \hat O_S}{\Delta t}\Bigr\rangle+i \,\langle [\hat H_S,\hat O_S]\rangle+i\sum_\nu\,g_\nu \langle{{\hat B}_\nu}\rangle \,\langle [  \hat A_\nu,\hat O_S]\rangle\nonumber\\
	&+\sum_{\nu\mu}\,\gamma_{\nu\mu}\langle {\hat B}_\mu {\hat B}_\nu\rangle\langle\hat A_\mu\hat O_{S}\hat A_\nu-\tfrac{1}{2}\,[\hat A_\mu\hat A_\nu,\hat O_S]_+\rangle\,.\label{dOS}
\end{align} 
Likewise, in light of \cref{deta2,Hn-diss}, the expectation value of an operator on ancilla $n$ evolves at the $n$th step as\marginnote{This holds only at the $n$th step. At any other step, since ancilla $n$ does not change its state, ${\Delta \langle \hat O_n\rangle}/{\Delta t}=\langle {\de \hat O_n}/{\Delta t}\rangle$.}
\begin{align}
	\frac{\Delta \langle \hat O_n\rangle}{\Delta t}&=\Bigl\langle \frac{\de \hat O_n}{\Delta t}\Bigr\rangle+i \,\langle [\hat H_n,\hat O_{n}]\rangle+i\sum_\nu\,g_\nu \langle{\hat A_\nu}\rangle\langle [  {\hat B}_\nu,\hat O_{n}]\rangle\nonumber\\
	&+\sum_{\nu\mu}\,\gamma_{\nu\mu}\langle \hat A_\mu \hat A_\nu\rangle\langle{\hat B}_\mu\hat O_n{\hat B}_\nu-\tfrac{1}{2}\,[{\hat B}_\mu{\hat B}_\nu,\hat O_n]_+\rangle\,.\label{dOn}
\end{align}
On the right hand sides of \cref{dOS,dOn}, expectation values of operators on $S$ are computed on state $\rho_{n-1}$ and those on the ancilla on $\eta_n$.
Note that here subscript $n$ must be intended as the ancilla index, not the time step. Accordingly,  the changes are understood as $\Delta \hat O_n=\hat O_n^{(n)}-\hat O_n^{(n-1)}$ and likewise for $\Delta \langle \hat O_n\rangle$, where each subscript denotes the time step.

\section{Continuous-time limit via coarse graining}
\label{sec-coarse2}

So far we have considered finite-difference equations of motion, which reflects a stroboscopic description of the dynamics at the discrete times $t_n=n\de t$ with $\de t$ \textit{short} enough that $\hat U_n$ can be replaced with its 2nd-order expansion in $\de t$. Clearly, if one observes the system evolution on a time scale much larger than $\de t$, then the dynamics will look like effectively time-continuous. 

This is illustrated in a simple case study in  \cref{fig-CTL}, where the open dynamics of the all-qubit CM of  \cref{section-all-qubit} is considered for $g_z=0$ with $S$ initially in state $\tfrac{1}{\sqrt{2}}(\ket{0}_S+\ket{1}_S)$ and each ancilla prepared in $\ket{0}_n$. 
Making $\de t$ too large (compared to $g^{-1}$) results in a generally abrupt change of the state of $S$ after each time step, which rules out a continuous interpolation [see  \ref{fig-CTL}(a)]. 
This change is instead negligible by setting a small collision time $\de t$ in a way that, for evolution times much longer than $\de t$, the dynamics will appear effectively continuous [see  \ref{fig-CTL}(b)].

Accordingly, if the collision time is small and for evolution times much larger than $\de t$, one can replace the elapsed time (after $n$ collisions) $t_n=n\de t$ with a continuous time variable, \ie $t_n\rightarrow t$, substituting at the same time the incremental ratio in \cref{drho2} with a continuous derivative,
\begin{equation}
	t_n=n\de t\rightarrow t\,,\,\,\,\,\,\frac{\de\rho_n}{\de t}\rightarrow \frac{d \rho}{dt}\,\,\,\,\,\,\,(\rm{ coarse graining})\,\,.\label{CG-op}
\end{equation}
Of course, all the discrete functions depending on the step number $n$ (such as $\rho_{n-1}$) become now continuous functions of time $t$.
This procedure is carried out after choosing a short enough but finite collision time $\de t$ (coarse graining time) which is then kept always fixed [which sets rates \ref{rates}]. This coarse graining procedure turns the finite-difference ME into a continuous-time ME. A prominent instance is the micromaser dynamics, which we will discuss in the next subsection.
In physical terms, the coarse-graining procedure means that we give up keeping track of the dynamics in fine detail (\ie on a time scale shorter than $\de t$) and are happy with a coarse description on a small but \textit{finite} time scale $\de t$.
   
We point out that different choices of $\de t$ (but still small) will result in generally \textit{different} rates \ref{rates}, hence the coarse-grained ME and associated dynamics are $\de t$-dependent. Notably, as rates \ref{rates} are proportional to $\de t$, if this is very short then the dissipator ${\mathcal{D}}_S$ will become in fact negligible with the only effect of the bath reducing to Hamiltonian $\hat H'_S$ [\cf\cref{drho2,HS-diss}]. In this extreme regime of ultra-short collision times, the open dynamics is thus unitary.

\begin{figure}[!h] 
	\raggedright
	\begin{floatrow}[1]
		\ffigbox[\FBwidth]{\caption[Continuous-time limit of the collisional dynamics]{\textit{Continuous-time limit of the collisional dynamics.} We consider the all-qubit CM in  \cref{section-all-qubit} for $g_z=0$ with the ancillas prepared in $\eta_n=\ket{0}_n\bra{0}$ and $S$ initially in $\ket{\psi_0}=\tfrac{1}{\sqrt{2}}(\ket{0}_S+\ket{1}_S)$ [thus $p=c=1/2$ according to \cref{qubit-state}]. The probability to find $S$ still in the initial state (survival probability) at the $n$th step is given by $\langle \psi_0\rvert  \rho_n\vert \psi_0\rangle =\tfrac{1}{2}(1{+}\cos ^{n}(g\Delta t))$ [\cf\cref{pncn}]. This is plotted in panel (a) for $g\Delta t=0.8\pi$, while in panel (b) we set $g\Delta t=10^{-1}$ (the inset shows the first 20 steps). Clearly, the dynamics cannot be approximated as continuous in the case (a) due to the generally non-negligible change of $\rho_{n}$ at each step, $\Delta\rho_n{=}\rho_n{-}\rho_{n-1}$. Note that setting an ultra-short collision time, \eg $g\de t=10^{-3}$ (not shown here), and keeping the same total simulated time $n_{\rm max}\de t$ as (a) or (b) would yield $\langle \psi_0\rvert  \rho_n\vert \psi_0\rangle \simeq 1$.}\label{fig-CTL}}%
		{\includegraphics[width=\textwidth]{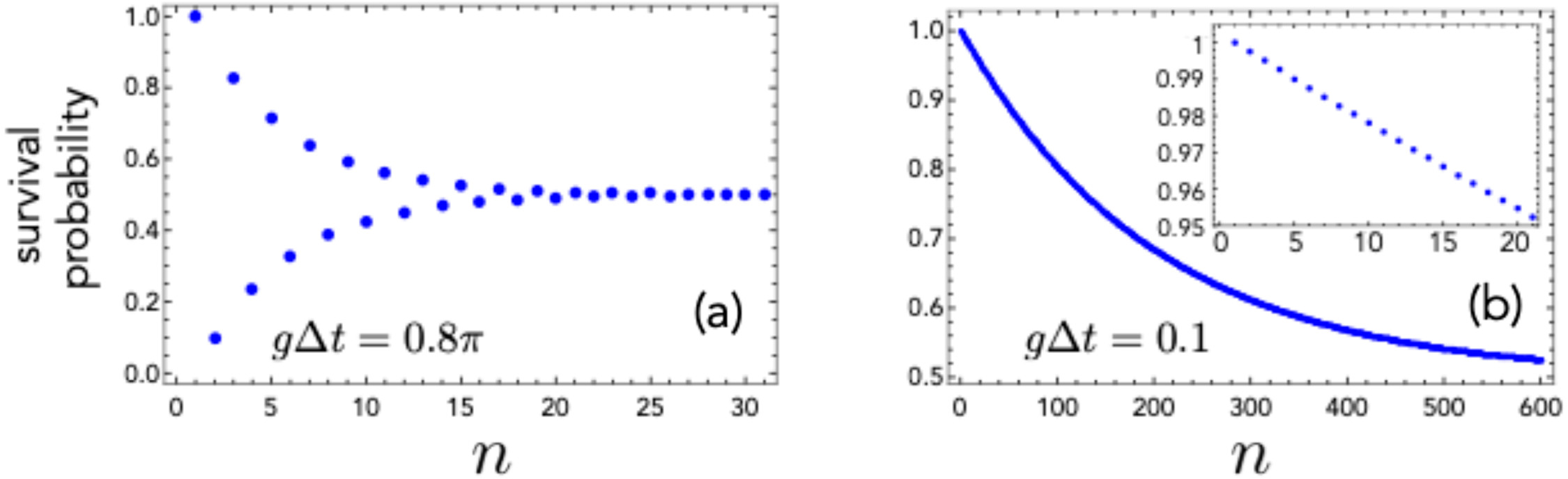}}
	\end{floatrow}
\end{figure}

\section{Micromaser}
\label{section-maser}

The micromaser~\cite{haroche_exploring_2006} is a system of utmost importance in
CMs theory as it is an experimental setup whose dynamics is, in fact by definition,
described by a CM. 
The paradigm of micromaser features a lossy cavity pumped by a beam of atoms which drive the cavity field into a lasing-like state.\footnote{This is the reason for the name ``micromaser", where ``maser" is intended as ``microwave laser" since the cavity frequency is in the range of microwaves.} More specifically, as sketched in  \ref{fig-maser}(a), a flux of Rydberg atoms ejected from an oven is directed through a velocity selector toward a high-finesse cavity where the atoms interact resonantly with a single normal mode of the cavity (the interaction with the other modes is off-resonant and thus can be neglected). In the ideal model, the atomic beam is monochromatic (fixed velocity) and the rate of injection $r$ is low enough that the atoms cross the cavity one by one (i.e.~there are never two atoms in the cavity at the same time). 

We have therefore a CM dynamics with $S$ embodied by the cavity mode and  ancillas by the flying atoms. In realistic conditions, atoms can be assumed as non-interacting and initially uncorrelated with each other, hence assumptions (1)--(3) in  \ref{conds} are all satisfied meaning that the dynamics is described by a basic Markovian CM.
For simplicity, we will neglect the cavity loss so that the atomic beam is the only environment driving the cavity open dynamics.
The interaction between the $n$th atom and the cavity mode is
well-described by the Jaynes and Cummings (JC) model~\cite{haroche_exploring_2006} in
which a two-level atom (qubit) with ground state $\ket{0}_n$, excited state
$\ket{1}_n$ and energy spacing $\omega_0$ [see  \ref{fig-maser}(c)] couples to a
cavity mode of frequency $\omega_c$.

\begin{figure}[!h] 
	\raggedright
	\begin{floatrow}[1]
		\ffigbox[\FBwidth]{\caption[Micromaser]{\textit{Micromaser.} (a): Basic micromaser setup. Atoms are heated in an oven (on the left). As atoms are ejected from the oven, a velocity selector filters only those of desired velocity $v$. Each selected atom then travels at speed $v$ towards the cavity (of length $L$) until it crosses it. (b): Characteristic times. If $L$ is the cavity length, each atom interacts with the cavity mode for a time $\tau=v/L$. Since $\tau\le \Delta t$, where $\Delta t$ is the time between two consecutive atomic injections, there are never two atoms in the cavity at the same time meaning that the dynamics is naturally described by a basic CM (atoms interact with the cavity mode one at a time). In the interaction picture, during the interval $[t_{n-1}+\tau,t_n]$ when the $n$th atom is out of the cavity, the system does not change its state. (c): Atomic and cavity-mode levels involved the interaction.}\label{fig-maser}}%
		{\includegraphics[width=\textwidth]{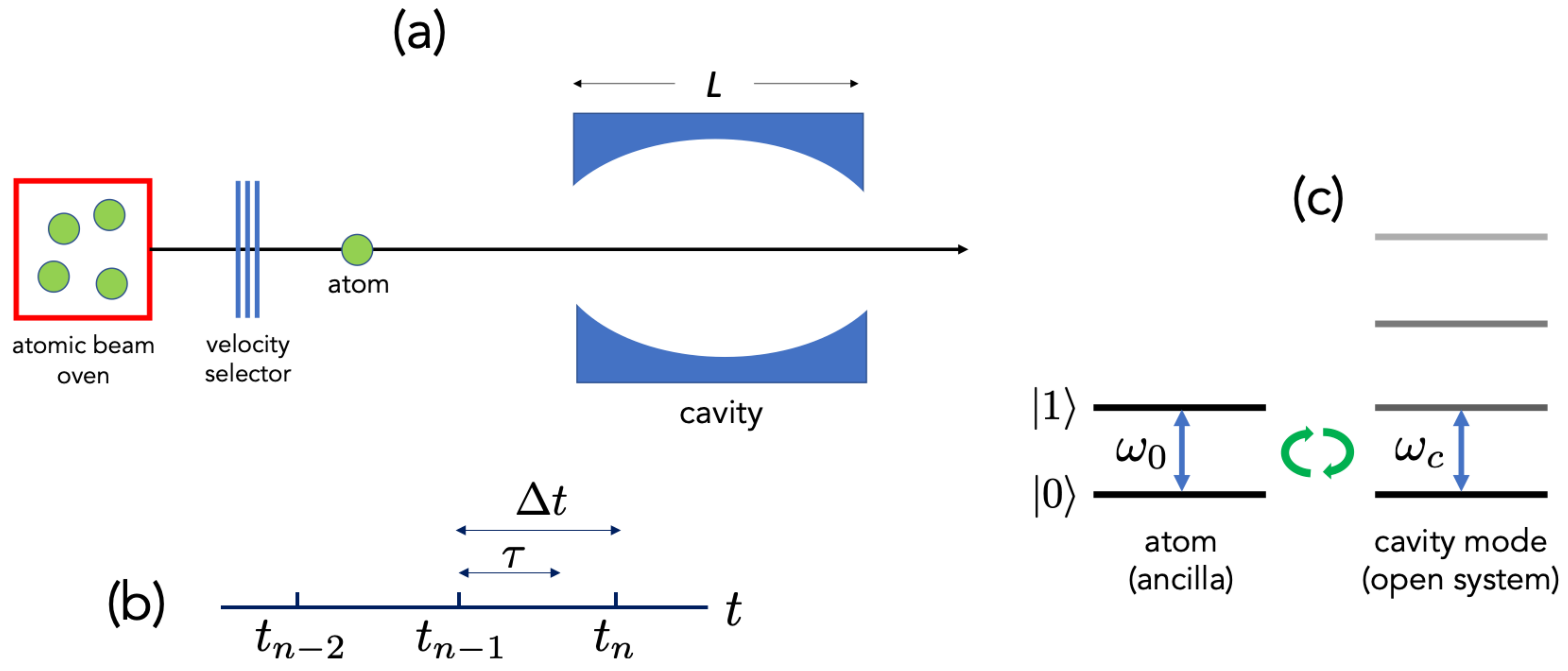}}
	\end{floatrow}
\end{figure}

On resonance ($\omega_a = \omega_c$), the JC
Hamiltonian reads $\hat H_{\rm JC}= \hat H_S+\hat H_n+ \hat V_n$ with
\begin{equation}
	\hat H_S=\omega_c \hat{a}^\dagger\hat{a}\,,\qquad\hat H_n= \omega_c\hat\sigma_{n+}\hat\sigma_{n-}\,,\qquad\hat V_n= g \left(\hat a\hat \sigma_{n+} +\hat a^\dag\hat\sigma_{n-}\right)\label{HJC}\,,
\end{equation}
where $\hat a$ and $\hat a^\dagger$ are bosonic annihilation and creation operators of the mode such that $[\hat a,\hat{a}^\dag]=1$ while (as usual) $\hat\sigma_{n-}=\hat\sigma_{n+}^\dag=\vert 0\rangle _n\langle 1\rvert  $ are pseudo-spin operators of the $n$th atom.

It is convenient to move to the \textit{interaction picture} with respect to the free Hamiltonian $\hat H_0=\hat H_S+\hat H_n$. Accordingly, the field and atomic operators are transformed as $\hat a \rightarrow \hat a e^{-i \omega_c t}$, $\hat \sigma_{n\pm}\rightarrow \hat \sigma_{n\pm} e^{\pm i \omega_c t}$ in a way that the coupling Hamiltonian $\hat V_n$ is unaffected. We note that expansion \ref{USn2} trivially holds here simply because the free Hamiltonians of $S$ and $n$ are zero in the interaction picture.

For the sake of argument, let us assume a constant atomic injection rate $r = 1/\Delta t$. Here, $\Delta t$ is the time elapsed between two consecutive injections in terms of which we discretize time as $t_n=n \Delta t$, hence $\Delta t$ embodies the CM time step [see  \ref{fig-maser}(b)].

It can be shown that the collision unitary at each step [\cf\cref{USn1}] takes
the form~\cite{englert2002five}
\begin{equation}
	{\hat U}_n {=} \exp\left[-i g\tau \left(\hat a\,\hat \sigma_{n+} {+}\hat a^\dag\hat\sigma_{n-}\right)\right] =\hat{\mathcal{C}} \ket{1}_n\langle {1}\rvert  +\hat{\mathcal{C}'} \ket{0}_n\langle {0}\rvert   -i\,(\hat{\mathcal{S}} \hat{a} \,\hat\sigma_{n+}+\hat{a}^\dagger\hat{\mathcal{S}}  \,\hat\sigma_{n-})\nonumber
\end{equation}
where for convenience we defined the nonlinear field operators
\begin{equation}
	\label{CCS}
	\hat{\mathcal{C}} = \cos \left(g\tau\sqrt{{\hat n} + 1}\right)\,, \hspace{0.7cm} 
	\hat{\mathcal{C}'} = \cos \left(g\tau\sqrt{{\hat n}}\right)\,, \hspace{0.7cm} 
	\hat{\mathcal{S}} = \frac{\sin \left(g\tau\sqrt{{\hat n} + 1}\right)}{\sqrt{{\hat n} + 1}}\,\,.\nonumber
\end{equation}
Here, $\tau$ is the time spent by each atom inside the cavity which is generally shorter than the injection time $\Delta t$ [see  \ref{fig-maser}(b)].

Let the atoms be prepared each in the same incoherent superposition of ground and excited states
\begin{equation}
	\eta_n =(1-p)\, \vert 0\rangle _n\langle 0\rvert  +p\, \vert 1\rangle _n\langle 1\rvert  \label{eta-maser}
\end{equation}
with $p$ a probability. Then the collision map, which fully describes the cavity open dynamics [\cf\cref{coll-map,eps}], is given by
\begin{align}
	\rho_{n}=&{\mathcal{E}}[\rho_{n-1}]=  {\rm Tr}_n\, \{{\hat U}_n\,\rho_{n-1}\,\eta\,{\hat U}_n^{\dagger}\}= \nonumber\\
	&(1-p) \,(\hat{\mathcal{C}'}\rho_{n-1}\,\hat{\mathcal{C}'} + \hat{\mathcal{S}}\,\hat{a}\,\rho_{n-1}\,\hat{a}^\dagger\hat{\mathcal{S}})+p \, (\hat{\mathcal{C}}\,\rho_{n-1} \,\hat{\mathcal{C}} + \hat{a}^\dagger\hat{\mathcal{S}}\rho_{n-1} \,\hat{\mathcal{S}}\,\hat{a} )\nonumber\,\\
\end{align}
with ${\rm Tr}_n$ the trace over the $n$th flying atom.

\subsection{Master equation of micromaser}

We note that, in the interaction picture, $\hat H_S=\hat H_n=0$ while $\hat V_n$ is just the same as in the Schr\"odinger picture (thus time-independent). 

Using \cref{HJC}, index $\nu$ in the expansion \ref{AB} here  takes values $\nu=\pm$ while $\hat A_{-}=\hat A_+^\dag=\hat a$, ${\hat B}_{-}={\hat B}_+^\dag=\hat \sigma_{n-}$ and $g_\pm=g$. In light of \cref{HS-diss} and given the initial state \ref{eta-maser}, the only non-zero moments of ancilla (\ie atomic) operators entering the finite-difference ME \ref{drho2} are $\langle \hat 0\rvert  \hat\sigma_- \hat \sigma_+\vert 0\rangle =\langle \hat 1\rvert  \hat\sigma_+ \hat \sigma_-\vert 1\rangle =1$ (the first-order Hamiltonian $\hat H'_S$ is zero since first moments vanish). Taking next the \textit{coarse-grained} continuous-time limit [\cf\cref{CG-op}], one finds the ME [\cf\cref{drho2}]
\begin{equation}
	\label{micro-ME}
	\dot \rho = (1-p)\,\Gamma\,\left(\hat{a}\rho\hat{a}^{\dagger}-\tfrac{1}{2}[\hat{a}^{\dagger }\hat{a},\rho]\right) + 
	p\, \Gamma\, \left(\hat{a}^{\dagger}\rho\hat{a}-\tfrac{1}{2}[\hat{a}\hat{a}^{\dagger },\rho]\right)\,\,,
\end{equation}
where we defined the rate\marginnote{To achieve this, a slight generalization of  \cref{section-EMs} is required since the injection time $\Delta t$ (time step) here can be generally larger than the collision time $\tau$. This leads to \ref{HS-diss} but with rates $\gamma_{\nu\mu}$ redefined as $\gamma_{\nu\mu}=g_{\nu} g_{\mu} \,\tau^2/\Delta t$.} 
$\Gamma=g^2\tfrac{\tau^2}{\Delta t}$. For $\tau=\de t$, this reduces to the simpler expression $\Gamma=g^2 \de t$.

Eq.~\ref{micro-ME} shows that atoms in the excited state $\ket1$ act as an incoherent pump (gain) on the cavity mode (corresponding to jump operator $\hat a^\dag$), while atoms in the ground state (jump operator $\hat a$) deplete the cavity (loss).

We note that a full micromaser description must account for fluctuations affecting the injection rate and, notably, cavity damping between atomic transits (neglected above). In such a case, we have an interesting example of a quantum system (cavity mode) in contact with \textit{two} baths, namely the atomic beam plus the external environment into which the cavity leaks out. 
Indeed, the cavity field steady state depends crucially on the balance between gain (due to the atomic pumping) and losses (due to cavity leakage). This leads to an extremely rich physics in the nonlinear strong-coupling regime $g\tau\gg1$,\marginnote{In this regime, operators \ref{CCS} entering $\hat U_n$ cannot be approximated as linear as done above.} where trapping states can arise. In general, the micromaser can produce non-classical light. 

\section{Continuous-time limit by introducing a diverging coupling strength}
\label{sec-div}

As discussed in  \cref{sec-coarse2}, the coarse-graining procedure returns a continuous-time ME with $\Delta t$-dependent rates [\cf\cref{rates}], where $\de t$ is small but finite. 

In some contexts, one may want to define a rigorous mathematical limit $\de t\rightarrow 0$ yielding a continuous-time ME where any dependence on $\de t$ is lost. Clearly, in order for this ME to feature a non-vanishing dissipator ${\mathcal{D}}_S$ (see final remarks of  \cref{sec-coarse2}), the price to pay is introducing $\de t$-dependent coupling strength(s) $g_\nu$. These must \textit{diverge} in such a way that rates $\gamma_{\mu\nu}$ (hence ${\mathcal{D}}_S$) keep finite [\cf\cref{rates}]. Yet, this may still be insufficient to get a well-defined continuous-time limit as illustrated by the next example.

Consider the all-qubit CM (\cf  \cref{section-all-qubit}) with $g_z=g$. Using \cref{Vn-qubits}, index $\nu$ in the expansion \ref{AB} here takes values $\nu=\pm,z$ while
$\hat A_{-}=\hat A_+^\dag=\sigma_-$, $\hat A_{z}=\hat \sigma_{z}$, ${\hat B}_{-}={\hat B}_+^\dag=\hat \sigma_{n-}$, ${\hat B}_{z}=\hat\sigma_{nz}$ and $g_\pm=g_z=g$ [\cf\cref{drho2,HS-diss}]. Since $\eta_n=\ket{0}_n\bra{0}$, the only non-zero moments of ancilla operators entering the finite-difference ME \ref{drho2} are $\langle 0\rvert   \hat \sigma_{nz}\vert 0\rangle =-1$ and $\langle \hat 0\rvert  \hat\sigma_{n-} \hat \sigma_{n+}\vert 0\rangle =1$. Hence, the first-order Hamiltonian and dissipator [\cf\cref{HS-diss}] explicitly read
\begin{align}
	\label{HS_CS}
	&\hat H'_S= -g\, \hat\sigma_z\,,\\
	&{\mathcal{D}}_S[\rho_n]=\gamma \left(\hat\sigma_{-}\,\rho_{n-1}\,\hat\sigma_{+}-\tfrac{1}{2}[\hat\sigma_{+}\hat\sigma_{-},\rho_{n-1}]_+\right)+\gamma \left(\hat\sigma_{z}\,\rho_{n-1}\,\hat\sigma_{z}-\rho_{n-1}\right)\,,\nonumber
\end{align}
where we set [\cf\cref{rates}]
\begin{equation}
	\gamma=g^2\Delta t\label{rate-ggg}\,.
\end{equation}
In order for the dissipator to survive the $\de t\rightarrow 0$ limit one can define a diverging coupling strength as
\begin{equation}
	g=\sqrt{\frac{\gamma}{\Delta t}}\,\,\,\,\,\,\,(\rm{diverging coupling strength})\label{gsqrt}\,\,.
\end{equation}
Such a scaling $\sim 1/\sqrt{\de t}$ of the coupling rate is a distinctive feature of many quantum CMs. 

However, while \cref{gsqrt} fixes the issue of the vanishing dissipator, it has a potential drawback. Indeed, as the coupling strength is also the characteristic rate of the 1st-order Hamiltonian $\hat H_S'$ [\cf \cref{HS_CS}], its divergence may cause $\hat H_S'$ to diverge as well for $\Delta t\rightarrow0$. 

Thereby, in general, in cases such as the present instance the introduction of a diverging coupling strength does not allow to perform a well-defined continuous-time limit fulfilling the double constraint that the dissipator \textit{and} $\hat H_S'$ must remain finite. Whether or not such a problem arises depends on the system--ancilla coupling Hamiltonian $\hat V_n$ as well as the initial ancilla's state. For instance, if in the considered example we set $g_z=0$ and $g=\sqrt{\gamma/{\Delta t}}$ [\cf\cref{Vn-qubits}] then of course $\hat H'_S=0$ for any $\Delta t$. Thus, in the limit $\Delta t\rightarrow 0$, the finite-difference Eq.~\ref{drho2} is turned into the well-defined continuous-time Lindblad ME
\begin{equation}
	\dot\rho=\gamma \left(\hat\sigma_{-}\,\rho\,\hat\sigma_{+}-\tfrac{1}{2}\,[\hat\sigma_{+}\hat\sigma_{-},\rho]_+\right)\,,\label{ME-SE}
\end{equation}
which is identical to the well-known ME describing \textit{spontaneous emission} of a two-level atom.\marginnote{Indeed, it is easily checked that, if $p=1$ and $c=0$ [\cf\cref{qubit-state}], then \cref{ME-SE} entails $p(t)=e^{-\gamma t}$, $c(t)=0$ namely the (initially excited) atom decays to the ground state with emission rate $\gamma$.} This is not accidental: in  \cref{section-CMQO}, we will show that the all-qubit CM with the diverging coupling strength \ref{gsqrt} (leading to this ME) can be directly derived from a microscopic atom-field model (see in particular  \cref{section-vacuum} discussing the field vacuum state).


As anticipated, however, also the initial state of ancillas matters. 
For instance, considering the above example for $g=\sqrt{\gamma/{\Delta t}}$ and $g_z=0$ but with the ancillas now initially in $\eta_n=\ket{+}_n\bra{+}$ will result again (for $\Delta t\rightarrow 0$) in a diverging Hamiltonian in this case given by $\hat H'_S=\sqrt{\gamma/{\Delta t}}\,\hat \sigma_x$.

It is natural to ask whether ensuring that $\hat H'_S=0$ is the only way for $\hat H'_S$ not to diverge (for $\Delta t\rightarrow 0$) due to \ref{gsqrt}. We show next that both $\hat H'_S$ and the dissipator can remain finite if one allows for the ancilla's state itself to depend on $\Delta t$. As a representative example in the all-qubit CM,  consider the initial ancilla's state $\eta_n=\ket{\chi}_n\bra{\chi}$ with
\begin{equation}
	\ket{\chi}_n=\frac{1}{1+\vert \alpha_n\vert ^2\Delta t}\left(\vert 0\rangle_n+\alpha_n \sqrt{\Delta t}\,\vert 1\rangle _n\right)\,,\label{superp}
\end{equation}
where $\alpha_n$ is generally complex. Setting again $g=\sqrt{\gamma/{\Delta t}}$ and $g_z=0$, the only non-zero ancilla moments [\cf\cref{moments}] in this case are $\langle\hat\sigma_{n-}\rangle=\langle\hat\sigma_{n+}\rangle^{*}=\alpha_n\sqrt{\Delta t}$ and $\langle \hat\sigma_{n-} \hat \sigma_{n+}\rangle=1$, where we neglected terms of order $\Delta t$ or higher. These entail the 1st-order Hamiltonian and dissipator
\begin{equation}
	\label{Hdiss-super}
	\hat H'_S= g\,(\alpha_n\hat\sigma_{-}+\alpha_n^{*}\hat\sigma_{+})\,,\,\,\,{\mathcal{D}}_S[\rho_n]=\gamma \left(\hat\sigma_{-}\,\rho_{n-1}\,\hat\sigma_{+}-\tfrac{1}{2}[\hat\sigma_{+}\hat\sigma_{-},\rho_{n-1}]_+\right)\,\,.
\end{equation}
Neither $\hat H'_S$ nor ${\mathcal{D}}_S$ depends on $\Delta t$, hence both remain finite for $\de t\rightarrow 0$. This happens because the $\sqrt{\Delta t}$ on the denominator of the coupling strength is canceled by that coming from the initial state with the latter not affecting the dissipator to leading order. 

In the case $\alpha_n=A\, e^{-i \omega_L t_n}$ with $A>0$, by taking the continuous-time limit of \ref{Hdiss-super} we get the ME
\begin{equation}
	\dot\rho=-i \left[\,g\,A\, (e^{-i\omega_L t}\hat\sigma_{-}+e^{i\omega_L t}\hat\sigma_{+}),\rho\right]+\gamma \left(\hat\sigma_{-}\,\rho\,\hat\sigma_{+}-\tfrac{1}{2}\,[\hat\sigma_{+}\hat\sigma_{-},\rho]_+\right)\,.\label{ME-BLOCH}
\end{equation}
This generalizes \ref{ME-SE} to the case where a driving Hamiltonian is added.
This ME is equivalent to the well-known optical Bloch equations describing the
evolution of an atom driven by a classical oscillating field while undergoing
spontaneous emission at the same time~\cite{meystre2007elements}. 

The assumption that we made of having a $\Delta t$-dependent ancilla state may appear somewhat artificial. In  \cref{section-CMQO}, we will show in detail that state \ref{superp} arises from an initial \textit{coherent} state of the electromagnetic field.

Before concluding the discussion on the continuous-time limit, it is worth noting that a diverging coupling strength [\cf\cref{gsqrt}] allows the condition underlying expansion \ref{USn2} (\ie $\hat H_S$, $\hat H_n$ much weaker than $\hat V_n$) to be satisfied for $\de t$ short enough.

In the following subsections, we will consider equations of motion for two important collisional dynamics: multiple baths and cascaded CMs.   

\section{Multiple baths}
\label{section-multiB}

In many realistic problems, the open system is in contact with many baths at once. Accordingly, it is useful to define CMs where $S$ collides with $M\ge 1$ baths of ancillas, as shown in  \cref{fig-BB}(a) for the case of two baths ($M=2$). At each step, $S$ collides with $M$ ancillas, one for each bath $i=1, \ldots ,M$. To make contact with previous theory, it is convenient to view the CM as featuring a single bath of $M$\textit{-partite} ancillas, each initially in state
\begin{equation}
	\eta_n=\eta_n^{(1)}\otimes\eta_n^{(2)}\otimes\cdots\otimes \eta_n^{(M)}+\chi_n^{({\rm corr})}\,.\label{chi-corr}
\end{equation}

\begin{figure}[!h] 
	\raggedright
	\begin{floatrow}[1]
		\ffigbox[\FBwidth]{\caption[Collision model with two baths of ancillas]{\textit{Collision model with two baths of ancillas}. (a): System $S$ collides with two baths of ancillas, labeled with 1 and 2. This CM can be formally seen  as basic CM [see  \ref{fig-CMdyn}] where each ancilla is bipartite and initially in state $\eta^{(1)}\otimes\eta^{(2)}+\chi^{({\rm corr})}$ (in the panel $\chi^{({\rm corr})}=0$). (b): Same as (a) except that now system $S$ is itself bipartite, comprising subsystems $S_1$ and $S_2$. Collisions with ancillas of bath $i$ involve only subsystem $S_i$.}\label{fig-BB}}%
		{\includegraphics[width=\textwidth]{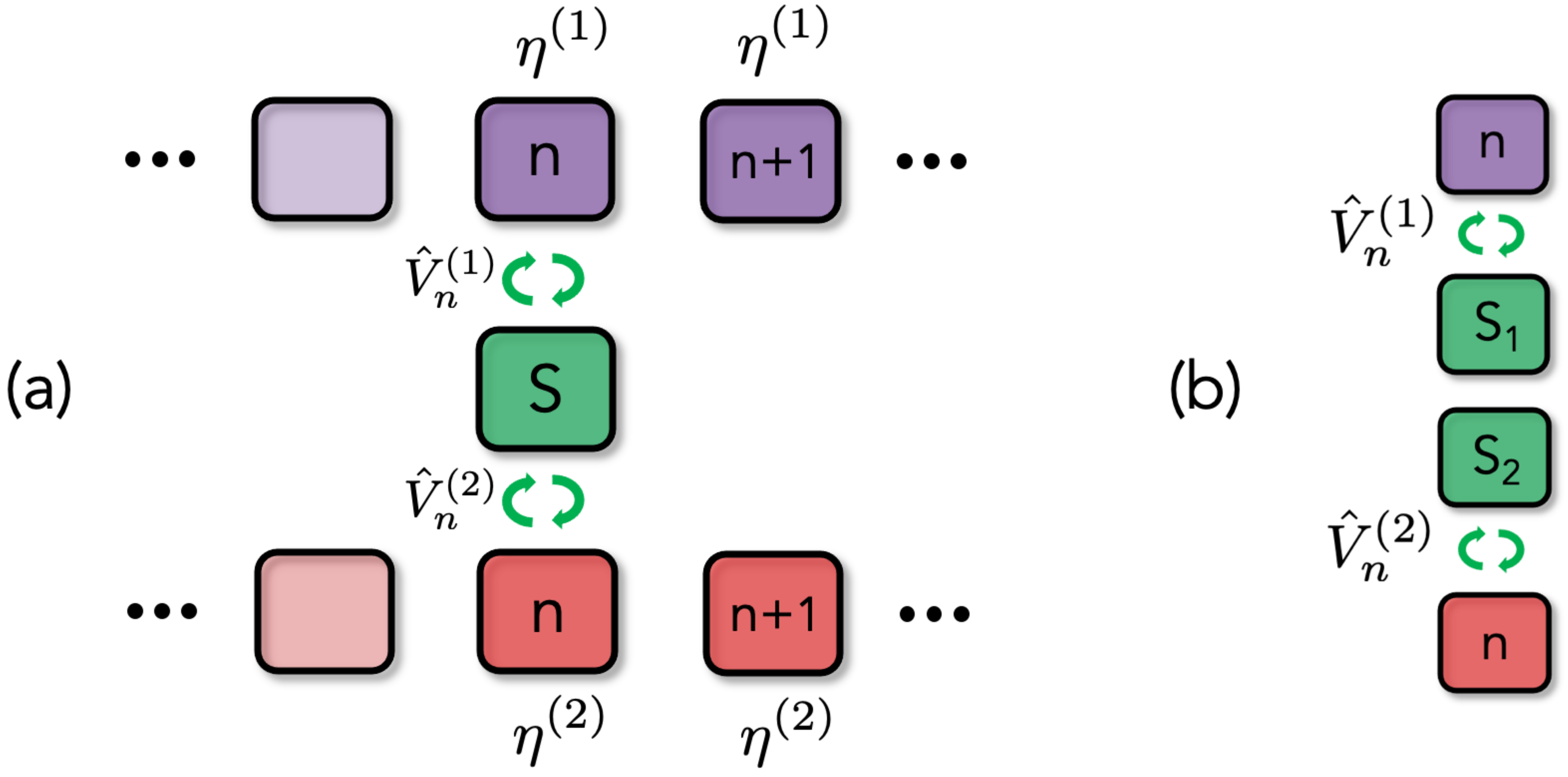}}
	\end{floatrow}
\end{figure}

Here, $\eta_n^{(i)}$ is the reduced state of ancilla of bath $i=1, \ldots ,M$. Note we allowed ancillas of different baths to share initial correlations described by term $\chi_n^{({\rm corr})}$. Thus when the $M$ baths are uncorrelated, $\chi_n^{({\rm corr})}=0$.   
The interaction Hamiltonian reads
\begin{equation}
	\hat V_n=\hat V_{n}^{(1)}+\hat V_{n}^{(2)}+\cdots+\hat V_{n}^{(M)}\,\,\,\,\,{\rm with}\,\,\,\hat V_{n}^{(i)}=\sum_{\nu} g_{\nu i}\hat A_{\nu i}\hat B_{\nu i} \label{Vn-many}\,\,,
\end{equation}
where as usual we expanded each $\hat V_n^{(i)}$ (coupling Hamiltonian between $S$ and an ancilla of bath $i$) in the form \ref{AB}. Here, $\hat B_{\nu i}$ is an operator acting on the ancilla of bath $i$ while $\hat A_{\nu i}$ is an operator of $S$ which we allow to be generally $i$-dependent.
Note that $\hat V_n$ can be written as
\begin{equation}
	\hat V_n=\sum_{i,\nu}  g_{\nu i} \hat A_{\nu i}\hat B_{\nu i} \label{Vn-many2}\,\,.
\end{equation}
This is still of the form \ref{AB} with the role of index $\nu$ now embodied by the double index $(\nu,i)$, hence all the theory in  \cref{section-moments,section-ev-moments} applies with the replacements $\nu\rightarrow( \nu,i)$, $\mu\rightarrow (\mu,j)$.

For \textit{uncorrelated} baths, \ie $\chi_n^{({\rm corr})}=0$ [\cf\cref{chi-corr}], all crossed second moments of the bath factorize as
\begin{equation}
	\langle {\hat B}_{\mu j} {\hat B}_{\nu i}\rangle= \langle {\hat B}_{\mu j} \rangle \langle{\hat B}_{\nu i}\rangle\,\,\,\,\,\,\,\,\,\,{\rm for}\,\,\,i\neq j\label{factor}
\end{equation}
with $\langle {\hat B}_{\nu i} \rangle={\rm Tr}_i \{\hat B_{\nu i}\eta_i\}$. This implies that when all the first moments vanish, \ie $\langle {\hat B}_{\nu i} \rangle=0$ for any $\mu$ and $i$, so do all the crossed second moments. In this case, based on \cref{drho2,HS-diss}, we get that
\begin{equation}
	\frac{\de \rho_n}{\de t}=\sum_{i=1}^M{\mathcal{D}}_S^{(i)}[\rho_{n-1}] \label{addi}
\end{equation}
with ${\mathcal{D}}_S^{(i)}$ the dissipator that would arise if $S$ were in contact only with bath $i$. We can thus say that the dissipative effects of uncorrelated baths are \textit{additive}. We point out that this holds as well (for $\de t$ short enough) when $\langle {\hat B}_{\nu i} \rangle\propto \sqrt{\de t}$ since in such a case \ref{factor} can be neglected. This can happen with states like \ref{superp} as we discussed in  \cref{sec-div}.

For \textit{correlated} baths, namely $\chi^{\rm (corr)}\neq 0$ [\cf\cref{chi-corr}], crossed second moments are generally non-zero. An interesting consequence of this occurs when $S$ itself is made out of $M$ subsystems $S_1$, \ldots, $S_M$ such that the collisions with ancillas of the $i$th bath involve only subsystem $S_i$ [see  \cref{fig-BB}(b)]. In this case, therefore, 
operator $\hat A_{\nu i}$ in \cref{Vn-many2} acts only on $S_i$. Then, based on \ref{HS-diss}, we see that the dissipator entering the ME will in particular contain terms of the form
\begin{equation}
	\propto \langle {\hat B}_{\mu j} {\hat B}_{\nu i}\rangle(\hat A_{\nu i}\rho_{n-1}\hat A_{\mu j}-\tfrac{1}{2}[\hat A_{\mu j}\hat A_{\nu i},\rho_{n-1}]_+)\,\,\,\,{\rm for}\,\,\,i\neq j\,\,.\label{jump-corr}
\end{equation}
These represent incoherent interactions between subsystems $S_i$ and $S_j$ mediated by the baths. Thus, correlations between the baths enable the establishment of correlations between the subsystems of $S$ even if these are not directly coupled.

\section{Cascaded master equation}
\label{sec-cascME}

As another important instance, we next derive the ME of the cascaded CM of  \cref{section-cascaded}. Recall that the collision unitary is given by [\cf\cref{Un-casc}] $\hat U_n=\hat U_{2,n}\hat U_{1,n}$ with $\hat U_{j,n}$ describing the sub-collision with subsystem $S_j$ (see  \ref{fig-chiral}). Equivalently, one can think of a single collision with a \textit{time-dependent} interaction that reads
\begin{equation}
	\label{V1V2}
	\hat V_n(t)=\begin{cases}
	\hat V_{1,n}& \quad t \in \left[t_{n-1}, t_{n-1}+\Delta t/2\right[\\
	\hat V_{2,n}& \quad t \in \left[t_{n-1}+\Delta t/2,t_n\right[\end{cases}
\end{equation} 
with $\hat V_{j,n}$ the interaction Hamiltonian between $n$ and $S_j$ such that $\hat U_{j,n}=e^{-i\hat V_{j,n}\frac{\Delta t}{2}}$. Thus $\hat V_n$ suddenly switches from $\hat V_{1n}$ to $\hat V_{2n}$ after the first subcollision.

The framework that we developed previously (in particular  \cref{section-EMs,section-moments}) holds for a time-independent $\hat V_n$, hence it cannot be directly applied for deriving the ME. We thus start over by expanding each subcollision unitary $\hat {U}_{jn}$ to the second order in ${\Delta t}/{2}$, eventually discarding terms of order higher than $\sim\Delta t^2$. This yields the overall collision unitary
\begin{equation}
	\hat U_n=\hat {U}_{2n}\hat {U}_{1n}\,\simeq\, \hat{\mathbb I}-i \left(\hat V_{1n}+\hat V_{2n}\right)\,\Delta t' -\left(\tfrac{1}{2}\,{\hat V}_{1n}^2+\tfrac{1}{2}\,{\hat V}_{2n}^2+\hat V_{2n}\hat V_{1n}\right)\,\Delta t^{\prime 2}.\label{U-casc}
\end{equation} 
with $\Delta t'=\Delta t /2$
Note that this is not invariant under the swap $1\leftrightarrow 2$, which is due to the intrinsic CM unidirectionality discussed in  \cref{section-cascaded}.
To gain a better physical insight, we note that \ref{U-casc} can be equivalently arranged as
\begin{equation}
	\hat {U}_{n}\,\simeq\, \hat{\mathbb I}-i \,\left(\hat V_{1n}+\hat V_{2n}+i\,\tfrac{\Delta t'}{2}\left[\hat V_{1n},\hat V_{2n}\right]\right)\,\Delta t' -\tfrac{1}{2}\left({\hat V}_{1n}+\hat V_{2n}\right)^2\,\Delta t^{\prime 2}\label{U-casc2}
\end{equation} 
Now observe that, if each ancilla collided with $S_1$ and $S_2$ \textit{at once} during the time $\Delta t'$, then one would get the usual collision unitary \ref{USn2} (for $\Delta t\rightarrow \Delta t'$) with the natural replacement $\hat V_n\rightarrow \hat V_{1n}+\hat V_{2n}$. This matches all the terms in \ref{U-casc2} but the \textit{unitary} contribution coming from the effective Hamiltonian $\hat {\mathcal{H}}_{Sn}$. Hence, the intrinsic system's unidirectionality, due to the fact that ancillas collide \textit{first} with $S_1$ and \textit{then} with $S_2$, is fully condensed in the appearance of the effective Hamiltonian $\hat {\mathcal{H}}_{Sn}$.\marginnote{Note that this is indeed the only term in \ref{U-casc2} which is not invariant under the exchange $S_1\leftrightarrow S_2$. Instead, it transforms as $\hat {\mathcal{H}}_{Sn}\rightarrow -\hat {\mathcal{H}}_{Sn}$}    
To work out the ensuing ME of $S$, let us expand $\hat V_{j,n}$ as $\hat V_{j,n}= \sum_{\nu} g_{\nu} \hat A_{j\nu} \, {\hat B}_{\nu}$ [\cf \cref{AB}].
Plugging this into \ref{U-casc2} and proceeding analogously to  \cref{section-EMs,section-moments}, we get the discrete ME
\begin{equation}
	\frac{\Delta \rho_n}{\Delta t}=-i \,[\hat H'_S+\hat H''_S,\rho_{n-1}]+ \,{\mathcal{D}}_S[\rho_{n-1}]\label{drho-chir}
\end{equation}
with
\begin{align}
	&\hat H'_S =\sum_\nu\,g'_\nu\, \langle{{\hat B}_\nu}\rangle  \hat A_{\nu}\,\nonumber\\
	&\hat H''_S={\rm Tr}_n \{\hat{\mathcal{H}}_{Sn}\eta_n\}=i\,\tfrac{\Delta t}{2}\sum_{\nu\mu}g'_{\nu} \,g'_{\mu}\,\bigl\langle[{\hat B}_{\nu} ,{\hat B}_{\mu}]\bigr\rangle\,\hat A_{2\mu}\hat A_{1\nu}\,  \label{H2}\\
	&{\mathcal{D}}_S[\rho_{n-1}]=\sum_{\nu\mu}\,g'_{\nu} g'_{\mu} \,\Delta t\langle {\hat B}_\mu {\hat B}_\nu\rangle\left(\hat A_{\nu}\rho_{n-1}\hat A_{\mu}-\tfrac{1}{2}\left[\hat A_{\mu}\hat A_{\nu},\rho_{n-1}\right]_+\right)\nonumber
\end{align}
where we set $g'_\nu=g_\nu/2$ and defined the collective operators $\hat A_{\nu}=\hat A_{1\nu}{+}\hat A_{2\nu}$.
Here, $\hat H'_S$ and ${\mathcal{D}}_S$ have the same form as \cref{HS-diss} with $\hat A_{\nu}$ now intended as collective operators. Notably, the second-order Hamiltonian $\hat {\mathcal{H}}_{Sn}$ upon partial trace results in an effective \textit{coherent} coupling between $S_1$ and $S_2$ (mediated by the ancillas) described by Hamiltonian $\hat{H}_S''$.
We point out that this is an effective \textit{second-order} Hamiltonian of $S$, in contrast to $\hat H'_S$ [this being the analogue of the Hamiltonian in \cref{HS-diss}], which in particular explains the notation we adopted.

As a significant example, let each ancilla be a qubit initially in state $\ket{0}_n$ with $\hat V_{j,n}$ of the form
\begin{equation}
	\hat V_{j,n}= \sqrt{\tfrac{2\gamma}{\Delta t}}\left(\hat A_j \,\hat \sigma_{n+}+\hat A_j^\dagger \,\hat\sigma_{n-}\right)\,.\label{Vjn}
\end{equation}
Then the only non-vanishing ancilla moments entering the ME are $\langle \hat\sigma_{n-}\hat\sigma_{n+} \rangle=1$. This yields
\begin{align}
	&\hat H'_S=0\nonumber\\
	&\hat H''_S=\tfrac{\gamma}{2}(i\hat A_2 \hat A_1^\dag-i\hat A_1 \hat A_2^\dag)\nonumber\\
	&{\mathcal{D}}_S[\rho_{n-1}]=\gamma \left(\hat A\,\rho_{n-1}\,\hat A^\dag-\tfrac{1}{2}\,[\hat A^\dag\hat A,\rho_{n-1}]_+\right)
\end{align}
with $\hat A=\hat A_1+\hat A_2$, hence in the CTL we end up with the ME
\begin{equation}
	\dot\rho=-i \left[\,\tfrac{\gamma}{2}(i\hat A_2 \hat A_1^\dag-i\hat A_1\hat A_2^\dag),\rho\right]+\gamma \left(\hat A\,\rho\,\hat A^\dag-\tfrac{1}{2}\left[\hat A^\dag\hat A,\rho\right]_+\right)\,.\label{ME-casc}
\end{equation}

\section{Equations of motion: state of the art}

Explicit derivations of the Lindblad master equation through the
continuous-time limit of a CM were given in \rref~\cite{brun_simple_2002,ziman_description_2005}. See
also \rref~\cite{ziman_all_2005} by the same authors of \rref\cite{ziman_description_2005}, which includes a general
characterization of decoherence channels of a qubit and their implementation via
suitably defined CMs. 

The dynamics most intuitively associated with a CM is arguably the dissipative
interaction of a system with a dilute \textit{gas} of particles (ancillas). In such a
case, the time between two next system--ancilla collisions is random, at variance with
the assumption of time-periodic collisions (one for each $\de t$) made in our
discussion. Yet, as shown in \rref~\cite{ScaraniPRE2019}, a Lindblad ME can be worked
out in this case as well even with strong collisions,\marginnote{For a strong collision,
	the collision unitary $\hat U_n$ cannot be approximated to the lowest orders.} the
associated rate $\gamma$ (entering the dissipator) being now the number of
collisions per unit time (similar CM and ME appeared in \rref~\cite{BrunnerNJP15}).
Note that, if the gas particles are quantum then a CM description relies on
approximating their motion as semiclassical. \rref~\cite{FilippovPRA20} showed that
this is equivalent to the low-density, fast-particle limit of a fully quantum
treatment.

The \textit{micromaser} theory (\cf  \cref{section-maser}) was first introduced by
Javanainen and
Meystre~\cite{filipowicz_theory_1986,filipowicz1986quantum,filipowicz1986microscopic}.
{Works that use explicitly the CM approach in particular for deriving the cavity
	field's master equation are \eg \rrefs~\cite{BergouPRA89,BriegelPRA95}}. An
introduction to micromaser can be found in the textbook by Meystre and
Sargent~\cite{meystre2007elements}. See also \rref~\cite{englert2002five}, which
includes the master equation.
Basics of cavity QED and JC model, which we referred to in  \cref{section-maser},
can be found \eg in the textbook by Haroche~\cite{haroche_exploring_2006}.   
Issues closely related to the continuous-time limit via diverging coupling strength
(see  \cref{sec-div}) were carefully investigated in
\rrefs~\cite{altamirano_unitarity_2017,grimmer_open_2016} (see also a
previous paper by Milburn~\cite{milburn2012decoherence}). Particular attention was
given to the regime of ultra-short collision times yielding a unitary dynamics (as we
discussed). This paradigm of unitary CM was proposed to carry out indirect quantum
control~\cite{layden_universal_2016} and universal two-qubit quantum gates in
spintronics systems~\cite{sutton2015manipulating}.

Cascaded master equations like \ref{ME-casc} were independently introduced in
1993 by Carmichael~\cite{CarmichaelPRL93} and Gardiner~\cite{GardinerPRLcascaded}
using the input--output formalism~\cite{meystre2007elements}. They were later derived
through a CM in \rrefs~\cite{giovannetti_master_2012,giovannetti_master_2012-1}
(introducing an internal bath dynamics as well)
although with a treatment somewhat different than the one in  \cref{sec-cascME}. 
Note that, for the sake of argument, we considered only a bipartite system
$S$. The generalization to more than two subsystems is straightforward,
leading to an interesting many-body Hamiltonian $\hat H''_S$. Multipartite cascaded
CMs can be advantageously applied to work out MEs of complex cascaded networks where
interference effects can
occur~\cite{cusumano_interferometric_2017,cusumano_interferometric_2018}.
From a more general perspective, cascaded systems are currently receiving large
attention in quantum optics also due to recent experimental realizations of chiral
emission (\eg in photonic crystals or fibers)~\cite{LodahlReviewNature17}.

\setchapterpreamble[u]{\margintoc}
\chapter{Quantum trajectories}
\label{section-qtraj}

The possibility of interpreting the Lindblad master equation as the result of an
ensemble average of different stochastic quantum trajectories, each corresponding to
a particular sequence of measurement outcomes on the environment, is a pillar of open
quantum systems dynamics with important applications in various fields such as
quantum optics and quantum
transport~\cite{breuer2007,haroche_exploring_2006,wiseman2009quantum,jacobs2014quantum,carmichael2009open}. 

Quantum trajectories emerge very naturally from a CM as soon as one imagines to measure each ancilla right after its collision with $S$. This and related concepts are the subject of the present section.

\section{Collision model unraveling}
\label{section-unrav}

Let us come back to the basic CM in  \cref{section-def} and assume for the sake or argument that $S$ and ancillas are initially in the pure states $\ket{\psi_0}$ and $\{\ket{\chi_n}\}$, respectively (thus $\eta_n=\ket{\chi_n}\bra{\chi_n}$). Accordingly, the initial joint state is $\sigma_{0}=\ket{\Psi_0}\bra{\Psi_0}$ with $\ket{\Psi_0}=\ket{\psi_0}\otimes_n \ket{\chi_n}$.\footnote{In the present section, we use a compact notation such that $\ket{\chi_n}=\ket{\chi_n}_n$ (and similarly for bras). This convention simplifies the formalism without affecting clarity.} At step $n$, this is turned into
\begin{equation}
	\ket{\Psi_n}=\hat{ U}_n\cdots\hat U_1\ket{\psi_0} \ket{\chi_1}\cdots\ket{\chi_n}\,.\label{psin}
\end{equation}
Let now $\{\ket{k_n}\}$ be  a single-ancilla orthonormal basis. Using the basis completeness, \eq\ref{psin} can be equivalently arranged by putting $\sum_{k_m}\ket{k_m}\bra{k_m}$ in front of each collision unitary $\hat U_m$ as
\begin{align}
	\label{psin-2}
	\ket{\Psi_n}&=\left(\sum_{k_n}\ket{k_n}\bra{k_n}\right)\hat{ U}_n\,
	 \ldots\,\left(\sum_{k_1}\ket{k_1}\bra{k_1}\right)\hat U_1\ket{\psi_0} \ket{\chi_1}\cdots\ket{\chi_n}\nonumber\\
	&=\sum_{k_n}
	\cdots\sum_{k_{1}}\,\,\,\left(\ket{k_n}\bra{k_n}\hat{ U}_n\right)
	\, \ldots\,\left(\ket{k_1}\bra{k_1}\hat U_1\right)\ket{\psi_0} \ket{\chi_1}\cdots\ket{\chi_n}\,.
\end{align}
Each ancilla state $\ket{\chi_m}$ can now be moved to the left and placed to the immediate right of the corresponding unitary $\hat U_m$, while kets $\ket{k_1}, \ldots ,\ket{k_n}$ can be moved to the right of $\ket{\psi_0}$. This allows to arrange $\ket{\Psi_n}$ as
\begin{equation}
	\ket{\Psi_n}=\sum_{k_n}
	\cdots\sum_{k_{1}}\,\,\bra{k_n}\hat{ U}_n\ket{\chi_n}\,
	 \ldots\,\bra{k_1}\hat U_1\ket{\chi_1}\,\,\ket{\psi_0}\, \ket{k_1}\cdots\ket{k_n}\,.\label{psin-3}
\end{equation}

\begin{figure}[!h] 
	\raggedright
	\begin{floatrow}[1]
		\ffigbox[\FBwidth]{\caption[Quantum trajectories in the all-qubit collision model]{\textit{Quantum trajectories in the all-qubit collision model}. Like the basic CM of  \ref{fig-CMdyn}, ancillas are prepared in $\otimes_n \ket{0_n}$ (thus uncorrelated) and $S$, initially in state $\ket{+}$, collides with each sequentially. After the collision [shown in panel (a)], the ancilla gets correlated with $S$ and (prior to the next collision) is measured in the basis $\{\ket{0_n},\ket{1_n}\}$ (b). If outcome 0 is recorded (c) no jump takes place and the state of $S$ is only slightly affected. Instead, if outcome 1 is recorded (d) then $S$ abruptly jumps to state $\ket{0}$. Note that, in either case, the ancilla is eventually uncorrelated with $S$, this being left in a pure state. We assumed that ancillas from 1 to $n-1$ were all measured in $\ket{0}$.}\label{fig-jumps}}%
		{\includegraphics[width=\textwidth]{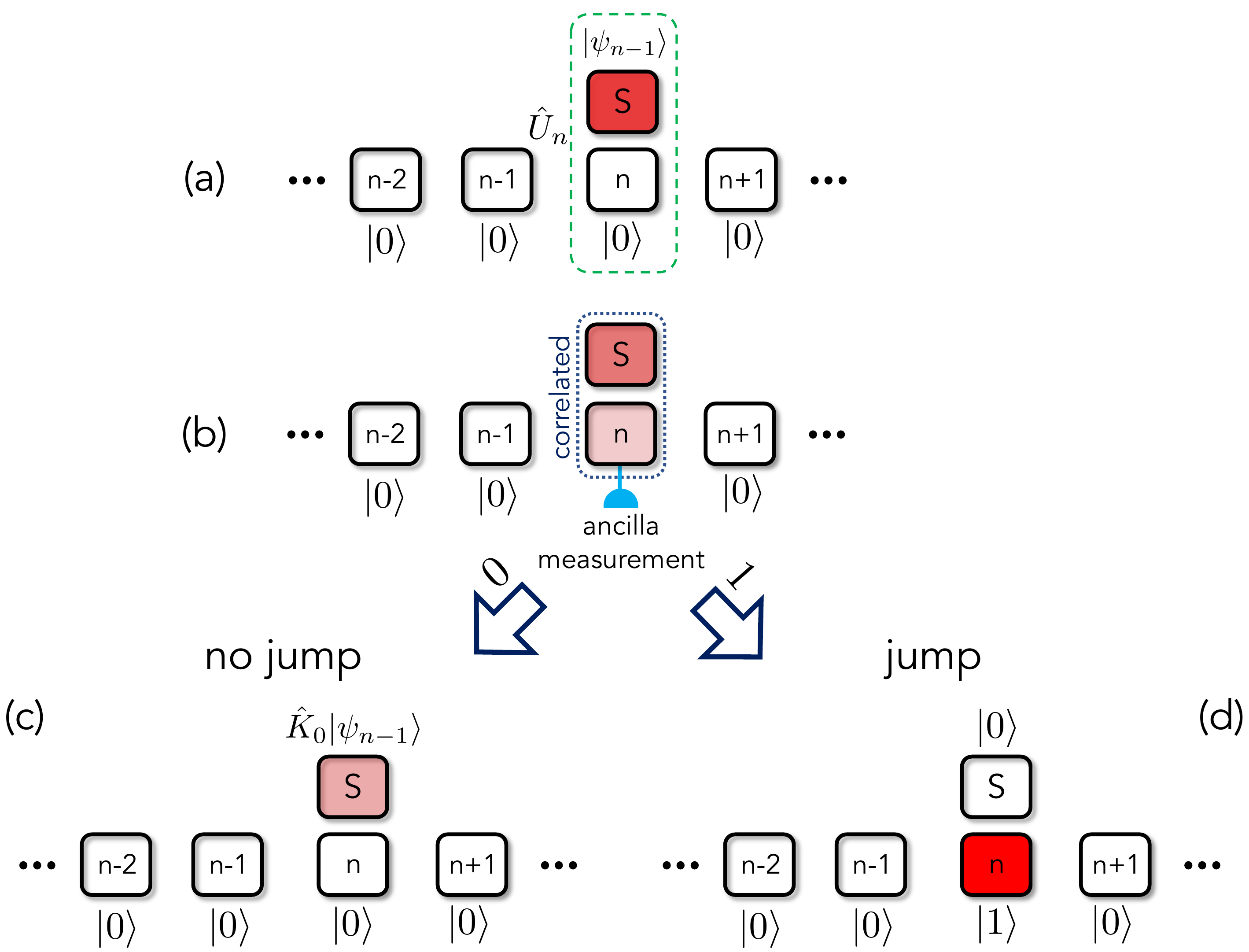}}
	\end{floatrow}
\end{figure}

Each sandwich on the left of $\ket{\psi_0}$ is effectively an operator on $S$
\begin{equation}
	\hat {K}_{k_m}=\bra{k_m}\hat{ U}_m\ket{\chi_m}\,,\label{cal-K}
\end{equation} 
in terms of which \cref{psin-3} is compactly expressed as
\begin{equation}
	\ket{\Psi_n}=\sum_{k_n}
	\cdots\sum_{k_{1}}\,\,\left(\hat {K}_{k_n}\cdots\hat{K}_{k_1}\ket{\psi_0}\right)\, \ket{k_1}\cdots\ket{k_n}\,.\label{psin-4}
\end{equation}
Note that operators \ref{cal-K} are generally {non}-unitary. Thereby, the state of $S$ between brackets $(\ldots )$ is {not} normalized. We thus rearrange \cref{psin-4} in the equivalent form
\begin{equation}
	\ket{\Psi_n}=\sum_{k_n}
	\cdots\sum_{k_{1}}\,\,\sqrt{p_{k_1\cdots k_n}}\,\,\,\,\left(\frac{\hat {K}_{k_n}\cdots\hat{K}_{k_1}\ket{\psi_0}}{\sqrt{p_{k_1\cdots k_n}}}\right)\, \ket{k_1}\cdots\ket{k_n}\,.\label{psin-5}
\end{equation}
with
\begin{equation}
	p_{k_1\cdots k_n}=\Vert \,\hat {K}_{k_n}\cdots\hat{K}_{k_1}\ket{\psi_0}\Vert ^2=\bra{\psi_0}\hat {K}_{k_1}^\dag\cdots\hat {K}_{k_n}^\dag\,\,\hat {K}_{k_n}\cdots\hat{K}_{k_1}\ket{\psi_0}\,\,,\label{pkn}
\end{equation}
where $\sum_{k_n}\cdots\sum_{k_{1}}\ p_{k_1\cdots k_n}=1$. Here, we used that \eq\ref{cal-K} defines a set of Kraus operators (see \ref{app-qmaps}) which thus fulfill $\sum_{k_m}\hat K_{k_m}^\dag\hat K_{k_m}=\mathbb{I}$.

The above shows that the CM dynamics can be seen as an {average} over a (very large) ensemble of ``histories" that result from projective measurements on the ancillas. 

Right after colliding with $S$ [see  \cref{fig-jumps}(a)], each ancilla is measured in the basis $\{\vert k_m\rangle \}$ [\cf\eq\ref{psin-2}] and the measurement outcome recorded, as sketched in  \cref{fig-jumps}(b). If this takes the specific value $k_m$, then operator $\hat {K}_{k_m}$ is applied on $S$. A \textit{specific} sequence of measurements results $\{k_1, \ldots ,k_n\}$ thus determines a particular history (realization), at the end of which $S$ is in state $\hat {K}_{k_n}\cdots\hat{K}_{k_1}\ket{\psi_0}$ (up to a normalization factor), this history occurring with probability $p_{k_1\cdots k_n}$. Remarkably, in each history, the state of $S$ remains pure at each step.

Note that the dynamics of histories does depend on the measurement basis $\{\ket{k_m}\}$. Different choices of this basis will result in different \textit{unravelings} of the same average dynamics (using a common jargon).
What we called histories so far usually go under the name of \textit{quantum trajectories}. The way the system evolves in a specific quantum trajectory is said \textit{conditional dynamics}: which Kraus operator \ref{cal-K} is to be applied on $S$ at each step is conditioned to the specific outcome of the measurement on the ancilla (recall that in quantum mechanics measurement is an intrinsically \textit{stochastic} process). A quantum circuit representation of the conditional CM dynamics is shown in \cref{fig-circ-qt}.

\begin{figure}[!h] 
	\raggedright
	\begin{floatrow}[1]
		\ffigbox[\FBwidth]{\caption[Quantum circuit representation of a CM conditional dynamics]{\textit{Quantum circuit representation of a CM conditional dynamics}.
		Compared to a basic CM (unconditional) dynamics [see  \ref{fig-CMdyn}(c)], each
		ancilla undergoes a projective measurement right after it collided with
		$S$.  The double wire indicates that the measurement outcome can be encoded
		as classical information~\cite{nielsen2002quantum}. The usual CM (unconditional)
		dynamics can be equivalently seen as an ensemble average over all possible
		conditional evolutions, each corresponding to a possible sequence of measurement
		outcomes.}\label{fig-circ-qt}}%
		{\includegraphics[width=\textwidth]{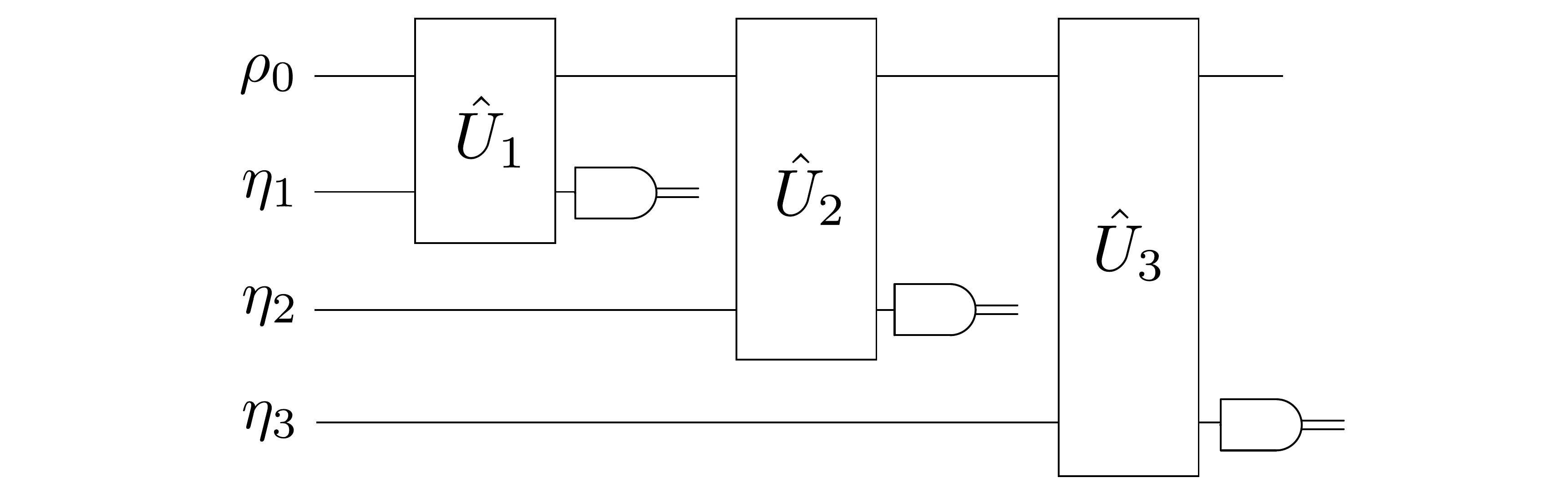}}
	\end{floatrow}
\end{figure}

\section{POVM and weak measurements}
\label{section-povm}
The above framework, when the interaction of $S$ with each ancilla is very weak, in fact defines the concept of \textit{weak measurements} in quantum mechanics.

Introductory textbooks to quantum mechanics usually describe measurements on a quantum system in terms of an orthonormal basis $\{\ket{k}\}$, each being the eigenstate of a certain observable with associated eigenvalue $k$ (assume for now that these are non-degenerate). According to the wavefunction collapse axiom, a measurement with outcome $k$ projects $S$ (initially in state $\ket{\psi}$) in the eigenstate $\ket{k}$ with probability $p_k=\vert \langle k\vert   \psi\rangle\vert ^2$. In the density--matrix language, this is expressed (and generalized at the same time) by saying that the act of measurement forces the state of $S$ to transform as
\begin{equation}
	\label{measur}
	\rho\rightarrow \frac{\hat \Pi_k \rho \hat \Pi_k}{p_k}\,\,\,\,\,{\rm with}\,\,\,\,\sum_k \hat \Pi_k=\mathbb{I}\,,
\end{equation}
whose associated probability is given by
\begin{equation}
	p_k={\rm Tr}_S \{\hat \Pi_k \rho\}\,.\label{prob-vn}
\end{equation}
Here, $\hat \Pi_k$ is the projector onto the eigenspace of eigenvalue $k$.\footnote{If $k$ is non-degenerate, $\hat P_k=\ket{k}\bra{k}$. Also, note that the expression of $p_k$ was obtained from ${\rm Tr} \{\hat \Pi_k \rho\hat \Pi_k\}$ by using the cyclic property of trace and $\hat \Pi_k^2=\hat \Pi_k$ (as $\hat \Pi_k$ is a projector).}  The $\hat \Pi_k$'s are a set of \textit{orthogonal} projectors, \ie $\hat \Pi_k\hat \Pi_{k'}=\delta_{k,k'}\hat \Pi_k$. Measurements of this kind are called Von Neumann measurements.

One can now define a generalized quantum measurement as
\begin{equation}
	\label{POVM}
	\rho\rightarrow \frac{\hat K_k \rho \hat K_k^\dag}{p_k}\,\,\,\,\,{\rm with}\,\,\,\sum_k \hat \Pi_k=\mathbb{I}\,,\,\,\,{\rm where}\,\,\,\hat \Pi_k=\hat K_k^\dag\hat K_k\,\,,
\end{equation}
the associated probability being $p_k={\rm Tr}_S \{\hat \Pi_k \rho\}$.  Here, the $\hat \Pi_k$'s are a set of positive operators (due to the constraint $p_k\ge 0$), which are not constrained to be orthogonal (at variance with Von Neumann measurements discussed before). Such a generalized quantum measure is usually referred to as \textit{positive operator-valued measure} (POVM).

Upon comparison of \ref{POVM} with the framework discussed in the last section, it should be clear that measuring each ancilla after the collision effectively performs a sequence of POVMs on $S$, one at each step. In this sense, the collisional dynamics is like continuously ``watching" the system. More specifically, when  the system--ancilla coupling is weak [as we assumed in  \cref{section-eqs}, \cf \eq\ref{USn2}] one talks about \textit{weak measurements}. The essential idea is that, instead of abruptly interrupting the dynamics through an instantaneous Von-Neumann measurement, one performs a gentle measurement that is yet diluted in time. Nevertheless, occasionally, this may still result in sudden changes of state (quantum jumps), as shown in the next section.

\section{Quantum trajectories in the all-qubit collision model and quantum jumps}
\label{section-qt-qubits}

To illustrate the framework in a concrete case, consider the (by now usual) all-qubit model of  \cref{section-all-qubit} for $g_z=0$ and $g=\sqrt{\gamma/\Delta t}$. There, we had already computed the Kraus operators \ref{cal-K} in the ancilla basis $\{\ket{0_n},\ket{1_n}\}$ [see \eq\ref{kraus-qubits}]. Assume that, right before a collision, qubit $S$ is in a superposition state $\ket{\psi}=c_{0} \ket{0}+c_{1}\ket{1}$ with $\vert c_0\vert ^2+\vert c_1\vert ^2=1$ (we omit the step index $n$ for a while).
The collision with ancilla $n$ and a subsequent measurement on $n$ in the basis $\{\ket{0_n},\ket{1_n}\}$ with outcome $\vert 0_n\rangle$ projects $S$ into the (unnormalized) state
\begin{equation}
	\hat K_0\vert \psi\rangle =\left(\ket{0}\bra{0}+\cos{\sqrt{\gamma \Delta t}}\,\ket{1}\bra{1}\right)\ket{\psi}=c_0\ket{0}+\cos{\sqrt{\gamma \Delta t}}\,\, c_1 \ket{1} \,,\label{K0-qt}
\end{equation}
and, if the measurement outcome is $\vert 1_n\rangle$, into the (unnormalized) state
\begin{equation}
	\hat K_1\vert \psi\rangle =\left(-i\,\sin\sqrt{\gamma \Delta t}\,\ket{0}\bra{1}\right)\,\ket{\psi}=-i\,\sin\sqrt{\gamma \Delta t}\,\,c_1\ket{0}\,\,.\label{K1-qt}
\end{equation}

\begin{figure}[!h] 
	\raggedright
	\begin{floatrow}[1]
		\ffigbox[\FBwidth]{\caption[Four sampled quantum trajectories in the all-qubit collision model]{Four sampled quantum trajectories in the all-qubit collision model of  \cref{section-all-qubit} for $g_z{=}0$ and $g{=}\sqrt{\gamma/\Delta t}$ [\cf\eq\ref{Vn-qubits}] when $S$ starts in state $\vert +\rangle _S$ and ancillas are all prepared in $\ket{0_n}$, each being measured in the basis $\{\ket{0_n},\ket{1_n}\}$ right after the collision with $S$. We plot the survival probability $\vert \langle +\vert \psi_n\rangle\vert ^2$ against the step number $n$, where each blue (red) dot stands for the measurement outcome $\ket{0_n}$ ($\ket{1_n}$). In each case, the survival probability tends to $\vert \langle +\vert 0\rangle\vert ^2{=}{1}/{2}$ witnessing that $S$ eventually converges to $\ket{0}$. Throughout we set $g\Delta t{=}\sqrt{\gamma \Delta t}{=}0.2$. The plots were obtained through a simple Monte Carlo simulation, where probabilities \ref{probs} are updated at each step and used to randomly select a measurement outcome and hence the corresponding state in \eq\ref{norm}. No jump occurs in trajectory (b), which exhibits a smooth exponential decay.}\label{fig-qt1}}%
		{\includegraphics[width=\textwidth]{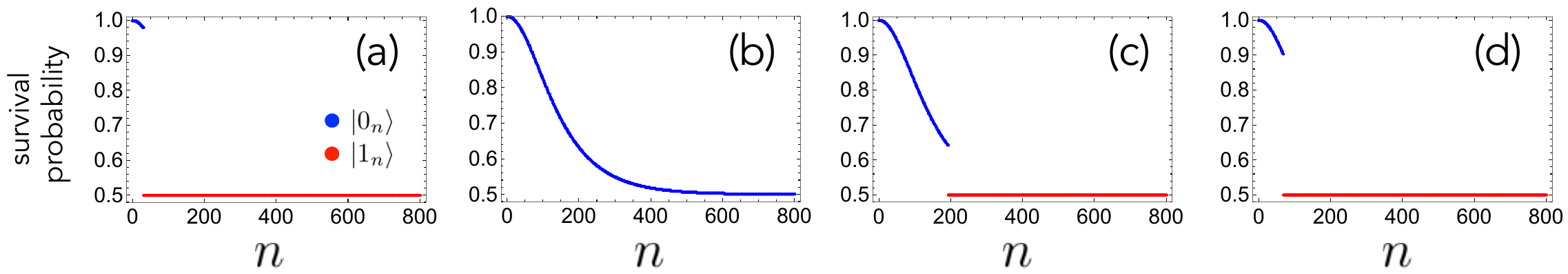}}
	\end{floatrow}
\end{figure}

These outcomes occur with probabilities $p_k=\bra{\psi}\hat K_k^\dagger \hat K_k\vert \psi\rangle $, which are explicitly worked out as
\begin{equation}
	p_0=\vert c_0\vert ^2+\cos^2\sqrt{\gamma \Delta t}\,\,\vert c_1\vert ^2\,,\,\,\,\,\,\,p_1=\sin^2\sqrt{\gamma \Delta t}\,\,\vert c_1\vert ^2\,\label{probs}
\end{equation}
(note that we correctly get $p_0+p_1=1$). Accordingly, the normalized version of \cref{K0-qt,K1-qt} reads
\begin{equation}
	\frac{\hat K_0\vert \psi\rangle }{\Vert \hat K_0\vert \psi\rangle \Vert }=\frac{c_0\ket{0}+\cos\sqrt{\gamma \Delta t}\, c_1 \ket{1} }{\sqrt{p_0}}\,,\,\,\,\,\frac{\hat K_1\vert \psi\rangle }{\Vert \hat K_1\vert \psi\rangle \Vert }=\frac{-i\,\sin\sqrt{\gamma \Delta t}\,c_1\ket{0}}{\sqrt{p_1}}\equiv\ket{0}\,\label{norm}
\end{equation}
up to an irrelevant phase factor in the last identity.
Thus both $\hat K_0$ and $\hat K_1$ have the effect of enhancing the $\ket{0}$'s component of $\ket{\psi}$. This entails that $\ket{\psi_n}$ asymptotically converges to $ \vert 0\rangle $. Therefore, we get that $S$ eventually ends up in $\ket{0}_S$ (\cf  \cref{section-homo}) even along single trajectories. 

\cref{probs,norm} can be used to simulate quantum trajectories through a random number generator. Some samples are shown in  \ref{fig-qt1}, where we plot the survival probability $\langle +\vert \psi_n\rangle \langle \psi_n\rvert +\rangle$
for $g\Delta t=\sqrt{\gamma\Delta t}=0.2$ when $S$ starts in state $\vert +\rangle$. Trajectories typically exhibit a continuous evolution, corresponding to repeated measurement outcomes $\ket{0_n}$ [recall sketch in  \ref{fig-jumps}(c)] interrupted by a sudden \textit{jump} when outcome $\ket{1_n}$ is recorded [recall sketch in  \cref{fig-jumps}(d)]. In the latter case, $S$ abruptly collapses to $\ket{0}$ in agreement with \ref{K1-qt} (signaled by the survival probability which jumps to 1/2) and then no longer changes its state. The precise step at which a jump occurs is {unpredictable} [\eg compare jumps in  \cref{fig-qt1}(a), (c) and (d)]. Note that jumps may even not occur at all, as in  \cref{fig-qt1}(b) where no ancilla is detected in $\ket{1_n}$.\marginnote{All no-jump trajectories have just the same evolution as that in  \ref{fig-qt1}(b).}

The reason why in the considered example only outcome $\ket{1_n}$ produces a sudden jump is that we set a relatively short collision time such that $g\Delta t\ll1$.\marginnote{If $\Delta t\sim g^{-1}$ both outcomes will generally produce a sudden change in the state of $S$ as is clear from \cref{K0-qt,K1-qt}.} Indeed, in this limit, \cref{K0-qt,K1-qt} reduce to
\begin{equation}
	\hat K_0\vert \psi\rangle \simeq c_0 \ket{0}+\left(1-\tfrac{1}{2}  \gamma\Delta t \right) c_1 \ket{1}\,,\,\,\,\,\hat K_1\vert \psi\rangle \simeq-i\, \sqrt{\gamma}\,\sqrt{\Delta t}\,c_1\ket{0} \,,\label{K01}
\end{equation}
the associated probabilities being
\begin{equation}
	p_0\simeq 1-\gamma \Delta t\,\vert c_1\vert ^2 \,,\,\,\,\,\,\,p_1\simeq \gamma\Delta t\,\vert c_1\vert ^2\,.\label{probs-exp1}
\end{equation}
We see that outcome $\ket{0_n}$ is very likely and, when occurring, it causes a tiny shrinking of the $\ket{1}$'s component. In contrast, outcome $\ket{1_n}$ is rather unlikely. However, \textit{if} occurring, it causes a \textit{dramatic} change of the state of $S$ which is projected to $\ket{0}$ altogether in one shot.

\section{Stochastic \se}
\label{section-stoca}

As seen thus far, during the conditional dynamics the state of $S$ remains pure all the time. However, its evolution is generally non-deterministic due to the occurrence of quantum jumps. In the previous instance, we saw that outcome $\ket{1_n}$ causes a sudden jump, in contrast to $\ket{0_n}$ producing only a small change in the state of $S$. We would like now both these behaviors to be incorporated into a \textit{single equation} that governs the \textit{stochastic }time evolution of state $\ket{\psi}$, like the usual \se \,does for conventional unitary (deterministic) dynamics. We next show how to achieve this for the CM and associated coupling Hamiltonian $\hat V_n$ considered in the previous section when $S$ starts in a pure state (a generalization will be presented in  \cref{section-SE-general}).

To this aim, we first express the low-order expansion of the Kraus operators [s\cf\cref{K0-qt}--\ref{K1-qt}] in the more compact form
\begin{equation}
	\hat K_0\vert \psi\rangle =\left(\mathbb{I}- \tfrac{1}{2}  \gamma\Delta t \,\hat\sigma_+\hat\sigma_{-} \right)\ket{\psi}\,,\,\,\,\,\hat K_1\vert \psi\rangle =-i\, \sqrt{\gamma}\,\sqrt{\Delta t}\,\,\hat\sigma_{-}\ket{\psi} \,,\label{K01-exp}
\end{equation}
(we used that $\hat\sigma_{+}\hat\sigma_{-}\vert \psi\rangle =c_1 \ket{1}$ and $\hat\sigma_{-}\vert \psi\rangle =c_1 \ket{0}$), the associated probabilities being
\begin{equation}
	p_1=1-p_0={\gamma}\,\langle \hat\sigma_{+}\hat\sigma_{-}\rangle\,{\Delta t}\,,\label{probs-exp}
\end{equation}
where $\langle\hat\sigma_{+}\hat\sigma_{-}\rangle{=}\langle\psi\rvert \,\hat\sigma_{+}\hat\sigma_{-}\vert \psi\rangle$.

The normalized state of $S$ for each measurement outcome is thus\marginnote{Using $1/\sqrt{1-x}\simeq 1+x/2$, the normalization factor of $\hat K_0\ket{\psi}$ [\cf\cref{K01-exp,probs-exp}] is $1/\sqrt{p_0}\simeq 1+\tfrac{\gamma}{2}\langle\hat\sigma_{+}\hat\sigma_{-}\rangle\Delta t$. Neglecting terms in $\Delta t^2$, we thus get the first identity in \ref{psip}. Also, note that, in the $1_n$ case, we could simply write $\ket{\psi_{n+1}}=\ket{0}$.}
\begin{align}
	&\ket{\psi_{n+1}}=\Bigl(\mathbb{I}- \tfrac{1}{2}  \gamma\Delta t \,(\hat\sigma_+\hat\sigma_{-}{-}\langle\hat\sigma_{+}\hat\sigma_{-}\rangle) \Bigr)\ket{\psi_n}\quad &({\rm for}\,\,\, 0_n)\nonumber\\
	&\ket{\psi_{n+1}}= \frac{\hat\sigma_{-}\ket{\psi_n}}{\sqrt{\langle\hat\sigma_{+}\hat\sigma_{-}\rangle}}\quad &({\rm for}\,\,\,1_n)\label{psip}
\end{align}
The corresponding changes in the state of $S$, $\Delta\ket{\psi_n}=\ket{\psi_{n+1}}-\ket{\psi_n}$, read
\begin{align}
	&\Delta\ket{\psi_{n}}=- \tfrac{1}{2}  \gamma\Delta t \,(\hat\sigma_+\hat\sigma_{-}{-}\langle\hat\sigma_{+}\hat\sigma_{-}\rangle) \ket{\psi_n}\quad&({\rm for}\,\, 0_n)\nonumber\\
	&\Delta\ket{\psi_{n}}= \left(\frac{\hat\sigma_{-}}{\sqrt{\langle\hat\sigma_{+}\hat\sigma_{-}\rangle}}-\mathbb{I}\right)\ket{\psi_n}\quad&({\rm for}\,\, 1_n)\label{dpsip}
\end{align}

We next define a binary \textit{random variable} $\Delta N$, which can take on values 0 or 1 with probabilities $p_0$ and $p_1$, respectively. Clearly, $(\Delta N)^2\equiv \Delta N$ and $\overline {\Delta N}=0 \cdot p_0+1 \cdot p_1=p_1$. Hence, in light of \eq\ref{probs-exp},
\begin{equation}
	\overline {\Delta N}=\overline {(\Delta N)^2}= p_1={\gamma}\,\langle \hat\sigma_{+}\hat\sigma_{-}\rangle\,{\Delta t}\,.
\end{equation}
The meaning of $\Delta N$ should be clear: $\Delta N=1$ when outcome $1_n$ is recorded and $S$ thereby evolves as in the second identity \ref{dpsip}.  Now, we combine together the two increments \ref{dpsip} as
\begin{equation}
	\Delta\ket{\psi_{n}}=- \tfrac{1}{2}  \gamma\Delta t \,(\hat\sigma_+\hat\sigma_{-}{-}\langle\hat\sigma_{+}\hat\sigma_{-}\rangle) \ket{\psi_n} + \left(\frac{\hat\sigma_{-}}{\sqrt{\langle\hat\sigma_{+}\hat\sigma_{-}\rangle}}-\mathbb{I}\right)\ket{\psi_n}\,\Delta N\label{SSE}\,.
\end{equation}
When $\Delta N=0$, $\Delta\ket{\psi_{n}}$ reduces to that for outcome $0_n$. When $\Delta N=1$, instead, we would get the sum of the two possible increments.  However, for $\Delta t$ short enough (as we are assuming), the term $\sim\Delta N$ dominates [plots such as those in  \ref{fig-qt1} could have been generated using \eq\ref{SSE}]. Now, we naturally take the continuous-time limit $\de t\rightarrow0$,\marginnote{This continuous-time limit corresponds to the one discussed in  \cref{sec-div} [indeed the coupling strength was chosen here in agreement with \eq\ref{gsqrt}].} obtaining
\begin{equation}
	{d}\ket{\psi}=- \tfrac{1}{2}  \gamma\,(\hat\sigma_+\hat\sigma_{-}{-}\langle\hat\sigma_{+}\hat\sigma_{-}\rangle) \ket{\psi} {d}t+ \left(\frac{\hat\sigma_{-}}{\sqrt{\langle\hat\sigma_{+}\hat\sigma_{-}\rangle}}-\mathbb{I}\right)\ket{\psi}\,{d} N\label{SSEc}\,,
\end{equation}
where
\begin{equation}
	\overline {{d} N}=\overline {({d} N)^2}={\gamma}\,\langle \hat\sigma_{+}\hat\sigma_{-}\rangle\,{{d} t}\,.\label{dN}
\end{equation}
\eq\ref{SSEc} fully describes the stochastic evolution of $S$ and indeed usually goes under the name of \textit{stochastic \se}. Note that, in contrast to the usual (deterministic) \se, this is highly \textit{nonlinear}. 
An equivalent way to write it is
\begin{equation}
	{d}\ket{\psi}=-i\hat H_{\rm eff}\,\ket{\psi}dt +i\tfrac{\gamma}{2}  \,\langle\hat\sigma_{+}\hat\sigma_{-}\rangle\ket{\psi} {d}t+ \left(\frac{\hat\sigma_{-}}{\sqrt{\langle\hat\sigma_{+}\hat\sigma_{-}\rangle}}-\mathbb{I}\right)\ket{\psi}\,{d} N\label{SSEc-3}\,,
\end{equation}
where
\begin{equation}
	\hat H_{\rm eff}=-i\tfrac{\gamma}{2}  \,\hat\sigma_+\hat\sigma_{-}
\end{equation}
is an effective non-Hermitian Hamiltonian.

\section{Unconditional dynamics: recovering the master equation}
\label{section-uncond}

Based on the discussion in  \cref{section-unrav}, the ensemble average of \ref{SSEc}, namely the average over all possible outcomes of the random variable ${d}N$, must return the Lindblad ME (recall   \cref{section-lind,section-moments,sec-div}). To prove this, we first work out the density--matrix  version of \eq\ref{SSEc}. The differential increment of $\rho=\ket{\psi}\bra{\psi}$ is
\begin{equation}
	{d}\rho={d}\left(\ket{\psi}\bra{\psi}\right)=\left({d}\ket{\psi}\right)\bra{\psi}+\ket{\psi}{d}\bra{\psi}+{d}\ket{\psi}{d}\bra{\psi}\,.\label{drhoS}
\end{equation}
As a point of utmost importance, note that, although of second order with respect to $d\psi$, the last term must be retained since $\overline{({d}N)^2}$ is in fact of \textit{first} order in ${d}t$ [\cf\eq\ref{dN}]. After plugging \ref{SSEc} and its bra in ${d}\ket{\psi}$ and ${d}\bra{\psi}$, respectively,  we replace ${d}N$ and $({d}N)^2$ with their common average \ref{dN}. To first order in ${d}t$, this yields as expected the Lindblad ME (see \ref{appendix-sse} for details)
\[
{d}\rho=\gamma \,\left(\hat\sigma_{-}\rho\,\hat\sigma_{+}-\tfrac{1}{2}\left[\hat\sigma_{+}\hat\sigma_{-},\rho\right]_+\right){d}t\,,\]
which we formerly derived in a different way in  \cref{sec-div}.

Consistently with the previous terminology (see end of  \cref{section-unrav}), any reduced dynamics discussed in  \cref{section-1,section-eqs} -- in particular that of $S$ -- is referred to as \textit{unconditional dynamics}. In real experiments, unconditional dynamics are usually not directly measurable but rather inferred by averaging over a large enough number of quantum trajectories. In this sense, although inherently stochastic, quantum trajectories reflect more closely the experimental reality. In contrast, the unconditional dynamics has a somewhat more indirect relationship with experiments but is fully deterministic.

{We mention that the connection with the Lindblad master equation just discussed
	has major computational applications in that it provides the basis for the widely
	used \textit{quantum jump method} or \textit{Monte Carlo wave
		function}~\cite{molmer1993monte,PlenioKnightRMP,haroche_exploring_2006,carmichael2009open}.
	This allows to work out the dynamics of open quantum  systems, especially of  large
	dimension, by keeping track of their \textit{wavefunction} over simulated quantum
	trajectories (and then averaging), thus bypassing the computationally demanding use of
	the density matrix.}

\section{A more general stochastic \se}
\label{section-SE-general}

Consider again the general (Markovian) CM in  \cref{section-moments} with coupling Hamiltonian \ref{AB} [which we assumed in the derivation \cref{drho2,deta2}]. The low-order collision unitary \ref{USn2} then explicitly reads
\begin{equation}
	\hat {U}_{n }\simeq \mathbb I-i \sum_{\nu}g_\nu  \hat A_\nu  {\hat B}_\nu \Delta t-\tfrac{1}{2}\,\sum_{\nu\mu} g_\nu g_\mu\hat A_\nu\hat A_{\mu} {\hat B}_\nu{\hat B}_{\mu}\, \Delta t^2\label{UAB}\,.
\end{equation} 
For simplicity we do not consider free Hamiltonian terms, which would simply result in an additional term $\sim \Delta t$ (we come back to this point at the end).

We will restrict to qubit ancillas (initially uncorrelated as usual), each prepared in state $\ket{\chi_n}$.\marginnote{For the sake of argument and to better highlight the physics,  as done throughout this Section, we will keep assuming that both the system and ancilla initial states are pure. The extension to mixed states is straightforward.} We also assume that first moments of the bath vanish [recall \eq\ref{moments}], \ie $\langle \hat B_\nu\rangle=\langle \chi_{n}\vert \hat B_\nu\vert \chi_n\rangle=0$ for all $\nu$ and $n$.

Based on  \cref{section-unrav} [see in particular \cref{cal-K,psin-4}], the $n$th collision transforms the state of $S$ and ancilla $n$ as
\begin{equation}
	\hat {U}_{n }\ket{\psi_{n-1}}\ket{\chi_n}=\hat K_0 \ket{\psi_{n-1}}\,\ket{0_n}+\hat K_1 \ket{\psi_{n-1}}\,\ket{1_n}\,,
\end{equation}
where, combining \cref{cal-K,UAB}, operators $\hat K_k$ are given by
\begin{equation}
	\hat K_k=\langle k\vert  \chi_n\rangle\,\mathbb{I}+\hat K_k^{(1)}\Delta t+\hat K_k^{(2)}\Delta t^2\,,\label{Kk-exp0}
\end{equation}
with
\begin{equation}
	\hat K_k^{(1)}=-i  \sum_\nu g_\nu \langle k\rvert  {\hat B}_\nu\vert \chi_n\rangle\, \hat A_\nu\,,\,\,\,\,\,\hat K_k^{(2)}=-\tfrac{1}{2}\sum_{\nu  \mu}\,g_\nu g_\mu\,\langle k\rvert  {\hat B}_\nu{\hat B}_\mu \vert \chi_n\rangle \,\hat A_\nu\hat A_{\mu}\label{Kk-exp2}\,.
\end{equation}
From now on, we drop index $n$. We consider next the case that $k=0,1$ with $\ket{\chi}=\ket{0}$, \ie we measure the ancilla  in a basis whose an element is just the initial state $\ket{\chi}$. This together with our initial assumption $\langle \hat B_\nu\rangle=0$ in particular yield
\begin{equation}
	\label{K-int}
	\langle k\vert  \chi\rangle=\delta_{k,0}\,\mathbb{I}\,,\,\,\,\hat K_0^{(1)}=0\,,\,\,\,\langle 0\rvert  {\hat B}_\nu{\hat B}_\mu \vert \chi\rangle =\langle 0\rvert  {\hat B}_\nu\vert 1\rangle \langle 1\rvert  {\hat B}_\mu \vert 0\rangle \,,\,\,\,\langle 1\rvert  {\hat B}_\nu{\hat B}_\mu \vert \chi\rangle =0\,,\nonumber
\end{equation}
where to compute the second moments we inserted $\ket{0}\bra{0}+\ket{1}\bra{1}=\mathbb{I}$ between ${\hat B}_\nu$ and ${\hat B}_\mu$.
Thereby
\begin{equation}
	\hat K_0^{(2)}=-\tfrac{1}{2}\,\underbrace{\sum_{\nu }g_\nu \,\langle 0\rvert  {\hat B}_\nu\vert 1\rangle \hat A_\nu}_{=\,\hat J^\dag} \,\underbrace{\sum_{ \mu}  g_\mu\langle 1\rvert  {\hat B}_\mu \vert 0\rangle  \hat A_{\mu}}_{=\,\hat J}\,,\,\,\,\hat K_1^{(2)}=0\,,
\end{equation}
where we suitably defined an operator $\hat J$ on $S$.\marginnote{The two quantities between brackets are easily shown to be mutually adjoint by recalling that $\sum_\nu g_\nu \hat A_\nu \hat B_\nu$ is Hermitian.} 

Putting together all the above and setting $g_\nu=\sqrt{\gamma_\nu/\Delta t}$, we conclude that
\begin{equation}
	\hat K_0=\mathbb{I}-\tfrac{1}{2}\hat L^\dag\hat L\,\Delta t\,,\,\,\,\hat K_1=-i  \hat L \,\sqrt{\Delta t}\label{Kk-exp}\,
\end{equation}
with associated probabilities
\begin{equation}
	p_1=1-p_0=\langle \hat L^\dagger \hat L\rangle\,{\Delta t}\,,\label{probs-exp-2}
\end{equation}
where the jump operator $\hat L$ is given by
\begin{equation}
	\hat L=\sqrt{\Delta t}\,\hat J=\sum_{\nu }\sqrt{\gamma_\nu}\,\langle 1\rvert  {\hat B}_\nu\vert 0\rangle \hat A_\nu \,.   
\end{equation}
In the example of  \cref{section-stoca} [see in particular \cref{K01-exp,probs-exp-2}], $\hat L=\sqrt{\gamma}\,\sigma_{-}$.

Since the structure of Kraus operators \ref{Kk-exp} is identical to \ref{K01-exp}, the reasoning followed in  \cref{section-stoca}  can be formally repeated leading to the general stochastic \se\,   [\cf\cref{SSEc,dN}]
\begin{equation}
	{d}\ket{\psi}=- \tfrac{1}{2}  \,(\hat L_+\hat L_{-}{-}\langle\hat L_{+}\hat L_{-}\rangle) \ket{\psi} {d}t+ \left(\frac{\hat\sigma_{-}}{\sqrt{\langle\hat L^\dag\hat L\rangle}}-\mathbb{I}\right)\ket{\psi}\,{d} N\label{SSEc2}\,,
\end{equation}
where $\overline {{d} N}=\overline {({d} N)^2}=\langle \hat L_{+}\hat L_{-}\rangle\,{{d} t}$. In the common case where an external drive or local field is applied on $S$ one simply needs to add the extra term $-i \hat H_S \ket{\psi}dt$, where Hamiltonian $\hat H_S$ could generally be time-dependent.

\section{Quantum trajectories: state of the art}
\label{soa-qt}

We already mentioned in the Introduction the seminal works by Caves and Milburn
(see in particular \rref~\cite{caves_quantum-mechanical_1987}). Therein, each ancilla
is modeled as a quantum harmonic oscillator which gets displaced due to the
interaction with $S$. Measuring the resulting displacement implements a
POVM. The corresponding unconditional dynamics is described by a characteristic ME,
whose dissipator (when $S$ is a harmonic oscillator itself) has the form
${\mathcal{D}}_S[\rho]=-K[\hat x,[\hat x,\rho]]$ with $K>0$ and $\hat x$ the position
operator~\cite{caves_quantum-mechanical_1987}. A bipartite generalization of this
collision model (with additional feedback) has been used more recently in some
gravitational decoherence theories to construct a classical channel that accounts for
Newtonian interaction 
~\cite{kafri2013noise,kafri2014classical}. These are critically reviewed in
\rref~\cite{altamirano_unitarity_2017}, which encompasses as well a general
presentation of metrological aspects of CMs (another one can be found in an
introductory section of \rref~\cite{gross_qubit_2018}).

A significant part of the discussion we developed relies on the seminal paper by
Brun~\cite{brun_simple_2002} already mentioned  in the Introduction. At variance with
Caves and Milburn, Brun employs qubit ancillas taking advantage of the quantum
information approach~\cite{nielsen2002quantum}.

It is important to note that in the considered instances we always measured the
ancillas in a basis containing the initial state $\ket{\chi_{n}}$. If this is not the
case, then two different outcomes could have comparable probabilities [unlike \eg
\eq\ref{probs-exp-2} where $p_0\ll p_1$]. The treatment in
\cref{section-SE-general} up to \eq\ref{Kk-exp2} would still apply, but the
stochastic \se\, would be different. A case of this kind is presented in the Brun's
paper and shown to lead to a quantum state diffusion
equation~\cite{brun_simple_2002}.

{We point out that micromaser (see  \cref{section-maser}) is a setup enabling direct
	measurement of the state of each ancilla (embodied by a flying atom). The related
	statistics of detections thus supplies informations on the cavity field and has been
	extensively studied, see \eg \rrefs~\cite{BriegelPRA94}.}

Finally, we mention that the collisional picture of quantum trajectories can be
profitably applied to quantum steering~\cite{beyer_collision-model_2018} and
engineering of quantum jump statistics~\cite{cilluffo2020microscopic}. Important
applications to stochastic quantum thermodynamics and quantum optics will be
discussed in  \cref{se-soa-thermo,soa-cmqo}, respectively.

\setchapterpreamble[u]{\margintoc}  
\chapter{Non-equilibrium quantum thermodynamics}
\label{section-thermo}

We now address the thermodynamics of quantum CMs in non - equilibrium transformations, this being arguably the area in which CMs (also known in this context as \textit{repeated interaction} schemes) occur most frequently. As the field is growing fast, the related body of literature is already considerable enough that several relevant topics cannot be covered here. Thus, given the pedagogical attitude of our paper, the present section aims to provide the reader with some basic tools for applying CMs in quantum thermodynamics problems. A number of topics that we do not discuss, e.g. exploiting CMs as a resource for improving thermodynamic performances,  are mentioned in the state of the art  \ref{se-soa-thermo} and related references supplied therein.

Before formulating general definitions and laws, we discuss a specific but quite paradigmatic non-equilibrium process: the relaxation to an equilibrium state.

\section{Relaxation to thermal equilibrium}
\label{section-QT1}

In  \cref{section-homo}, we introduced \textit{mixing} collision maps, namely those dynamics such that $S$ reaches a state $\rho^{*}$ no matter what initial state it started from (\ie $\rho_n\rightarrow \rho^{*}$ for any $\rho_{0}$) . If so, then $\rho^{*}$ is necessarily the only possible steady state, \ie the unique fixed point of the collision map (${\mathcal{E}}[\rho^{*}]=\rho^{*}$). It is natural to ask whether, by converging to $\rho^{*}$, $S$ inherits some intensive property of the bath. The most natural one is \textit{temperature}: if the bath is in an equilibrium state at a given temperature, will $S$ asymptotically end up in a Gibbs state at the same temperature? In other words, we wonder whether $S$ will \textit{thermalize} with the ancillas. 

We can formally define thermalization in terms of a basic CM (\cf  \cref{section-def}) where each ancilla is initially in the Gibbs state (henceforth referred to as {thermal state})
\begin{equation}
	\eta_{\rm th}= \frac{e^{-\beta H_n}}{Z_n}\,\label{eta-th1}
\end{equation}
with $\beta=1/(KT)$ the inverse temperature and $Z_n={\rm Tr}_n\left\{e^{-\beta H_n}\right\}$ the partition function. We say that thermalization occurs when $\rho_n\rightarrow \rho^{*}$ for any $\rho_{0}$ such that the asymptotic state $\rho^{*}$ is a thermal state of $S$ at the same temperature as each ancilla, \ie
\begin{equation}
	\rho^{*}=\frac{e^{-\beta H_S}}{Z_S}\,.\label{rho-th1}
\end{equation}
This definition can be generalized in many ways. For instance, one can conceive a generalized thermalization whose steady state is given by \ref{rho-th1} but $\beta$ generally differs from the bath's one. If so then equilibrium is never reached. Another possibility is that $S$ ends up in a thermal state like \ref{rho-th1} even though the bath is not in a thermal state (this would again entail lack of equilibrium).

\section{System thermalizing with a bath of quantum harmonic oscillators}
\label{section-harm-th}

A typical instance to illustrate thermalization is the basic CM in  \cref{section-def} in the case that ancillas are quantum harmonic oscillators (with associated bosonic ladder operators $\hat b_n$ and $\hat b_n^\dag$ such that $[\hat b_n,\hat b_{n'}^\dag]=\delta_{n,n'}$). The free Hamiltonian of $S$ (ancilla $n$) is $\hat H_S{=}\omega_{0}\hat A_{+}\hat A_{-}$ ($\hat H_n{=}\omega_0 \,\hat b_n^\dagger \hat b_n$), while  for the interaction Hamiltonian we take $\hat V_n{=} \sqrt{\gamma/\Delta t}\,(\hat A_{+}\hat b_n+{\rm H.c.})$. The nature of $\hat A_\pm$, which are ladder operators of $S$ fulfilling $[\hat H_S,\hat A_\pm]{=}\pm \omega_0 \hat A_\pm$, will be left unspecified for a while.

Each ancilla is initially in the Gibbs state [\cf\eq\ref{eta-th1}]
\begin{equation}
	\eta_{\rm th}= \frac{e^{-\beta H_n}}{Z_n}=\sum_k \frac{e^{-\beta \omega_0 k}}{Z_n}\,\ket{k}_n\bra{k}\,,\label{eta-th}
\end{equation}
with $\{\ket{k}_n\}$ the basis of Fock states.\marginnote{A Fock or number state $\ket{k}_n$ (for $k{=}0,1,2, \ldots $) fulfills $\hat H_n \ket{k}_n{=}\omega_0 k \ket{k}_n$. Hence, $e^{-\beta \hat H_n} =\sum_k e^{-\beta \omega_{0}k}\ket{k}_n\bra{k}$.} Recalling \cref{drho2,HS-diss}, we see that $\hat H'_S=0$ while the dissipator is given by
\begin{align}
	{\mathcal{D}}_S[\rho_{n-1}]=
	&\gamma\, \langle\hat b_n\hat b_n^\dag\rangle(\hat A_{-}\rho_{n-1}\hat A_{+}-\tfrac{1}{2}\,[\hat A_{+}\hat A_{-},\rho_{n-1}]_+)+\nonumber\\
	&\gamma\, \langle\hat b_n^\dagger \hat b_n\rangle(\hat A_{+}\rho_{n-1}\hat A_{-}-\tfrac{1}{2}\,[\hat A_{-}\hat A_{+},\rho_{n-1}]_+)\,.
\end{align}
Replacing $\hat b_n\hat b_n^\dag=\hat b_n^\dagger \hat b_n+1$ and introducing the thermal number of excitations
\begin{equation}
	\bar{n}_{\omega_0}= \langle\hat b_n^\dag\hat b_n\rangle={\rm Tr}_n \{\hat b_n^\dag\hat b_n \,\eta_n\}=\frac{1}{e^{\beta \omega_0}-1}\,\,,\label{nav0}
\end{equation}
the dissipator is written as
\begin{align}
	{\mathcal{D}}_S[\rho_{n-1}]=
	&\gamma_-(\,\hat A_{-}\rho_{n-1}\hat A_{+}-\tfrac{1}{2}\,[\hat A_{+}\hat A_{-},\rho_{n-1}]_+)+\nonumber\\
	&\gamma_+(\hat A_{+}\rho_{n-1}\hat A_{-}-\tfrac{1}{2}\,[\hat A_{-}\hat A_{+},\rho_{n-1}]_+)\,,\label{diss-th}
\end{align}
where we defined the emission and absorption rates
\begin{equation}
	\gamma_-=\gamma\, (\bar n_{\omega_0}+1)\,,\,\,\,\,\,\gamma_+=\gamma\,\bar n_{\omega_0}\,.\label{rates-ea1}
\end{equation}
This is a well-known master equation describing a system in contact with a thermal
bath, where we can recognize the \textit{Einstein
	coefficients}~\cite{loudon2000quantum} $A_E=\gamma$ (spontaneous emission rate) and
$ B_{\rm E}=\gamma\, \bar n_{\omega_0}$ (stimulated emission/absorption rate). These are related to rates
\ref{rates-ea1} according to $\gamma_-= A_{\rm E}+ B_{\rm E}$ and $\gamma_+= B_{\rm E}$.  Note that
\cref{nav0,rates-ea1} entail
\begin{equation}
	\frac{\gamma_+}{\gamma_-}=e^{-\beta \omega_{0}}\,\,.\label{gpgm}
\end{equation}
This identity connects rates (associated with relaxation, thus a non-equilibrium process) to temperature (defined for equilibrium states).

Similar conclusions hold when ancillas are qubits (instead of harmonic oscillators), \ie $\hat H_n=\omega_0 \hat\sigma_{n+}\hat\sigma_{n-}$ and $\hat V_n= \sqrt{\gamma/\Delta t}\, \left(\hat A_{+}\hat \sigma_{n-}+{\rm H.c.}\right)$. The resulting ME dissipator is identical to \ref{diss-th} except that the thermal number of excitations of each ancilla is now given  by $\bar{n}_{\omega_0}=1/(e^{\beta\omega_0}+1)$ [instead of \ref{nav0}]. This is just ME \ref{micro-ME} which we encountered in  \cref{section-maser}, describing the cavity dynamics of a micromaser with the atomic population given by $p=\bar{n}_{\omega_0}$ and for $\tau=\Delta t$, $g=\sqrt{\gamma/\Delta t}$ (where in that case $S$ is a harmonic oscillator such that $\hat A_-=\hat A_+^\dag=\hat a$). Note that this rules out atomic initial states such that $p>1/2$, \ie that cannot be regarded as thermal states at any temperature (unless one defines a negative temperature such that $\beta<0$).

Mostly for the sake of argument, in all the forthcoming instances we will refer to the case that $S$ is a qubit (ancillas being still harmonic oscillators), thus we will set $\hat A_\pm=\hat \sigma_{\pm}$.

In the basis $\{\ket{0},\ket{1}\}$ of $S$, ME \ref{diss-th} translates into a pair of differential equations for the excited-state population $p$ and coherences $c$ [recall \eq\ref{qubit-state}], which read
\begin{equation}
	\label{dp-dc}
	\dot p=\gamma_+(1-p)-\gamma_- p\,,\,\,\,\dot c=-\tfrac{1}{2}\, (\gamma_+{+}\gamma_-)c\,\,.
\end{equation}
Under stationary conditions the derivatives vanish, yielding $c=0$ and
\begin{equation}
	p=\frac{1}{1+(\gamma_+/\gamma_-)^{-1}}\,\,.\label{p-eq}
\end{equation} 
Using \ref{gpgm} this means that, regardless of the initial state, $S$ eventually ends up in
\begin{equation}
	\rho_{\rm th}= \frac{e^{-\beta H_S}}{{\rm Tr}_S \left\{e^{-\beta H_S}\right\}}=\frac{1}{1+e^{-\beta\omega_0}}\ket{0}_S\bra{0}+\frac{e^{-\beta\omega_{0}}}{1+e^{-\beta\omega_0}}\ket{1}_S\bra{1}\,,\label{rho-th}
\end{equation}
namely the thermal state at the same temperature of ancillas (defined by $\beta$). Thus thermalization occurs.

Although very common, the thermalization process considered here regards a specific class of systems. In the next section, we consider a general situation where $S$ and ancillas are unspecified, shedding some light at the same time on the reason why thermalization may take place.

\section{Thermalization and energy conservation}
\label{section-balance}

Occurrence of thermalization depends, in particular, on the form of system--ancilla Hamiltonian. For example, let us consider the last instance of the previous section and simply add a detuning to $S$ such that $\hat H_S=(\omega_0+\delta)\hat\sigma_{+}\hat\sigma_{-}$. The ancilla thermal state $\eta_n$ and rates \ref{rates-ea1} are unaffected by $\delta$ and thus ME \ref{diss-th} continues to hold unchanged, hence $S$ still asymptotically converges to \ref{rho-th}. Yet, this is \textit{not} the thermal state of $S$ at the ancilla temperature, thus thermalization now does not take place.\marginnote{The state can still be arranged as a thermal state but at an effective temperature different from the ancilla's one.}

An important necessary (although generally not sufficient) condition for thermalization to occur is that collisions be \textit{energy-conserving}. This means that $\hat H_S+\hat H_n$ (total free Hamiltonian of $S$ and $n$th ancilla) is a constant of motion in the $n$th collision, \ie it commutes with the collision unitary
\begin{equation}
	[\hat U_n,\hat H_S+\hat H_n]=0\,\label{EC1}.
\end{equation}
This is because if this is true then the $S$-ancilla state
\begin{equation}
	\frac{e^{-\beta H_S}}{Z_S}\otimes\frac{e^{-\beta H_n}}{Z_n}\propto e^{-\beta( \hat H_S+\hat H_n)}
\end{equation}
is clearly unaffected by the $n$th collision. It follows that state \ref{rho-th1} is a fixed point of the collision map (\ie a steady state), this being a necessary condition for thermalization as we discussed in  \cref{section-homo,section-QT1}.

Based on the form of the collision unitary [\cf \cref{USn1,USn2}], energy conservation can be equivalently expressed as
\begin{equation}
	[\hat V_n,\hat H_S+\hat H_n]=0\,.\label{EC}
\end{equation}
In the example mentioned at the beginning of this subsection, when $\delta\neq 0$ \ref{EC} does not hold thus thermalization cannot occur.

Physically, conservation of $\hat H_S+\hat H_n$ means that if the free energy of $S$ decreases then that of the colliding ancilla grows by exactly the same amount (and viceversa). This intuition can be made formally rigorous as follows. 

Let us first define an eigenoperator $\hat A_\nu$ of $\hat H_S$ with eigenvalue $\omega_\nu$ as an operators on $S$ fulfilling
\begin{equation}
	[\hat H_S, \hat A_\nu]=-\omega_\nu \hat A_\nu\,.\label{comm}
\end{equation}
Likewise, eigenoperators of $\hat H_n$  are defined as
\begin{equation}
	[\hat H_n, \hat B_\nu]=-w_\nu \hat B_\nu\,\label{eigB}
\end{equation}
with $w_\nu$ the associated eigenvalues. Here, $\hat A_\nu$ and $\hat B_\nu$ are defined as dimensionless operators. Note that the values taken by index $\nu$ in \cref{comm,eigB} are generally different.

Now, for given $\hat H_S$ and $\hat H_n$, it can be shown (see \ref{appendix-equi}) that the most general class of interaction Hamiltonians $\hat V_n$ satisfying \ref{EC} has the form
\begin{equation}
	\hat V_n=\sum_\nu  g_\nu \left(\hat A_\nu^\dagger \,{\hat B}_\nu+\hat A_\nu \,{\hat B}_\nu^\dag\right)\,\,\,\,{\rm with}\,\,\,\omega_\nu=w_\nu\,\,.\label{V-EC}
\end{equation}
It can be immediately checked that \ref{V-EC} fulfills \ref{EC}.

Many coupling Hamiltonians appearing throughout this paper can be recognized as falling within this class.
Note that $\hat V_n\neq 0$ only provided that there exist eigenvalues common to both $\hat H_S$ ad $\hat H_n$.  To make clear the physical meaning of \ref{V-EC}, it suffices to consider a generic eigenstate $\ket{E}$ of $\hat H_S$ with energy $E$ and note that $\hat A_\nu \ket{E}$ is another eigenstate but with energy $E-\omega$, while $\hat A^\dagger_\nu \ket{E}$ is an eigenstate with eigenvalue $E+\omega$.\marginnote{From \eq\ref{comm}, $\hat H_S\hat A_\nu=\hat A_\nu\hat H_S-\omega_\nu \hat A_\nu$. Hence, $\hat H_S\hat A_\nu  \ket{E}=\hat A_\nu\hat H_S  \ket{E}-\omega_\nu \hat A_\nu \ket{E}=E\hat A_\nu \ket{E}-\omega_\nu\hat A_\nu   \ket{E}=(E-\omega_\nu)\hat A_\nu \ket{E}$, showing that $\hat A_\nu \ket{E}$ is eigenstate of $\hat H_S$. The property for $\hat A_\nu^\dag$ is proven likewise by noting that $[\hat H_S, \hat A^\dagger_\nu]=\omega_\nu \hat A^\dagger_\nu$. Note that $\hat A_\nu \ket{E}$ (or $\hat A_\nu^\dagger \ket{E}$) could be zero: \eg for a qubit of Hamiltonian $\omega_0\ket{1}\bra{1}$ we have $\hat \sigma_{-}\ket{0}=\hat \sigma_{+}\ket{1}=0$.} Analogous properties hold for $\hat B_\nu$. Thereby, according to $\hat V_n$, if $S$ undergoes a transition $\ket{E_i}\rightarrow\ket{E_f}$ changing its energy by the amount $\omega=E_f-E_i$ then the ancilla will make a simultaneous transition with energy change $-\omega$.
For instance, in  \cref{section-QT1}, if $S$ is a qubit making the transition $\ket{0}\rightarrow \ket{1}$ with energy gain $\omega_{0}$ then a harmonic-oscillator ancilla can only decay from a Fock state $\ket{k}$ to $\ket{k-1}$ losing the same amount of energy $\omega_{0}$.

The general ME corresponding to interaction \ref{V-EC} can be calculated in terms of the ancilla's moments [see  \cref{section-moments} and \eq\ref{drho2}]. Since $\eta_n$ is a thermal state (mixture of eigenstates of $\hat H_n$), each $\hat {\hat B}_\nu$ (in light of the aforementioned properties) has vanishing expectation value. Thus $\hat H'_S=0$ [\cf\eq\ref{HS-diss}]. Regarding the dissipator ${\mathcal{D}}_S$, we note that $\langle {\hat B}_{\nu'} {\hat B}_\nu\rangle=\langle {\hat B}_{\nu'}^\dagger {\hat B}_\nu^\dag\rangle=0$ for all $\nu$ and $\nu'$. Therefore,
\begin{align}
	{\mathcal{D}}_S[\rho_{n-1}]=
	&\sum_{\nu,\nu'}\,\gamma_{\nu,\nu'}\langle {\hat B}_{\nu'} {\hat B}_\nu^\dag\rangle(\hat A_\nu\rho_{n-1}\hat A_{\nu'}^\dagger -\tfrac{1}{2}[\hat A_{\nu'}^\dag\hat A_\nu,\rho_{n-1}]_+)+\nonumber\\ &\sum_{\nu,\nu'}\,\gamma_{\nu,\nu'}\langle {\hat B}_{\nu'}^\dagger {\hat B}_\nu\rangle(\hat A_\nu^\dag\rho_{n-1}\hat A_{\nu'} -\tfrac{1}{2}[\hat A_{\nu'}\hat A_\nu^\dag,\rho_{n-1}]_+)\,.
\end{align}

\section{Non-equilibrium steady states with  baths at different temperatures}
\label{section-QT1T2}

We have dealt so far with a non-equilibrium process where however $S$ eventually ends up in an equilibrium state. We next consider a dynamics where $S$ never attains equilibrium although it reaches a (non-equilibrium) steady state. This is the simultaneous interaction with $\mathit{many}$ thermal baths at different temperatures, which is a paradigmatic dynamics to illustrate \eg thermal conduction, where it is known that $S$ can reach an effective thermal state at a temperature which is a weighted average of those of the reservoirs. CMs are very effective in handling multiple baths as discussed in  \cref{section-multiB}.

We thus focus on a CM comprising $M=2$ baths of ancillas labeled with 1 and 2 as shown in  \cref{fig-BB}(a). Ancillas of bath $i=1,2$ are in a thermal state $\eta^{(i)}=\eta_{\rm th}^{(i)}$ with inverse temperature $\beta_i$ [\cf\eq\ref{chi-corr} for $\chi_n^{({\rm corr})}=0$] where in general $\beta_1\neq \beta_2$. The coupling Hamiltonian ruling each collision has the form \ref{Vn-many2}. As in the instance in  \cref{section-QT1}, we assume that first moments vanish for each bath, \ie $\langle \hat B_{\nu i}\rangle=0$. Hence, $\dot{\rho}={\mathcal{D}}_S[\rho]$ with [\cf\eq\ref{addi}]
\begin{equation}
	{\mathcal{D}}_S[\rho]={\mathcal{D}}_S^{(1)}[\rho]+{\mathcal{D}}_S^{(2)}[\rho]\,,\label{addi2}
\end{equation}
where ${\mathcal{D}}_S^{(i)}$ is the dissipator that would arise if $S$ collided only with ancillas of bath $i$. 

As an illustrative instance, fully in line with  \cref{section-QT1}, we model each ancilla of bath $i$ as a harmonic oscillator of frequency $\omega_0$ initially in a thermal state like \ref{eta-th} with inverse temperature $\beta_i$. The coupling with $S$ has the same form as in  \cref{section-QT1} with coupling strength $\sqrt{\gamma_i/\Delta t}$. Note that $S$ and all ancillas. have the same frequency $\omega_{0}$ in a way that, if $\gamma_2$ were zero (meaning that bath 2 is decoupled from $S$), then $S$ would reach thermal equilibrium with bath 1 (and viceversa).

Due to \ref{addi2} we see that the dissipator is analogous to \ref{diss-th} under the replacements $\gamma_\pm\rightarrow \gamma'_\pm$ with the effective emission and absorption rates given by
\begin{equation}
	\gamma'_\pm=\gamma_\pm^{(1)}+\gamma_\pm^{(2)}\,,\label{gpp}
\end{equation}
where
\begin{equation}
	\gamma_-^{(i)}=\gamma_i\, (\bar n^{(i)}+1)\,,\,\,\,\,\,\gamma_+^{(i)}=\gamma_i\,\bar n^{(i)}\,\,\,\,\,\,{\rm with}\,\,\,\,\,\bar n_i=(e^{\beta_i \omega_{0}}-1)^{-1}\label{rates-ea}\,\,.
\end{equation}
As the ME is formally identical to that in  \cref{section-QT1}, $S$ asymptotically converges to an effective thermal state of the form \ref{rho-th} with inverse temperature $\beta_{\rm eff}$ given by\marginnote{We write down the analogue of \ref{gpgm} under the replacements $\beta\rightarrow\beta_{\rm eff}$ and $\gamma_\pm\rightarrow \gamma'_{\pm}$ and then solve for $\beta_{\rm eff}$.}
\begin{equation}
	\beta_{\rm eff}=\frac{1}{\omega_{0}}\log\frac{\gamma'_-}{\gamma'_+}=\frac{1}{\omega_{0}}\log \,\frac{(\gamma_1+\gamma_2) e^{
			(\beta_1+\beta_2)\omega_0}-\gamma_1 e^{\beta_1
			\omega_0}-\gamma_2 e^{\beta_2 \omega_0}}{\gamma_2
		\left(e^{\beta_1 \omega_0}-1\right)+\gamma_1 \left(e^{\beta_2
			\omega_0}-1\right)}\,.\label{beta12}
\end{equation}
This entails that $\beta_{\rm eff}$ is generally different from both $\beta_1$ and $\beta_2$ (confirming that a non-equilibrium steady state is reached), reducing to $\beta_1$ for $\gamma_2=0$ and to $\beta_2$ for $\gamma_1=0$. Thermal equilibrium is retrieved when the two baths have the same temperature, in which case \ref{beta12} predicts (as expected) $\beta_{\rm eff}=\beta_1=\beta_2$ regardless of $\gamma_1$ and $\gamma_2$.

Since CMs can keep track of the bath dynamics in a relatively straightforward way, they are an advantageous tool for calculating the rate of change (or flux) of thermodynamic quantities in non-equilibrium transformations (such as thermalization) even beyond the weak coupling regime [\ie when the collision unitary cannot be approximated with the lowest-order expansion \ref{USn2}]. The general definition and calculation of these, as well as the basic laws governing them, will be a main subject of the following subsections.

\section{Time dependence of the total system--bath Hamiltonian}
\label{section-HSBt}

We allow the free Hamiltonian of the open system $S$ to be generally \textit{time-dependent}. This allows to encompass situations where $S$ is subject to an external classical drive such that one or more parameters of $\hat H_S$ can be deterministically modulated in time according to an assigned protocol. For instance, in the CM considered in  \cref{section-QT1}, we could have $\hat H_S(t)=\omega_0(\lambda_t) \,\vert e\rangle \langle e\rvert  $, describing a time-modulated detuning with $\lambda_t$ some smooth function of time defining the protocol. We also assume that the characteristic time over which $\hat H_S(t)$ changes is much larger than $\Delta t$, hence during the $n$th collision we can approximate $\hat H_{S}(t)\simeq \hat H_{S}^{(n)}$ so that $\hat H_S$ becomes step-dependent.\marginnote{More in detail, $\hat H_{S}^{(n)}$ can \eg be defined as the time average of $\hat H_{S}(t)$ during the $n$th time interval.}

Accordingly, the total $S$-bath Hamiltonian at an arbitrary time $t$ has the general expression
\begin{equation}
	\hat H_{SB}(t)=\hat H_S(t)+\hat H_B+\hat V(t)\,,\label{HSBt-1}
\end{equation}
with the (time-independent) bath Hamiltonian given by
\begin{equation}
	\hat H_B=\sum_n \hat H_n\label{HBn}
\end{equation}
and the $S$-$B$ coupling Hamiltonian by
\begin{equation}
	\hat V(t)= \sum_n \Theta_n (t) \,\hat V_n\,,\label{V-td}
\end{equation}
where $\Theta_n (t)=1$ for $t_{n-1}\le  t< t_n$ and zero otherwise. 

Notably, besides the possible time dependence coming from $H_S(t)$, the total Hamiltonian has an \textit{intrinsic} time dependence due to the sudden replacement of the bath ancilla interacting with $S$ at times $t=t_n$. This time dependence, due to the periodic \textit{switching} (on and off) of the interaction with ancillas, is a distinctive feature of CMs not present in conventional microscopic system--bath models. This generally introduces a contribution to the work as we will see in  \cref{section-work}.

\section{Rate of change of energy of $S$}
\label{section-free}

We generally define the internal energy (or simply energy) of $S$ as the quantum expectation value $E_{S}=\langle \hat H_S\rangle={\rm Tr}_S\{\hat H_S \rho\}$. Since in general both the operator $\hat H_S$ itself and the state of $S$ evolve in time, the change of $E_{S}$ at each step has two contributions
\begin{equation}
	\label{dees}
	\Delta E_{S}={\rm Tr}_S\{\Delta \hat H_S\, \rho_{n-1}\}+{\rm Tr}_S\{\hat H_S\,\Delta \rho_{n}\}
\end{equation}
with $\Delta \hat H_S=\hat H_S^{(n)}-\hat H_S^{(n-1)}$ (subscripts between brackets denote the step number).
Using \eq\ref{dOS}, in terms of the usual decomposition \ref{AB} of $\hat V_n$, the rate of change of $E_S$ at each collision (\ie during the time interval $t_{n-1}\le t<t_n$)  is generally given by
\begin{equation}
	\frac{\Delta E_S}{\Delta t}=\Bigl\langle \frac{\de \hat H_S}{\Delta t}\Bigr\rangle+i\sum_\nu\,g_\nu \langle{{\hat B}_\nu}\rangle \,\langle [  \hat A_\nu,\hat H_S]\rangle+\sum_{\nu\mu}\,\gamma_{\nu\mu}\langle {\hat B}_\mu {\hat B}_\nu\rangle\langle\hat A_\mu\hat H_{S}\hat A_\nu-\tfrac{1}{2}\,[\hat A_\mu\hat A_\nu,\hat H_S]_+\rangle.\label{dOSqt}
\end{equation}

\section{Heat flux}
\label{section-heat}

Analogously to $S$, the energy of the $n$th-ancilla is defined as $E_{n}=\langle \hat H_n\rangle$. As ancillas are uncoupled to one another, $E_{n}$ can change only during the $n$th collision. Accordingly, $\de E_{n}$ at the $n$th step is also the change of energy of the entire bath $B$, \ie $\Delta E_{n}=\Delta E_B^{(n)}$. This in fact gives the exchanged \textit{heat} whose definition reads
\begin{equation}
	\delta Q=-\Delta E_B^{(n)}=-\de E_{n}\,.\label{dQ}
\end{equation}
Therefore, using \eq\ref{dOn}, the heat flux (exchanged heat per unit time) is given by
\begin{equation}
	\frac{\delta Q}{\Delta t}=-i\sum_\nu\,g_\nu \langle{\hat A_\nu}\rangle\langle [  {\hat B}_\nu,\hat H_{n}]\rangle -\sum_{\nu\mu}\,\gamma_{\nu\mu}\langle \hat A_\mu \hat A_\nu\rangle\langle{\hat B}_\mu\hat H_n{\hat B}_\nu-\tfrac{1}{2}\,[{\hat B}_\mu{\hat B}_\nu,\hat H_n]_+\rangle\label{dOnqt}\,
\end{equation}
(note that, unlike $\hat H_S$, $\hat H_n$ is time-independent).

\section{Work rate}
\label{section-work}

Work is the contribution to the change of total energy $E_{SB}=\langle \hat H_{SB}\rangle$ due to the time dependence of the total Hamiltonian operator $\hat H_{SB}(t)$ [\cf\eq\ref{HSBt-1}]. Thus a natural definition of the work performed in each time step $t_{n-1}\le  t< t_n$ reads
\begin{equation}
	\delta W={\rm Tr}_{SB}\,\{{\Delta \hat H_{SB}\,\sigma_{n-1}}\}\label{work1}\,    
\end{equation}
with $\Delta \hat H_{SB}$ the change of operator $\hat H_{SB}$ in the considered time interval. Since the only time-dependent terms in $\hat H_{SB}(t)$ are (in general) $\hat H_S(t)$ and $\hat V(t)$, we can split $\delta W$ into a pair of corresponding terms
\begin{equation}
	\delta W=\delta W_d+\delta W_{\rm sw}\label{dw2}
\end{equation}
with
\begin{equation}
	\delta W_d= {\rm Tr}_{S}\,\{{\Delta \hat H_{S}\,\rho_{n-1}}\}\,,\,\,\,\delta W_{\rm sw}= {\rm Tr}_{SB}\,\{{\de\hat V\,\sigma_{n-1}}\}\,,\label{work-2}
\end{equation}
where subscript $d$ stands for ``drive" (we used that $\hat H_{S}$ acts only on $S$).
Here, $\delta W_{\rm sw}$ is the contribution due to the time dependence of $\hat V(t)$. We call it \textit{switching work} since, physically, it is the work (generally) required for replacing an ancilla which completed its collision with a fresh one.

\begin{figure}[!h] 
	\raggedright
	\begin{floatrow}[1]
		\ffigbox[\FBwidth]{\caption[Redefinition of the time step]{\textit{Redefinition of the time step}. (a): The interaction with ancilla $n$ (yellow area) is switched on at time $t_{n-1}$ and then turned off at $t_{n}=t_{n-1}+\Delta t$, at which time interaction $V_{n+1}$ is switched on (green). To correctly take into account the work required for the switching (if any), we redefine the time interval as $\left[t_{n-1},t_{n}\right[\,\,\rightarrow\,
				\left[t_{n-1}+\varepsilon,t_{n}+\varepsilon\right[$ with $\varepsilon\rightarrow0^+$. (b): The redefined time step in turn can be split into a pair of consecutive intervals:
				$\left[t_{n-1}+\varepsilon,t_{n}-\varepsilon\right[$ (interval I) and $\left[t_{n}-\varepsilon, t_{n}+\varepsilon\right[$ (interval II). In I, $\hat V(t)=\hat V_n$ (constant). During II, instead, $\hat V(t)$ jumps as $\hat V_n\rightarrow \hat V_{n+1}$ at $t=t_n$.}\label{fig-sw}}%
		{\includegraphics[width=\textwidth]{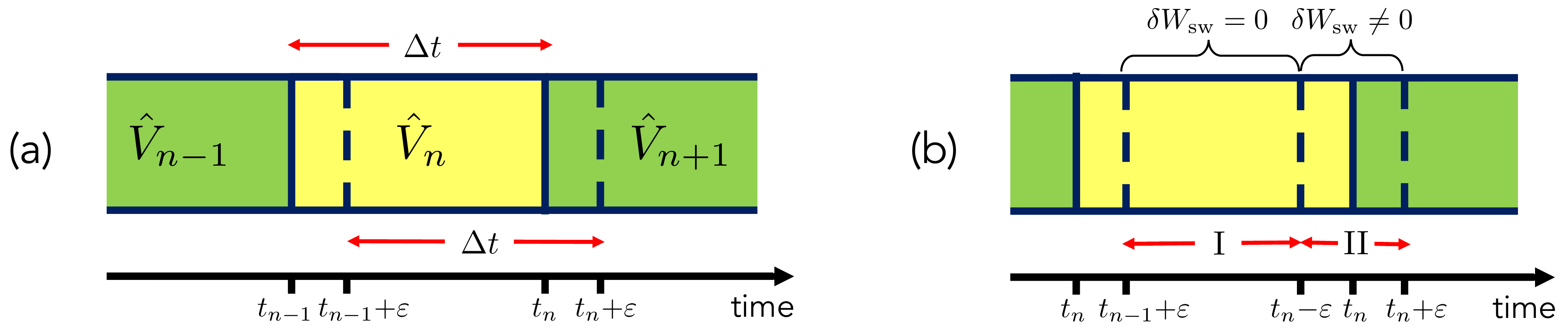}}
	\end{floatrow}
\end{figure}

Now a subtle but relevant issue arises  since the time derivative of $\Theta_n(t)$ [\cf \eq\ref{V-td}] is singular at times $t=t_{n}$ (for any $n$). At these times, $\hat V(t)$ undergoes the instantaneous switch $\hat V_{n}\rightarrow \hat V_{n+1}$.
To take this switch into due account, all the changes throughout must be intended as computed over the time interval $\left[t_{n-1}+\varepsilon,t_{n}+\varepsilon\right[$ as sketched in  \ref{fig-sw}(a) with the understanding that $\varepsilon\rightarrow0^+$. As the singularity of \ref{HSBt-1} at time $t=t_n$ comes only from $\hat V(t)$, this slight change of time interval does not affect all the thermodynamic quantities other than $\delta W_{\rm sw}$ (in particular $\delta W_d$) with the only exception of $E'_S=\langle \hat H'_S\rangle$ which will be analyzed in  \cref{section-1st}.

Now, the redefined time step $\left[t_{n-1}+\varepsilon,t_{n}+\varepsilon\right[$ can be conveniently decomposed into a pair of consecutive intervals  [see  \ref{fig-sw}(b)]: $\left[t_{n-1}+\varepsilon, t_{n}-\varepsilon\right[$ (interval I) and $\left[t_{n}-\varepsilon,t_{n}+\varepsilon\right[$ (interval II). 
As $\hat V(t)$ is constant all over interval I, the switching work $\delta W_{\rm sw}$ is performed only in the very short time interval II [within which $\hat V(t)$ undergoes the sudden jump $\hat V_n\rightarrow \hat V_{n+1}$]. Accordingly, the switching work is correctly worked out as $\delta W_{\rm sw}={\rm Tr}_{SB}\{(\hat V_{n+1}-\hat V_n)\hat \sigma(t_n{-}\varepsilon)\}$. 
More explicitly, using that $\hat V_{n+1}$ and $\hat V_n$ respectively involve ancillas $n$ and $n+1$, we get
\begin{equation}
	\delta W_{\rm sw}= {\rm Tr}_{S,n+1}\,\{{\hat V_{n+1}\,\rho_{n}\eta_{n+1}}\}-{\rm Tr}_{S,n}\,\{{\hat V_{n}\,\varrho_{Sn}}\}\,,\label{Wsw}
\end{equation}
where $\varrho_{Sn}$ (\cf  \cref{section-EMs}) is the joint state of $S$ and ancilla $n$ \textit{right after} they collided with one another.\marginnote{Note that states must be continuous functions of time, hence in particular $\sigma(t_n{-}\varepsilon)=\sigma(t_n)=\sigma_n$.}

\section{First law of thermodynamics}
\label{section-1st}

During a \textit{single} collision, the dynamics of $S$ and the involved ancilla is governed by the total Hamiltonian
\begin{equation}
	\hat H_{\rm coll}=\hat H_S+\hat H_n+\hat V_n\,.\label{HSn-coll}
\end{equation}
Since operators $\hat H_n$ and $\hat V_n$ are time-independent, $\Delta \hat H_{\rm coll}=\Delta \hat H_S$. Making now the replacements $\langle \Delta \hat H_{\rm coll}\rangle=\Delta E_S+\Delta E_n+ {\rm Tr}_{S,n}\,\{{\hat V_{n}\,\de\varrho_{S,n}}\}$ and [\cf\eq\ref{work-2}] $\langle\Delta \hat H_S\rangle=W_d$, we get
\begin{equation}
	\Delta E_S-\delta Q+ {\rm Tr}_{S,n}\,\{{\hat V_{n}\,\de\varrho_{Sn}}\}=\delta W_d\,,\label{1stlaw-coll}
\end{equation}
where we used $\langle\Delta \hat H_{\rm coll}\rangle=\langle\Delta \hat H_S\rangle=W_d$ and $\Delta E_n=-\delta Q$ [\cf\eq\ref{dQ}].

To connect the last identity with the switching work, in \eq\ref{Wsw} we replace $\varrho_{Sn}=\rho_{n-1}\eta_n+\Delta \varrho_{Sn}$ obtaining
\begin{equation}
	\delta W_{\rm sw}= {\rm Tr}_{S,n+1}\,\{\hat V_{n+1}\,\rho_n\eta_{n+1}\}- {\rm Tr}_{S n}\,\{\hat V_{n}\,\rho_{n-1}\eta_{n}\}- {\rm Tr}_{S,n}\,\{{\hat V_{n}\,\Delta\varrho_{Sn}}\}\,.\label{Wsw2}
\end{equation}
Combining this with \ref{1stlaw-coll} and recalling the definition of $\hat H'_S$ [\cf\eq\ref{HS-diss}] and total work \ref{dw2}, we finally end up with the \textit{1st law} of thermodynamics
\begin{equation}
	\Delta E_S+\Delta E'_S=\delta Q+\delta W\,,\label{1stlaw}
\end{equation}
where $\Delta E_S+\Delta E'_S$ can be identified as the total energy change of $S$ when also the bath-induced Hamiltonian $\hat H'_S$ is accounted for. 

The analogous law for instantaneous rates/fluxes (in the continuous-time limit) reads $\dot E_S+\dot  E'_S=\dot  Q+\dot  W$.

An important case occurs for energy-conserving interactions [see  \cref{section-balance} and \cref{EC}, \ref{HSn-coll}]. In this case, in the absence of drive \ie for $\delta W_d=0$, we get $[\hat H_S,\hat H_{\rm coll}]=-[\hat H_n,\hat H_{\rm coll}]$. Hence,
\begin{equation}
	\Delta E_S=\delta Q\,.\label{ESQ}
\end{equation}
This formalizes energy conservation in thermodynamic terms: energy lost (gained) by $S$ is absorbed from (released to) the bath of ancillas in the form of heat. Note that \ref{1stlaw} in this case reduces to $\Delta E'_S=\de W_{\rm sw}$, namely the work (done by some external agent) for switching on and off the interaction with ancillas is entirely converted into extra energy of $S$ which adds to $E_S$.\marginnote{This is reasonable since $\delta W_{\rm sw}$ is in fact the contribution to the change of $\langle \hat H'_S\rangle$ coming from a step dependence of operator $\hat H'_S$ [\cf\cref{HS-diss,work-2}].} This work yet vanishes for interaction Hamiltonians and ancilla states such that ${\rm Tr}_n\{ \hat V_n\, \eta_n\}=0$, as in  \cref{section-QT1}.

\section{Qubit coupled to baths of harmonic oscillators}
\label{qubit-sect}

To illustrate the thermodynamic quantities introduced so far and the 1st law, let us reconsider the CM of  \cref{section-QT1} when $S$ is a qubit and each ancilla a quantum harmonic oscillator. Since the interaction is energy conserving, \eq\ref{ESQ} holds. Moreover, $\delta W_d=0$ (no drive) and ${\rm Tr}_n\{ \hat V_n\, \eta_n\}=0$, hence $\Delta E_S'=0$. Consistently, the switching work vanishes since, using \eq\ref{1stlaw}, $W_{\rm sw}=\Delta E_S{-}\de Q=0$. Thus overall no work is performed. Hence, in this thermalization process, \eq\ref{ESQ} coincides with the 1st law.

Using \eq\ref{dOnqt} or the opposite of \ref{dOSqt}, after simple calculations we get that in the continuous-time limit the heat flux is given by\marginnote{First-order terms vanish since in this case $[ {\hat B}_\nu,\hat H_{n}]\propto \hat b_n, \hat b_n^\dag$ whose expectation value on $\eta_n$ is zero. We also used $\vert 0\rangle _S\langle 0\rvert   =\hat \sigma_- \hat \sigma_+=\mathbb{I}_S-\hat \sigma_+ \hat \sigma_-$ and $[\hat b_n,\hat b_n^\dag]=1$.}
\begin{equation}
	\frac{dQ}{dt}=  \omega_0\,\left[\gamma_+\,(1-p)-\gamma_- p\,\right]\,.\label{hf1}   
\end{equation}
where the introduced the excited-state probability $p=\langle \hat \sigma_+ \hat \sigma_-\rangle=1-\langle \hat \sigma_- \hat \sigma_+\rangle$ and the previously defined rates \ref{rates-ea1}. 
Using \eq\ref{dp-dc}, we get as expected that $\dot Q=\omega_0 \dot p=\dot E_S$. We see that the heat flux undergoes an exponential decay (in magnitude) until it stops when $S$ reaches thermal equilibrium.

Next, as in the beginning of  \cref{section-balance}, we add a detuning to $S$ such that $\hat H_S=(\omega_0+\delta)\hat\sigma_{+}\hat\sigma_{-}$. As the evolution of $p$ is just the same, heat flux \ref{hf1} is identical. However, since now $[\hat H_S,\hat H_n]\neq 0$, $\dot Q$ no longer matches $\dot E_S$. Indeed, applying \eq\ref{dOSqt} in the continuous-time limit yields
\begin{equation}
	\frac{d E_S}{d t}=(\omega_0+\delta)\left[\gamma_+\,(1-p)-\gamma_-p\right]\,.\label{daaa}   
\end{equation}
Upon comparison with \ref{hf1}, this shows that $\dot E_S$ differs from $\dot Q$ whenever $\delta\neq 0$. Their difference, using the 1st law \ref{1stlaw} and $\hat H'_S=0$, is the switching work per unit time
\begin{equation}
	\dot W_{\rm sw}=\dot E_S-\dot Q=\delta\left[\gamma_+\,(1-p)-\gamma_-p\right]\,.
\end{equation}
This provides the complete energy balance at each instant, showing that in order for $S$ to reach the asymptotic state work must be performed by an external agent.

Note that, in the situation just analyzed, $\dot p=0$ entails $\dot Q=\dot E_S=\dot W_{\rm sw}=0$, meaning that no energy flux occurs throughout the system once the steady state is reached. This is true regardless of $\delta$ since $\gamma_+(1-p)-\gamma_- p=\dot p$.

Differently from the case just seen, let us now illustrate an instance featuring an uninterrupted heat flux. This is the dynamics of  \cref{section-QT1T2} featuring system $S$ is contact with two baths at different temperatures, in which case (as explained at that time) the open dynamics of $S$ is formally the same (so that $S$ reaches a steady state) except that the absorption and emission rates are replaced by $\gamma_\pm\rightarrow \gamma'_\pm$ [\cf\cref{gpp,rates-ea}]. Thus in particular
\begin{equation}
	\dot p= \gamma'_+(1-p)-\gamma'_- p\,,\label{dotpp}
\end{equation}

Accordingly,
\begin{equation}
	\frac{dE_S}{dt}=  \omega_0\,\left[\gamma'_+\,(1-p)-\gamma'_- p\,\right]\,\label{hf5}\,.
\end{equation}
Instead, the heat flux of bath $i$ [\cf\eq\ref{dOnqt}] is given by
\begin{equation}
	\frac{dQ_i}{dt}= \omega_0\,\left[\gamma^{(i)}_+\,(1-p)-\gamma^{(i)}_- p\,\right]\,\label{hf8}\,.
\end{equation}
Since $\gamma'_\pm=\gamma_\pm^{(1)}\pm \gamma_\pm^{(2)}$ we get the energy balance
\begin{equation}
	\frac{dQ_1}{dt}+\frac{dQ_2}{dt}= \frac{dE_S}{dt}
\end{equation}
(the switching work vanishes). This embodies a continuity equation for heat [see \cref{fig-balance}].

\begin{figure}[!h] 
	\raggedright
	\begin{floatrow}[1]
		\ffigbox[\FBwidth]{\caption[Stationary heat flux in a CM with two baths]{\textit{Stationary heat flux in a CM with two baths.} System $S$ collides with two baths of thermal ancillas, one at temperature $T_1$ one at $T_2$ with $T_1\neq\ T_2$. In general, the continuity equation for heat current reads $\dot Q_1+\dot Q_2=\dot E_S$, meaning that the net energy entering/exiting from the dashed region must balance the change of energy of $S$. As stationary conditions are reached, the energy of $S$ no longer changes and a permanent heat current $\dot Q_1=-\dot Q_2$ flows from the hot to the cold bath.}\label{fig-balance}}%
		{\includegraphics[width=\textwidth]{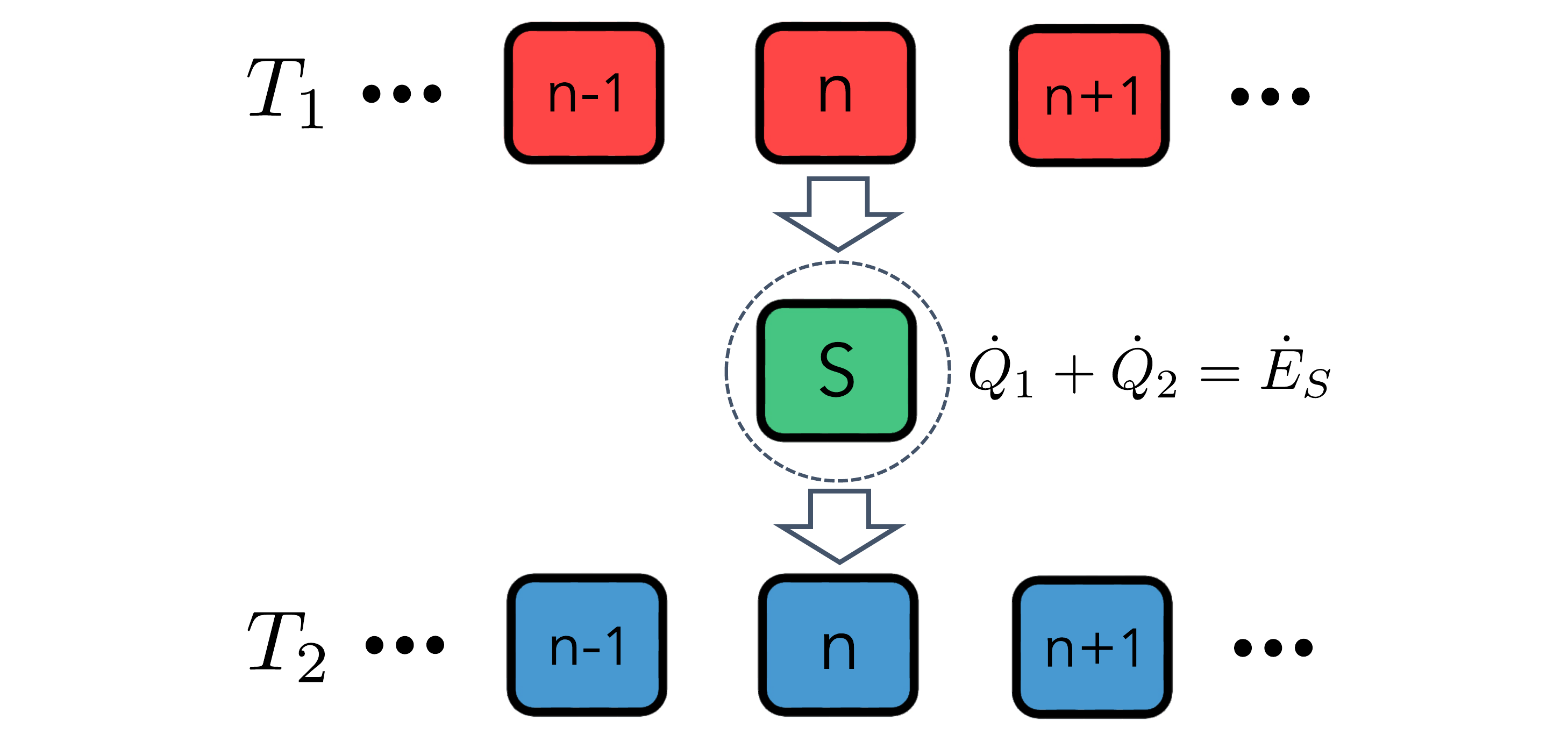}}
	\end{floatrow}
\end{figure}

Asymptotically, $\dot E_S=0$ so that
\begin{equation}
	\frac{dQ_1}{dt}=-\frac{dQ_2}{dt}\,\label{hf3}\,,
\end{equation}
showing that stationary heat current flows from one bath to the other. In these conditions, by deducing from \ref{dotpp} the steady value of $p$ and using  \ref{gpp}--\ref{rates-ea}, we get the heat current
\begin{equation}
	\frac{dQ_1}{dt}=-\frac{dQ_2}{dt}=\frac{\gamma_1\gamma_2\,({\bar{n}_1}-{\bar n}_2)}{\gamma_1+\gamma_2+2(\gamma_1 {\bar n}_1+\gamma_2 {\bar n}_2)}\,.
\end{equation}
As expected, for ${\bar n}_1>{\bar n}_2$ that is $T_1>T_2$, $\dot Q_1>0$ meaning that heat flows from bath 1 towards bath 2.

\section{Second law of thermodynamics}
\label{section-2nd}

Each collision changes the joint state of $S$ and the involved ancilla, which evolves from $\rho_{n-1}\eta_n$ (uncorrelated) to $\varrho_{Sn}$ (generally correlated). The relative entropy of these two states, which we call \textit{entropy production} $\Sigma$ for reasons that will be clear shortly, fulfills
\begin{equation}
	\Sigma ={\mathcal{S}}(\varrho_{Sn}\parallel \rho_{n-1}\eta_n)\ge 0\,,
\end{equation} 
which simply follows from the property that relative entropy is always non-negative (see \ref{app-ent}).
If, due to the interaction during the collision, $\varrho_{Sn}$ is a correlated state then it must be different from the initial state $\rho_{n-1}\otimes\eta_n$, entailing $\Sigma>0$. 
Thus the strict positivity of $\Sigma $ witnesses establishment of system--ancilla correlations at each collision.

It can be shown\footnote{This is worked out as
	\begin{align}
		{\mathcal{S}}(\varrho_{Sn}\parallel \rho_n\otimes\eta_n)=&
		{\rm Tr}\{\varrho_{Sn}\log\varrho_{Sn}\}-
		{\rm Tr}\{\varrho_{Sn}\log\rho_n\otimes\eta_n\}=\nonumber\\&
		{\rm Tr}\{\varrho_{Sn}\log\varrho_{Sn}\}-
		{\rm Tr}\{\rho_n\log\rho_n\}-
		{\rm Tr}\{\eta'_n\log\eta_n\}=\nonumber\\&
		{\rm Tr}\{\varrho_{Sn}\log\varrho_{Sn}\}-
		{\rm Tr}\{\rho_n\log\rho_n\}-
		{\rm Tr}\{\eta'_n\log\eta_n\}\nonumber\,.
	\end{align}
	Now, adding and subtracting ${\rm Tr}\{\eta'_n\log\eta'_n\}$ yields
	${\mathcal{S}}(\varrho_{Sn}\parallel \rho_n\otimes\eta_n)=
	{\mathcal{I}}\{\varrho_{Sn}\}-{\rm Tr}\{\eta'_n\log\eta_n\}+{\rm Tr}\{\eta'_n\log\eta'_n\}=
	{\mathcal{I}}\{\varrho_{Sn}\}+{\mathcal{S}}(\eta'_n\parallel \eta_n)\ge 0\,.$
} that $\Sigma$ can be split into the two contributions
\begin{equation}
	\Sigma ={\mathcal{I}}_{Sn}+{\mathcal{S}}(\eta'_n\parallel \eta_n)\ge 0\,,\label{Sig2}
\end{equation} 
where ${\mathcal{I}}_{Sn}$ stands for the mutual information (see \ref{app-ent}) of $S$ and ancilla $n$ at the end of the collision, while ${\mathcal{S}}(\eta'_n\parallel \eta_n)$ is the relative entropy (see \ref{app-ent}) between the final and initial states of the ancilla. Now, since $S$ and $n$ are initially uncorrelated (mutual information zero), we have
\begin{equation}
	{\mathcal{I}}_{Sn}=\Delta {\mathcal{S}}_S+\Delta {\mathcal{S}}_n\,\label{Isn}
\end{equation}
with $\Delta {\mathcal{S}}_S={\mathcal{S}}(\rho_n)-{\mathcal{S}}(\rho_{n-1})$ and $\Delta {\mathcal{S}}_n={\mathcal{S}}(\eta'_n)-{\mathcal{S}}(\eta_{n})$ the change of entropy of $S$ and ancilla, respectively.
Here, we used that the $S$-$n$ dynamics during the collision is globally unitary, hence it cannot change the entropy of the joint state, \ie $\Delta {\mathcal{S}}_{Sn}={\mathcal{S}}(\varrho_{Sn})-{\mathcal{S}}(\rho_{n-1}\eta_{n})=0$.

While the above holds for any ancilla state $\eta_n$, we now focus on a thermal bath of ancillas, \ie we take $\eta_n=\eta_{\rm th}$ [\cf\eq\ref{eta-th}]. In this case, recalling \eq\ref{Srel}, the second term of \ref{Sig2} is given by,\marginnote{Inside the trace, we added and subtracted a term $\eta_{\rm th} \log \eta_{\rm th}$ and used ${\rm Tr}_n \{\de \eta_{n}\}=0$ (since the state of ancilla of course remains normalized). We finally used $\delta Q=-{\rm Tr}_n\{\hat H_n (\eta'_n-\eta_{\rm th})\}$.}
\begin{equation}
	{\mathcal{S}}(\eta'_n\parallel \eta_{\rm th})=-\VNS{\eta'_n\,\log{\eta_{\rm th}}-\eta'_n \log \eta'_n}=-\de {\mathcal{S}}_n-\beta \delta Q\,.\label{See}
\end{equation} 
Replacing \cref{Isn,See} in \eq\ref{Sig2}, we end up with the 2nd law in the form
\begin{equation}
	\de {\mathcal{S}}_S\ge \beta\, \delta Q\label{2ndlaw}\,,
\end{equation}
which in terms of instantaneous rates (in the continuous-time limit) reads $\dot{\mathcal{S}}_S\ge \beta\, \dot Q$.

In particular, note that we get an identity connecting a thermodynamic quantity to an information-theoretical one. Hence, production of entropy in (the thermodynamics sense) results from creation of system--ancilla correlations as well as perturbation of the ancilla thermal state (caused by the interaction with $S$).

We point out that the above derivation of the 2nd law for each time \textit{step} relies crucially on having used a CM, this allowing to decompose the bath into distinct uncorrelated units which $S$ interacts with one at a time. In particular, we exploited that $S$ at each step is initially uncorrelated with the involved ancilla and this is still in the respective thermal state. The analogue of \eq\ref{Sig2} for the entire bath $B$ holds only if it is referred to the entire evolution up to the considered step (\ie replacing $t_{n-1}\rightarrow t_{0}$). This is because $S$ is uncorrelated with all the ancillas and these are all in a thermal state only at the initial time $t=t_0$ [see  \cref{fig-CMdyn}(a) and (d)]. From this viewpoint, it is remarkable that we got inequality \ref{2ndlaw} connecting the entropy change of system $S$ with the heat exchanged with the full bath $B$. This highlights particularly well a major advantage of employing a collisional description of non-equilibrium processes.

\section{Landauer's principle}
\label{section-Land}

Let us define $\tilde {\mathcal{S}}=-{\mathcal{S}}$ and $\tilde Q=-Q$ in a way that $\de\tilde {\mathcal{S}}$ represents the decrease of entropy while $\tilde Q>0$ is positive when heat flows from $S$ to $B$. Then \ref{2ndlaw} yields
\begin{equation}
	\beta \,\delta \tilde Q\ge  \de  \tilde{\mathcal{S}}\,.\label{land}
\end{equation}
This is the quantum version of the so called Landauer's
principle~\cite{Landauer61}, stating that the heat dissipated into the bath is
lower-bounded by the entropy decrease of system $S$. It entails that, in
order to decrease the entropy of the open system so as to gain more information about
it (see \ref{app-ent}), a finite amount of heat must be dissipated into the
reservoir. In the continuous-time limit, the corresponding statement in terms of heat
flux and instantaneous entropy decrease per unit time reads $\beta \dot Q\ge \dot {\tilde{\mathcal{S}}}$.

As an illustration, consider once again the CM analyzed at the beginning of  \cref{section-2nd}. The dissipated heat per unit time is given by the opposite of \ref{hf1}. The entropy  instead reads $S_S=- (1-p)\log p-p \log p$ (we assume zero coherences $c$ for simplicity). Hence, $\dot{\tilde S}_S=-\dot p \log \left(\tfrac{1-p}{p}\right)$ and we get
\begin{equation}
	\beta \,\dot{\tilde Q}-\dot {\tilde{ S}}=\gamma\left(\beta \omega_0+\log\frac{p}{1-p}\right)\bigl((1+e^{\beta\omega_0})p-1\bigr)\,\,.
\end{equation}
Both factors between brackets on the right hand side change their sign when $p$ becomes greater than $1/(1+e^{\beta\omega_0})$, meaning that the product is indeed non-negative at any time $t$.

\section{Non-equilibrium quantum thermodynamics: state of the art}
\label{se-soa-thermo}

The definition of thermodynamic quantities and derivation of thermodynamics laws
are largely based on
\rrefs~\cite{barra_thermodynamic_2015,Esp17,de2018reconciliation} (see also
\rref~\cite{kosloff2019quantum} where some aspects concerning the use of CMs in
quantum thermodynamics are discussed). {Note that \eq\ref{Sig2} was first derived
	for bath thermal states in \rref~\cite{Esposito_2010} and then generalized in
	\rrefs~\cite{Esp17,ManzanoPRX18}.}

We present next an overview of the quantum thermodynamics literature focusing on works that make explicit use of a collisional approach (our concern being mostly the methodological relevance for CMs theory).

The use of a CM to gain insight into the thermalization of a quantum system (see
Sections \cref{section-QT1}, \cref{section-harm-th},\cref{section-balance})
appeared in a seminal work published in 2002~\cite{scarani_thermalizing_2002}
(related to \rref~\cite{ziman_diluting_2002} mentioned in  \ref{soa-def}). This
linked together dissipation, fluctuations (by deriving a CM-based version of the
fluctuation--dissipation theorem~\cite{landau1980statistical}) and maximal
system--ancilla entangling power. Notably, the CM approach allowed the authors to
explicitly show how, due to entanglement, a dissipative (thus irreversible) process
can result from a jointly unitary system--bath dynamics (see also
\rref~\cite{scarani_entanglement_2007}). {Roughly in the same period, a similar CM
	was used by Diosi, Feldmann and Kosloff~\cite{diosi_exact_2006}, where however the
	joint dynamics is made irreversible by randomizing identities of the ancillas.}

Deviations from thermalization, in particular because of lack of energy-conserving
interactions (see  \cref{section-balance}), were investigated in
\rrefs~\cite{chimonidou_relaxation_2008,GrimmerBomb,GrimmerMischief}.

In the context of resource theories, \rref~\cite{Lostaglio2018elementarythermal}
introduced a resource theory called ``elementary thermalization operations" (ETOs)
and showed that Markovian ETOs are closely linked to memoryless CMs.
\rref~\cite{Almost_thermalPRA} instead studied almost thermal operations by relaxing
the constraint of having identical ancillas all in the same thermal state.

Since only the reduced state of $S$ is involved in the definition of
thermalization, an interesting question is whether or not $S$ can share
correlations with the ancillas even after reaching thermalization. Strong evidence
that $S$ gets asymptotically uncorrelated with the bath was provided in
\rref~\cite{cusumanoPRA18}. 

Note that not only a CM can model thermal baths, but can even implement an
effective thermometer as proposed in \rrefs~\cite{ThermometryPRL,ThermometryPRA}
showing that collective measurements on the ancillas can provide quantum metrological
advantages (an extension to stochastic collisions has been recently put forward in
\rref~\cite{o2021stochastic}) .

A class of problems where the collisional approach is very helpful are
non-equilibrium dynamics in the presence of multiple, usually thermal, baths (see \cref{section-multiB,section-QT1T2}). A standard case typically features a
multipartite open system $S$ [\cf \cref{fig-BB}(b)] comprising a
generally large number of subsystems $\{S_1, \ldots ,S_N\}$ which are coupled to one another
(modeled \eg as a spin
chain)~\cite{de2018reconciliation,karevski_quantum_2009,LandiPRE2014,barra_thermodynamic_2015,PereiraChainPRE,heineken2020quantum,ShaoPRE18}.
Note that switching work (see  \cref{section-work}) was first identified in a
system of this kind by Barra in \rref~\cite{barra_thermodynamic_2015} and then
further investigated in \rrefs~\cite{Esp17,de2018reconciliation}.

As seen in  \cref{section-multiB}, uncorrelated multiple baths typically result in MEs of the form \ref{addi} featuring only local dissipators. 
The thermodynamic consistency of such \textit{local MEs} (regardless of the way
they are derived) was disputed~\cite{levy2014local}. In this context,
\rref~\cite{de2018reconciliation} considered a CM with multiple baths and coupled
subsystems yielding a local ME. By highlighting the key role of switching work (see
\cref{section-work}), full consistency with both laws of thermodynamics (see Sections
\Cref{section-1st,section-2nd}) was demonstrated.

Note that, while the baths are commonly assumed to be uncorrelated,
\rref~\cite{de2020quantum} studied how inter-bath correlations affect thermal machine
performances. This corresponds to a CM with multiple baths  where in
\eq\ref{chi-corr} $\chi_n^{\rm corr}\neq 0$, resulting in ME terms that couple the subsystems
to one another [\cf\eq\ref{jump-corr}]. The corresponding ME can then be arranged
in terms of collective jump operators as first demonstrated in
\rref~\cite{daryanoosh2018quantum}. The effect of correlated ancillas was also
recently studied in the derivation of quantum Onsager relations via a collision
model~\cite{OnurPRR21}.

Multiple baths naturally enter thermal machines (see next) as these usually operate between reservoirs at different temperatures.

In 2003, Scully, Zubairy, Agarwal and Walther~\cite{scully2003Science} proposed a
heat engine based on the micromaser setup of  Sections \cref{section-maser}
with the difference that each thermal atom is a three-level system featuring a nearly
two-fold-degenerate ground state (doublet). They showed that coherences stored in the
doublet can work as an added control parameter to extract work from a single heat
bath with some features unattainable by classical engines~\cite{scully2003Science}.
This established a paradigm of proposed engines/thermal machines whose working
principle exploits some genuine quantum property (such as
entanglement)~\cite{SunPRE2006,ShaoPRE14,KurizkiEntropy2016,TurpenkePRE16,BaragiolaPRE19,LutzEPL09,dag2019temperature,HardalSciRep2015}.

CMs have become a routine description tool to investigate thermal machines, mostly
in the quest for quantum-enhanced
performances~\cite{Uzdin_2014,StrunzThermoPRL,ManzanoPRE19,BarraPRL2,dechiaraPRE2020,PiccionePRA21}
and/or with the aim to explore quantum non-Markovian effects (see  \ref{soa-nm}).
Note in particular the possibility of using CMs to model processes with
\textit{partial} thermalization, which was investigated in
\rrefs~\cite{Baumer2019imperfect,quadeer2021work,Landi2stroke}.

A topical research line is investigating thermodynamics laws in the presence of
\textit{non-thermal} reservoirs, mostly motivated by the hope that bath in
non-classical states could enable improved thermodynamic performances.
\rref~\cite{LandiWeak19} considered a CM with each ancilla prepared in a thermal
state with added \textit{coherences} of the order of $\sim\sqrt{\de t}$ quite like state
\ref{superp} in  \cref{section-EMs}. A bound was derived demonstrating explicitly
that the consumption of bath quantum coherences can convert heat into work on
$S$. \rref~\cite{ManzanoNJP21} showed that coherences in the energy basis
can both enhance (or deteriorate in some cases) the performance of thermal machines
and let them operate in otherwise forbidden regimes. \rref~\cite{Rom_n_Ancheyta_2019}
showed that coherences in the bath can cause a thermalization to an apparent
temperature which could be spectroscopically inferred~\cite{Rom_n_Ancheyta_2019}.

A major class of bath quantum states with promising thermodynamic advantages are
\textit{squeezed states}. A broadband (white-noise) squeezed reservoir can be
simulated via a CM featuring identical harmonic oscillator ancillas each prepared in
the same one-mode squeezed state, which could be implemented through an array of beam
splitters as proposed in the 90s in \rref~\cite{MSKPRA95} (see also
\cref{section-gaussian}). Such scheme can be generalized by considering non-identical
ancillas each initially in a squeezed-thermal state (so as to encompass a thermal
reservoir as a special case). Baths of ancillas prepared in squeezed-thermal states
were used in \rrefs~\cite{ManzanoPRE2016,ManzanoSOLOPRE,manzano2020non}.

The collisional approach to the Landauer's bound for fluxes (see
\cref{section-Land})  was introduced in \rref~\cite{lorenzo_landauers_2015}, where a
major focus was exploring the bound when $S$ is part of a larger
multipartite system which causes deviations from the Markovian behavior. One of the
considered case studies was the cascaded configuration of
\Cref{section-cascaded,sec-cascME}, where the dependence of heat fluxes in the
transient regime on intra-system correlations was formerly studied in
\rref~\cite{lorenzo_fluxes}. We also note that, although not explicitly connected
with CMs, a pertinent basic reference on the Landauer's principle adopting the
language of quantum maps is a 2014 paper by Reeb and Wolf~\cite{Reeb_2014}.

An intensively investigated topic in quantum thermodynamics is the possibility to
define thermodynamic quantities and non-equilibrium laws at the level of single
\textit{quantum trajectories} (instead of unconditional dynamics as assumed
throughout the present section) in a way that the resulting thermodynamics acquires
an intrinsically stochastic nature {(see the recent review in
	\rref~\cite{manzano2021quantum}).} As discussed in  \cref{section-qtraj}, CMs are the
natural microscopic framework for describing quantum trajectories, which explains their
use as an advantageous tool in studies of \textit{stochastic quantum
	thermodynamics}~\cite{HorowitzPRE12,horowitz_entropy_2013,BarraPRE,ManzanoPRX18,StrasbergPRE2019,StrasbergPRL2019,Cresser_Crooks,landi2021informational}.

A major appeal of CMs in quantum thermodynamics (and beyond) is that they allow
relaxing the standard weak-coupling assumption and thus exploring the ``ultra-strong"
coupling regime where counter-rotating terms cannot be neglected as done \eg in
\rrefs~\cite{de2018reconciliation,benentiPRA2015,PlastinaPLA20,roman2020enhanced}.

CMs can be used to introduce decoherence for extending \textit{fluctuations
	theorems} to quantum \textit{non-unitary} transformations~\cite{Smith_2018}.

Although not discussed in  \cref{section-work}, the work on $S$ can be seen
as resulting from collisions with a bath of ancillas in the case that the unitary
collision is approximated to first order, resulting only in $\hat H'_S$
[\cf\Cref{drho2,HS-diss}]. This was used for proposing a definition of work
independent of the $S$ free Hamiltonian~\cite{StrunzWork20}

\setchapterpreamble[u]{\margintoc}    
\chapter{Non-Markovian collision models}
\label{section-NM}

So far we have been focusing on memoryless (\ie Markovian) CMs. Yet, an important application of CMs is the description of \textit{non-Markovian} (NM) dynamics. This will be the subject of the present section.

Corresponding to assumptions (1)--(3) (\cf  \ref{conds}) underpinning the basic, Markovian, CM (see  \ref{conds}), one can identify three main classes of NM extensions of CMs:
\begin{itemize}
	\item[(i)] CMs with added ancilla--ancilla collisions;
	\item[(ii)] CMs with initially-correlated ancillas;
	\item[(iii)] CMs with multiple collisions.
\end{itemize}
It is understood that each class relaxes the corresponding hypothesis in  \cref{section-def} without breaking the other two. Of course mixed cases relaxing two or all of the hypotheses are also possible, an instance being the so called \textit{composite} CMs (which will be introduced in  \cref{section-comp}) which have connections with both classes (1) and (3).

In the following, we introduce each of the above three classes discussing some related basic properties.

\section{Ancilla--ancilla collisions}
\label{section-aacu} 

Introducing ancilla--ancilla collisions is physically motivated since it natural to think that ancillas can generally interact with one another.
In its (arguably) simplest formulation (see  \ref{fig-AA}), such a CM is obtained from the basic CM of  \cref{section-def} by adding extra pairwise ancilla--ancilla (AA) collisions between system--ancilla (SA) collisions.
As sketched in  \ref{fig-AA}, the CM dynamics starts with a standard collision between $S$ and ancilla 1 (unitary $\hat{U}_1$). Then ancillas 1 and 2 collide together (unitary $\hat W_{12}$). This is followed by an SA collision between $S$ and ancilla 2 (unitary $\hat U_2$), then an AA collision 2--3, then $S$-3 and so on. As a key feature, AA collisions are interspersed with SA collisions: for instance, \textit{prior to} the collision with $S$, ancilla 2 interacts with ancilla $1$ (with which $S$ is correlated due to the previous collision). As a result of this AA collision, $S$ and ancilla 2 are thus already correlated \textit{before} collision $S$-2 starts. Hence, regarding the open dynamics of $S$, the second step (ending with $S$-$2$ collision) cannot be described by a CPT map and so cannot all the remaining steps. The CP-divisibility condition [\cf\eq\ref{DMCM-2}] thereby does not hold, making the dynamics non-Markovian.

Calling $\hat W_{n,n-1}$ the unitary describing the AA collision between ancillas $n-1$ and $n$, the joint $S$-$B$ dynamics is given by
\begin{equation}
	\sigma_n=\hat U'_{n} \cdots\hat U'_{2}\, \hat{ U}'_{1}\,\sigma_{0}\,(\hat{ U}'_{1})^\dag\,(\hat{ U}'_{2})^\dagger \cdots\,,(\hat U'_{n})^\dag\label{sigmaAA}
\end{equation}
with the step unitary $\hat U'_{n} $ defined as
\begin{equation}
	\hat U_n'=\hat U_{n} \hat W_{n,n-1}\,\,\,({\rm for}\,\,\,n\ge 2) \,,\,\,\,\,\,\,\hat U'_1=\hat U_1\,,\label{Unp}
\end{equation}
hence (except for $n=1$) $\hat U'_{n}$  describes an AA collision followed by a SA one.
This can be contrasted with \eq\ref{sigman} holding for a basic CM. As usual, we take as initial state $\sigma_{0}=\rho_{0}\otimes_n \eta_n$ featuring no correlations.

\begin{figure}[!h] 
	\raggedright
	\begin{floatrow}[1]
		\ffigbox[\FBwidth]{\caption[Non-Markovian collision model with ancilla--ancilla collisions]{\textit{Non-Markovian collision model with ancilla--ancilla collisions}. Just like the basic CM of  \ref{fig-CMdyn}, all ancillas are initially uncorrelated and in the first step $S$ collides with ancilla 1 (a), getting correlated with it (not shown here). Yet, before $S$ collides with $n=2$, ancillas 1 and 2 collide together (b). As a result of this AA collision, $S$, ancilla 1 and ancilla 2 are jointly correlated (c). Now, $S$ collides with 2 (d) with which it is however correlated already \textit{before} collision $S$-2 starts. Collisions with ancillas $m>3$ are obtained by iteration.}\label{fig-AA}}%
		{\includegraphics[width=\textwidth]{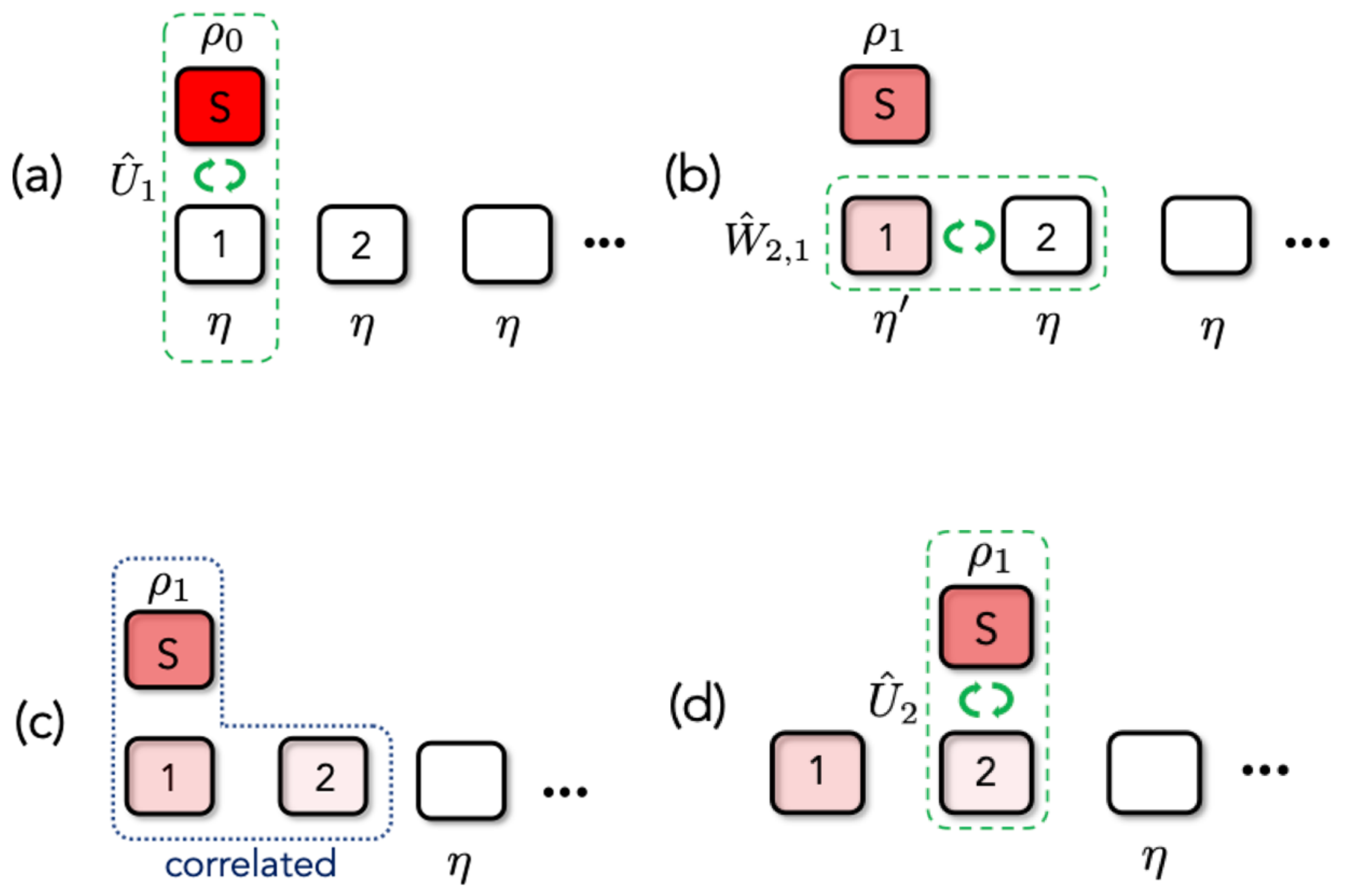}}
	\end{floatrow}
\end{figure}

To understand the main features of the open dynamics entailed by this CM, it is helpful to take each AA collision unitary in the form of a partial SWAP [\cf\eq\ref{p-swap}]
\begin{equation}
	\hat W_{n,n-1}=\sqrt{q}\,\,\mathbb{I}+\sqrt{p} \,\hat S_{n,n-1}\,\label{Wswap}
\end{equation}
with $q=1-p$, where we recall that unitary $\hat S_{n,n-1}\equiv \hat S_{n,n-1}^\dag$ [\cf\eq\ref{part-s}] swaps the states of ancillas $n-1$ and $n$. Here, the swap probability $p$ can be regarded as a measure of the effectiveness of AA collisions. 

For $p=0$, $\hat W_{n,n-1}=\mathbb{I}$, thus AA collisions are fully ineffective. We retrieve in this case the standard memoryless CM [\cf  \cref{section-def}] where $S$ undergoes the usual Markovian dynamics given by
\begin{equation}
	\rho_n={\mathcal{E}}^n [\rho_0]\label{EAA}
\end{equation}
with ${\mathcal{E}}$ the usual collision (CPT) map [\cf\eq\ref{coll-map}].

Let us now study the other extreme case $p=1$, when $\hat W_{n,n-1}$ is just a swap and AA collisions have the maximum effect. First note that the unitary transformation defined by $\hat S_{n,n-1}$ turns an operator acting on $S$ and $n$ into its analogue on $S$ an ancilla $n-1$
\begin{equation}
	\hat O_{S,n-1}=\hat S_{n,n-1}\hat O_{S,n}\hat S_{n,n-1}\,.\label{SOS}
\end{equation}
Using this and $\hat S_{2,1}\hat S_{2,1}=\mathbb{I}$, the overall unitary at step $n=2$ can be arranged as
\begin{equation}
	\hat U'_{2}\hat U'_1=\hat U_{2} \hat S_{2,1} \hat{ U}_{1}= (\hat S_{2,1}  \hat S_{2,1}) \,\hat U_{2} \hat S_{2,1} \hat{ U}_{1}= \hat S_{2,1}(  \hat S_{2,1} \hat U_{2} \hat S_{2,1}) \hat{ U}_{1}=\hat S_{2,1}\hat U_1^2\,,
\end{equation}
where we used that $\hat S_{2,1} \hat U_{2} \hat S_{2,1}=\hat U_{S,1}$ [due to \eq\ref{SOS}].

Upon iteration,\marginnote{In the case  $n=3$, we get $\hat U'_{3}\hat U'_{2}\hat U'_{1}=\hat U_{3} \hat S_{3,2} \hat U_{2} \hat S_{2,1} \hat{ U}_{1}=\hat U_{3} \hat S_{3,2}(\hat S_{2,1}\hat U_1^2)=\hat S_{2,1}\hat S_{3,2}\,\hat U_1^3$, where we used that $(\hat S_{2,1}\hat S_{3,2})\hat U_{3}(\hat S_{3,2}\hat S_{2,1}){=}\hat S_{2,1}\hat U_{2}\hat S_{2,1}=\hat U_1$ and $(\hat S_{2,1}\hat S_{3,2})^2=\mathbb{I}$.} at step $n$
\begin{equation}
	\hat U'_{n}\cdots\hat U'_{1}=\hat S_{2,1}\cdots\hat S_{n-1,n-2}\hat S_{n,n-1}\,\hat U_1^n\,.\label{U1nrho}
\end{equation}
Thereby, we get that the CM dynamics can be equivalently seen as the usual collision between $S$ and ancilla 1 yet repeated $n$ times, followed by a sequence of AA swaps. 
This property, along with the assumption that ancillas start all in the \textit{same} state $\eta_n$, allows to work out the evolution of $S$ as (see \ref{app-AA})
\begin{equation}
	\rho_n={\mathcal{F}}_n [\rho_{0}]={\rm Tr}_{1}\{\hat U_1^n\rho_{0}\,\eta_{1}\hat U_1^{\dag n}\}\,.\label{mapAA1}
\end{equation}
This can be contrasted with the case $p=0$ [see \eq\ref{EAA}] which, since $\eta_n$ is the same for all ancillas, can be written as
\begin{equation}
	\rho_n={\mathcal{E}}^n [\rho_{0}]\,\,\,\, {\rm with}\,\,\,\,{\mathcal{E}}[\rho]={\rm Tr}_1\{\hat U_1\, \rho \,\eta_1 \,\hat U_1^\dag\}\equiv {\mathcal{F}}_1[\rho]\,.\label{En3}
\end{equation}
Interestingly, from a formal viewpoint, maps \cref{mapAA1,En3} differ for the fact that, while in \ref{mapAA1} the exponentiation to power $n$ involves the collision \textit{unitary}, in \ref{En3} the exponentiation is instead over the collision \textit{map} (\ie the exponentiation is carried out after the partial trace). 

Physically, \eq\ref{mapAA1} describes just the same open dynamics which $S$ would undergo if it were interacting all the time with the \textit{same} ancilla.\marginnote{Indeed, $(\hat U_n)^n{=}(e^{-i \hat H_{\rm coll} \de t})^n{=}e^{-i \hat H_{\rm coll} t_n}$ (with $t_n=n\de t$ as usual).} Notably, adopting such a viewpoint, even \eq\ref{En3} could be seen as resulting from an everlasting interaction with the same ancilla, yet with the crucial difference that the ancilla state is \textit{periodically reset} to $\eta_1$ at each time step $\de t$.

The reason why, when $p=1$, the open dynamics effectively results from a non-stop interaction always with the same ancilla [\cf\ref{mapAA1}] is easily grasped. As pictured in  \ref{fig-swap}, at the end of collision $S$-1 [see  \ref{fig-swap}(a)], $S$ and ancillas 1--2 are in state $\varrho_{S,1}\otimes \eta_2$ with $\varrho_{S,1}$ a correlated state. Swap $\hat S_{2,1}$ is now applied [see  \ref{fig-swap}(b)], yielding
\begin{equation}
	\hat S_{2,1} \varrho_{S,1}\otimes \eta_2\hat S_{2,1}=\eta_1 \otimes  \varrho_{S,2}\,,\label{transfer}
\end{equation}
which transfers altogether the \textit{joint} $S$-1 state to $S$ and ancilla 2, while 1 returns to state $\eta$ uncorrelated with $S$ and 2 [see  \ref{fig-swap}(c)].
This entails that the $S$-2 collision [see  \ref{fig-swap}(d)] is seen by $S$ (open dynamics) just as if the collision with ancilla 1 resumed and then continued up to time $t=t_2$. 
We point out that, while the above \textit{in particular} implies that 1 and 2 swap their respective reduced states (during the AA collision), this alone would not be sufficient for \eq\ref{mapAA1} to hold. The transfer of system--ancilla correlations from $S$-1 to $S$-2 brought about by \ref{transfer} is thus essential. Analogous considerations apply at any step with $S$-$n$ correlations transferred to $S$ and ancilla $n+1$.

\begin{figure}[!h] 
	\raggedright
	\begin{floatrow}[1]
		\ffigbox[\FBwidth]{\caption[Fully swapping ancilla--ancilla collisions]{\textit{Fully swapping ancilla--ancilla collisions}. The unitary describing each AA collision is a full swap, $\hat W_{n,n-1}=\hat S_{n,n-1}$. At the end of the first SA collision (a), a swap is applied on ancillas 1 and 2 (b). Thereby, in particular, state $\eta'$, is transferred to ancilla $2$ with 1 thus returning to the initial state $\eta$ (c). Actually, it is the joint (correlated) state of $S$ and 1 which is transferred altogether to $S$ and 2 (c). Thus, in terms of {\textit open} dynamics, it is just as if the first SA collision resumed with the same ancilla, lasting a further time $\de t$ until $t=t_2$ (d).} \label{fig-swap}}%
		{\includegraphics[width=\textwidth]{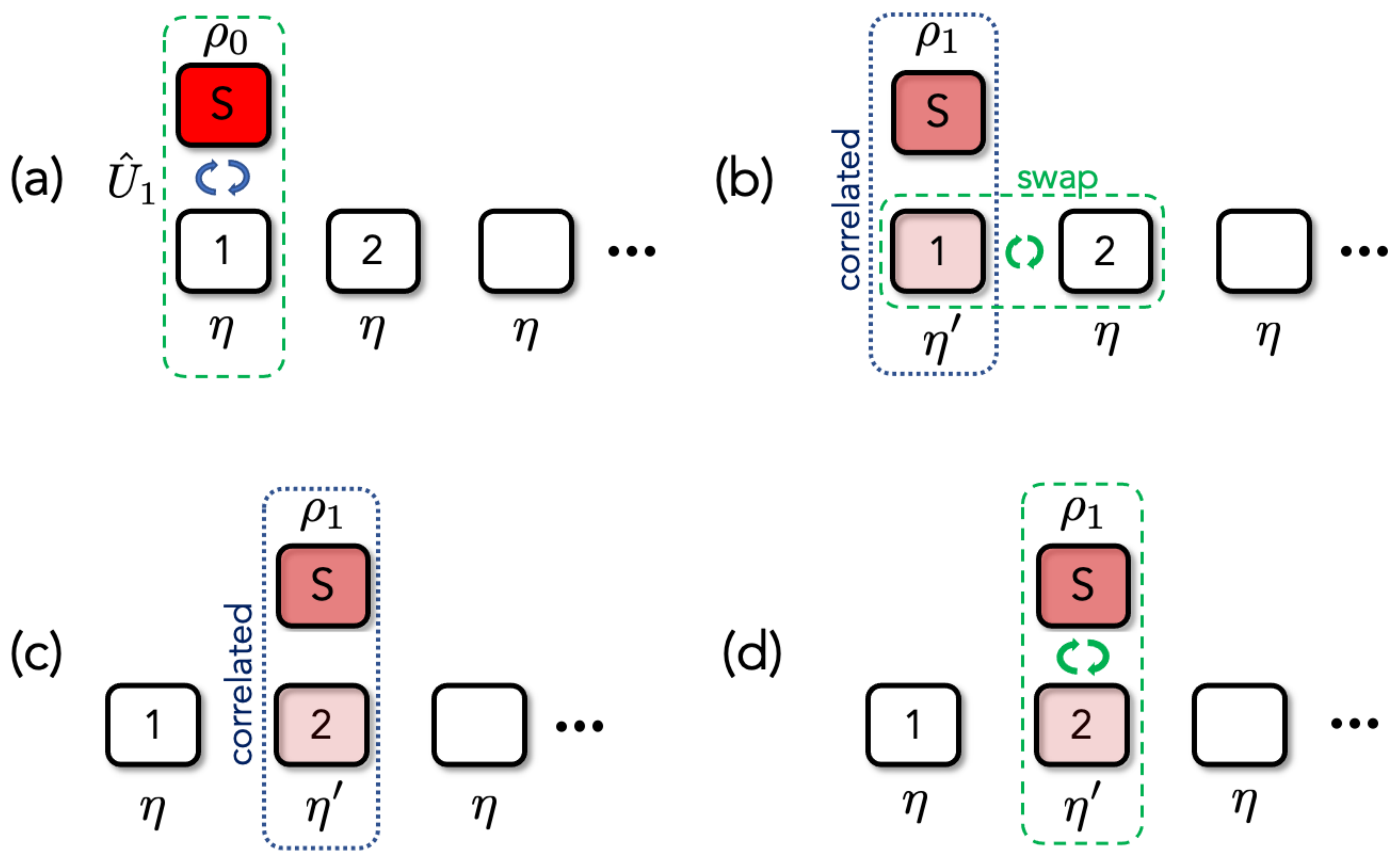}}
	\end{floatrow}
\end{figure}

It is worth stressing that the mapping into a continuous interaction with the same ancilla does not hold for the \textit{joint} dynamics.   
A major appeal of the CMs with ancilla--ancillas collisions as defined here is that the open dynamics can be analytically described, as will be shown in  \cref{section-comp} by connecting these models with composite CMs. Moreover, under an appropriate redefinition of AA collisions, one can even work out a closed ME for $S$ as discussed next.

\section{Non-Markovian master equation in the presence of ancilla--ancilla collisions}
\label{section-NMME}

In the previous CM when AA collisions are full swaps ($p=1$), the dynamics is strongly \textit{non-Markovian}. Formally, this is because there is no way of decomposing map \ref{mapAA1} into a sequence of CPT maps, one for each step, thus the CP divisibility condition \ref{DMCM-2} is not satisfied. To understand the physical reason behind NM behavior, 
think of a continuous coherent interaction between $S$ and another system $A$. If this dynamics were memoryless, the knowledge of the reduced state $\rho(t')$ at an intermediate time $t'$, such that $t_0<t'<t$, would be sufficient for determining the evolution of $\rho$ between $t'$ and $t$ (if the Hamiltonian is known). This cannot be the case as during the evolution the two systems are generally in a \textit{correlated} state $\varrho_{SA}(t)$ such that $\rho(t)={\rm Tr}_A\{\varrho_{SA}(t)\}$: knowing only $\rho(t)$ does not allow reconstructing the joint state $\varrho_{SA}(t)$.\marginnote{Note that the same statement applies to the dynamics of each single collision even for a basic memoryless CM. Yet, this lasts only a short time $\de t$, so that on a time scale far larger than $\de t$ the dynamics is Markovian.}

To sum up, if $p=0$, to get $\rho_n$ it is enough knowing the state of $S$ at the previous step and apply map ${\mathcal{E}}={\mathcal{F}}_1$, \ie $\rho_{n}={\mathcal{F}}_1[\rho_{n-1}]$. In contrast, if $p=1$, we need to know in which state $S$ ultimately started at $t=t_0$ and apply map ${\mathcal{F}}_n$, \ie $\rho_{n}={\mathcal{F}}_n[\rho_{0}]$. We might expect these two evolutions to be special cases of a recurrence rule, valid for any swap probability $p$, expressing $\rho_n$ generally in terms of $\rho_{0}$, $\rho_{1}$, \ldots, $\rho_{n-1}$ in a way that, as $p$ tends to 1, the number of previous steps which $\rho_n$ in fact depends on grows up. Unfortunately, it is not possible to work out such a closed relationship unless one introduces a little \textit{modification} in the CM, as shown next.

First of all, it is convenient to introduce a compact formalism for unitary operators and partial traces expressing them as quantum maps\marginnote{Despite we use the same symbol, map ${\mathcal{T}}_i$ here is different from map ${\mathcal{T}}_n$ introduced in  \ref{AD-section}.}
\begin{equation}
	\mathcal U[\sigma]=\hat U \sigma\hat U^\dag\,,\,\,\,\,{\mathcal T}_i [\sigma]={\rm Tr}_i [\sigma]\,,\label{UT-maps}
\end{equation}
where $i$ can be any subsystem of the joint system which state $\sigma$ generically refers to (here $\hat U$ is intended as a generic unitary). For instance, in terms of \ref{UT-maps}, the usual open dynamics of a basic CM of Section [\cf  \cref{section-def}] could be expressed as
\begin{equation}
	\rho_{n}={\mathcal{T}}_n\cdots{\mathcal{T}}_1\,\mathcal{U}_n\, \ldots\,\mathcal{U}_1[\sigma_{0}]={\mathcal{T}}_n\,\mathcal{U}_n\cdots{\mathcal{T}}_1\,\mathcal{U}_1[\sigma_{0}]\,\,\label{coll-maps}
\end{equation}
(any ${\mathcal{T}}_n$ commutes with ${\mathcal{U}}_{n'\neq n}$). When AA collisions are added, each $\mathcal{U}_m$ is replaced by $\mathcal{U}_m{\mathcal{W}}_{m,m-1}$.

It is immediate to check that an AA collision in the form of a partial swap [\cf\eq\ref{Wswap}] is described by the map
\begin{equation}
	{\mathcal{W}}_{n,n-1}[\sigma]=q\sigma+p\hat S_{n,n-1}\sigma\hat S_{n,n-1}+\sqrt{q p}\,\,[\sigma, \hat S_{n.n-1}]_+\,.\label{W1}
\end{equation}
The aforementioned modification of the CM with partial swaps consists in removing terms $\sim \sqrt{qp}$, namely we replace \ref{W1} with the new map
\begin{equation}
	{\mathcal{W}}_{n,n-1}=q\,{\mathcal{I}}+p \,{\mathcal{S}}_{n,n-1}\,\label{Winc}\,.
\end{equation}
This is a well-defined CPT map, having $\sqrt{q}\,{\mathbb I}$ and $\sqrt{p}\,\hat S_{n,n-1}$ as Kraus operators (see \ref{app-qmaps}).
Note that, while the removal of such terms affects the collisional dynamics, all the salient features discussed so far hold. In particular, map ${\mathcal{W}}_{n,n-1}$ swaps the states of ancillas with probability $p$ or leave them unchanged.

To get a closed ME for $\rho_{n}$, we note that the joint state at each step evolves as
\begin{equation}
	\sigma_{n}=\mathcal{U}_n\,({q\,\mathcal I}+p \,{\mathcal{S}}_{n,n-1})\,[\sigma_{n-1}]=q\,\,\mathcal{U}_n[\sigma_{n-1}]+p\, \,\mathcal{U}_n{\mathcal{S}}_{n,n-1}[\sigma_{n-1}] \label{rec5}
\end{equation}
for $n\ge2$ and $\sigma_1={\mathcal{U}}_1[\sigma_{0}]$. 

For $n=2$, we explicitly get $\sigma_{2}=q\,\mathcal{U}_2\,[\sigma_1]+p\, \,\mathcal{U}_2\,{\mathcal{S}}_{2,1}[\sigma_{1}]$. Replacing next $\sigma_1={\mathcal{U}}_1[\sigma_0]$ only in the second term yields
\begin{equation}
	\sigma_{2}=q\,\mathcal{U}_2\,[\sigma_1]+p\, \,\mathcal{U}_2\,{\mathcal{S}}_{2,1}\,{\mathcal{U}}_1[\sigma_{0}] =q\,\mathcal{U}_2\,[\sigma_1]+p \,\mathcal{U}_{2}^2[\sigma_{0}]\,,\label{step2}
\end{equation}
where we used the identity ${\mathcal{S}}_{n,n-1}\,\mathcal{U}_{n-1}={\mathcal{U}}_n{\mathcal{S}}_{n,n-1}$\marginnote{Using \eq\ref{SOS}, we get $\hat S_{n,n-1}\hat U_{n}\, \ldots \,\hat U_{n}^\dag\hat S_{n,n-1}=\hat S_{n,n-1}\hat U_{n}\hat S_{n,n-1}^2\, \ldots \,\hat S_{n,n-1}^2\hat U_{n}^\dag\hat S_{n,n-1}=\hat U_{n-1}\hat S_{n,n-1}\, \ldots \,\hat S_{n,n-1}\hat U_{n-1}^\dag$, proving the identity.} along with the invariance of the initial state $\sigma_{0}$ under any swap of ancillas (see  \cref{section-aacu}). 
Notably, \eq\ref{step2} is now arranged so as to feature only powers of ${\mathcal{U}}_2$. We can accomplish an analogous task at step $n=3$ starting from $\sigma_{3}=q\,\mathcal{U}_2\,[\sigma_2]+p\, \,\mathcal{U}_2\,{\mathcal{S}}_{2,1}[\sigma_{2}]$. Similarly to what done in the previous step, we replace $\sigma_2$ with \ref{step2} only in the second term, obtaining
\begin{align}
\sigma_{3}=&q\,\mathcal{U}_3[\sigma_2]+qp\, \mathcal{U}_3\mathcal{S}_{3,2}\mathcal{U}_2[\sigma_{1}]+p^2\mathcal{U}_3\mathcal{S}_{3,2} \mathcal{U}_{2}^2[\sigma_0]=\nonumber\\
&q\left(\mathcal{U}_3[\sigma_2]+p\,\mathcal{U}_3^2[\sigma_1]\right)+p^2 \mathcal{U}_{3}^3[\sigma_{0}],
\end{align}
which now features only powers of ${\mathcal{U}}_3$.

Upon induction, at the $n$th step we get
\begin{equation}
	\sigma_n= q\sum_{j=1}^{n\meno1}p^{j-1}{\mathcal{U}}_{n}^j\,[\sigma_{n\meno j}] \; \piu \;  p^{n-1} {\mathcal U_n}^n[\sigma_0]\,\,,\label{sigma-kernel}
\end{equation}
containing only powers of ${\mathcal{U}}_{n}$ (collision unitary corresponding to the last SA collision).
Note that the larger the power of ${\mathcal{U}}_{n}$ the older is the state it acts on. This property is remarkable since, given that ${\mathcal{U}}_{n}$ does not act on ancillas different from the $n$th one, the trace over all ancillas yields an equation formally analogous to \ref{sigma-kernel} with $\sigma_{n}$ replaced by $\rho_{n}$ and each power ${\mathcal{U}}_{n}^j$ by map ${\mathcal{F}}_j$ [\cf\eq\ref{mapAA1}]
\begin{equation}
	\rho_n\ug q\,\sum_{j=1}^{n\meno1}p^{j-1}{\mathcal{F}}_j\,[\rho_{n\meno j}] \; \piu \;  p^{n-1} {\mathcal{F}}_n[\rho_0]\,.\label{rho-kernel}
\end{equation}
As promised, we thus end up with a closed equation for the reduced state of $S$, which holds for arbitrary swap probability $p$. The corresponding dynamics interpolates between the memoryless case for $p=0$ and the strongly NM dynamics for $p=1$ [\cf\eq\ref{mapAA1}]. For arbitrary $p$, note that, due to the exponential weights $p^{j-1}$ and $p^{n-1}$, the current state is more affected by the latest steps. This formalizes the property that the system keeps \textit{memory} of its past evolution, whose memory length ranges from 1 (Markovian case occurring for $p=0$) to $n$ (strongly NM case occurring for $p=1$).

Most remarkably, by defining a memory rate $\Gamma$ through $p=e^{{-\Gamma \Delta t}}$ in a way that, for $\Delta t\ll \Gamma^{-1}$, $p\simeq 1-\Gamma \Delta t$, one can convert \eq\ref{rho-kernel} into a corresponding ME in the continuous-time limit (see  \ref{app-swap}) which reads
\begin{equation}
	\dot{\rho} =\Gamma\int_{0}^t dt'e^{-\Gamma t'} \;\mathcal{F}(t')\left[ \dot{\rho}(t{-}t')
	\right]+e^{-\Gamma t}\; \dot{\mathcal{F}}(t) [\rho_0]\;.\label{ME-kernel}
\end{equation}
Here, $\mathcal{F}(t)$ is the continuous-time version of map
\ref{mapAA1}.\marginnote{This is obtained by replacing in the definition
	[\cf\eq\ref{mapAA1}] $\hat U_n=(e^{-i \hat H_{\rm coll}\de t})^n=e^{-i \hat H_{\rm coll}t_n}$ with $\hat U(t)=e^{-i \hat H_{\rm coll}t}$ (with $t_n\rightarrow t$).}

This
kind of integro-differential non-Markovian MEs are called memory-kernel MEs.
Independently of its derivation as the continuous-time limit of an intrinsically CPT
discrete dynamics, it can be shown that \eq\ref{ME-kernel} correctly entails a
continuous-time CPT dynamics for any
$\Gamma>0$~\cite{ciccarello_collision-model-based_2013}.

\section{Initially-correlated ancillas}
\label{section-ica} 

Consider the basic CM of  \cref{section-def} where the initial state of the ancillas is generalized as
\begin{equation}
	\rho_B= \sum_{m=1}^M p_m \,\chi_{m}\,,\label{corr-state}
\end{equation}
where probabilities $\{p_m\}$ fulfill $\sum_{m=1}^Mp_m=1 $ while
\begin{equation}
	\chi_m=\eta_{m1}\otimes\eta_{m2}\otimes\, \ldots\,\,.
\end{equation}
Here, $\{\eta_{mn}\}$ are an arbitrary set of $M$ states of ancilla $n$.   
When all the $p_m$'s but one are zero, we recover the memoryless CM [\cf\eq\ref{sigma0}]. In the general case, however, \ref{corr-state} is a not a product state and thus describe initially \textit{correlated} ancillas [see panel (a) of  \ref{fig-ic-anc}].   
After $n$ collisions, the joint initial state $\sigma_{0}=\rho_{0}\otimes \rho_{B}$ evolves into (\cf\eq\ref{sigman})
\begin{equation}
	\sigma_n=\sum_{m=1}^M p_m\, \hat U_n \cdots \hat U_1 \,\rho_{0} \otimes \chi_{m}\,\hat U_1^\dagger \cdots\hat U_n^\dag\,.   
\end{equation}
The corresponding open dynamics of $S$ is given by
\begin{equation}
	\label{rhon-nm}
	\rho_n=\sum_{m=1}^M p_m\, {\rm Tr}_B\left\{\hat U_n \cdots \hat U_1 \,\rho_{0} \otimes \chi_{m}\,\hat U_1^\dagger \cdots\hat U_n^\dag\right\}=\sum_{m=1}^M p_m \,\Lambda_{mn}[\rho_{0}]\,,
\end{equation}
where
\begin{equation}
	\Lambda_{mn}=\left({{\mathcal E} _m}\right)^n\label{Ln}
\end{equation}
with the CPT map ${\mathcal{E}}_m$ defined by
\begin{equation}
	\rho'={\mathcal{E}}_m[\rho]={\rm Tr}_{n}\left\{\hat {U}_ {n}\rho\,\eta_{mn}\,\hat {U}_{ {n} }^\dag\right\}\,. \label{coll-map-m}
\end{equation}
The evolution is thus a \textit{mixture} of $M$ dynamics, each described by a dynamical map $\Lambda_{mn}$ (\cf  \ref{markovianity-section}) with associated collision map ${\mathcal{E}}_m$. As shown by \ref{Ln}, each $\Lambda_{mn}$ alone describes a fully Markovian collisional dynamics [\cf \eq\ref{DMCM}]. 

According to \ref{rhon-nm}, the dynamical map of the present collision model reads
\begin{equation}
	\Lambda_n=\sum_{m=1}^M p_m \,\Lambda_{mn}\,.
\end{equation} 
Remarkably, while each $\Lambda_{mn}$ can be divided into elementary CPT collision maps [\cf \eq\ref{Ln}] thus being Markovian [\cf  \ref{markovianity-section}] this is generally \textit{not} possible for $\Lambda_n$ despite it results from a seemingly innocent mixture of $\Lambda_{mn}$'s. This is best illustrated with a simple counterexample, which is discussed next.

\begin{figure}[!h] 
	\raggedright
	\begin{floatrow}[1]
		\ffigbox[\FBwidth]{\caption[Non-Markovian collision model with initially-correlated ancillas]{\textit{Non-Markovian collision model with initially-correlated ancillas}. Before interacting with $S$, ancillas are initially correlated with one other (a). Thereby, after colliding with ancilla 1, $S$ gets correlated with all the bath ancillas. Thus each collision (starting from the second one) is generally not described by a CPT map on $S$, making the dynamics non-Markovian.}\label{fig-ic-anc}}%
		{\includegraphics[width=\textwidth]{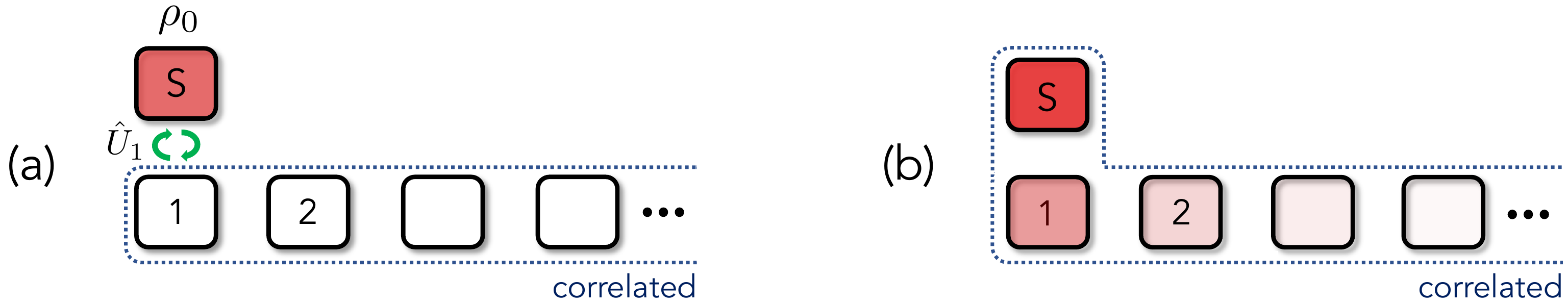}}
	\end{floatrow}
\end{figure}

Consider the all-qubit CM [see  \cref{section-all-qubit}]  with the ancillas starting in the correlated state
\begin{equation}
	\rho_B=p \ket{00\cdots}_B\bra{00\cdots}+q  \ket{11\cdots}_B\bra{11\cdots}\,,
\end{equation}
where $\ket{ii\cdots}=\otimes_n \ket{i}_n$ with $i=0,1$ and with $p=1-q$ a probability. Assuming that $S$ starts in state $|1\rangle_S$, at the end of the first collision the joint state reads
\begin{align}
	\sigma_1=&p\,\left(\hat U_{1}\ket{10}_{S1}\bra{01}\hat U_{1}^\dag\right)\,\ket{00\cdots}_{23\cdots}\bra{00\cdots}+\nonumber\\
	&q\,\left(\hat U_{1}\ket{11}_{S1}\bra{11}\hat U_{1}^\dag\right)\,\ket{11\cdots}_{23\cdots}\bra{11\cdots}\,.\label{jo}
\end{align}
Taking for simplicity $g_z=0$ [\cf \eq\ref{Vn-qubits}] and based on \ref{Un-qubits}, we have
\begin{align}
	&\hat U_{1}\ket{10}_{S1}= \cos(g\dt) \ket{1}_S\ket{0}_1  -i\sin(g\dt)\ket{0}_S\ket{1}_1 \nonumber \\
	&\hat U_{1}\ket{11}_{S1}=\ket{1}_S\ket{1}_1\,.
\end{align}
By replacing these in \ref{jo} and tracing over ancilla 1, we get the reduced state of $S$ and ancillas 2,3, \ldots
\begin{align}
	{\rm Tr}_{1}\{\sigma_{1}\}=&p\left(c^2 \ket{1}_S\bra{1}+s^2\ket{0}_S\bra{0}\right)\otimes\ket{00\cdots}_{23\cdots}\bra{00\cdots}+\nonumber\\
	&q\ket{1}_S\bra{1}\otimes\ket{11\cdots}_{23\cdots}\bra{11\cdots}\,\,,
\end{align}
where we set $c=\cos(g\dt)$ and $s=\sin(g\dt)$.

For $0<p<1$, this is a correlated state between $S$ and all ancillas 2,3, \ldots \,. This means that each collision starting from the second one is generally not described a CPT map. It follows that the overall dynamical map $\Lambda_{n}$ does not satisfy the CP-divisibility condition \ref{DMCM-2}, which witnesses the non-Markovian nature of the dynamics.

We note that, since a dynamics like \ref{coll-map-m} is a mixture of Markovian dynamics, if each of these admits a continuous-time limit then one can work out as many Lindblad master equations $\dot \rho_m= {\mathcal L}_m [\rho]$ having a form like \eq\ref{drho2}. Solving these, the overall dynamics then results from the mixture of the respective solutions $\rho(t)=\sum_m p_m \rho_m(t)$. Due to non-Markovianity, however, $\rho(t)$ generally cannot be expressed as the solution of a well-defined Lindblad master equation.

It is worth pointing out that, while state \ref{corr-state} is not entangled as
it is a mixture of product states,\footnote{Yet, one such state can still feature	non-classical correlations in the form of so called quantum discord~\cite{modiRMP2012}.} the essential conclusions on the non-Markovian nature
of the dynamics apply to entangled states as well as is for instance the case of
single-photon wavepackets to be discussed in  \cref{section-singleph}.

\section{Multiple collisions}
\label{section-multiple}

Another mechanism for introducing memory in a CM is allowing each ancilla to collide with $S$ at many distinct, non-consecutive, steps, instead of only one [as in the basic CM of  \cref{section-def}]]. 

\begin{figure}[!h] 
	\raggedright
	\begin{floatrow}[1]
		\ffigbox[\FBwidth]{\caption[Non-Markovian collision model with non-local collisions]{\textit{Non-Markovian collision model with non-local collisions}. Like in a basic CM, ancillas are non-interacting and initially uncorrelated. Yet, system $S$ interacts with the bath \textit{non-locally} in the following sense: at the $n$th step, $S$ collides at once with ancillas $n-d$ and $n$ (bi-local collision). As a major consequence, ancilla $n$ collides with the system \textit{twice}: the first time at step $n$ (a) and then again at step $n+d$ (b). Thus $d$ is the delay between the two collisions with the same ancilla. Before the second collision starts, ancilla $n$ and $S$ are already correlated so that the CP-divisibility condition \ref{SG-2} does not hold, making the dynamics NM. Notice that, until step $n=d{-}1$ (c), no memory effect can occur as each ancilla underwent at most one collision with $S$ [the dotted square in (c) is a phantom ancilla].}\label{fig-delay}}%
		{\includegraphics[width=\textwidth]{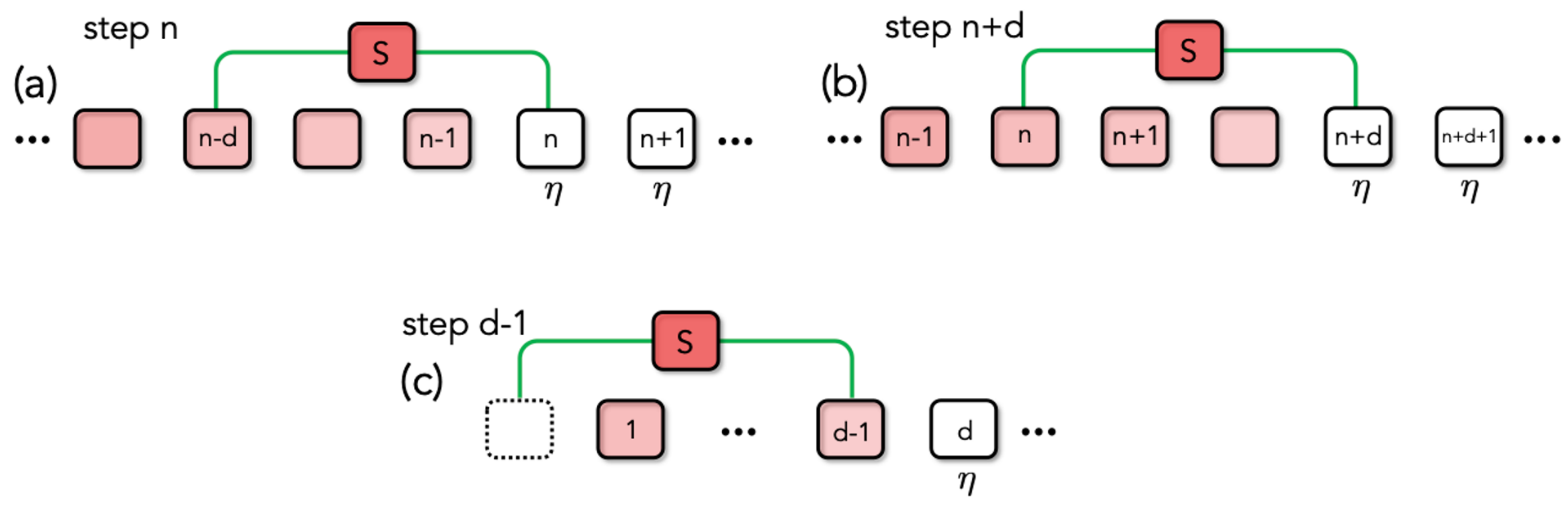}}
	\end{floatrow}
\end{figure}

A simple instance is a CM featuring a sequence of collisions like 
\begin{align}\hat U_1\,,\hat U_2\,,\hat U_3\,,\hat U_1\,,\hat U_4\,,\hat U_5\,,\hat U_1, \ldots \nonumber
\end{align}
such that $S$-1 collision takes places every three steps. 

While several possible multiple-collision schemes can be conceived (see also  \ref{soa-nm}), here we focus on CMs with \textit{non-local} collisions that naturally arise in quantum optics dynamics where delay times (light retardation) are non-negligible. The paradigm of such dynamics is shown in  \ref{fig-delay}: at each step, $S$ simultaneously collides with \textit{many} ancillas (two in the simplest case, as in the figure). More in detail, at the $n$th step, $S$ collides with ancillas $n-d$ and $d$, where $d$ is an integer such that $d\ge 1$ [see  \ref{fig-delay}(a)]. Accordingly, at step $n+d$, the collision will involve ancillas $n$ and $n+d$ [see  \ref{fig-delay}(b)]. It follows that a generic ancilla labeled with $n$ collides with $S$ \textit{twice}: the first time at step $n$ and then again at step $n+d$. The resulting collisional dynamics is evidently non-Markovian: before the second collision starts (step $n+d$), $S$ is already correlated with ancilla $n$ due to the first collision (step $n$), hence the dynamical map will generally not fulfill the CP divisibility condition [\cf\eq\ref{SG-2}].
As a paradigmatic, analytically solvable, instance consider the usual all-qubit model of \cref{section-all-qubit} with $\hat H_S=\hat H_n=0$ and the interaction Hamiltonian describing the $n$th collision now replaced by
\begin{equation}
	\hat V^{(n)}=\sqrt{\tfrac{\gamma}{\Delta t}}\,\,\hat\sigma_{+}\,(\hat \sigma_{n,-}+e^{i\phi}\hat \sigma_{n-d,-})+{\rm H.c.}\label{Vt5}\,
\end{equation}
Here and throughout the present subsection, superscript ``$(n)$" refers to the time step.\marginnote{Since more than one ancilla collides with the system at each step, we can longer use a common index for labeling the colliding ancilla and time step.} 
For completeness, we allowed for a phase shift $\phi$ between the couplings to the two ancillas. As initial state, we take $S$ in state $\ket{1}$ and each ancilla in state $\ket{0}$. Thus the joint initial state is $\ket{\Psi^{(0)}}=\ket{1}_S\otimes_m\ket{0}_m$. 

Defining the total number of excitations as $\hat N=\vert 1\rangle _S\langle 1\rvert  +\sum_m \ket{1}_m\langle 1\rvert  $, we note that this is conserved at all steps since $[\hat V^{(n)},\hat N]=0$. The eigenspace of $\hat N$ with eigenvalue $N=1$ (single-excitation sector) is spanned by $\ket{e_S}=\ket{1}_S\otimes_m\ket{0}_m$ (excitation on $S$) and $\ket{e_{m'}}=\ket{0_S}\otimes_{m\neq m'}\ket{0}_m \otimes \ket{1}_{m'}$ (excitation on ancilla $m'$). Thereby, since $\ket{\Psi^{(0)}}=\ket{e_S}$, the joint dynamics remains at all steps within the single-excitation sector. Accordingly, the joint state at step $n$ can be expanded as
\begin{equation}
	\ket{\Psi^{(n)}} = \alpha^{(n)}\ket{e_S} + \sum_m \lambda^{(n)}_m  \ket{e_m}\,.
	\label{st}
\end{equation}
In terms of excitation amplitudes $\alpha^{(n)}$ and $\lambda^{(n)}_m$, the initial state $\vert \Psi^{(0)}\rangle$ reads $\alpha^{(0)}=1$ and $\lambda_m^{(n)}=0$ for any $m$. For formal convenience, we will assume that the excitation amplitudes are defined also for negative values of the step index $n$, taking values $\alpha^{(n\le 0)}=1$, $\lambda_m^{(n\le 0)}=0$ (for any $m$).
The evolution of the joint state at each step reads $\ket{\Psi^{(n+1)}} =  \hat{U}_{n+1} \ket{\Psi^{(n)}}$\marginnote{Due to the specific type of calculations involved, in this subsection we define the generic step as the time interval between times $t_{n}$ and $t_{n+1}$, instead of $t_{n-1}$ and $t_{n}$ as usually done throughout the paper. This helps keeping notation relatively light.} where the collision unitary is defined by $\hat U_{n}=\exp[-i\hat V^{(n)}\de t]$ with $\hat V_n$ having the form \ref{Vt5}. 

At short enough  $\de t$ (we limit the analysis to this regime only), we expand $\hat U_{n+1}$ to the 2nd order in $\hat V^{(n)}$ and then apply it to \ref{st}. Projecting the resulting state on $\vert e_S\rangle$ yields a recurrence relation for amplitudes $\alpha^{(n)}$ and $\lambda_{m}^{(n)}$, which reads
\begin{equation}
	\alpha^{(n+1)}= \alpha^{(n)} -\gamma \Delta t \,\alpha^{(n)}-i \sqrt{\gamma\Delta t}  \,e^{i \phi} \lambda_{n-d}^{(n)} \,.
	\label{eps-delay}
\end{equation}
Here, we used that $\lambda_n^{(n)}=0$ since, at step $n$, the $n$th ancilla is still in the initial state $\ket{0}_n$ [see  \ref{fig-delay}(a)].
Our goal is expressing $\lambda_{n-d}^{(n)}$ in terms of $\{\alpha^{(n)}\}$ so as to end up with a closed equation for $\{\alpha^{(n)}\}$ (which fully describes the open dynamics). 

Considering first the case $n\ge d$, note that ancilla $n-d$ collides with $S$ the first time at step $n-d$ and then at step $n$. Thus the corresponding amplitude at step $n-d+1$ cannot change any more until step $n$
\begin{equation}
	\lambda_{n-d}^{(n)} =\lambda_{n-d}^{(n-1)}=\dots= \lambda_{n-d}^{(n-d+1)}\,\,.\label{chain}
\end{equation}
Amplitude $\lambda_{n-d}^{(n-d+1)}$ can be worked out, similarly to $\alpha^{(n+1)}$ in \eq\ref{eps-delay}, by applying the collision unitary $\hat U_{n-d+1}$ to $ \ket{\Psi^{(n-d)}}$ and projecting next to $\vert 1_{n-d}\rangle$. This yields
\begin{equation}
	\lambda_{n-d}^{(n-d+1)} =   -i\sqrt{\gamma\Delta t} \,\alpha^{(n-d)} +\tfrac{1}{2}\gamma \Delta t \,\lambda_{n-2d}^{(n-d)}\,.
\end{equation}
Due to \eq\ref{chain}, this coincides with $\lambda_{n-d}^{(n)}$ so that \eq\ref{eps-delay} becomes     
\[
\alpha^{(n+1)}-\alpha^{(n)}=-\gamma \Delta t\, \alpha^{(n)}-\gamma \Delta t \,e^{i \phi}\,\alpha^{(n-d)}\,,\]
where the term $\sim\lambda_{n-2d}^{(n-d)}$ was neglected being of order $\sim\Delta t^{3/2}$. We thus get
\begin{equation}
	\frac{\Delta \alpha^{(n)}}{\Delta t}=-\gamma \alpha^{(n)}-\gamma  e^{i \phi}\alpha^{(n-d)}\,\,\,\,\,\,\,\,{\rm for}\,\,\,n\ge d\,.\label{DDE-fin0}   
\end{equation}
where $\Delta \alpha^{(n)}=\alpha^{(n+1)}-\alpha^{(n)}$.

We are left with the case $0\le n< d$. For these values of $n$, \eq\ref{eps-delay} misses the last term because of the initial conditions [see below \eq\ref{st}], thus reducing to ${\Delta \alpha^{(n)}}/{\Delta t}=-\gamma \alpha^{(n)}$.

To sum up, we thus conclude that the dynamics of $S$ is governed by the finite-difference equation
\begin{equation}
	\label{DDE-fin}
	\frac{\Delta \alpha^{(n)}}{\Delta t}=
	\begin{cases}
		-\gamma\,\alpha^{(n)}\qquad\qquad\qquad\,\,\,\,\,{\rm for}\;\;\;\,\, n< d\ \\ 
		-\gamma\, \alpha^{(n)}-\gamma  e^{i \phi}\alpha^{(n-d)}\;\;\,\,{\rm for}\;\;\,\,\,\, n\ge d
	\end{cases}\,\,.
\end{equation}
We can understand this equation as follows. Until step $n=d-1$ [see  \ref{fig-delay}(c)], each ancilla undergoes at most one collision with $S$: in this initial stage, the dynamics is identical to a memoryless basic CM with $\alpha^{(n)}$ undergoing a standard exponential decay just like for spontaneous emission [\cf\eq\ref{ME-SE}]. Step $n=d$ is the first featuring an ancilla  undergoing a second collision with $S$ (this is ancilla $m=1$). From this step on, thereby, memory effects come into play as witnessed by the presence of term $\alpha^{(n-d)}$.

In the continuous-time limit, such that $t_n\rightarrow t$ and $d\Delta t\rightarrow \tau$ with $\tau$ a characteristic \textit{delay time}, \eq\ref{DDE-fin} is turned into\marginnote{This equation usually appears in the literature with phase  $\phi\rightarrow\phi+\pi$, \ie without the minus sign in the second term.}
\begin{equation}
	\dot{\alpha}=-\gamma  \alpha(t)-\gamma e^{i \phi}\alpha(t-\tau)\theta(t-\tau)\,,\label{DDE-c}
\end{equation}
which is a so called delay differential equation [here $\theta(x)=1$ for $x>0$ and $\theta(x)=0$ otherwise].

\section{Composite collision models}
\label{section-comp}

Besides the three non-Markovian generalizations of CM discussed so far, each constructed so as to directly break one of the assumptions (1)--(3) in  \ref{conds}, there is a further natural scheme to endow a CM with memory. 

\begin{figure}[!h] 
	\raggedright
	\begin{floatrow}[1]
		\ffigbox[\FBwidth]{\caption[Composite collision model]{\textit{Composite collision model}. The composite system $S$ [see panel (a)] is made out of subsystems ${\mathcal{S}}$ (the open system under study) and $M$ (``memory"). System $S$ undergoes collisions with the ancillas (just like in a memoryless CM) which however involve only subsystem $M$. Before each $M$-ancilla collision [see panel (c)], ${\mathcal{S}}$ and $M$ collide with one another [see (b)] through unitary $\hat U_{{\mathcal{S}}M}$, hence they are generally correlated. Due to these correlations, the open dynamics of ${\mathcal{S}}$ cannot be divided into a sequence of CPT maps, one for each step, and is thus non-Markovian. In contrast, the dynamics of $S$ (i.e, ${\mathcal{S}}$ plus $M$) is fully Markovian since no correlations with ancilla $n$ exists prior to the $\hat U_{Mn}$ collision.}\label{fig-comp}}%
		{\includegraphics[width=\textwidth]{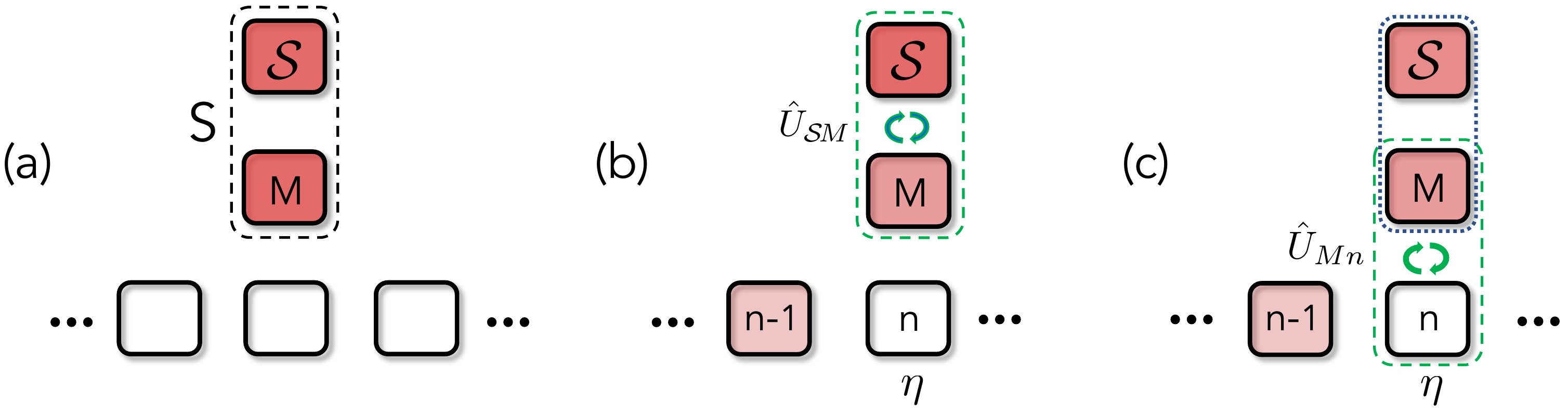}}
	\end{floatrow}
\end{figure}

Consider a memoryless CM where $S$ is {bipartite} as sketched in  \ref{fig-comp}(a). Its two susbsytems are ${\mathcal{S}}$ and $M$, the latter referred to as the ``memory". The former, namely ${\mathcal{S}}$, is the \textit{open system} we are concerned with.
By hypothesis, ancillas collide only with memory $M$ [see  \cref{fig-comp}(a) and (c)] through unitaries $\hat U_{Mn}$. Between two next collisions, however, ${\mathcal{S}}$ undergoes a collision with $M$ described by unitary $\hat U_{{\mathcal{S}}M}$ [see  \cref{fig-comp}(b)]. 
Now, while the reduced dynamics of $S$ is of course fully Markovian, so is not that of ${\mathcal{S}}$ which will be generally correlated with $M$ before each internal collision $\hat U_{{\mathcal{S}}M}$. 
More explicitly, if $\varrho_{{\mathcal{S}}M}^{(n-1)}$ is the (generally correlated) state of ${\mathcal{S}}$ and $M$ at the end of collision $\hat U_{M,n-1}$, the state of ${\mathcal{S}}$ at step $n$ is given by\footnote{In the present subsection, $\rho_n$ denotes the state of ${\mathcal{S}}$ not $S$}
\begin{equation}
	\rho_{n}={\rm Tr}_{M}{\rm Tr}_{n} \{\hat U_{Mn}\hat U_{{\mathcal{S}}M}\,\varrho_{{\mathcal{S}}M}^{(n-1)}\otimes \eta_{n}\,\hat U_{{\mathcal{S}}M}^\dag\hat U_{Mn}^\dag\}\,.
\end{equation}
This is not a CPT map on ${\mathcal{S}}$ because unitary $\hat U_{Mn}\hat U_{{\mathcal{S}}M}$ acts on a state featuring correlations between ${\mathcal{S}}$ and $M$-$n$ (since $\varrho_{{\mathcal{S}}M}^{(n-1)}$ is not a product state).

We see that in this dynamics the effective environment in contact with ${\mathcal{S}}$ in fact comprises the ancillas \textit{plus} ${M}$. Only the former are still ``fresh" when colliding with $S$. In contrast, $M$ is continuously recycled, thus keeping memory of the evolution at previous steps. 
Note that, in contrast to ${\mathcal{S}}$, the reduced dynamics of $S$ is always fully memoryless (in this specific respect similarly to the cascaded CM of  \cref{section-cascaded}). One can thus describe the non-Markovian system ${\mathcal{S}}$ as ``embedded" into the Markovian system $S$, in line with a common jargon in the open quantum systems literature. Indeed, this way of endowing a dynamics with memory ultimately is a typical one in the theory of open quantum systems. We also note that, as anticipated, a composite CM does not originate from breaking a single hypothesis among (1)--(3) (see beginning of the section). Indeed, as the effective bath seen by ${\mathcal{S}}$ comprises in fact both $M$ and ancillas in a way that $M$ could be seen itself as an additional ancilla, we could say that both hypotheses (1) and (3) do not hold (since $M$ keeps interacting with the other ancillas and because $S$ collides with $M$ more than once). We will yet see in the next subsection that, so long as only the open dynamics is concerned, one can establish a precise mapping between CMs with ancilla--ancilla collisions and composite CMs.

In order to express the open dynamics in terms of the compact notation for unitaries and partial traces defined in \eq\ref{UT-maps}, let us define the collision map on $M$ (corresponding to the $M$-$n$ collision) as\marginnote{We assume $\eta$ to be the same for all ancillas. If not, ${\mathcal{M}}$ simply becomes $n$-dependent.}
\begin{equation}
	{\mathcal{M}}[\ldots ]={\rm Tr}_n\{{\hat U}_{Mn}\ldots \otimes\eta_{n}\hat U_{Mn}^\dag\} = {\mathcal{T}}_n\,{\mathcal{U}}_{Mn}[\, \ldots \otimes\eta_{n}]\,.\label{map-Mn}
\end{equation}
Accordingly, the initial state of $S$ after $n$ steps turns into
\begin{equation}
	\rho_{n}={\mathcal{T}}_M\left({\mathcal{M}}\,\mathcal{U}_{{\mathcal{S}}M}\right)^n[\rho_0\otimes\eta_M]\,\,\label{coll-maps1}\,,
\end{equation}
where the leftmost partial trace returns the final reduced state of ${\mathcal{S}}$ (we assumed that the system starts in state $\sigma_{0}=\rho_{0}\otimes \rho_{M}\otimes_n \eta_n$).

As an illustrative instance, consider a qubit ${\mathcal{S}}$, a memory qubit $M$ and a bath of qubit ancillas, whose pseudo-spin ladder operators are respectively denoted as $\hat \sigma_{\pm}$, $\hat \sigma_{M\pm}$ and $\{\hat \sigma_{n\pm}\}$. The ${\mathcal{S}}$-$M$ and $M$-ancilla collisions are described by unitaries\marginnote{One could define yet more general composite CMs featuring non-unitary collisions.}
\begin{equation}
	\hat U_{{\mathcal{S}}M}=\exp{[-i \hat V_{{\mathcal{S}}M} \Delta t]}\,,\qquad\hat U_{Mn}=\exp{[-i \hat V_{Mn} \Delta t]}\,\label{UnWn}
\end{equation}
with
\begin{equation}
	\hat V_{{\mathcal{S}}M}= G\,  (\hat \sigma_{+} \hat \sigma_{M-} +\hat \sigma_{-} \hat \sigma_{M+})\,,\qquad\hat V_{Mn}= g \, (\hat \sigma_{M+} \hat \sigma_{n-} +\hat \sigma_{M-} \hat \sigma_{n+} )\,.\label{HnWn}
\end{equation}
Both unitaries \ref{UnWn} conserve the total number of excitations $\hat N=\vert 1\rangle _S\langle 1\rvert  +\vert 1\rangle _M\langle 1\rvert  +\sum_n \ket{1}_n\langle 1\rvert  $. Accordingly, if all ancillas and $M$ are initially in state $\ket{0}$ and ${\mathcal{S}}$ is in state $\ket{1}$, a reasoning analogous to that in  \cref{section-multiple} [around \eq\ref{st}] entails that the joint state at each step necessarily has the form
\begin{equation}
	\ket{\Psi^{(n)}}= \alpha^{(n)} \ket{e_{\mathcal{S}}}+ \beta^{(n)} \ket{e_{M}}+\sum_{m=1}^n  \lambda_m^{(n)} \ket{e_{m}}\,,\label{Psi-n}
\end{equation}
where, in analogy to  \cref{section-multiple}, $\ket{e_i}$ with $i={\mathcal{S}}, M, m$ is the state where subsystem $i$ is in the excited state $\ket{1}$ and all the others in $\ket{0}$. 
Here, the subscript on each amplitude denotes the time step.

Using \eq\ref{Un-qubits} with the replacements $g_z=0$ and $S\rightarrow M$, the effective representation of unitary $\hat U_{{\mathcal{S}}M}$ in the present dynamics reads
\begin{align}
	\label{USMn}
	\hat U_{{\mathcal{S}}M}=&\ket{00}_{{\mathcal{S}}M}\bra{00}+ \cos (G \Delta t) (\ket{10}_{{\mathcal{S}}M}\bra{10}+\ket{01}_{{\mathcal{S}}M}\bra{01})\\
	&-i \sin G \Delta t) (\ket{01}_{{\mathcal{S}}M}\bra{10}+\ket{10}_{{\mathcal{S}}M}\bra{01})\,\nonumber,
\end{align}
where we used that state $\ket{11}$ is never involved in the dynamics. The form of $\hat U_{Mn}$ is identical provided that $G$ is replaced by $g$ and ${\mathcal{S}}M\rightarrow Mn$.

Based on \eq\ref{Psi-n}, applying $\hat U_{Mn}\hat U_{{\mathcal{S}}M}$ on $\ket{\Psi^{(n-1)}}$ yields for $\alpha^{(n)}$ and $\beta^{(n)}$ the recurrence relation (see Appendix \ref{recurr} for details)
\begin{equation}
	\label{recur}
	\left(\begin{array}{c}    \alpha^{(n)} \\
		\beta^{(n)} 
	\end{array}
	\right)={\mathbf{D}}\,{\mathbf .}\left(\begin{array}{c}    \alpha^{(n-1)} \\
		\beta^{(n-1)} 
	\end{array}
	\right)\,\,\,\,\,\,{\rm with}\,\,\,\,{\mathbf{D}}=\left(\begin{array}{cc}    C& -ic S  \\
		-i S&c\, C  
	\end{array}
	\right)\,,
\end{equation}
where for brevity we set $c=\cos (g \Delta t)$, $s=\sin (g \Delta t)$, $C=\cos (G \Delta t)$, $S=\sin (G \Delta t)$. 
The solution of this equation is simply given by
\begin{equation}
	\left(\begin{array}{c}    \alpha^{(n)} \\
		\beta^{(n)} 
	\end{array}
	\right)={\mathbf{D}}^n\mathbf{.}\left(\begin{array}{c}    \alpha^{(0)} \\
		\beta^{(0)} 
	\end{array}
	\right)
\end{equation}
with $\alpha^{(0)}=1$ and $\beta^{(0)}=0$.

\begin{figure}[!h] 
	\raggedright
	\begin{floatrow}[1]
		\ffigbox[\FBwidth]{\caption[Dynamics of a composite collision model]{\textit{Dynamics of a composite collision model}. System ${\mathcal{S}}$, memory $M$ and all ancillas are qubits, while the collision unitaries have the form \ref{UnWn}. Initially, ${\mathcal{S}}$ is in the excited state with $M$ and each ancilla in the ground state. Each panel shows $p_S=\vert \alpha^{(n)}\vert ^2$ with the corresponding $p_M=\vert \beta^{(n)}\vert ^2$ in the inset. Throughout we set $g =\sqrt{G/\de t}$. The first three panels [(a)--(c)] report the dynamics in the case $G = 1$ for $\de t=2$ (a), $\de t=1$ (b) and $\de t=0.1$ (c), while in panel (d) we set $G = 0.1$, $\de t=0.1$.}\label{fig-comp2}}%
		{\includegraphics[width=\textwidth]{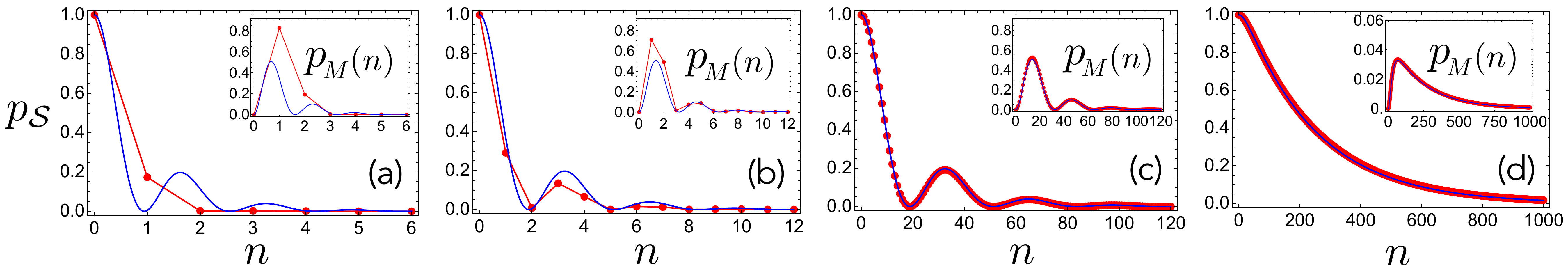}}
	\end{floatrow}
\end{figure}

In  \ref{fig-comp2}, we plot the evolution of the excited-state population of ${\mathcal{S}}$ and $M$, respectively denoted with $p_{\mathcal{S}}=\vert \alpha^{(n)}\vert ^2$ and $p_M=\vert \beta^{(n)}\vert ^2$, for $g=\sqrt{\gamma/\de t}$ and $G=1$ [panels (a)--(c)], $G=0.1$ (d), where energies are expressed in units of $\gamma=g^2\de t$. As $\de t$ decreases, the curves become more and more continuous as shown (for the case $G=1$) by panels (a)--(c).
We see that when $G$ is large [(a)--(c)], ${\mathcal{S}}$ and $M$ keep exchanging an excitation which eventually leaks out and gets dissipated into the bath of ancillas. In particular, ${\mathcal{S}}$ undergoes damped oscillations, exhibiting revivals which fade away for $n$ large enough. For $G$ small enough, however, the excitation of ${\mathcal{S}}$ monotonically decays and no revivals show up, while $p_M$ reaches a maximum and then decays. In the latter regime (small $G$), the interaction of $M$ with ancillas dominates over the ${\mathcal{S}}$-$M$ coupling so that as an excitation is transferred from ${\mathcal{S}}$ to $M$ this is immediately released into the bath before being reabsorbed by ${\mathcal{S}}$. 

The above behavior is analogous to the dynamics of an atom coupled to a lossy
cavity mode, a longstanding paradigm of non-Markovian
dynamics~\cite{haroche_exploring_2006}. Specifically, the regimes of damped
oscillations [see  \cref{fig-comp2}(a)--(c)] and monotonic decay [see \
\ref{fig-comp2}(d)] respectively correspond to the so called strong and weak
coupling regimes of cavity QED. This link with cavity-QED dynamics can be formulated
as an explicit mapping if we assume
\begin{equation}
	G \dt \ll g\de t\ll 1
\end{equation}
and expand accordingly the overall unitary for short $\de t$ as [\cf\cref{UnWn}--\ref{HnWn}]
\begin{equation}
	\hat U_{{\mathcal{S}}M}\hat U_{Mn}\simeq \mathbb{I}-i (\hat V_{{\mathcal{S}}M}+\hat V_{Mn})\de t-\tfrac{1}{2}\hat V_{Mn}^2\de t^2\,.
\end{equation}
This expression is now identical to \eq\ref{USn2} of  \cref{section-EMs} with $\hat H_0=\hat V_{{\mathcal{S}}M}$ and $\hat V_n=\hat V_{Mn}$. It follows that the coarse-grained ME of the composite ${\mathcal{S}}$-$M$ system is given by [\cf\cref{drho2,HS-diss}]
\begin{equation}
	\dot\rho_{{\mathcal{S}}M}{=}-i \,[ G(\hat \sigma_{+} \hat \sigma_{M-} +\hat \sigma_{-} \hat \sigma_{M+}),\rho_{{\mathcal{S}}M}]+\gamma \left(\hat\sigma_{M-}\rho_{{\mathcal{S}}M}\hat\sigma_{M+}{-}\tfrac{1}{2}\,[\hat\sigma_{M+}\hat\sigma_{M-},\rho_{{\mathcal{S}}M}]_+\right)\,,\label{ME-CQED}
\end{equation}
where as usual $\gamma=g^2\de t$.\marginnote{Given the approximations made, $\hat U_{{\mathcal{S}}M}$ and $\hat U_{Mn}$ commute. Hence, we can replace $\hat U_{{\mathcal{S}}M}\hat U_{Mn}\simeq e^{-i (\hat V_{{\mathcal{S}}M}+\hat V_{Mn})\de t}$, which can now be effectively thought as a single collision of duration $\de t$.} This is the bipartite ME of a two-level atom coupled to a leaky cavity mode initially in the vacuum state.\marginnote{As no Fock states with more than one photon are involved in such dynamics, the bosonic ladder operators of the cavity can be equivalently replaced with ladder spin operators (here denoted as $\hat \sigma_{M\pm}$).} Now, if $\alpha(t)$ [$\beta(t)$] is the excited-state amplitude of ${\mathcal{S}}$ ($M$) at time $t$, it can be shown (see Appendix \ref{lin-sys}) that ME \ref{ME-CQED} is equivalent to the pair of coupled equations
\begin{equation}
	\label{syst}
	\dot{\alpha}= -iG\beta\,,\,\,\,\dot{\beta}=-i G\alpha  -\tfrac{\gamma}{2}\beta\,.
\end{equation}
Solving the latter equation for $\beta(t)$\marginnote{Specifically, this yields $ \beta(t)=-i\, G \int_0^t {\rm d}t'\exp[{- \frac{\gamma}{2}(t-t')}]\alpha(t')$.} and replacing in the former yields the integro-differential equation
\begin{equation}
	\dot{\alpha}=- \,G^2 \int_0^t {\rm d}t'\,e^{-\frac{\gamma}{2}(t-t')}\alpha(t')\label{eps-int}\,,
\end{equation}
whose solution is
\begin{equation}
	\alpha(t)=e^{-\frac{\gamma}{4}t}  \left[\cos
	\left(\tfrac{\delta t}{2}\right)+\tfrac{\gamma}{2\delta}\sin
	\left(\tfrac{\delta t}{2}\right)\right]\label{cossin}\,\,\,\,\,\,{\rm with}\,\,\,\delta\ug\sqrt{4
		G^2-\tfrac{1}{4}\gamma^2}
\end{equation}

\section{Mapping ancilla--ancilla collisions into a composite collision model}
\label{section-mapCCM}

In  \cref{section-aacu}, we saw that the open dynamics of a CM with fully-swapping ancilla--ancilla (AA) collisions reduces to a continuous interaction between $S$ and the same ancilla. Note that this can be seen as a special case of a composite CM with the memory trivially decoupled from the bath.
Accordingly, one could guess that, when it comes to \textit{arbitrary} AA collisions, the open dynamics is effectively described by a suitably defined composite CM. We will show next that this is indeed the case and, remarkably, it is true no matter the form of AA unitaries.\marginnote{We will consider unitary AA collisions, yet the property can be extended to non-unitary collisions (as those in  \cref{section-NMME}).}

Let $\hat R_n=\hat W_{n+1,n}\hat U_{Sn}  \cdot\cdot\cdot\hat W_{3,2}\hat U_{S2} \hat W_{2,1}\hat U_{S1}$ be the unitary describing the joint dynamics at the $n$th step.\marginnote{Note that the definition of step adopted here is slightly different from \cref{sigmaAA,Unp}. This yet has no effect on the open dynamics, which is our focus. Also, at variance with  \cref{section-aacu}, here we explicitly show the $S$-dependence of SA collision unitaries, which facilitates establishing the connection with the notation used for introducing composite CMs.} Unitaries $\hat R_n$ then fulfill
\begin{equation}
	\hat {R}_n=\hat W_{n+1,n}\hat U_{Sn} \,\hat {R}_{n-1}\,.\label{Unn1}
\end{equation}
Let us also define for convenience a pairwise unitary on ancillas $n$ and $n-1$ as
\begin{equation}
	\hat W'_{n,n-1}=\hat S_{n,n-1}\hat W_{n,n-1}\,\label{Wp}
\end{equation}
with $\hat S_{n,n-1}$ the usual swap operator.

At step $n=2$, we can arrange the total unitary as
\begin{equation}
	\hat {R}_2=\hat W_{3,2}\hat U_{S2}\hat W_{2,1}\hat {U}_{S1}=\hat W_{3,2}(\hat S_{2,1}\hat U_{S1}\hat S_{2,1})\hat W_{2,1}\hat U_{S1}=\hat W_{3,2}\hat S_{2,1}\hat U_{S1}\hat W'_{2,1}\hat U_{S1}\,,\label{UU2}
\end{equation}
where we expressed $\hat U_{S2}$ in terms of $\hat U_{S1}$ via \ref{SOS} and used definition \ref{Wp}.

At step $n=3$, using \Cref{Unn1,UU2}, we get
\begin{align}
	\hat {R}_3&=\hat W_{4,3}\hat U_{S3}\hat W_{3,2}\hat S_{2,1}\hat U_{S1}\hat W'_{2,1}\hat U_{S1}=\hat W_{4,3}(\hat S_{3,2}\hat U_{S2}\hat S_{3,2})\hat W_{3,2}\hat S_{2,1}\hat U_{S1}\hat W'_{2,1}\hat U_{S1}\nonumber\\
	&=\hat W_{4,3}\hat S_{3,2}\hat U_{S2}\hat W'_{3,2}\hat S_{2,1}\hat U_{S1}\hat W'_{2,1}\hat U_{S1}.\nonumber
\end{align}
Now, recalling that $\hat O_n\hat S_{n,n-1}=\hat S_{n,n-1}\hat O_{n-1}$, we move swap $\hat S_{2,1}$ to the left until it is placed to the right of $\hat S_{3,2}$. This turns $\hat U_{S2}\hat W'_{3,2}$ into $\hat U_{S1}\hat W'_{3,1}$, hence we get
\begin{equation}
	\hat {R}_3=\hat W_{4,3}\hat S_{3,2}\hat S_{2,1}\hat U_{S1}\hat W'_{3,1}\hat U_{S1}\hat W'_{2,1}\hat U_{S1}=\hat W_{4,3}(\hat S_{3,2}\hat S_{2,1})\hat U_{S1}\hat W'_{3,1}\hat U_{S1}\hat W'_{2,1}\hat U_{S1}\,.\nonumber
\end{equation}
By induction, at step $n$
\begin{equation}
	\hat {R}_n=\hat W_{n+1,n}(\hat S_{n,n-1}\hat S_{n-1,n-2}\, \ldots \,\hat S_{2,1})\,\hat U_{S1}\hat W'_{n,1}\hat U_{S1}\hat W'_{n-1,1}\, \ldots \,\hat U_{S1}\hat W'_{2,1}\hat U_{S1}\,.\nonumber
\end{equation}
To get the reduced dynamics of $S$, we evolve the initial state via unitary $\hat {R}_n$ and trace off the ancillas as usual. In doing so, we add a further $\hat S_{n+1,n}$ to get  another $\hat W'$ operator and move all the swaps to the left. Due to the partial trace, the sequence of swaps is eventually eliminated\marginnote{This is because, if $\{\vert k_1,k_2,\ldots,k_n\rangle \}$ is an ancillas' basis, any given sequence of two-ancilla unitaries applied to all basis states $\vert k_1,k_2,\ldots,k_n\rangle$ yields another valid basis for computing the partial trace over the ancillas.} so that we end up with
\begin{align}
	\rho_n=&{\rm Tr}_{1,2, \ldots ,n}\{\hat {R}_n \,\rho_{0}\otimes\eta_{n}\,\, \hat {R}_n^\dag\}=\label{rhon-com} \\
	&{\mathcal{T}}_1{\mathcal{T}}_2\, \ldots \,{\mathcal{T}}_n\,\,{\mathcal{W}}'_{n+1,1}\mathcal U_{S1}{\mathcal{W}}'_{n,1}\mathcal U_{S1}{\mathcal{W}}'_{n-1,1}\, \ldots \,\mathcal U_{S1}\mathcal W'_{2,1}\mathcal U_{S1}[\rho_0\otimes_m \eta_m]\nonumber
\end{align}
where as usual
${\mathcal{U}}_{S1}$ and ${\mathcal{W}}'_{n,1}$ are  respectively the unitary maps associated with $\hat U_{S1}$ and $\hat W'_{n,1}$ [\cf\eq\ref{UT-maps}]. This open dynamics is identical to that of a composite CM as can be seen more explicitly by introducing a collision map on ancilla $1$ as ${\mathcal{M}}={\mathcal{T}}_n {\mathcal{W}}'_{n,1}$ so that \eq\ref{rhon-com}  can be written as
\begin{equation}
	\rho_{n}={\mathcal{T}}_1\,\left({\mathcal{M}}\,\mathcal{U}_{{\mathcal{S}}1}\right)^n\,[\rho_0\otimes\eta_1].
\end{equation}
Upon comparison with \eq\ref{coll-maps1}, we see that the open dynamics is indeed that of a composite CM where ancilla 1 embodies the memory. In this equivalent picture, the original SA collision unitary turns into the unitary describing the collision internal to the composite $S$-1 system, while the original AA unitary now embodies the collision describing memory--ancilla collisions.

The fact that it is enough to consider a single ancilla in order to get a composite system jointly undergoing Markovian dynamics clearly follows from the pairwise nature of each AA collision. For instance, if between two next SA collisions there occurred AA collisions overall involving \textit{three} ancillas, then the composite Markovian system would comprise \textit{two} ancillas (besides $S$). Thus the size of the effective composite system somehow measures how big is the portion of bath which we have to keep track in detail in order to describe our non-Markovian open dynamics. This effectively illustrates a distinctive feature of many non-Markovian dynamics, namely the impossibility to trace off the entire bath dynamics even if one is interested solely in the open dynamics.\marginnote{In this respect, the collisional dynamics in  \cref{section-NMME} is a remarkable exception which relies crucially on the non-unitary nature of AA collisions [\cf\eq\ref{Winc}].}

\section{Non-Markovian collision models: state of the art}
\label{soa-nm}

Non-Markovian CMs with ancilla--ancilla collisions (see
\cref{section-aacu,section-NMME}) were first introduced in
\rrefs~\cite{ciccarello_collision-model-based_2013,ciccarello_quantum_2013} in the
form of incoherent partial swaps [\cf\ref{Winc}] alongside ME \ref{ME-kernel}.
CMs of the same class but with unitary ancilla--ancilla collisions, typically in the
form of partial swaps [\cf\eq\ref{Wswap}], were considered in
\rrefs~\cite{mccloskey_non-markovianity_2014,cakmak_non-markovianity_2017,campbellPRA18,CampbellPrecursors2019}
mostly with the goal of investigating the relationship between non-Markovianity and
system--environment correlations (and changes in the bath state). Notably, this type
of CMs are a convenient tool to introduce non-Markovian effects in quantum
thermodynamics studies (see  \cref{section-thermo}), which was applied in particular
to investigate the Landauer principle of  \cref{section-Land} in the presence of
baths with memory~\cite{pezzutto2016implications,man2019validity,xiaPRA21}, the
temperature dependence of non-Markovianity~\cite{LoFrancoPRA18}, quantum engine
performances~\cite{pezzutto2019out,abah2020implications} and a non-Markovian generalization of quantum homogenization (see Section \cref{section-homo}) \cite{Ghosh}. Remarkably, collisional
dynamics with ancilla--ancilla collisions can be experimentally implemented in
all-optical setups~\cite{stroboscopicPRA15,cuevas2019}. While most of these works
considered qubits, continuous-variable versions of CMs with ancilla--ancilla
collisions were proposed and studied in \rref~\cite{jin_non-markovianity_2018},
featuring multipartite ancillas (environmental blocks), and
\rref~\cite{ramirez2020memory}, where both beam-splitter-like and two-mode-squeezing
ancilla--ancilla interactions were investigated.
It is also worth mentioning that ME \ref{ME-kernel} stimulated the study of a
corresponding class of well-defined memory-kernel
MEs~\cite{vacchini_non-markovian_2013,vacchini_general_2014,vacchini_generalized_2016,vacchini2020quantum,chruscinski_sufficient_2016,chruscinski_generalized_2017,siudzinska_memory_2017}.

A CM with initially-correlated ancillas (see  \cref{section-ica}) was introduced in
\rref~\cite{rybar_simulation_2012}. The authors showed that any CPT map on a qubit
$S$ can be simulated by a CM where $S$ collides with qutrits
(i.e.,~three-level ancillas) initially prepared in a suitable, generally correlated,
state. This includes the so called indivisible quantum
channels~\cite{wolf2008dividing}, namely CPT maps that cannot be decomposed into
infinitesimal CPT maps, thus violating in particular \eq\ref{SG-2}. The link
discussed in  \cref{section-ica} between correlated ancillas and mixtures of
dynamical maps was extensively studied in \rref~\cite{filippov_divisibility_2017}
within a broader framework connected with concepts such as eternal CP
indivisibility~\cite{megier2017eternal} and pictorially illustrated through Pauli
maps (see also \rref \cite{pathak2019non}). Note that in a condensed-matter scenario it is natural to consider ancillas as
coupled spins described by a many-body Hamiltonian and, as such, initially
correlated~\cite{mascarenhas_quantum_2017}. While one might expect that for growing
inter-ancillary correlations the dynamics becomes more and more non-Markovian,
correlations alone are yet insufficient to ensure non-Markovian behavior which
indeed depends as well on the specific features of system--ancilla interaction. This
was shown by Bernardes \etal~\cite{bernardes_environmental_2014} in terms of the
non-Markovianity measure of \rref~\cite{RHP} and then experimentally tested in
all-optical~\cite{bernardes_experimental_2015} and NMR
settings~\cite{bernardes_high_2016}. The CM in
\rref~\cite{bernardes_environmental_2014} was used as well to investigate the
relationship between coarse-graining time and correlation
time~\cite{bernardes_coarse_2017}.
A collisional dynamics with initially-correlated ancillas was also experimentally
implemented through the IBM Q Experience processors~\cite{garcia2020ibm}. We also
quote the use of such class of CMs in \rref~\cite{ModiMSK} investigating the
relationship between CP divisibility and non-Markovianity.

CMs with multiple collisions (see  \cref{section-multiple}) were proposed as a
paradigm of non-Markovian quantum chain~\cite{bodor_structural_2013} (see also
\rref~\cite{pellegrini2009non}) and recently applied in the study of quantum Markov
order~\cite{taranto2019} and quantum cooling~\cite{taranto2020exponential}.

The derivation of \eq\ref{DDE-fin} follows \rref~\cite{cilluffo2019quantum}. The
equation is usually derived without resorting to the collisional approach, see \eg
\rrefs~\cite{dorner_zoller,tufarelli2013}. Note that the phase $\phi$ which we
included for completeness in the coupling Hamiltonian \ref{Vt5} significantly
affects the emission.

This class of CMs with multiple, non-local, collisions were introduced in quantum
optics by \rrefs~\cite{pichler_photonic_2016,grimsmo_time-delayed_2015} which
considered quantum emitters under a continuous drive [a dynamics considerably more
involved than \eq\ref{DDE-fin}]. \rref~\cite{pichler_photonic_2016} showed that
the problem can be efficiently solved numerically using Matrix Product States (MPS),
while \rref~\cite{grimsmo_time-delayed_2015} proposed an elegant diagrammatic
technique mapping the non-Markovian dynamics of the emitter into the Markovian
dynamics of a cascade of fictitious emitters. An algorithm for describing
non-Markovian quantum trajectories based on such CMs was proposed in
\rref~\cite{whalen2019}, while a thorough comparison between the collisional and MPS
approach to time-delayed quantum optics dynamics was recently carried out in
\rref~\cite{regidor2020modelling}. 
We note that this class of CMs with non-local collisions describe the dynamics of
so called giant atoms~\cite{Kockum5years} (a new paradigm of quantum optics) in the
regime of non-negligible time delays~\cite{cilluffo2020}. 

Another type of CMs with multiple collisions was considered in
\rref~\cite{gennaro_entanglement_2008} (see also~\cite{gennaro_structural_2009})
considering an open system $S$ undergoing random collisions with a
two-ancilla bath. At each step, both the ancilla colliding with $S$ and the
collision unitary are selected randomly. It was found that the purity of
$S$ as well as bipartite and tripartite entanglement reach time averaged
equilibrium values characterized by large fluctuations.

Composite collision models of  \cref{section-comp}, whose theory was formulated in
\rref~\cite{lorenzo_composite_2017}, are used as a versatile tractable model for
investigating non-Markovian
problems~\cite{lorenzo_landauers_2015,LoFrancoPRA18,filippovPRL,filippovPRL2,xiaPRA21,karpat2020synchronization},
including generalized versions where each subsystem is in contact with a different
bath of ancillas~\cite{ShaoPRE18}. The descriptive power of composite CMs
(generalized to multiple baths) was studied in \rref~\cite{cattaneo2020collision},
where it was shown that they can simulate efficiently the Markovian dynamics of any
multipartite open quantum system, \ie with an error and resources (in terms of size
and number of memory systems $M$) that scale polynomially with the size of
$S$ and simulation time.

The mapping of a CM with ancilla--ancilla collisions into a composite CM (see
\cref{section-mapCCM}) was introduced in \rref~\cite{kretschmer_collision_2016}. In
\rref~\cite{campbellPRA18}, the mapping was further developed and used for defining
the concept of ``{memory depth}". 
These works consider unitary ancilla--ancilla collisions, yet even when these are
incoherent partial swaps (as in  \cref{section-NMME}) a mapping into a
suitably-defined composite CM is still possible as shown in
\rref~\cite{lorenzo_class_2016}.

\setchapterpreamble[u]{\margintoc}  
\chapter{Collision models from conventional models}
\label{section-CMQO}

We saw in  \cref{section-maser} that the micromaser is naturally described by a CM. The micromaser is an instance of engineered, intrinsically discrete dynamics. In the present section, we discuss another major scenario (common in quantum optics) that admits a CM description.    The paradigmatic model is a system $S$ -- in typical cases a cavity mode or atom(s) -- coupled to a white-noise bosonic field (we clarify later what ``white-noise" means). 

The present section is conceptually important in that it shows how CMs are related to conventional system--bath microscopic models. The latter ones typically describe the bath as a continuum of modes which interact with $S$, in general, \textit{all at the same time} [see  \ref{fig-intro}(b)]. 
This is in stark contrast with (memoryless) CMs [see  \ref{fig-intro}(a)], where $S$ interacts with the bath units (ancillas) \textit{one at a time} (a major reason why CMs are an advantageous theoretical tool). Another key difference between the two frameworks is that, while in a CM the total Hamiltonian of $S$ and all the ancillas is intrinsically {time-dependent} (as we discussed in particular in  \cref{section-HSBt,section-work}), conventional microscopic models usually feature a \textit{time-independent} total Hamiltonian. The latter case matches the physical expectation that, since $S$ and the bath form a closed system, no intrinsic time-dependence is expected to arise in the total Hamiltonian. These issues (in particular) will be clarified in what follows, from which the CM will emerge as an effective \textit{picture} to study a dynamics originally formulated in a conventional microscopic model. Notably, this will provide physical intuition about a number of properties of CMs postulated on a rather abstract ground in  \cref{section-def,section-EMs}.

\section{White-noise bosonic bath and weak-coupling approximation}
\label{wn}

Let $S$ be a quantum system of frequency $\omega_0$ coupled to a continuum of bosonic modes $f$ (field), whose normal-mode ladder operators $\hat b_\omega$ and $\hat b_\omega^\dag$ fulfill the commutation rules $[\hat b_\omega,\hat b_{\omega'}^\dag]=\delta(\omega{-}\omega')$, $[\hat b_\omega,\hat b_{\omega'}]=[\hat b_\omega^\dag,\hat b_{\omega'}^\dag]=0$. The total Hamiltonian reads
\begin{equation}
	\hat H=\hat H_S+\hat H_f+\hat V\label{H-bos}
\end{equation}
with
\begin{align}
	&\hat H_{S}= \omega_0 \,{\hat A}^\dag{\hat A}\,,\,\,\,\hat{H}_{f}= \int_{-\infty}^\infty d\omega\, (\omega_0+\omega)\, \hat{b}^\dagger_\omega\, \hat{b}_\omega\nonumber\\
	&\hat V=\sqrt{\tfrac{\gamma}{2\pi}}\int_{-\infty}^\infty d\omega\,\left({\hat A}^\dag\hat b_\omega+{\hat A}\,\hat b_\omega^\dag\right)\,\,.\label{HQO}
\end{align}
The $S$ operators ${\hat A}$ and ${\hat A}^\dag$ could be fermionic or bosonic, the essential requirement being only that  $\hat A$ is an eigenoperator of $\hat H_S$, \ie $[{\hat H}_S,{\hat A}]=-\omega_0{\hat A}$ [\cf\eq\ref{comm}]. Three major features of the Hamiltonian model \ref{H-bos} stand out:
\begin{itemize}
	\item [(a)] The coupling strength is $\omega$-independent (white coupling);
	\item  [(b)] $\hat V$ does not contain counter-rotating terms ${\sim}{\hat A}\hat b_\omega$, $\hat A^\dag\hat b_\omega^\dag$;
	\item [(c)] Frequency $\omega$ takes values on the entire real axis.
\end{itemize}

\begin{figure}[!h] 
	\raggedright
	\begin{floatrow}[1]
		\ffigbox[\FBwidth]{\caption[Sketch of involved frequencies]{\textit{Sketch of involved frequencies}. Here, $\omega_0$ is the frequency of $S$ while the blue strip represents the spectrum of normal frequencies of the bath (\ie the field $f$; we consider a single frequency band for simplicity). The open system $S$ significantly couples only to field modes with frequency lying within a narrow window of width $\gamma$ centered at $\omega_0$. Accordingly, once can extend the field spectrum to the entire $\omega$-axis (light blue strip) by introducing fictitious modes (including in particular frequencies $\omega<0$).}\label{fig-window}}%
		{\includegraphics[width=\textwidth]{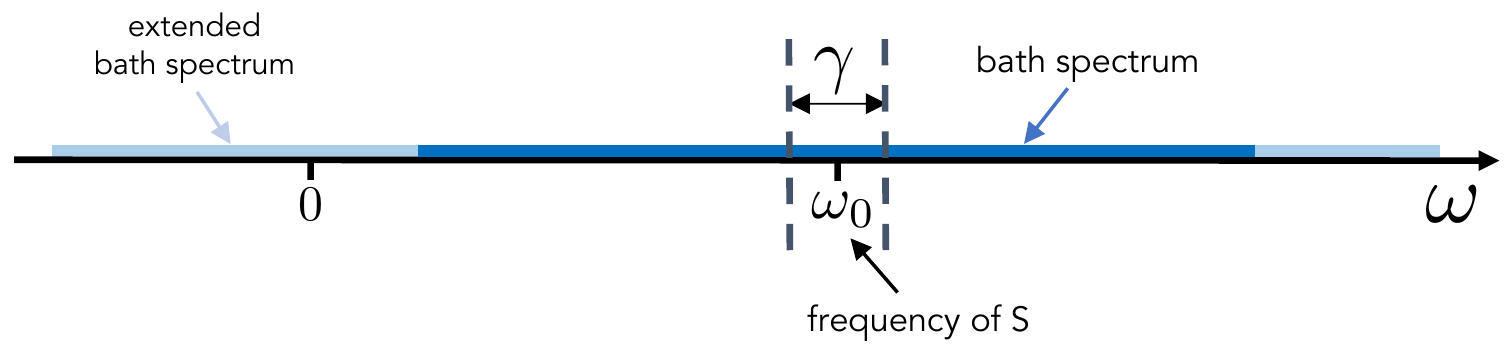}}
	\end{floatrow}
\end{figure}

These are all idealizations: in reality, the coupling depends on $\omega$,
counter-rotating terms are present and $\omega$ is lower-bounded. The validity
of (a)--(c) relies on the \textit{weak coupling} approximation, namely the weakness
of $S$-$B$ interaction (a usual situation, e.g.~in quantum optics).\footnote{For
	a derivation of Hamiltonian \ref{H-bos}--\ref{HQO} through the weak-coupling
	approximation see \eg Appendix A of \rref~\cite{cilluffo2020}.} Because of it,
$S$ undergoes a significant energy exchange only with field modes whose
frequency $\omega$ lies within a narrow window around $\omega_0$ of width
$\sim \gamma$ such that $\gamma\ll \omega_0$ (see  \ref{fig-window}). Accordingly, it
makes no difference if the coupling rate at any frequency $\omega$ is replaced
with its value at $\omega_0$, which we called $\sqrt{\gamma/2\pi}$ in \eq\ref{HQO}, at
the same time extending  integrals over $\omega$ to the entire real axis (see \
\ref{fig-window}) by introducing in particular negative-frequency fictitious modes
(these remain uncoupled to $S$ in fact). Moreover, counter-rotating terms
rotate fast compared to the time scale $\gamma^{-1}$ and are thus discarded
(rotating wave approximation or RWA). Note that, for self-consistency, introduction
of negative frequencies and RWA must be  performed \textit{together}: without the
latter, an unphysical resonance at $\omega=-\omega_0$ would arise.

\section{Time modes}
\label{section-tm}

Instead of normal modes (ladder operators $\hat b_\omega$), the bosonic bath can be equivalently represented in terms of \textit{time modes} (henceforth all integrals are intended to run from $-\infty$ to $\infty$)
\begin{equation}
	\hat b _s  = \tfrac{1}{\sqrt{2 \pi}} \int d\omega  \,\hat{b}_\omega e^{-i \omega s}\,,\label{tm}
\end{equation}
which are thus related to $\hat b_\omega$ through Fourier transform. 
As is easily checked, time modes fulfill bosonic commutation rules
\begin{equation}
	[\hat b_s, \hat b_{s'}^\dag]=\delta(s-s')\,,\,\,\,\,[\hat b_s,\hat b_{s'}]=[\hat b_s^\dag,\hat b_{s'}^\dag]=0\,.\label{bt-comm}
\end{equation}
Despite having dimensions of time, $s$ should be regarded for now as just a label and time modes as an alternative way to represent the field (the connection with true time $t$ will become clear shortly). 

\section{Interaction picture}
\label{section-int-pic}

In the interaction picture with respect to $\hat H_0=\hat H_{S}+\hat H_f$, ladder operators transform as $\hat A{\rightarrow} \hat A e^{-i\omega_0 t}$ and $\hat b_{\omega}\rightarrow \hat b_{\omega} e^{-i(\omega_{0}+\omega) t}$ so that the joint $S$-field state $\sigma$ evolves as $\dot \sigma=-i \,[\hat V_t ,\sigma]$ with
\begin{equation}
	\hat V_t=\sqrt{\gamma}\,\hat A^\dagger \,{\hat b}_{s=t}+ {\rm H.c.}\,\,,\label{Vt1}
\end{equation}
hence, in the interaction picture: (i) time modes are non-interacting with each other,\marginnote{In the Schr\"odinger picture, time modes do couple to one another since $\hat H_f$ clearly cannot have a diagonal form when expressed in terms of time modes (note that these are \textit{not} normal modes).} (ii) at time $t$, $S$ only couples to the time mode $\hat b_{s=t}\equiv\hat b_t$. Note that (i) and (ii) strongly recall, respectively, assumptions (1) and (3) of  \ref{conds}, representing in fact a continuous version of these. 

A consequence of the interaction picture is that $\hat V_t$ becomes \textit{time-dependent}, hence the time evolution operator (propagator) is given by
\begin{equation}
	\label{Ut2}
	\U_t=\hat {\mathcal{T}} \,e^{-i\int_{t_{0}}^{t} ds\, \V(s)}   
\end{equation}
with $\hat {\mathcal{T}}$ the time-ordering operator.

\section{Time discretization and coarse graining}
\label{section-tdcg}

Let us next consider a mesh of the time axis defined by $t_n=n\Delta t$ with $n$ an integer and $t_0=0$. In terms of this mesh, the propagator \ref{Ut2} can be split as\marginnote{We assume that $t/\Delta t$ is an integer. If not, the error committed becomes negligible for vanishing $\Delta t$.}
\begin{equation}
	\label{Magnus}
	\U_t=\hat U_{t/\Delta t} \cdots\,\hat U_2\,\hat U_1\qquad {\rm with}\,\,\,\, \hat U_n=\hat {\mathcal{T}}\,e^{-i\int_{t_{n-1}}^{t_n} ds\, \V_s}\,.
\end{equation}
We take a time step much shorter than the characteristic interaction time, i.e.,~$\Delta t\ll \gamma^{-1}$. This allows us to expand each $\hat U_n$ to second order in
$\Delta t$, which yields\marginnote{This perturbative expansion of the propagator is
	known as Magnus expansion~\cite{Magnus1954}.}
\begin{equation}
	\hat U_n\simeq \mathbb{I} -i \,(\hat V_n  + \hat V'_n )\, \Delta t -\tfrac{1}{2} \hat V_n^2 \,\Delta t^2\label{Un-app1}
\end{equation}
with
\begin{equation}
	\hat V_n=\tfrac{1}{\Delta t}\int_{t_{n-1}}^{t_n} ds \,\V_s\,,\,\,\,\, \hat V_n'=\tfrac{i}{2\Delta t}\int_{t_{n-1}}^{t_n} ds \int_{t_{n-1}}^{s} ds' \,[\V_{s'},\V_{s}]\,.\label{Mag2}
\end{equation}

\section{Emergence of the collision model}
\label{section-emer}

It can be shown (see Appendix B of \rref~\cite{cilluffo2020}) that term
$\hat V_n'$ gives negligible contribution for $\Delta t$ short enough. Thus each elementary
unitary \ref{Un-app1} reduces to
\begin{equation}
	\hat U_n=\mathbb{I} -i \,\hat V_n  \, \Delta t -\tfrac{1}{2} \hat V_n^2\, \Delta t^2\label{Un-app}\,,
\end{equation}
where, using \eq\ref{Vt1}, $\hat V_n$ has the explicit form
\begin{equation}
	\hat V_n=\sqrt{\tfrac{\gamma}{\Delta t}}\left(\hat A^\dagger \hat b_{n}+\hat A \,\hat b_{n}^\dag\right)\,,\label{HBn2}
\end{equation}
where we defined
\begin{equation}
	\hat b_n=\tfrac{1}{\sqrt{\Delta t}}\int_{t_{n-1}}^{t_n} dt ~ \hat b _t \,.\label{bn0}
\end{equation}
It is easily verified that $\hat b_n$ fulfill standard bosonic commutation rules
\begin{equation}
	[\hat b_{n}, \hat b_{n'}^\dag]=\delta_{n,n'}\,,\,\,\,[\hat b_{n}, \hat b_{n'}]=[\hat b^\dagger_{n}, \hat b^\dagger_{n'}]=0\,.\label{comm-bn}
\end{equation}
This is precisely the basic CM of  \cref{section-def} in the case that each ancilla is a quantum harmonic oscillator of frequency $\omega_0$.  A number of comments follow.
\begin{itemize}
	\item [(1)]  Note how the characteristic $1/\sqrt{\Delta t}$ dependence of the coupling strength  -- which we assumed repeatedly in this paper (see \eg  \cref{sec-div}) -- here in fact results from the model's white coupling [\cf\eq\ref{HQO}] combined with the need for well-defined bosonic commutation rules of the $\hat b_n$'s.\marginnote{Commutation rules \ref{comm-bn} crucially rely on having incorporated a factor $1/\sqrt{\Delta t}$ in the definition of $\hat b_n$ [\cf\eq\ref{bn0}].}
	\item [(2)] The CM arises in the interaction picture [recall \eq\ref{Vt1}], which explains the time-dependent nature of the collisional Hamiltonian.
	\item [(3)] The interaction picture is key in order for $S$ to collide with a new ancilla $\hat b_n$ at each time step and for the ancillas to be mutually non-interacting. In the Schr\"odinger picture, $S$ would be interacting all the time with the same ancilla and the ancillas would be coupled to one another (reflecting an analogous properties of continuous time modes).
	\item [(4)] Among the three hypotheses in  \cref{section-def} which ensure lack of memory, the CM that we derived fulfills (1) and (3). Whether or not (2) holds (initially-uncorrelated ancillas) depends on the field initial state, as shown next.
\end{itemize}

\section{Initial state of ancillas and condition for Markovian dynamics}
\label{in-field}

We assume throughout that $S$ and the bosonic bath are initially uncorrelated, that is $\sigma_0=\rho_0\otimes \rho_f$. The field initial state $\rho_f$ is usually expressed in terms of the continuous normal modes (frequency domain) or through the time modes (time domain). Thus, in order to derive the corresponding initial state of ancillas, one first needs to express $\rho_f$ in terms of modes $\hat b_{n}$. At this point, we observe that, for an unspecified $\Delta t$, modes $\hat b_n$ in \eq\ref{bn0} clearly embody only part of the field degrees of freedom. This can be formally seen by Fourier-expanding $\hat b_t$ can in each time interval $\left[t_{n-1},t_n\right[$ as
\begin{equation}
	\hat b_t=\sum_{n=-\infty}^{\infty}\sum_{k=-\infty}^{\infty}\Theta_n(t)\,\tfrac{1}{\sqrt{\Delta t}}\,e^{-i \frac{2\pi k}{\Delta t}t}\,\hat b_{n,k}\,\,\,\,\,\,\,\,\,{\rm with}\,\,\,\,\,\,\hat b_{n,k}=\tfrac{1}{\sqrt{\Delta t}}\int_{t_{n-1}}^{t_n} dt ~ e^{i \frac{2\pi k}{\Delta t}t}\hat b _t\label{fourier}\,\,
\end{equation}
[recall that $\Theta_n(t)=1$ inside interval $t_{n-1}\le t<t_n$ while $\Theta_n(t)=0$
elsewhere]. Here, ladder operators $\hat b_{n,k}$ are defined so as to obey bosonic
commutation rules, $[\hat b_{n,k}, \hat b_{n'k'}^\dag]=\delta_{n,n'}\delta_{k,k'}$, $[\hat b_{n,k}, \hat b_{n'k'}]=0$. Note that for $k=0$ we
retrieve ancillas' modes \ref{bn0}, that is $\hat b_n\equiv \hat b_{n.0}$. It is easily shown that
modes $\hat b_{n,k\neq 0}$ contain only field frequencies $\omega$ that diverge in the
limit $\Delta t\rightarrow 0$~\cite{gross_qubit_2018}. Accordingly, it is reasonable to assume
that for all practical purposes these modes remain always unexcited, that is one in
fact always deals with field initial states of the form
\begin{equation}
	\rho_f =\rho_{B}\bigotimes_{n }\bigotimes_{k\neq0 }\vert 0\rangle _{n,k}\langle 0\rvert  \,\,,
\end{equation}
where $\rho_B$ stands for the state of modes $\hat b_n=\hat b_{n,0}$ (our \textit{ancillas}) while $\vert 0\rangle _{n,k}$ is the vacuum state of each mode $\hat b_{n,k}$.\marginnote{Note that the approximation according to which modes $\hat b_{n,k\neq0}$ remain unexcited all the time is consistent with \cref{Un-app,HBn2} where only modes $\hat b_n\equiv\hat b_{n,0}$ appear.}

Based on the above, the initial state of ancillas (modes $\hat b_n$) is generally inferred from $\rho_f$ (initial state of the bosonic bath) by decomposing the field into modes $\hat b_{n,k}$ through the inverse of transform \ref{tm} followed by \ref{fourier} (or only the latter when $\rho_f$ is already expressed in terms of time modes).

Notably, besides properties (1) and (3) of  \ref{conds} (always matched as discussed before), property (2) will be fulfilled whenever $\rho_f$ is such that
\begin{equation}
	\rho_B=\bigotimes_n \eta_n\,\,\,\,\,\,\,\rm{ (condition for Markovian dynamics)}\label{prod}
\end{equation}
with $\eta_n$ the initial state of mode $\hat b_n$. In this case, the emerging CM is memoryless
(see  \cref{section-def,section-inhomCP}). 
It turns out that condition \ref{prod} is fulfilled by a number of relevant classes of field states, some of which are illustrated next.

\section{Vacuum state}
\label{section-vacuum}

The field vacuum state $\ket{{\rm vac}}$ is defined as the state such that $\hat b_\omega \ket{{\rm vac}}=0$ for any $\omega$. Since the analogous statement clearly holds for time modes, \eq\ref{fourier} entails that $\hat b_{n,k} \ket{{\rm vac}}=0$ for any $n$, $k$. Hence, $\rho_B$ is of the form \ref{prod} -- meaning that the dynamics is Markovian -- with
\begin{equation}
	\eta_n=\ket{0}_{n}\bra{0}\,.
\end{equation}
In the case that $S$ is a qubit, namely  $\hat A=\hat \sigma_-$ [\cf\eq\ref{HQO}], conservation of the total number of excitations $\hat \sigma_+\hat \sigma_-{+}\sum_n \hat b_n^\dag\hat b_n$ entails that the state of each ancilla must lie in the subspace spanned by the pair of Fock states $\ket{0}_n$ and $\ket{1}_n$ with $\ket{1}_n=\hat b_n^\dagger \ket{{\rm vac}}$. Thus ancillas behave as effective qubits.\marginnote{In passing, this justifies the convention to define $\ket{1}$ such that $\hat\sigma_{z}\ket{1}=\ket{1}$, which we followed throughout the paper.} We thus recover the all-qubit CM of  \cref{section-all-qubit} (when $g_z=0$ and each ancilla is prepared in $\ket{0}$), which we used in particular to derive the spontaneous-emission ME \ref{ME-SE}.

\section{Thermal states}
\label{section-thermal}

Formally, a thermal state of the bosonic bath at inverse temperature $\beta=(K{T})^{-1}$ would read
\begin{equation}
	\rho_{f}=Z^{-1}e^{-\beta{\hat H_{f}}}\label{rho-th-f}
\end{equation} 
with $Z={{\rm Tr}_f\{e^{-\beta{\hat H_f}}\}}$ the field partition function. In our case, replacing $\hat H_{f}$ with the expression in \eq\ref{HQO} would yield an unphysical thermal state due to the absence of a lower bound of the field spectrum. To get around this difficulty, it is customary to make the brute-force approximation consisting in replacing $\hat H_f$ in \ref{rho-th-f} with
\begin{equation}
	\label{app-Hs}
	\hat H_f\simeq \omega_{0}\int_{-\infty}^\infty d\omega\,  \hat{b}^\dagger_\omega\, \hat{b}_\omega\,.
\end{equation}
Upon comparison with $\hat H_f$ in \eq\ref{HQO}, we see that this is equivalent to stating that the field normal modes are perfectly resonant with $S$ (neglecting the dispersion). This again relies on weak coupling according to which only field normal modes within a narrow bandwidth around $\omega_0$ (\cf\ \ref{fig-window}) exchange a significant amount of energy with $S$. Under approximation \ref{app-Hs}, by noting that $\int d\omega\,\hat{b}^\dagger_\omega\, \hat{b}_\omega$ is the total number of bosonic excitations, which can be equivalently expressed as $\int d t\,\hat{b}^\dagger_t\, \hat{b}_t=\sum_{n,k}\hat b_{n,k}^\dag\hat b_{n,k}$ [\cf\eq\ref{fourier}], we have
\begin{align}
	\rho_{f}\simeq &Z^{-1}e^{-\beta \omega_{0}\int d\omega\,\hat{b}^\dagger_\omega\, \hat{b}_\omega}=Z^{-1}e^{-\beta \omega_{0}\int d t\,\hat{b}^\dagger_t\, \hat{b}_t}=\nonumber\\	&Z^{-1}e^{-\beta \omega_0 \sum_{n,k}\hat b_{n,k}^\dag\hat b_{n,k}}=\bigotimes_{n,k} Z_{n,k}^{-1}\,\,e^{-\beta \omega_0 \hat b_{n,k}^\dag\hat b_{n,k}}\,.\label{rho-th-2}
\end{align} 
with $Z_{n,k}={\rm Tr}_{n,k}\{e^{-\beta \omega_0 \hat b_{n,k}^\dag\hat b_{n,k}}\}$. Thereby, \eq\ref{prod} holds with
\begin{equation}
	\eta_n=Z_{n,0}^{-1}\,\,e^{-\beta \omega_0 \hat b_{n}^\dag\hat b_{n}}\,.\label{therm}
\end{equation}
It follows that $S$ is governed by the same finite-temperature master equation that we obtained in  \cref{section-QT1} to describe thermalization.\marginnote{Unlike  \cref{section-QT1}, here ancillas do not have a free Hamiltonian since in the interaction picture chosen above the only Hamiltonian term is that describing the $S$-field interaction. Yet, the reduced dynamics of $S$ is the same as in  \cref{section-QT1} because the $S$-ancilla coupling and the ancilla initial state are identical.}

\section{Coherent states}
\label{section-coh}

A generic coherent state of the bosonic bath field has the form $\rho_f=\ket{\alpha}\bra{\alpha}$
with\marginnote{For a discrete bosonic field, a multimode coherent state has the form
	$\bigotimes_j\exp(\alpha_j\,\hat b_j^\dagger {-}\alpha_j^{*} \hat b_j)\,\vert {\rm vac}\rangle =\exp[\sum_j(\alpha_j\,\hat b_j^\dagger {-}\alpha_j^{*} \hat b_j )]\,\vert {\rm vac}\rangle $, whose \eq\ref{coher} represents the continuous
	version~\cite{loudon2000quantum}.}
\begin{equation}
	\vert \alpha\rangle =e\,^{\int {d}\omega\,\left(\alpha_\omega \hat b^\dagger_\omega-\alpha^{*}_\omega \hat b_\omega\right)}\,\vert {\rm vac}\rangle \,\label{coher}
\end{equation}
with $\alpha_\omega$ the pulse shape in the frequency domain. The standard continuous-wave case occurs for $\alpha_\omega\propto \delta(\omega-\omega_d)$ with $\omega_d$ the drive frequency. The state can be equivalently expressed in terms of time modes as
\begin{equation}
	\vert \alpha\rangle =e\,^{\int {d}t\,\left(\alpha_t \hat b^\dagger_t-\alpha^{*}_t \hat b_t\right)}\,\vert {\rm vac}\rangle \,,\label{coher-t}
\end{equation}
where $\alpha_t={1}/{\sqrt{2 \pi}} \int d\omega  \,\alpha_\omega e^{-i \omega t}$ encodes the pulse shape in the time domain. By decomposing $\hat b_t$ through \ref{fourier}, the exponent of \ref{coher} becomes
\begin{equation}
	\int {d}t\,\left(\alpha_t \hat b^\dagger_t-{\rm H.c.}\right)=\sum_{n,k}\tfrac{1}{\sqrt{\Delta t}}\left(\int_{t_{n-1}}^{t_n} {d}t\,\alpha_t \,e^{i \frac{2\pi k}{\Delta t}t}\right)\hat b^\dagger_{n,k}-{\rm H.c.}\,.
\end{equation}
Accordingly, condition \ref{prod} for Markovian dynamics is matched for $\eta_n=\ket{\alpha_n}_n\bra{\alpha_n}$, where
\begin{equation}
	\vert \alpha_n\rangle = e^{ \alpha_n\sqrt{\Delta t}\,\hat b^\dagger_n-\alpha_n^{*}\sqrt{\Delta t}\, \hat b_n}\,\vert 0_n\rangle \,\,\,\,{\rm with}\,\,\,\,\alpha_n=\tfrac{1}{\Delta t}\int_{t_{n-1}}^{t_n} {d}t\,\alpha_t\,\,\,
\end{equation}
($\alpha_n$ is the mean value of $\alpha_t$ on interval $[t_{n-1},t_n]$). 

Thus each ancilla is initially in a (single-mode) coherent state of amplitude $\alpha_n\sqrt{\Delta t}$ (note the $\sqrt{\Delta t}$-proportionality). For $\Delta t$ small enough this can be approximated to the lowest order as
\begin{equation}
	\vert \alpha_n\rangle = e^{-\tfrac{1}{2}\vert \alpha_n\vert ^2 \Delta t}\sum_{k=0}^\infty\,\frac{(\alpha_n\sqrt{\Delta t})^k}{\sqrt{k!}}\ket{k_n}\simeq \frac{1}{1+\vert \alpha_n\vert ^2\Delta t}\left(\vert 0\rangle_n+\alpha_n \sqrt{\Delta t}\,\vert 1\rangle _n\right)\,,\label{alphan}
\end{equation}
which is normalized to the first order in $\Delta t$ (here $\ket{k_n}=(\hat b_n^\dag)^k/\sqrt{k!}\,\vert {\rm vac}\rangle $). We thus retrieve state \ref{superp}, which we considered in  \cref{sec-div} for the all-qubit CM showing that it leads to optical Bloch Eqs.~\ref{ME-BLOCH}.\marginnote{Strictly speaking, when $S$ is a qubit each ancilla behaves as an effective three-level system (with Hilbert space spanned by $\{\ket{0_n},\ket{1_n},\ket{2_n}\}$) due to the possible transition $\ket{1_S}\ket{1_n}\rightarrow\ket{0_S}\ket{2_n}$. Yet, in the limit of short $\Delta t$, this has negligible probability compared to $\ket{0_S}\ket{1_n}\rightarrow\ket{1_S}\ket{1_n}$ since the $\ket{1_n}$'s component of state \ref{alphan} is of order $\sim \sqrt{\Delta t}$ so that the all-qubit CM is effectively retrieved (as usual, $\ket{0_S}$ and $\ket{1_S}$ are respectively the ground and excited states of $S$).}

\section{General white-noise Gaussian state}
\label{section-gaussian}

By definition, a \textit{Gaussian} state of the field is fully specified by the
knowledge of first and second moments $\langle \hat b_t \rangle$ and $\langle \hat b^\dagger_t \hat b_{t'} \rangle$, $\langle \hat b_t \hat b_{t'} \rangle$
with $\langle \ldots  \rangle={\rm Tr}_f\{\ldots \,\rho_f\}$. For $\delta$-correlated second moments, namely \eg
$\langle \hat b_t \hat b_{t'} \rangle\propto \delta(t-t')$, $\rho_f$ is a so called \textit{white-noise} Gaussian state. The
standard way to express its general form is~\cite{wiseman2009quantum}
\begin{equation}
	\langle d\hat B_t \rangle =\beta_t \,dt\,,\,\,\,\langle d\hat B^\dagger_t  \,d\hat B_t\rangle=N \,dt\,,\,\,\,\langle d\hat B_t  \,d\hat B_t\rangle=M \,dt\label{dB}\,.
\end{equation}
with $N\ge 0$ and where $\beta_t$ and $M$ are complex coefficients subject to the constraint $\vert M\vert ^2\le N(N+1)$. Here, $M$ measures the amount of \textit{squeezing} of the field, while $d\hat B_t=\int_t^{t+dt}ds\,\hat b_s$ is the so called quantum noise increment fulfilling  the commutation rule $[d\hat B_t,d\hat B_t^\dag]=dt$ [following from $[\hat b_t,\hat b^\dagger_{t'}]=\delta(t-t')$]. Thus \eq\ref{dB} gives first and second moments of noise increments at the same time, while those at different times vanish (meaning, in particular, that time modes are initially uncorrelated). Using \ref{bn0} this entails that first and second moments of ancillas are given by
\begin{equation}
	\label{dbn}
	\langle \hat b_n \rangle =\beta_n \,\sqrt{\Delta t}\,,\,\,\,\langle \hat b_n^\dagger   \,\hat b_{n'}\rangle=\delta_{n,n'}\,N \,,\,\,\,\langle \hat b_n  \,\hat b_{n'}\rangle=\delta_{n,n'}\,M \,
\end{equation}
with $\beta_n$ the mean value of $\beta_t$ on the $n$th interval. Second moments vanish for $n\neq n'$, guaranteeing that condition \ref{prod} holds.\marginnote{Any two-mode Gaussian state $\rho_{12}$ such that $\langle \hat b^\dagger _1 \hat b_2\rangle=\langle \hat b_1 \hat b_2\rangle=0$ is necessarily a product state, i.e.,~$\rho_{12}=\rho_{1}\otimes\rho_2$ (third- or higher-order correlation functions are zero since Gaussian states are by definition fully specified by first and second moments). This is naturally generalized to more than two modes.} Corresponding to the continuous field state [\cf\eq\ref{dB}], here $N$ is the average number of excitations of each ancilla while $M$ measures its squeezing. 

The states discussed in the previous sections are special cases of \ref{dbn}: $\beta_n=N=M=0$ (vacuum), $\beta_n=M=0$ and $N=\bar n_{\omega_0}$ (thermal state), $\beta_n=\alpha_n$, $N=\vert \alpha_n\vert ^2$ and $M=0$ (coherent state) [recall definition \ref{nav0}].

In light of \cref{drho2,HS-diss}, the above in fact provides the most general master equation of $S$ for an arbitrary white-noise Gaussian state of the field. Note that the continuous-time limit [\cf  \cref{sec-div}] is always well-defined since $\langle \hat b_n \rangle\propto \sqrt{\Delta t}$ [\cf\eq\ref{dbn}].

\section{Initially-correlated ancillas}
\label{section-singleph}

There are a variety of field states such that condition \ref{prod} does not hold, which makes the dynamics non-Markovian. The simplest instance is probably a single-photon state like
\begin{equation}
	\ket{\Psi}_f=\int  dt \,\Psi_t \,\hat b_t^\dagger \ket{{\rm vac}}\,,
\end{equation}
where $\Psi_t$ is a photonic wavepacket.
Using \ref{fourier}, the corresponding initial state of the ancillas reads $\rho_B=\ket{\psi}_B\langle \psi\rvert  $ with
\begin{equation}
	\ket{\psi}_B=\sum_n c_n \vert 1_n\rangle \,\,\,\,\,\,{\rm with}\,\,\,\,\,c_n=\tfrac{1}{\sqrt{\Delta t}}\int_{t_{n-1}}^{t_n}dt\,\Psi_t\,,
\end{equation}
which is a generally entangled, thus \textit{correlated}, state [\cf  \cref{section-ica}].

\section{Connection with input--output formalism}
\label{link-io}

The collisional picture of the dynamics (see  \cref{section-emer}) was defined above in terms of evolution of \textit{states}. Yet, one can let equivalently evolve operators so that each collision is governed by the operatorial equation\marginnote{In the present subsection, time arguments appear in the standard form (not as subscripts or superscripts).}
\begin{equation}
	\frac  {d}{dt}\hat b_n(t)=i\, [\hat V_n, \hat b_n(t)]=-i \sqrt{\tfrac{\gamma}{\Delta t}}\,\hat A(t)\,.
\end{equation}
where we used \eq\ref{HBn2}.
Since $\Delta t$ is very short we can replace the derivative with $\Delta\hat b_n/\Delta t$, where $\Delta\hat b_n=\hat b_n(t_{n})-\hat b_n(t_{n-1})$  (recall that the $n$th collision occurs in the time interval $t_{n-1}\le t< t_n$). This yields
\begin{equation}
	\label{io-d}
	\hat b_n(t_{n})=\hat b_n(t_{n-1})-i \sqrt{\gamma\Delta t}\,\,\hat A(t_{n-1})\,.
\end{equation}
This equation can be understood by interpreting $\hat b_n(t_{n-1})$ as an \textit{input}
discrete field, whose interaction with $S$ produces an \textit{output}
field $\hat b_n(t_{n})$. Indeed, \ref{io-d} can be seen as the discrete version of the
central equation underpinning the so called input--output formalism of quantum optics
(see e.g.~\rref~\cite{wiseman2009quantum})
\begin{equation}
	\label{io}
	\hat b^{({\rm out})}(t)=\hat b^{({\rm in})}(t)-i \sqrt{\gamma}\,\hat A(t)
\end{equation}
with $\hat b^{({\rm in})}(t)$ and $\hat b^{({\rm out})}(t)$ being the continuous limits of $\hat b_n(t_{n-1})/\sqrt{\Delta t}$ and $\hat b_n(t_{n})/\sqrt{\Delta t}$, respectively.

\section{Collision models from conventional models: state of the art}
\label{soa-cmqo}

The above derivation of the CM from the microscopic bosonic model is largely based on
\rrefs~\cite{ciccarello_collision_2017,gross_qubit_2018,cilluffo2020} (see also
\rref~\cite{fischer2017derivation}). In particular, \rref~\cite{cilluffo2020}
encompasses the extension to a multipartite system $S$ that can couple to
the field non-locally. This brings about a new feature in that, relaxing the
hypothesis that $S$ is point-like (as assumed throughout in the above),
term $\hat V_n'$ in the elementary unitary \ref{Un-app1} has a contribution due
to vacuum fluctuations that yields an effective (second-order) induced Hamiltonian
for $S$~\cite{cilluffo2020}. In the case of systems each interacting with a
waveguide field at multiple coupling points (such as ``giant
atoms"~\cite{Kockum5years} or oscillators in looped geometries~\cite{Hammerer2019}),
this effective Hamiltonian can be made decoherence-free~\cite{carollo2020mechanism}.
This phenomenon was predicted in \rref~\cite{KockumPRL2018} (through methods not
based on CMs) and then experimentally observed in a circuit-QED
setup~\cite{kannan2020waveguide}. Mapping the dynamics into an effective CM allows
for a full-fledged interpretation of the physical mechanism underlying  such class of
decoherence-free Hamiltonians, which was shown in \rref~\cite{carollo2020mechanism}.

Note that, while for vacuum and coherent states (\cref{section-vacuum,section-coh})
the field time bins naturally behave as effective qubits, this is generally not the
case (for instance for thermal or squeezed states). However, as shown in
\rref~\cite{gross_qubit_2018}, one can always replace the time bins with suitably
defined qubits yielding the same open dynamics of $S$. 

In the model we considered, $\hat H_S$ is time-independent. One can yet extend
the framework so as to account for an external drive on $S$, an approach
that was successful in studying directional emission into a waveguide from a quantum
emitter subject to a pulsed laser~\cite{fischer_scattering_2018}.

Relying on its tight link with the input--output formalism (see  \ref{link-io}),
the CM mapping was recently exploited to infer equations of motion and input--output
relations of cavity-waveguide systems~\cite{JacobsPRA20,JacobsPRL20}, carry out
quantum simulations of coherent light--matter interactions~\cite{Bouten2019,Bouten2},
design qubit-oscillator circuits for implementing quantum error correction
codes~\cite{GirvinPRL20} and investigate non-equilibrium thermodynamics (see
\cref{section-thermo}) in waveguide QED~\cite{maffei2021probing}.

The CM mapping discussed here can be extended to a system $S$ coupled to
the field at many points in the regime of non-negligible \textit{delays}. This
results in non-Markovian CMs with multiple \textit{non-local} collisions (see
\cref{section-multiple}), which were applied in
\rrefs~\cite{pichler_photonic_2016,grimsmo_time-delayed_2015,cilluffo2019quantum}. 

Due to the natural connection of CMs with quantum trajectories (see
\cref{section-qtraj}), another promising application of the collisional mapping are
non-Markovian extensions of photon counting and quantum trajectories (usually
formulated for Markovian dynamics~\cite{wiseman2009quantum,carmichael2009open}).
Examples are non-Markovian dynamics induced by single-photon states (see
\cref{section-singleph})~\cite{dabro2017,dabroJPA,dabrowska2020posteriori},
{superposition of coherent states}~\cite{dkabrowska2019quantum} and delayed coherent
feedback~\cite{whalen2019,regidor2020modelling}.

We finally mention that, formally, even in the case of micromaser (\cf
\cref{section-maser}) one can define an effective quantum field whose the two-level
atoms are the corresponding quanta~\cite{CresserPRA92}.

\setchapterpreamble[u]{\margintoc}  
\chapter{Conclusions}
\label{section-concl}

In this paper, by adopting a pedagogical approach we presented the theory of quantum collisions models (CMs), reviewing at the same time the related state of the art. In line with  \ref{fig-struc}, our discussion analyzed first the basic properties of CMs in  \cref{section-1,section-eqs} and then considered the major areas of application of CMs to date: quantum trajectories/weak measurements (\cref{section-qtraj}), non-equilibrium quantum thermodynamics (\cref{section-thermo}), non-Markovian extensions of CMs \ref{section-NM} and white-noise microscopic models (\cref{section-CMQO}), the latter being recurrent in quantum optics.

Besides those featured in the previous state-of-the-art sections, there exist further interesting applications of CMs (and new ones keep being proposed). One of these is \textit{quantum Darwinism}
~\cite{SteveDarwinPRA,ChisholmPRR2020,LorenzoZenoPRR,lorenzo2019reading,ccakmak2021quantum},
where a CM description allows for a dynamical study of information spreading across
the bath. Very recently, CMs started being applied to \textit{quantum biology}
problems, mostly as a versatile tool for modeling decoherence including non-Markovian
effects (see  \cref{section-NM}). In particular, \rref~\cite{ChisholmNJP}
investigated quantum transport across a Fenna--Matthews--Olson complex, while
\rref~\cite{olaya} studied decoherence of an avian-inspired quantum magnetic sensor. 
Other recent applications include: {quantum
	classifiers}~\cite{turkpencce2019steady} simulation of the Unruh effect~\cite{unruh},
quantum friction~\cite{FrictionPRA19}, information scrambling~\cite{scramblingPRA20}, quantum batteries~\cite{landi2021} and quantum~metrology \cite{metro21}.

Needless to say, while the paper dealt with well-established theory, there are a number of problems which are still open some of which are mentioned next. 

\cref{section-NM} introduced various classes of non-Markovian CMs. The relationships between these classes are still unexplored, \eg whether or not it is possible to map one class into another, which was proven only for ancilla--ancilla collisions and composite CMs (see  \cref{section-mapCCM}). This is an interesting question also from a fundamental viewpoint since it would help clarifying the relationship between seemingly different memory mechanisms corresponding to the relaxation of one of assumptions (1)--(3) in  \ref{conds}.

Another open issue concerns the derivation of CMs from conventional microscopic
models, which was carried out in  \cref{section-CMQO} only for bosonic baths. The
procedure we followed there does depend on the bosonic commutation rules of the
field, allowing to define in a relatively natural way independent ancillas (in the
sense that operators of different ancillas are mutually commuting). A strictly
analogous procedure for fermionic fields would lead to non-commuting ancillas, hence
a suitable non-trivial extension is demanded. It appears reasonable to expect that a
CM mapping exists also in this case since Markovian dynamics and Lindblad master
equations occur for fermionic baths as well. This problem is arguably related to the
definition of input--output formalism for fermionic fields~\cite{gardiner2004input}.

While writing this paper, the interest in CMs keeps growing as \eg witnessed by
regular submissions of preprints to the Los Alamos archive. A natural question is to
what extent the field of application of CMs could be enlarged. Should one envisage
such approach becomes one day the conventional methodology? This is a non-trivial
question to answer. One of the key points is the ability of CMs to describe
non-Markovian dynamics. While research along this line is still in the early stages,
one can expect (see \eg  \cref{section-aacu,section-mapCCM}) that the higher is the
degree of non-Markovianity the larger will be the number of (effective) bath ancillas
one has to keep track with the same level of detail as the open system $S$
(see also \rref~\cite{filippovPRL}). Aside from the obvious difficulty to account for
many degrees of freedom, we note that at some point this might even question the very nature of the
collisional approach whose spirit is reducing complex dynamics to a sequence of
simple interactions. This is well-illustrated by the instance in  \ref{fig-delay}
to describe which we needed to cope somehow with all ancillas at each step (which was
possible only because a single excitation was involved in the problem).

What appears by now well-assessed is that the collisional approach performs extremely well in a number of problems such as derivation of well-defined master equations, both Markovian and non-Markovian, the calculation of thermodynamic rates in non-equilibrium processes (where handling conventional microscopic models is often beyond reach), the physical interpretation of complex dynamics, the study of non-Markovianity.

An interesting future direction would be to synergically combine CMs or CM-inspired
methods with other techniques (such as tensor network), as recently done in
\rref~\cite{purkayastha2020periodically}.

On a merely pedagogical ground, we envisage that CMs could become a standard strategy for introducing students to the basics of open quantum systems theory. In this respect, note that our discussion dealt with most main concepts of this field such as quantum maps, Lindblad master equation, steady states, POVMs, quantum trajectories, stochastic Schr\"odinger equation, Stinespring dilation theorem. The required background  is in fact some familiarity with elementary quantum mechanics. Moreover, developing a physical intuition of the various topics (\eg the conditions for the Lindblad master equation to hold) is facilitated compared to conventional microscopic models (\cf\ref{app-lind}). 

We hope that the systematic settlement of the CMs theory that we tried to carry out here could spur an increasing use of CMs among students and researchers or at least stimulate a ``collisional thinking" of open quantum systems problems in addition to, or possibly in combination with, other methods.

\chapter{Acknowledgments}
What we learned about collision models over these years greatly benefited from discussions and collaborations with a number of valuable people to whom we are deeply grateful. Among these are (in alphabetical order) D.~Burgarth, G.~Benenti,  S.~Campbell, A.~Carollo, D.~Chisholm, D.~Cilluffo, G.~De Chiara, A.~Grimsmo, J.~A.~Gross,  G.~T.~Landi, S.~Maniscalco, R.~McCloskey, M.~Paternostro, T.~Tufarelli and B.~Vacchini.

We gratefully acknowledge D.~Cilluffo, S.~Campbell, G.~T.~Landi, B.~Vacchini, G.~De Chiara and G. Manzano for the critical reading of the manuscript. We are indebted to D.~Cilluffo for the help offered in the preparation of  \cref{section-MPS}.

We acknowledge financial support from MIUR through project PRIN (Project No.2017SRN-BRK QUSHIP).

\appendix 
\pagelayout{wide} 
\addpart{Appendix}
\pagelayout{wide} 
\chapter{Appendix\;}
\section{Density matrices}
\label{app-rho}

The most general state of a quantum system $S$ is described by a \textit{density operator} $\rho$ (often referred to as density matrix). This is a Hermitian, positive semi-definite operator of trace one. As such, it can always be expanded (``spectrally decomposed") as 
\def\theequation{A.\arabic{equation}}
\setcounter{equation}{0}
\begin{equation}
	\rho=\sum_\nu p_\nu \vert \nu\rangle \langle \nu\rvert   \label{SD}\,
\end{equation}
with $p_\nu\ge 0$ (positivity\footnote{Rigorously speaking, this expresses non-negativity, but we will refer to this property as ``positivity" to simplify the language.}) and ${\rm Tr}\rho=\sum_\nu p_\nu=1$ (normalization). Here, $\{\vert \nu\rangle \}$ are the eigenstates of $\rho$, \ie $\rho\vert \nu\rangle =p_\nu \vert \nu\rangle $ for all $\nu$, which form an orthonormal basis of the Hilbert space of $S$.  When all probability $p_\nu$ vanish but one, $\rho$ reduces to a simple projector, in which case we say that the state is \textit{pure}. In all other cases, we deal with a mixed state. While the usual description through kets is always possible for pure states, the density--matrix language is indispensable for representing \textit{mixed} states.

Spectral decomposition \ref{SD} expresses $\rho$ as a mixture of \textit{orthogonal} (pure) states. A density matrix can however be alternatively expressed as a mixture of non-orthogonal states, for instance a legitimate state for a qubit is $\rho=1/2 \vert 0\rangle \langle 0\rvert  +1/2 \vert +\rangle \langle +\rvert  $ with $\ket{\pm}=\tfrac{1}{\sqrt{2}}(\ket{0}+\ket{1})$, where $\ket{0}$ and $\ket{+}$ are non-orthogonal.

The density--matrix language is essential for describing subsystems.  Assume that $S$ is part of a larger bipartite system, the other subsystem being $E$ (no matter how big). Then, if $\sigma$ is the joint $S-E$ state, the state of $S$ is given by the partial trace over $E$
\begin{equation}
	\rho={\rm Tr}_E \,\sigma=\sum_{\mu} {\mbox{}_E} \langle \mu\rvert   \sigma\vert \mu\rangle _E\label{trace}\,,
\end{equation}
where $\{\vert \mu\rangle _E\}$ is an arbitrary orthonormal basis of $E$ (it is easily checked that this satisfies the definition of density operator).

\section{Von-Neumann entropy, mutual information and relative entropy}
\label{app-ent}

Given a (generally mixed) state $\rho$ the \textit{Von Neumann entropy} is
defined as~\cite{nielsen2002quantum}
\def\theequation{B.\arabic{equation}}
\setcounter{equation}{0}
\begin{equation}
	{\mathcal{S}}(\rho)=-\VNS{\rho\log{\rho}}\,.
\end{equation}
This is the natural quantum analogue of the Shannon entropy occurring in classical information theory. This can be seen by spectrally decomposing $\rho$ as in \eq\ref{SD}, which entails
\begin{equation}
	{\mathcal{S}}(\rho)=-\sum_\nu p_\nu \log p_\nu\,.
\end{equation}
Also, this shows that ${\mathcal{S}}(\rho)\ge 0$ for any $\rho$. Specifically, entropy vanishes for pure states and is non-zero for mixed states. This matches the picture of a mixed state as a statistical mixture of pure states. For instance, consider the qubit state $\rho=1/2 \ket{0}\bra{0}+1/2 \ket{1}\bra{1}=\tfrac{1}{2}\mathbb{I}$. This can be interpreted by saying that we are fully ignorant about whether $S$ is in $\ket{0}$ or $\ket{1}$. Entropy is a measure of such \textit{ignorance}. Indeed, in the considered instance, it takes its maximum value ${\mathcal{S}}=\log2$.\footnote{This is the maximum value for a qubit. In general, for a system with Hilbert space dimension $d$, the maximum entropy is $S=\log d$  (for a qubit, $d=2$)} 
In contrast, $\mathcal S(\ket{0}\bra{0})=0$ as we are fully sure that $S$ is in the pure state $\ket{0}$. 
An important property of the Von Neumann entropy is that it does not change under a unitary transformation, \ie
\begin{equation}
	\mathcal S(\rho)=\mathcal S(\hat U \rho\hat U^\dag)\label{SU}
\end{equation}
for any state $\rho$ and unitary $\hat U$. This is immediately seen from \ref{SD} by noting that $\hat U \rho \,\hat U^\dag$ has the same spectral decomposition as $\rho$ under the change of basis $\{\ket{\nu}\}\rightarrow\{\hat U\ket{\nu}\}$. 

The Von Neumann entropy underpins the definition of two useful quantities, quantum relative entropy and quantum mutual information.

Unlike Von Neumann entropy which is associated with a single state, the quantum \textit{relative entropy} depends on a pair of states, say $\rho$ and $\rho'$. It is defined as
\begin{equation}
	\mathcal S(\rho\parallel \rho')=-\VNS{\rho\,\log{\rho'}}-\mathcal S(\rho)={\rm Tr}\{\rho\,(\log \rho-\log\rho')\}\,.\label{Srel}
\end{equation} 
It can be shown that $\mathcal S(\rho\parallel \rho')\ge0$ (non-negativity) with $\mathcal S(\rho\parallel \rho')=0$ if and only if $\rho=\rho'$. Relative entropy is useful because it is a measure of the \textit{distinguishability} between two quantum states. Notably, it is not symmetric under swap of states, \ie $\mathcal S(\rho\parallel \rho')\neq \mathcal S(\rho'\parallel \rho)$.\footnote{This is a reason why relative entropy cannot be used to define a metric in the Hilbert space.}

Another entropic quantity is quantum \textit{mutual information}, the quantum version of mutual information (a longstanding measure of correlations). Given a pair of systems $S$ and $E$, it is defined as
\begin{equation}
	{\mathcal{I}}_{SE}=\mathcal S(\rho_S)+\mathcal S(\rho_E)-\mathcal S(\rho_{SE})\,
\end{equation}
with $\rho_{SE}$ the joint state and $\rho_{S(E)}={\rm Tr}_{E(S)} \{\rho_{SE}\}$ the reduced states. Mutual information fulfills ${\mathcal{I}}_{SE}\ge0$ with
\begin{equation}
	{\mathcal{I}}_{SE}=0\,\,\Leftrightarrow\,\,\rho_{SE}=\rho_S\otimes\rho_E\,.
\end{equation}
Thus ${\mathcal{I}}_{SE}>0$ witnesses the existence of $S$-$E$ correlations.

\section{Quantum maps}
\label{app-qmaps}

Transformations of density matrices are described by \textit{quantum maps}.  A quantum map transforms a state $\rho$ into another state $\rho'$, which is expressed as $\rho'=\mathcal M[\rho]$.
A major class of quantum maps is that defined by
\def\theequation{C.\arabic{equation}}
\setcounter{equation}{0}
\begin{equation}
	\rho'=\mathcal M[\rho]=\sum_m \hat K_m \,\rho\, \hat K_m^\dag\,\,\,\,\,\,{\rm with}\,\,\,\,\,\,\sum_m \hat K_m^\dag\hat K_m=\mathbb{I}\,\,.\label{kraus-map}
\end{equation}
These are called \textit{completely positive and trace-preserving (CPT) maps}.\footnote{We do not discuss here the concept of \textit{complete }positivity, using \ref{kraus-map} as the definition of a CPT map.} 

The rightmost expansion in \ref{kraus-map} is called \textit{Kraus decomposition} and $\hat K_m$ the Kraus operators. The Kraus decomposition (demonstrably) ensures that, if $\rho$ is a well-defined density matrix, then so is $\rho'$. The importance of CPT maps indeed relies on the fact that they describe physically-legitimate transformations, \eg due to a dynamical evolution or measurement, \ie they map physical states into physical states.

Note that, like any operator, a unitary transformation transforms a density matrix as $\rho'=\hat U \rho \,\hat U^\dag$ subject to $\hat U^\dag\hat U=\mathbb{I}$, which is a special case of quantum map \ref{kraus-map} having only one Kraus operator $\hat U$. Actually, a unitary transformation fulfills $\hat U\hat U^\dag=\mathbb{I}$ as well, while in general $\sum_m\hat K_m\hat K_m^\dag\neq\mathbb{I}$ this expressing the fact that a quantum map is generally \textit{non-unitary}. 

Non-unitarity most notably entails that the scalar product of two states is not invariant under a quantum map. The best instance to see this is probably the decay of a two-level atom: the excited state $\ket{e}$ eventually evolves into the ground state $\ket{g}$, while the ground state is unaffected. Thus $\ket{e}$ and $\ket{g}$ (which are orthogonal states) are both mapped into the same state $\ket{g}$, with the scalar product thereby changing from zero to one.

\section{Dynamical map}
\label{app-DM}

If $S$ is closed  (decoupled from anything else) its state $\rho$ evolves in time according to the Von Neumann (or quantum Liouville) equation (recall that we set $\hbar=1$) $\dot \rho=-i [\hat H_S,\rho]$. This is in fact just the \se\, expressed in the density--matrix formalism. Accordingly, the time evolution of $\rho$ is \textit{unitary}, ${\rho_t=\hat U_{St} \rho_0\hat U_{St}^\dag}$, with $\hat U_{St}=e^{-i \hat H_S t}$ the time-evolution operator.

If $S$ is open then its time evolution is generally \textit{non-unitary}. This can be seen in the case that $S$ and $E$ overall form a closed system so that they jointly evolve unitarily as $\sigma_t=\hat U_t\,\sigma_{0}\, \hat U_t^\dag$. Hence, tracing off $E$, the state of $S$ at time $t$ is given by
\def\theequation{D.\arabic{equation}}
\setcounter{equation}{0}
\begin{equation}
	\rho_t=\sum_{\mu} {\mbox{}_E\,} \langle \mu\rvert  \,  \hat U_t\,\rho_0\otimes\rho_E\, \hat U_t^\dag\,\vert \mu\rangle _E\,,
\end{equation}
where we assumed that $S$ and $E$ start in the \textit{uncorrelated} state $\sigma_{0}=\rho_0\otimes\rho_E$. Replacing now $\rho_E$ with its spectral decomposition $\rho_E=\sum_\lambda p_\lambda \vert \lambda\rangle _E\langle \lambda\rvert  $, $\rho$ can be arranged in the form
\begin{equation}
	\rho_t=\Lambda_t[\rho_0]=\sum_{\nu\lambda} \left(\sqrt{p_\lambda}\,_E\langle \mu\rvert  \, \hat U_t\,\vert \lambda\rangle _E\right)\rho_0 \left(\sqrt{p_\lambda}\,_E\langle \mu\rvert  \,  \hat U_t\,\vert \lambda\rangle _E\right)^\dag\,.\label{kraus1}
\end{equation}
\eq\ref{kraus1} defines the so called \textit{dynamical map}: for any given initial state of $S$, $\rho_0$, $\Lambda_t$ returns the dynamically evolved state at time $t$, $\rho_t$. The dynamical map $\Lambda_t$ can be seen as the open-system counterpart of the time-evolution operator. Remarkably, by comparing \eq\ref{kraus1} with \ref{kraus-map}, we see that $\Lambda_t$ is a CPT map whose generic Kraus operator, labeled by the double index ($\nu,\lambda$), reads
\begin{equation}
	\hat K_{\nu\lambda}=\sqrt{p_\lambda}\,_E\langle \mu\rvert  \, \hat U_t\,\vert \lambda\rangle _E\,.
\end{equation}

\section{Stinespring dilation theorem}
\label{app-stine}

We have just seen in \ref{app-DM} that, starting from an uncorrelated $S$-$E$ state, a global unitary dynamics results upon partial trace in a CPT map on $S$. According to the Stinespring dilation theorem, the converse property holds as well: given a CPT map $\mathcal M$ [\cf\eq\ref{kraus-map}] one can always find an ancillary system $A$, an initial state of $A$, $\rho_A$, and a global unitary $\hat U_{SA}$ (acting on $S$ and $A$) such that
\def\theequation{E.\arabic{equation}}
\setcounter{equation}{0}
\begin{equation}
	\rho'=\mathcal M[\rho]={\rm Tr}_A \left\{\hat U_{SA}\, \rho\otimes\rho_A \,\hat U_{SA}^\dag\right\}\,\,.\label{SDT}
\end{equation}
Note that in general there are infinite pairs  $(\rho_{A},\hat U_{SA})$ producing the same CPT map $\mathcal M$ through \ref{SDT}. We stress that the lack of initial correlations between $S$ and $A$ in \eq\ref{SDT} is essential for a CPT map to emerge.

For more detailed treatments of the topics from \ref{app-rho} to \ref{app-stine} see
\eg\rrefs~\cite{nielsen2002quantum,preskill1998lecture}.

\section{Lindblad master equation}
\label{app-lind}

Consider the class of dynamical maps such that
\def\theequation{F.\arabic{equation}}
\setcounter{equation}{0}
\begin{equation}
	\Lambda_{t}=\Lambda_{t-t'} \,\Lambda_{t'}\,.\label{SG-3}
\end{equation}
for any $t$ and $t'$ such that $0\le t'\le t$.  \eqref{SG-3} is called \textit{semigroup property} and can be regarded as a formal definition of a Markovian, \ie memoryless, dynamics.

It can be shown~\cite{breuer2007} that, if \ref{SG-3} holds, then $\rho=\Lambda_t [\rho_{0}]$
is the solution of a master equation (ME) having the general form
\begin{equation}
	\frac{d \rho}{dt}=-i \,[  \hat {\mathcal{H}},\rho]+\sum_ {\nu} \gamma_{\nu} \left(\hat L_{\nu} \rho  \hat L_{\nu}^\dag-\tfrac{1}{2}[\hat L_{\nu}^\dagger \hat L_{\nu} ,\rho ]_+\right)\,\label{lindblad}
\end{equation}
with $\hat {\mathcal{H}}$ a Hermitian operator and $\gamma_\nu\ge 0$ for each $\nu$. Here, $\hat L_{\nu}$ are a set of operators on $S$ called \textit{jump operators}. \eq\ref{lindblad} is the so called Gorini--Kossakowski--Sudarshan--Lindblad equation, more often referred to simply as \textit{Lindblad ME} (or ME in Lindblad form).

\subsection{Microscopic derivation from a conventional system--bath model}

We ask under what physical conditions the Lindblad ME correctly describes the open dynamics. Thus consider the generic system--bath Hamiltonian
\begin{equation}
	\hat H=\hat H_S+\hat H_B+\hat V\,.\label{H-mod}
\end{equation}
In the interaction picture with respect to $\hat H_0=\hat H_S+\hat H_B$, the joint $S$-$B$ state evolves according to the Von-Neumann equation $\dot\sigma=-i\,[\hat V_t,\sigma]$.
Solving it formally yields
\begin{equation}
	\sigma_t=\sigma_0-i\int_0^tdt'\,[\hat{V}_{t'},\hat{\sigma}_{t'}]\,.
\end{equation}. 
Plugging this back into the Von-Neumann equation one gets
\begin{equation}
	\frac{d {\sigma}}{dt}=-i\,[\hat V_t, \sigma_0] - \int_0^tdt'\,[\hat V_t,[\hat V_{t'},\sigma_{t'}] ]\,.\label{lindblad2}
\end{equation}
We assume no initial correlations between system and environment, \ie $\sigma_{0} = \rho_0\otimes\rho_B$, where $\rho_{0}$ and $\rho_B$ are respectively the initial reduced density operators of $S$ and $B$. Also, we assume ${\rm Tr}_B[\hat V_t,\sigma_{0}]=0$, which is the case \eg when $\rho_B$ is such that $[\hat H_B,\rho_B]=0$ (\eg in the case of a thermal state).
Using these and tracing off the bath $B$ in \eq\ref{lindblad2} yields
\begin{equation}
	\frac{d \rho}{dt}= - \int_0^tdt'\,{\rm Tr}_B \left\{ [\hat V_t,[\hat V_{t'},\sigma_{t'}] ]\right\}\,.\label{lindblad3}
\end{equation}
Although system and bath start in a product state, as a consequence of their interaction, mutual correlations between the two will build up. However if $B$ is a reservoir (very large number of degrees of freedom), one intuitively expects its reduced state to be little modified by the interaction with the system. Accordingly, in \eq\ref{lindblad3} one can approximate
\begin{equation}
	\sigma_t\simeq \rho_t \otimes \rho_B \label{Born}\,,
\end{equation}
which is known as \textit{Born approximation}.\footnote{We point out that approximation \ref{Born} is made only in \eq\ref{lindblad3} determining the reduced dynamics of $S$.} At the same time, we expand the interaction Hamiltonian as $\hat V_t= \sum_\nu g_\nu \hat A_{\nu t}  \hat B_{\nu t}$ (always possible), where $\hat A_{\nu t}=e^{i \hat H_S t}\hat A_\nu e^{-i \hat H_S t}$ and $\hat B_{\nu t}=e^{i \hat H_B t}\hat B_\nu e^{-i \hat H_B t}$ are a set of operators on respectively $S$ and $B$ in the interaction picture ($\hat A_\nu$ and $\hat B_\nu$ are Hermitian). With these replacements and the Born approximation \ref{Born}, \ref{lindblad3} takes the form
\begin{align}
	\frac{d \rho}{dt}=- \sum_{\mu ,\nu}g_\mu g_\nu \int_0^tdt'&\,{\rm Tr}_B\left\{\left[ \hat A_{\mu t}   \hat B_{\mu t},[ \hat A_{\nu t'}  \hat B_{\nu t'},\rho_{t'} \rho_B] \right]\right\}=\nonumber\\
	=- \sum_{\mu ,\nu}g_\mu g_\nu \int_0^tdt' &\left( (\hat A_{\mu t} \hat A_{\nu t'} \rho_{t'}  - \hat A_{\nu t'} \hat\rho_{t'} \hat A_{\mu t} )\langle \hat B_{\mu t}\hat B_{\nu t'} \rangle\right.\nonumber\\ &\left.+(\rho_{t'}\hat A_{\nu t'} \hat A_{\mu t}  - \hat A_{\mu t} \hat\rho_{t'}\hat A_{\nu t'} )\langle \hat B_{\nu t'} \hat B_{\mu t} \rangle\right)\,, \label{lindblad5}
\end{align}
where we defined 
\[
\langle \hat B_{\nu t'} \hat B_{\mu t} \rangle={\rm Tr}_B\{\hat B_{\nu t'} \hat B_{\mu t} \rho_B\}.
\]

For a large reservoir $B$, each two-time correlation function $\langle \hat B_{\mu t} \hat B_{\nu t'} \rangle$ is strongly peaked around $t-t'\simeq \tau_c$ with $\tau_c$ usually referred to as the \textit{correlation time}. This entails that any fluctuation in the bath state due to its interaction with the environment dies out on a time scale of the order $\sim \tau_c$. This time is typically very short, in particular when compared to the evolution timescale of $S$. Accordingly, in each integral over $t'$ appearing in \eq\ref{lindblad5}, we can approximate
\begin{equation}
	\rho_{t'}\simeq \rho_t \,,\label{mark-app}
\end{equation}
which is known as the \textit{Markov approximation}.

\subsection{Secular approximation}

For each $\hat A_\nu$, we now conveniently define $\hat A_{\nu\omega}=\sum'_{E,E'}\hat \Pi_{E}\hat A_\nu\hat \Pi_{E'}$ with $\hat \Pi_E$ the projector onto the eigenspace of $\hat H_S$ of energy $E$ and where the sum runs over all pairs $(E,E')$ such that $E'-E=\omega$. It is then easily checked that $\hat A_\nu=\sum_\omega \hat A_{\nu\omega}$ and, moreover, $[\hat H_S,\hat A_{\nu\omega}]=-\omega \hat A_{\nu\omega}$. It follows that, in the interaction picture, $\hat A_{\nu t}=\sum_{\omega}e^{-i \omega t}\hat A_{\nu\omega}$.  Replacing this and \ref{mark-app} in \eq\ref{lindblad5} this can be arranged in the form
\begin{equation}
	\frac{d \rho}{dt}=\sum_{\omega,\omega'}\sum_{\nu,\mu}e^{i (\omega'-\omega)t}\,\gamma_{\nu\mu}(\omega)\left(\hat A_{\mu\omega}\rho \hat A^\dagger_{\nu\omega'}-\hat A^\dagger_{\nu\omega'}\hat A_{\mu\omega}\rho\right)+{\rm H.c.}\,\label{lindblad4}
\end{equation}
with
\begin{equation}
	\gamma_{\nu\mu}(\omega)= \int_0^\infty dse^{i\omega s}\,\langle\hat B^\dagger_{\nu t}\hat B_{\mu (t-s)}\rangle\,,
\end{equation}
where in the last integral we approximated the upper limit of integration with $+\infty$ since the integrand function (see above) decays with a characteristic time $\tau_c$.
For $\rho_B$ such that $[\hat H_B,\rho_B]=0$ (\eg a thermal state), the above two-time correlation function actually depends only on the time difference $s$ and thus can be replaced with $\hat B^\dagger_{\nu s}\hat B_{\mu 0}$.

The secular approximation consists in throwing away all counter-rotating terms in \eq\ref{lindblad4}, \ie those corresponding to $\omega\neq \omega'$. This results in an equation with  time-independent coefficients, which reads
\begin{equation}
	\frac{d \rho}{dt}=\sum_{\omega}\sum_{\nu,\nu'}\gamma_{\nu\nu'}(\omega)\left(\hat A_{\nu'\omega}\rho \hat A^\dagger_{\nu\omega}-\hat A^\dagger_{\nu\omega}\hat A_{\nu'\omega}\rho\right)+{\rm H.c.}\label{lindblad55}
\end{equation}

\subsection{Master equation in Lindblad form}
Defining next
\begin{align}
	&S_{\nu\nu'}(\omega)=\tfrac{1}{2i}\left(\gamma_{\nu\nu'}(\omega)-\gamma^{*}_{\nu'\nu}(\omega)\right)\nonumber\\
	&\gamma_{\nu\nu'}(\omega)=\gamma_{\nu\nu'}(\omega)+\gamma^{*}_{\nu'\nu}(\omega)=\int_{-\infty}^\infty ds\,e^{i\omega s}\,{\rm Tr}_B\{\hat B_\nu^\dag(s)\hat B_{\nu'}(0)\}\,,
\end{align}
\eq\ref{lindblad5} takes the form
\begin{equation}
	\frac{d \rho}{dt}=-i[\hat {\mathcal{H}},\rho]+{\mathcal{D}}[\rho]\,\label{lindblad6}
\end{equation}
with
\begin{align}
	&\hat{\mathcal{H}}=\sum_{\omega}\sum_{\nu,\nu'}S_{\nu\nu'}(\omega)\hat A^\dagger_{\nu}(\omega)\hat A_{\nu'}(\omega)\\
	&{\mathcal{D}}[\rho]=\sum_{\omega}\sum_{\nu,\nu'}\gamma_{\nu\nu'}(\omega)\left(\hat A_{\nu'\omega}\rho\hat A^\dagger_{\nu\omega}-\tfrac{1}{2}\,[\hat A^\dagger_{\nu\omega}\hat A_{\nu'\omega},\rho]_+\right)\,.\nonumber
\end{align}
This master equation can be put in the standard Lindblad form \ref{lindblad} upon diagonalization of each matrix $\gamma_{\nu\nu'}(\omega)$.

The above derivation of the Lindblad master equation from the Hamiltonian model
\ref{H-mod} follows standard textbooks, in particular
\rrefs~\cite{breuer2007,carmichael2009open}, to which the reader is referred for
further details. In the context of the present paper, it serves the purpose of
illustrating that the derivation of the Lindblad ME from a standard microscopic model
is a relatively involved procedure which requires a number of non-trivial
approximations.

\section{Lindblad master equation from the stochastic  \se}
\label{appendix-sse}

Using \cref{SSEc,dN}, the three terms in \eq\ref{drhoS} are worked out as
\def\theequation{G.\arabic{equation}}
\setcounter{equation}{0}
\begin{align}
	\left({d}\ket{\psi}\right)\bra{\psi}&=- \tfrac{1}{2}  \gamma\,(\hat\sigma_+\hat\sigma_{-}{-}\langle\hat\sigma_{+}\hat\sigma_{-}\rangle) \ket{\psi}\langle \psi\rvert  \,{d}t+ \left(\frac{\hat\sigma_{-}}{\sqrt{\langle\hat\sigma_{+}\hat\sigma_{-}\rangle}}-\mathbb{I}\right)\ket{\psi}\bra{\psi}\,\overline{{d} N}\nonumber\\
	&=-\tfrac{\gamma}{2}\,\hat\sigma_{+}\hat\sigma_{-}\rho\,{d}t+ \tfrac{\gamma}{2} {\langle\hat\sigma_{+}\,\hat\sigma_{-}\rangle} \,\rho \,{d}t+\tfrac{\gamma}{\sqrt{\langle\hat\sigma_{+}\,\hat\sigma_{-}\rangle}} \,\hat\sigma_{-}\,\rho\,\,{d}t-\gamma {\langle\hat\sigma_{+}\,\hat\sigma_{-}\rangle} \,\rho \,{d}t\,,\label{id1}\\
	\ket{\psi}\left({d}\bra{\psi}\right)&=- \tfrac{1}{2}\gamma\ket{\psi} \bra{\psi}  (\hat\sigma_+\hat\sigma_{-}{-}\langle\hat\sigma_{+}\hat\sigma_{-}\rangle)\, {d}t+\ket{\psi}\bra{\psi} \left(\frac{\hat\sigma_{+}}{\sqrt{\langle\hat\sigma_{+}\hat\sigma_{-}\rangle}}-\mathbb{I}\right)\,\overline{{d} N}\nonumber\\
	&=-\tfrac{\gamma}{2}\,\rho\,\hat\sigma_{+}\hat\sigma_{-}\,{d}t+\tfrac{\gamma}{2} {\langle\hat\sigma_{+}\,\hat\sigma_{-}\rangle} \,\rho \,{d}t+\tfrac{\gamma}{\sqrt{\langle\hat\sigma_{+}\,\hat\sigma_{-}\rangle}}\, \rho\,\hat\sigma_{+}\,\,\,{d}t-\gamma {\langle\hat\sigma_{+}\,\hat\sigma_{-}\rangle} \,\rho \,{d}t\,,\label{id2}\\
	{d}\ket{\psi}{d}\bra{\psi}&= \left(\frac{\hat\sigma_{-}}{\sqrt{\langle\hat\sigma_{+}\hat\sigma_{-}\rangle}}-\mathbb{I}\right)\ket{\psi}\bra{\psi}\left(\frac{\hat\sigma_{+}}{\sqrt{\langle\hat\sigma_{+}\hat\sigma_{-}\rangle}}-\mathbb{I}\right)\,\overline{{d} N^2}\nonumber\\
	&=\gamma \,\hat\sigma_{-}\rho\,\hat\sigma_{+}\,{d}t-\tfrac{\gamma}{\sqrt{\langle\hat\sigma_{+}\,\hat\sigma_{-}\rangle}} \,\hat\sigma_{-}\,\rho\,\,{d}t-\tfrac{\gamma}{\sqrt{\langle\hat\sigma_{+}\,\hat\sigma_{-}\rangle}} \,\rho\,\hat\sigma_{+}\,\,\,{d}t+\gamma {\langle\hat\sigma_{+}\,\hat\sigma_{-}\rangle}  \,\rho \,{d}t\label{id3}\,,
\end{align}
where in the second line of \cref{id1,id3} we replaced  $\rho=\ket{\psi}\bra{\psi}$ and neglected  terms $\sim {d}t^2$ and $\sim{d}t\,\overline{{d}N}$ [note that instead $\overline{({\rm d}N)^2}\sim {d}t$]. Summing the three increments, it can be seen that many terms cancel out in a way that we are left with ${d}\rho=\gamma \,\left(\hat\sigma_{-}\rho\,\hat\sigma_{+}-\tfrac{1}{2}\left[\hat\sigma_{+}\hat\sigma_{-},\rho\right]_+\right){d}t$.

\section{Equivalence between \cref{EC} and \ref{V-EC}}\label{appendix-equi}

For simplicity, we assume here that both $S$ and ancilla $n$ are finite-dimensional systems (the derivation can yet be easily extended to infinite dimension). Let us introduce the spectral decompositions of $\hat{H}_S$ and $\hat{H}_n$ as
\def\theequation{H.\arabic{equation}}
\setcounter{equation}{0}
\begin{eqnarray}
	\hat{H}_S&=& \sum_j E_j \hat{\Pi}_S^j \;, \qquad \sum_j \hat{\Pi}_S^j  = {\mathbb I}_S \;, \\
	\hat{H}_n &=& \sum_i
	e_i \hat{\Pi}_n^i \;, \qquad\; \sum_i \hat{\Pi}_n^i  = {\mathbb I}_n\;,
\end{eqnarray}
where $E_j$ ($e_i$) is the generic eigenvalue of $\hat{H}_S$ ($\hat{H}_n$) and $\hat{\Pi}_S^j$ ($\hat{\Pi}_n^i$) the projector on the corresponding eigenspace. Projectors associated with different energies are orthogonal, \ie
\begin{equation}
	\hat \Pi_S^j\hat \Pi_{S}^{j'}=\delta(E_j-E_{j'})\hat \Pi_S^j\,,\,\,\, \hat \Pi_n^i\hat \Pi_{n}^{i'}=\delta(E_i-E_{i'})\hat \Pi_n^i\,\,.\label{proj-ort}
\end{equation}
Here, we conveniently defined $\delta(x)$ as a function taking value 1 for $x=0$ and $0$ otherwise.

Accordingly, by denoting with $\bar{E}_\ell$ the eigenvalues of $\hat{H}_S+\hat{H}_n$ this can be spectrally-decomposed as
\begin{equation}
	\label{SD-HSn}
	\hat{H}_S+\hat{H}_n=
	\sum_{\ell} {\bar{E}}_\ell \; \hat{\Pi}_{Sn}^{\ell}\;, 
\end{equation}
where  $\hat{\Pi}_{Sn}^{\ell}$ are the (complete) orthonormal projectors on the system--ancilla Hilbert space defined by
\begin{equation} 
	\hat{\Pi}_{Sn}^{\ell}= \sum_{j,i} \delta( E_j + e_{i} - {\bar{E}}_\ell ) \,
	\,\, \hat{\Pi}_S^{j} \otimes \hat{\Pi}_n^{i}\;.\label{pisn}
\end{equation} 

Now, we observe that the commutation between $\hat V_n$ and $\hat H_S+\hat H_n$ [\cf\eq\ref{EC}] is equivalent to stating that $\hat V_n$ can be spectrally decomposed in the same basis of projectors $\{\hat{\Pi}_{Sn}^{\ell}\}$ as $\hat H_S+\hat H_n$ [\cf\eq\ref{SD-HSn}], \ie
\begin{equation}
	\label{Vn66}
	\hat{V}_n=
	\sum_{\ell} 
	v_\ell  \; \hat{\Pi}_{Sn}^{\ell}=\sum_{j,i} \sum_{\ell}
	v_\ell  \;  \delta( E_j + e_{i} - {\bar{E}}_\ell )     \, \hat{\Pi}_S^{j}  \otimes \hat{\Pi}_n^{i} \;.
\end{equation}
Here, in the last step we replaced $\hat{\Pi}_{Sn}^{\ell}$ with \ref{pisn}.

Consider now the operator defined by
\begin{align}
	&\sum_{j'',j',i'',i'} \delta(E_{j''} +e_{i''}-(E_{j'} +e_{i'})) 
	\hat{\Pi}_S^{j''}  \otimes \hat{\Pi}_n^{i''} \; \hat{V}_n \; \hat{\Pi}_S^{j'} \otimes \hat{\Pi}_n^{i'}=\nonumber \\ 
	&\sum_{j,i} \sum_{\ell}
	v_\ell  \;  \delta( E_j + e_{i} - {\bar{E}}_\ell ) \quad\times\\ 
	&\quad\sum_{j'',j',i'',i'} \delta(E_{j''} +e_{i''}-(E_{j'} +e_{i'})) \,
	\hat{\Pi}_S^{j''}  \otimes \hat{\Pi}_n^{i''} \;   ( \hat{\Pi}_S^{j}  \otimes \hat{\Pi}_n^{i} ) \; \hat{\Pi}_S^{j'} \otimes \hat{\Pi}_n^{i'} \nonumber\,,
\end{align} 
where in the last step we replaced $\hat V_n$ with \ref{Vn66}. This operator coincides just with $\hat V_n$. Indeed, using the orthogonality relations \ref{proj-ort}, the last expression can be arranged as
\begin{align}
	&\sum_{j,i} \sum_{\ell}
	v_\ell  \;  \delta( E_j + e_{i} - {\bar{E}}_\ell ) \qquad \times   \nonumber \\
	&\sum_{j'',j',i'',i'} \delta(E_{j''} +e_{i''}-(E_{j'} +e_{i'})) \delta( E_{j''} -E_{j})  \delta( e_{i''} -e_{i}) 
	\delta( E_{j} -E_{j'})  \delta( e_{i} -e_{i'}) 
	\, \hat{\Pi}_S^{j}  \otimes \hat{\Pi}_n^{i} \nonumber \\
	&=\sum_{j,i} \sum_{\ell}
	\Delta_\ell  \;  \delta( E_j + e_{i} - {\bar{E}}_\ell )     \, \hat{\Pi}_S^{j}  \otimes \hat{\Pi}_n^{i}  = \hat{V}_n\;.
\end{align}
Thereby,
\begin{equation}
	\hat{V}_n   = 
	\sum_{j,j',i,i'} \delta(E_{j} +e_{i}-(E_{j'} +e_{i'})) \,
	\hat{\Pi}_S^{j}  \otimes \hat{\Pi}_n^{i} \; \hat{V}_n \; \hat{\Pi}_S^{j'} \otimes \hat{\Pi}_n^{i'}\;.
\end{equation} 
Plugging now $\hat{V}_n=\sum_\nu g_\nu \hat A'_\nu\hat B'_\nu$ [\cf\eq\ref{AB}] on the right-hand side yields
\begin{equation}
	\label{Vn3}
	\hat{V}_n=\sum_\nu g_\nu \sum_{j,j',i,i'} \,
	\delta(E_j- E_{j'} + e_i-e_{i'})\,\,
	\hat{\Pi}_S^j\,\hat{A}'_\nu\,\hat{\Pi}_S^{j'}\otimes
	\hat{\Pi}_n^{i} \,\hat{B}'_\nu\,\hat{\Pi}_n^{i'}\,.
\end{equation}
By defining $\hat A_\nu=\hat{\Pi}_S^j\,\hat{A}'_\nu\,\hat{\Pi}_S^{j'}$ for $j<j'$, we note that it is an eigenoperator of $\hat{H}_S$ with eigenvalue $\omega_\nu= E_j- E_{j'}$ [\cf\eq\ref{comm}].
Likewise, $\hat B_\nu=\hat{\Pi}_n^i\,\hat{B}'_\nu\,\hat{\Pi}_n^{i'}$ with $i<i'$ is an eigenoperator of $\hat{H}_n$ with eigenvalue $w_\nu= e_i- e_{i'}$ [\cf\eq\ref{eigB}]. Therefore \ref{Vn3} is exactly of the same form as \ref{V-EC}, which completes the proof.

\section{Fully swapping ancilla--ancilla collisions: proof of \eq\ref{mapAA1}}\label{app-AA}

Using \eq\ref{U1nrho}, the reduced state of $S$ at the $n$th step is given by
\def\theequation{I.\arabic{equation}}
\setcounter{equation}{0}
\begin{equation}
	\rho_n={\rm Tr}_{1,2,\ldots,n}\{(\hat S_{2,1}\cdots\hat S_{n-1,n-2}\hat S_{n,n-1})\,\,\hat U_1^n\,\rho_{0}\otimes_{m=1}^n\eta_{m}\,\hat U_1^{\dag n}\,(\hat S_{n,n-1} \hat S_{n-1,n-2}\, \ldots\hat S_{2,1})\}\,.\label{appAAeq}   
\end{equation}
Taking now advantage of the homogeneity of $\eta_n$, we can write
\begin{equation}
	(\hat S_{3,2}\cdots\hat S_{n-1,n-2}\hat S_{n,n-1})\,\eta_{2}\otimes\eta_{3}\otimes\cdots\otimes\eta_n\,(\hat S_{3,2}\cdots\hat S_{n-1,n-2}\hat S_{n,n-1})^\dag=\eta_{2}\otimes \eta_{3}\otimes\cdots\otimes\eta_n\,.\nonumber\label{homog}
\end{equation}
Replacing back in \ref{appAAeq}, this reduces to (we refer to a basis $\vert k_1,k_2\rangle _{12}$ for computing the partial trace)
\begin{eqnarray}
	\rho_{n}&=&{\rm Tr}_{1,2}\,\{\hat S_{2,1}\hat U_{1}^n\rho_{0}\,\eta_1\,\eta_2\,\hat U_1^{n\dag}\hat S_{2,1}\}=\sum_{k_1,k_2}\langle k_1,k_2\rvert  \hat S_{2,1}^\dag\,\hat U_{1}^n\rho_{0}\,\eta_1\,\eta_2\,\hat U_1^{n\dag}\hat S_{2,1}\vert k_1,k_2\rangle \nonumber\\
	&=&\sum_{k_1,k_2}\langle k_1,k_2\rvert  \hat U_{1}^n\rho_{0}\,\eta_1\,\eta_2\,\hat U_1^{n\dag}\vert k_1,k_2\rangle =\sum_{k_1}\langle k_1\rvert  \hat U_{1}^n\rho_{0}\,\eta_1\,\hat U_1^{n\dag}\vert k_1\rangle ={\rm Tr}_1\{ \hat U_{1}^n\rho_{0}\,\eta_1\,\hat U_1^{n\dag}\}\,,\nonumber
\end{eqnarray}
where we used that $\hat S_{2,1}\vert k_1,k_2\rangle $ is another valid basis for computing the \textit{partial trace} (this being invariant under a change of basis). This completes the proof of \eq\ref{mapAA1}.

\section{Ancilla--ancilla collisions: derivation of master equation~\ref{ME-kernel}
} \label{app-swap}

By subtracting from \eq\ref{rho-kernel} the analogous equation for $\rho_{n-1}$ we get 
\def\theequation{J.\arabic{equation}}
\setcounter{equation}{0}
\begin{equation}
	\label{fd}
	\Delta \rho_n= (1- p) \sum_{j=1}^{n-2}p^{j-1}\mathcal{F}_j[\Delta \rho_{n- j}]+ (1- p)p^{n-1} \mathcal{F}_{n-1}[\rho_1]
	+ \;\; \Delta \left(p^{n-1} \mathcal{F}_n\right)[\rho_0]\,,
\end{equation}
where, as usual, $\Delta A_n=A_n-A_{n-1}$ with $A$ a map or state.
By expressing each power of $p$ in the form of an exponential as $p^j=e^{-\Gamma (\Delta t j)}=e^{-\Gamma t'}$ with $t'=j \Delta t$ and likewise $p^n=e^{-\Gamma t}$, in the limit $\Delta t \ll \Gamma$ the three terms on the right hand side of \eq\ref{fd} become
\begin{align}\label{fd2}
	&\frac{(1-p) \sum_{j=1}^{n-2}p^{(j-1)}\mathcal{E}_j\left[{\rho_{n- j}-\rho_{n-1- j}}\right]}{{\Delta t}}\simeq\Gamma \int_{0}^t {}dt' e^{-\Gamma t'}\mathcal{E}(t')\left[\frac{d\rho(t- t')}{d (t- t')}\right]\,,\,\nonumber\\
	& \frac{(1-p)p^{n-1} \mathcal{E}_{n-1}}{{\Delta t}}\,[\rho_1]{}\simeq\Gamma e^{-\Gamma t}\mathcal{E}(t)\, [\rho_0]\,,\,\nonumber\\
	&\frac{\Delta (p^{n-1} \mathcal{E}_n)}{\Delta t}\,\ug\frac{p^{n-1} \mathcal{E}_n- p^{n-2} \mathcal{E}_{n-1}}{{\Delta t}}\simeq{}\frac{e^{-\Gamma (t+ 2{\Delta t})}\mathcal{E}(t+{\Delta t})- e^{-\Gamma (t+ {\Delta t})}\mathcal{E}(t)}{{\Delta t}} \ug \frac{d}{dt}{}\left(e^{-\Gamma t }\mathcal{E}(t)\right)\,.\nonumber
\end{align}
Thus in the continuous-time limit, \eq\ref{fd} reduces to \eq\ref{ME-kernel}.

\section{Composite CMs: derivation of the recurrence relation \ref{recur}} \label{recurr}

From \cref{Psi-n} (for $n\rightarrow n-1$) and \ref{USMn} it follows

\def\theequation{K.\arabic{equation}}
\setcounter{equation}{0}
\begin{equation}
	\label{umn}
	\hat U_{Mn}=\ket{00}_{Mn}\bra{00}+ \cos (g \Delta t) (\ket{10}_{Mn}\bra{10}+\ket{01}_{Mn}\bra{01})-i \sin (g \Delta t) (\ket{01}_{Mn}\bra{10}+\ket{10}_{Mn}\bra{01})\,,
\end{equation}
\begin{eqnarray}
	\hat U_{Mn}\hat U_{{\mathcal{S}}M}\ket{\Psi^{(n-1)}}&=& \hat U_{Mn}\left((C \alpha^{(n-1)} -i S\beta^{(n-1)}) \ket{e_{\mathcal{S}}}+(C\beta^{(n-1)}-i S\alpha^{(n-1)} ) \ket{e_{M}}+\sum_{m=1}^{n}  \lambda_m^{(n-1)}\ket{e_{m}}\right)\nonumber\\
	&= & (C \alpha^{(n-1)} -i S \beta^{(n-1)} )\ket{e_{\mathcal{S}}}+(c\,C\beta^{(n-1)}-i cS\alpha^{(n-1)} ) \ket{e_{M}}\nonumber\\
	&&\qquad+\sum_{m=1}^{n-1}  \lambda_m^{(n-1)}\ket{e_{m}}-i s (C\beta^{(n-1)}-i S\alpha^{(n-1)} ) \ket{e_{n}}\,.
\end{eqnarray}
Comparing with \eq\ref{Psi-n}, we get the recurrence relation \ref{recur} for the excitation amplitudes of ${\mathcal{S}}$ and $M$.

\section{Composite CMs: derivation of linear system \ref{syst}} \label{lin-sys}

By looking at \eq\ref{Psi-n} we see that, upon trace over the bath, the joint state of ${\mathcal{S}}$ and $M$ has the form
\def\theequation{L.\arabic{equation}}
\setcounter{equation}{0}
\begin{equation}
	\rho_{{\mathcal{S}}M}=\vert  \alpha_n\vert ^2\ket{10}\bra{10}+
	\vert  \beta_n\vert ^2\ket{01}\bra{01}+(\alpha_n\beta^{*}_n\ket{10}\bra{01}\piu{\rm H.c.})+(1\meno\vert \alpha_n\vert ^2\meno \vert \beta_n\vert ^2)\ket{00}\bra{00} \,.             
\end{equation}
This remains true when $\alpha_n\rightarrow \alpha(t)$ and $\beta_n\rightarrow \beta(t)$. Replacing $\rho_{{\mathcal{S}}M}(t)$ into master equation~\ref{ME-CQED} this is turned into the coupled differential equations
\begin{equation}
	\tfrac{{\rm d} }{{\rm d}t}\vert \alpha\vert ^2= i G (\alpha\beta^{*} \meno  \alpha^{*}\beta)\,,\,\,\tfrac{{\rm d} }{{\rm d}t}\vert \beta\vert ^2=-i G(\alpha \beta^{*}  \meno \alpha^{*} \beta ) \meno \gamma \vert \beta\vert ^2\,,\,\,\tfrac{{\rm d} }{{\rm d}t}\left(\alpha\beta^{*}\right)=-\tfrac{\gamma}{2}\alpha\beta^{*}+i \left[G (\vert \alpha\vert ^2 \meno\vert \beta\vert ^2 )\right]\,.
\end{equation}
It is easily checked that these are indeed equivalent to \ref{syst} (\eg $\tfrac{{\rm d} }{{\rm d}t}\vert \alpha\vert ^2$ is obtained from $\alpha^{*}\dot{\alpha}= -i G\alpha^{*}\beta$ by adding to either side the respective c.c.). This completes the proof.

\backmatter
\setchapterstyle{plain}
\printbibliography[heading=bibintoc, title=Bibliography]

\end{document}